\documentclass[11pt,a4paper]{article}
\usepackage[utf8]{inputenc}

\usepackage[left=3cm,right=3cm,top=3cm,bottom=3cm]{geometry}

\author{Andreas Deuchert, Christian Hainzl, Marcel Maier (born Schaub)}

\usepackage[english]{babel}

\usepackage{fancyhdr}

\usepackage{soul}
\setuldepth{0}

\usepackage{wallpaper}

\usepackage{anyfontsize}

\usepackage[absolute]{textpos}

\usepackage{upref}

\newcommand{\superscript}[1]{\ensuremath{^{\textrm{#1}}}}

\newcommand{\tho}[0]{\superscript{th}}

\usepackage{enumerate}

\usepackage{amsmath}
\numberwithin{equation}{section}
\usepackage{mathtools}

\renewcommand{\leq}{\leqslant}
\renewcommand{\geq}{\geqslant}

\usepackage{amsfonts}
\usepackage{amssymb}

\usepackage{amsthm}
\usepackage{thmtools, thm-restate} 

\usepackage{esvect}
\usepackage{cancel}
\usepackage{nicefrac}

\usepackage{fontawesome}
\newcommand{\speaker}{{\hspace{0.7pt}\text{\raisebox{-1.7pt}{\scalebox{1.4}{\faVolumeOff}}\hspace{0.7pt}}}}

\usepackage[T1]{fontenc}

\DeclareMathAlphabet{\mathbbs}{U}{bbold}{m}{n}
\newcommand{\Idbb}{\mathbbs 1}

\usepackage[arrow, matrix, curve]{xy}

\usepackage{ifthen}

\usepackage{oldgerm}

\usepackage{eurosym}

\usepackage{float}
\usepackage[rflt]{floatflt}
\usepackage{graphicx}
\usepackage{xcolor}

\usepackage{dlfltxbcodetips}

\newtheorem{thm}{Theorem}[section]
\newtheorem{bigthm}{Theorem}

\newtheorem{lem}[thm]{Lemma}

\newtheorem{kor}[thm]{Corollary}

\newtheorem{prop}[thm]{Proposition}

\newtheoremstyle{cited}{}{}{\itshape}{}{
}{\textbf{.}}{.5em}{\textbf{#1 #2} #3}
\theoremstyle{cited}

\theoremstyle{definition}
\newtheorem{defn}[thm]{Definition}

\newtheorem{bem}[thm]{Remark}

\newtheorem{bems}[thm]{Remarks}
\newtheorem*{varbems}{Remarks}

\newtheorem{asmp}[thm]{Assumption}

\newcommand{\ra}{\rightarrow}

\newcommand{\lk}{\left(}
\newcommand{\rk}{\right)}

\newcommand{\ov}[1]{\overline{#1}}

\newcommand{\dx}{\mathrm{d}x}

\newcommand{\dt}{\mathrm{d}t}
\newcommand{\dd}{\mathrm{d}}

\DeclareMathOperator{\Tr}{Tr}

\DeclareMathOperator{\divv}{div}

\DeclareMathOperator{\curl}{curl}

\DeclareMathOperator{\sgn}{sgn}

\DeclareMathOperator*{\esssup}{ess \, sup}

\DeclareMathOperator{\spec}{spec}
\DeclareMathOperator{\rank}{rank}

\newcommand{\loc}{\mathrm{loc}}

\newcommand{\symm}{\mathrm{symm}}

\newcommand{\Imm}{\mathrm{Im}}
\newcommand{\Rem}{\mathrm{Re}}
\let\Im\undefined
\let\Re\undefined
\DeclareMathOperator{\Im}{\Imm}
\DeclareMathOperator{\Re}{\Rem}

\renewcommand{\tilde}{\widetilde}
\renewcommand{\hat}{\widehat}

\DeclareFontFamily{U}{matha}{\hyphenchar\font45}
\DeclareFontShape{U}{matha}{m}{n}{
      <5> <6> <7> <8> <9> <10> gen * matha
      <10.95> matha10 <12> <14.4> <17.28> <20.74> <24.88> matha12
      }{}
\DeclareSymbolFont{matha}{U}{matha}{m}{n}
\DeclareFontSubstitution{U}{matha}{m}{n}

\DeclareFontFamily{U}{mathx}{\hyphenchar\font45}
\DeclareFontShape{U}{mathx}{m}{n}{
      <5> <6> <7> <8> <9> <10>
      <10.95> <12> <14.4> <17.28> <20.74> <24.88>
      mathx10
      }{}
\DeclareSymbolFont{mathx}{U}{mathx}{m}{n}
\DeclareFontSubstitution{U}{mathx}{m}{n}

\DeclareMathDelimiter{\vvvert}{0}{matha}{"7E}{mathx}{"17}

\DeclareFontFamily{U}{mathx}{\hyphenchar\font45}
\DeclareFontShape{U}{mathx}{m}{n}{
      <5> <6> <7> <8> <9> <10>
      <10.95> <12> <14.4> <17.28> <20.74> <24.88>
      mathx10
      }{}
\DeclareSymbolFont{mathx}{U}{mathx}{m}{n}
\DeclareFontSubstitution{U}{mathx}{m}{n}
\DeclareMathAccent{\widecheck}{0}{mathx}{"71}
\DeclareMathAccent{\wideparen}{0}{mathx}{"75}

\newcommand{\Hbb}{\mathbb{H}}

\newcommand{\Nbb}{\mathbb{N}}

\newcommand{\Rbb}{\mathbb{R}}

\newcommand{\Zbb}{\mathbb{Z}}

\newcommand{\Acal}{\mathcal{A}}

\newcommand{\Dcal}{\mathcal{D}}

\newcommand{\Gcal}{\mathcal{G}}
\newcommand{\Hcal}{\mathcal{H}}

\newcommand{\Ocal}{\mathcal{O}}

\newcommand{\Qcal}{\mathcal{Q}}
\newcommand{\Rcal}{\mathcal{R}}
\newcommand{\Scal}{\mathcal{S}}
\newcommand{\Tcal}{\mathcal{T}}

\newcommand{\Wcal}{\mathcal{W}}

\newcommand{\hfrak}{\mathfrak{h}}

\newcommand{\EGL}{\mathcal E^{\mathrm{GL}}_{D}}
\newcommand{\EGLh}{\mathcal E^{\mathrm{GL}}_{D, h}}
\newcommand{\EGLGSE}{E^{\mathrm{GL}}(D)}
\newcommand{\FBCS}{\mathcal F^{\mathrm{BCS}}_{
h, T
}}
\newcommand{\Abold}{\mathbf A}
\newcommand{\Asolo}{A}

\newcommand{\Bbold}{\mathbf B}

\newcommand{\Zbold}{\mathbf Z}
\newcommand{\sbold}{\mathbf s}
\newcommand{\Hmag}{H_{\mathrm{mag}}}
\newcommand{\Lmag}{L_{\mathrm{mag}}}

\newcommand{\Wvec}[1]{W^{#1, \infty}(\Rbb^3; \Rbb^3)}

\newcommand{\Lsymm}{{L^2(Q_h \times \Rbb_{\mathrm s}^3)}}
\newcommand{\Hsymm}{{H^1(Q_h \times \Rbb_{\mathrm s}^3)}}

\newcommand{\Tc}{{T_{\mathrm{c}}}}
\newcommand{\Dc}{{D_{\mathrm{c}}}}
\newcommand{\betac}{{\beta_{\mathrm{c}}}}
\newcommand{\gzfunction}{\rho}
\newcommand{\gzfunctiondiff}{\tau}

\newcommand{\Woperator}{\Wcal}

\usepackage{esint} 

\usepackage{hyperref}
\newcommand{\e}{\mathrm{e}}
\renewcommand{\i}{\mathrm{i}}

\usepackage[
	sorting=nyt, 
	backend=biber, 
    giveninits=true,	
	maxbibnames=99,
	backref=true,
	eprint=true,
	isbn=false,
	date=year,
	]{biblatex}

\DeclareLabelalphaTemplate{
   \labelelement{
     \field[uppercase,final]{shorthand}     
     \field[uppercase,strwidth=1,strside=left,names=-5,noalphaothers]{labelname}
   }
   \labelelement{
     \field[strwidth=2,strside=right]{year}
   }
   \labelelement{
		\field[uppercase]{label}
   }
}
\renewbibmacro*{note+pages}{%
  \printfield{note}%
  \setunit{\bibpagespunct}%
  \printfield{pages}%
  \setunit{\bibpagespunct}
  \printfield{pagetotal}
  \newunit}

\DefineBibliographyStrings{english}{%
  backrefpage = {page}, 
  backrefpages = {pages}, 
}

\renewbibmacro{in:}{} 
\DeclareFieldFormat{pages}{#1}
\DeclareFieldFormat{pagetotal}{#1~pages}
\DeclareFieldFormat[article]{volume}{\mkbibbold{#1}}
\DeclareFieldFormat[article,inbook,incollection,inproceedings,patent,thesis,unpublished]{title}{#1} 

\AtEveryBibitem{\clearfield{number}}
\AtEveryBibitem{\clearfield{issue}}

\usepackage{pdfpages}
\usepackage{chngcntr}
\counterwithin{footnote}{section}
\counterwithout{footnote}{subsection}

\usepackage{ifthen}
\newcommand{\masterfile}{DHS2}

\usepackage{lmodern}

\begin{document}

\title{
%
Microscopic Derivation of Ginzburg--Landau Theory \\ and the BCS Critical Temperature Shift \\ in the Presence of Weak Macroscopic External Fields
%
%
%
%
%
%
%
}

\maketitle

\begin{abstract}
%
We consider the Bardeen--Cooper--Schrieffer (BCS) free energy functional with weak and macroscopic external electric and magnetic fields and derive the Ginzburg--Landau functional. We also provide an asymptotic formula for the BCS critical temperature as a function of the external fields. This extends our previous results in \cite{DeHaSc2021} for the constant magnetic field to general magnetic fields with a nonzero magnetic flux through the unit cell.
\end{abstract}

\tableofcontents



\section{Introduction and Main Results}


\subsection{Introduction}

Ginzburg--Landau (GL) theory has been introduced as the first macroscopic and phenomenlogical description of superconductivity in 1950 \cite{GL}. The theory comprises a system of partial differential equations for a complex-valued function, the order parameter, and an effective magnetic field. Ginzburg--Landau theory has been highly influencial and investigated in numerous works, among which are \cite{Sigal1, Sigal2, Serfaty, SandierSerfaty, Correggi3, Correggi2, Giacomelli1, Giacomelli2, Giacomelli3} and references therein.

Bardeen--Cooper--Schrieffer (BCS) theory of superconductivity is the first commonly accepted and Nobel prize awarded microscopic theory of superconductivity \cite{BCS}. As a major breakthrough, the theory features a pairing mechanism between the electrons below a certain critical temperature, which causes the electrical resistance in the system to drop to zero in the superconducting phase. This effect is due to an effective attraction between the electrons, which arises as a consequence of the phonon vibrations of the lattice ions in the superconductor.

One way to formulate BCS theory mathematically is via the BCS free energy functional or BCS functional for short. As Leggett pointed out in \cite{Leg1980}, the BCS functional can be obtained from a full quantum mechanical description of the system by restricting attention to quasi-free states, see also \cite{de_Gennes}. Such states are determined by their one-particle density matrix and the Cooper pair wave function. The BCS functional has been studied intensively from a mathematical point of view in the absence of external fields in \cite{Hainzl2007, Hainzl2007-2, Hainzl2008-2, Hainzl2008-4, FreiHaiSei2012, BraHaSei2014, FraLe2016, DeuchertGeisinger} and in the presence of external fields in \cite{HaSei2011, BrHaSei2016, FraLemSei2017, A2017, CheSi2020}. The BCS gap equation arises as the Euler--Lagrange equation of the BCS functional and its solution is used to compute the spectral gap of an effective Hamiltonian, which is open in the superconducting phase. BCS theory from the point of view of its gap equation is studied in \cite{Odeh1964, BilFan1968, Vanse1985, Yang1991, McLYang2000, Yang2005}.

The present article continues a series of works, in which the \emph{macroscopic} GL theory is derived from the \emph{microscopic} BCS theory in a regime close to the critical temperature and for weak external fields. This endeavor has been initiated by Gor'kov in 1959 \cite{Gorkov}. The first mathematically rigorous derivation of the GL functional from the BCS functional has been provided by Frank, Hainzl, Seiringer, and Solovej for periodic external electric and magnetic fields in 2012 in \cite{Hainzl2012}. An important assumption of this work is that the flux of the external magnetic field through the unit cell of periodicity of the system vanishes. This excludes for example a homogeneous magnetic field. The techniques from this GL derivation have been further developed in \cite{Hainzl2014} to compute the BCS critical temperature shift caused by the external fields. The first important step towards overcoming the zero magnetic flux restriction in \cite{Hainzl2012,Hainzl2017} has been made by Frank, Hainzl, and Langmann, who considered in \cite{Hainzl2017} the problem of computing the BCS critical temperature shift for systems exposed to a homogeneous magnetic field within the framework of linearized BCS theory. Recently, the derivation of the GL functional and the computation of BCS critical temperature shift (for the full nonlinear model) could be extended to the case of a constant magnetic field by Deuchert, Hainzl, and Maier in \cite{DeHaSc2021}. The goal of the present work is to further extend the results in \cite{DeHaSc2021} to the case of general external magnetic fields with an arbitrary flux through the unit cell. 

GL theory arises from BCS theory when the temperature is sufficiently close to the critical temperature and when the external fields are weak and slowly varying. More precisely, if $0 < h \ll 1$ denotes the ratio between the microscopic and the macroscopic length scale, then the external electric field $W$ and the magnetic vector potential $\Abold$ are given by $h^2 W(hx)$ and $h\Abold(hx)$, respectively. Furthermore, the temperature regime is such that $T - \Tc = -\Tc Dh^2$ for some constant $D >0$, where $\Tc$ is the critical temperature in absence of external fields. When this scaling is in effect, it is shown in \cite{Hainzl2012} and \cite{DeHaSc2021} that the Cooper pair wave function $\alpha(x,y)$ is given by
\begin{align}
\alpha(x,y) = h\, \alpha_*(x - y) \, \psi \left( \frac{h(x+y)}{2}\right) \label{eq:intro1}
\end{align}
to leading order in $h$. Here, $\alpha_*$ is the microscopic Cooper pair wave function in the absence of external fields and $\psi$ is the GL order parameter. 

Moreover, the influence of the external fields causes a shift in the critical temperature of the BCS model, which is described by linearized GL theory in the same scaling regime. More precisely, it has been shown in \cite{Hainzl2014,Hainzl2017}, and \cite{DeHaSc2021} that the critical temperature shift in BCS theory is given by
\begin{align}
\Tc(h) = \Tc (1 - \Dc h^2) \label{eq:intro2}
\end{align}
to leading order, where $\Dc$ denotes a critical parameter that can be computed using linearized GL theory.

The present work is an extension of the paper \cite{DeHaSc2021}, where the case of a constant magnetic field was considered. In this article, we incorporate periodic electric fields $W$ and general vector potentials $\Abold$ that give rise to periodic magnetic fields. This, in particular, generalizes the results in \cite{Hainzl2012,Hainzl2014} to the case of general external magnetic fields with non-zero flux through the unit cell. We show that within the scaling introduced above, the Ginzburg--Landau energy arises as leading order correction on the order $h^4$. Furthermore, we show that the Cooper pair wave function admits the leading order term \eqref{eq:intro1} and that the critical temperature shift is given by \eqref{eq:intro2} to leading order. The main technical novelty of this article is a further development of the phase approximation method, which has been pioneered in the framework of BCS theory for the case of the constant magnetic field in \cite{Hainzl2017} and \cite{DeHaSc2021}. It allows us to compute the BCS energy of a class of trial states (Gibbs states) in a controlled way. This trial state analysis is later used in the proofs of the upper and of the lower bound for the BCS free energy. The proof of our lower bound additionally uses a priori bounds for certain low-energy BCS states that include the magnetic field and have been established in \cite{DeHaSc2021}.

\subsection{Gauge-periodic samples}
\label{Magnetically_Periodic_Samples}

Our objective is to study a system of three-dimensional fermionic particles that is subject to weak and slowly varying external electromagnetic fields within the framework of BCS theory. Let us define the magnetic field $\Bbold \coloneqq h^2 e_3$. It can be written in terms of the vector potential $\Abold_{\Bbold}(x) \coloneqq \frac{1}{2} \Bbold \wedge x$, where $x \wedge y$ denotes the cross product of two vectors $x,y \in \mathbb{R}^3$, as $\Bbold = \curl \Abold_\Bbold$. To the vector potential $\Abold_{\Bbold}$ we associate the magnetic translations
\begin{align}
	T(v)f(x) &\coloneqq  \e^{\i \frac {\Bbold} 2\cdot (v\wedge x)} f(x+v), & v &\in \Rbb^3, \label{Magnetic_Translation}
\end{align}
which commute with the magnetic momentum operator $-\i \nabla + \Abold_\Bbold$. The family $\{ T(v) \}_{v\in \mathbb{R}^3}$ satisfies $T(v+w) = \e^{\i \frac{\Bbold}{2} \cdot (v \wedge w)} T(v) T(w)$ and is therefore a unitary representation of the Heisenberg group. We assume that our system is periodic with respect to the Bravais lattice $\Lambda_h \coloneqq \sqrt{2\pi} \, h^{-1} \, \Zbb^3$ with fundamental cell
\begin{align}
Q_h &\coloneqq \bigl[0, \sqrt{2\pi} \, h^{-1}\bigr]^3 \subseteq \Rbb^3. \label{Fundamental_cell}
\end{align}
Let $b_i = \sqrt{2\pi} \, h^{-1} \, e_i$ denote the basis vectors that span $\Lambda_h$. The magnetic flux through the face of the unit cell spanned spanned by $b_1$ and $b_2$ equals $2 \pi$, and hence the abelian subgroup $\{ T(\lambda) \}_{\lambda \in \Lambda_h}$ is a unitary representation of the lattice group. 

Our system is subject to an external electric field $W_h(x) = h^2W(hx)$ with a fixed function $W \colon \Rbb^3 \ra \Rbb$, as well as a magnetic field defined in terms of the vector potential $\Abold_h(x) = h \Abold(hx)$, which admits the form $\Abold \coloneqq  \Abold_{e_3} + A$ with $A \colon \Rbb^3\ra \Rbb^3$ and $\Abold_{e_3}$ as defined above. We assume that $A$ and $W$ are periodic with respect to $\Lambda_1$. The flux of the magnetic field $\curl A_h$ through all faces of the unit cell $Q_h$ vanishes because $A_h$ is a periodic function. Accordingly, the magnetic field $\curl \Abold_h$ has the same fluxes through the faces of the unit cell as $\Bbold$. 

The above representation of $\Abold_h$ is general in the sense that any periodic magnetic field field $B(x)$ that satisfies the Maxwell equation $\divv B = 0$ can be written as the curl of a vector potential $A_B$ of the form $A_B(x)= \frac{1}{2} b \wedge x + A_{\mathrm{per}}(x)$, where $b$ denotes the vector with components given by the average magnetic flux of $B$ through the faces of $Q_h$ and $A_{\mathrm{per}}$ is a periodic vector potential. For more information concerning this decomposition we refer to \cite[Chapter 4]{Diss_Marcel}. For a treatment of the two-dimensional case, see \cite{Tim_Abrikosov}.


\subsection{The BCS functional}
\label{BCS_functional_Section}

In BCS theory a state is conveniently described by its generalized one-particle density matrix, that is, by a self-adjoint operator $\Gamma$ on $L^2(\Rbb^3) \oplus L^2(\Rbb^3)$, which obeys $0 \leq \Gamma \leq 1$ and is of the form
\begin{align}
\Gamma = \begin{pmatrix} \gamma & \alpha \\ \ov \alpha & 1 - \ov \gamma \end{pmatrix}. \label{Gamma_introduction}
\end{align}
Here, $\ov \alpha$ denotes the operator $\alpha$ with the complex conjugate integral kernel in the position space representation. Since $\Gamma$ is self-adjoint we know that $\gamma$ is self-adjoint and that $\alpha$ is symmetric in the sense that its integral kernel satisfies $\alpha(x,y) = \alpha(y,x)$. This symmetry is related to the fact that we exclude spin degrees of freedom from our description and assume that all Cooper pairs are in a spin singlet state. The condition $0 \leq \Gamma \leq 1$ implies that the one-particle density matrix $\gamma$ satisfies $0 \leq \gamma \leq 1$ and that $\alpha$ and $\gamma$ are related through the inequality
\begin{align}
\alpha \alpha^* \leq \gamma ( 1- \gamma). \label{gamma_alpha_fermionic_relation}
\end{align}

Let us define the magnetic translations $\mathbf T(\lambda)$ on $L^2(\Rbb^3)\oplus L^2(\Rbb^3)$ by
\begin{align*}
	\mathbf T(v) &\coloneqq  \begin{pmatrix}
		T(v) & 0 \\ 0 & \ov{T(v)}\end{pmatrix}, & v &\in \Rbb^3.
\end{align*}
We say that a BCS state $\Gamma$ is \emph{gauge-periodic} provided $\mathbf T(\lambda) \, \Gamma \, \mathbf T(\lambda)^* = \Gamma$ holds for any $\lambda\in \Lambda_h$. This implies the relations $T(\lambda) \, \gamma \, T(\lambda)^* = \gamma$ and   $T(\lambda)\,\alpha \,\ov{T(\lambda)}^* = \alpha$, or, in terms of integral kernels,
\begin{align}
\gamma(x, y) &= \e^{\i \frac \Bbold 2 \cdot (\lambda \wedge (x-y))} \; \gamma(x+\lambda,y+ \lambda), \notag\\
\alpha(x, y) &= \e^{\i \frac \Bbold 2 \cdot (\lambda \wedge (x+y))} \; \alpha(x+\lambda,y+ \lambda), & \lambda\in \Lambda_h. \label{alpha_periodicity}
\end{align}

We further say that a gauge-periodic BCS state $\Gamma$ is \emph{admissible} if 
\begin{align}
\Tr \bigl[\gamma + (-\i \nabla + \Abold_\Bbold)^2\gamma\bigr] < \infty \label{Gamma_admissible}
\end{align}
holds. Here $\Tr[\Rcal]$ denotes the trace per unit volume of an operator $\Rcal$ defined by
\begin{align}
\Tr [\Rcal] &\coloneqq  \frac{1}{|Q_h|} \Tr_{L^2(Q_h)} [\chi \Rcal \chi],  \label{Trace_per_unit_volume_definition}
\end{align}
where $\chi$ denotes the characteristic function of the cube $Q_h$ in \eqref{Fundamental_cell} and $\Tr_{L^2(Q_h)}[\cdot]$ is the usual trace over an operator on $L^2(Q_h)$. By the condition in \eqref{Gamma_admissible}, we mean that $\chi \gamma \chi$ and $\chi (-\i \nabla + \Abold_\Bbold)^2 \gamma \chi$ are trace-class operators. 
Eqs.~\eqref{gamma_alpha_fermionic_relation}, \eqref{Gamma_admissible}, and the same inequality with $\gamma$ replaced by $\overline{\gamma}$ imply that $\alpha$, $(-\i \nabla + \Abold_\Bbold)\alpha$, and $(-\i \nabla + \Abold_\Bbold) \ov \alpha$ are locally Hilbert--Schmidt. We will rephrase this property as a notion of $H^1$-regularity for the kernel of $\alpha$ in Section~\ref{Preliminaries} below.

Let $\Gamma$ be an admissible BCS state. We define the Bardeen--Cooper--Schrieffer free energy functional, or BCS functional for short, at temperature $T\geq 0$ by the formula
\begin{align}
\FBCS(\Gamma) &\coloneqq  \Tr\bigl[ \bigl( (-\i \nabla + \Abold_h)^2 - \mu + W_h \bigr)\gamma \bigr] - T\, S(\Gamma)  \notag \\
&\hspace{120pt} - \frac{1}{|Q_h|} \int_{Q_h} \dd X \int_{\Rbb^3} \dd r\; V(r) \, |\alpha(X,r)|^2,
\label{BCS functional}
\end{align}
where  $S(\Gamma)= - \Tr [\Gamma \ln(\Gamma)]$ denotes the von Neumann entropy per unit volume and $\mu\in \Rbb$ is a chemical potential. The interaction energy is written in terms of the center-of-mass and relative coordinates $X = \frac{x+y}{2}$ and $r = x-y$. Throughout this paper, we write, by a slight abuse of notation, $\alpha(x,y) \equiv \alpha(X,r)$. That is, we use the same symbol for the function depending on the original coordinates and for the one depending on $X$ and $r$.

The natural space for the interaction potential guaranteeing that the BCS functional is bounded from below is $V \in L^{\nicefrac 32}(\Rbb^3) + L_{\varepsilon}^{\infty}(\mathbb{R}^3)$, that is, the set of interaction potentials, for which $V$ is relatively form bounded with respect to the Laplacian. Under these assumptions it can be shown that the BCS functional satisfies the lower bound
\begin{align}
\FBCS(\Gamma) &\geq \frac 12 \Tr \bigl[ \gamma + (-\i \nabla + \Abold_\Bbold)^2 \gamma\bigr] - C \label{BCS functional_bounded_from_below}
\end{align}
for some constant $C>0$. In other words, the BCS functional is bounded from below and coercive on the set of admissible BCS states. 

The normal state $\Gamma_0$ is the unique minimizer of the BCS functional when restricted to admissible states with $\alpha = 0$ and reads
\begin{align}
\Gamma_0 &\coloneqq  \begin{pmatrix} \gamma_0 & 0  \\ 0 &  1-\ov\gamma_0 \end{pmatrix}, & \gamma_0 &\coloneqq  \frac 1{1 + \e^{ ((-\i\nabla + \Abold_h)^2 + W_h-\mu)/T}}.  \label{Gamma0}
\end{align}
Its name is motivated by the fact that it is also the unique minimizer of the BCS functional if the temperature $T$ is chosen sufficiently large.
We define the BCS free energy by
\begin{align}
F^{\mathrm{BCS}}(h, T) \coloneqq  \inf\bigl\{ \FBCS(\Gamma) - \FBCS(\Gamma_0) : \Gamma \text{ admissible}\bigr\} \label{BCS GS-energy}
\end{align}
and say that the system is superconducting at temperature $T$ if $F^{\mathrm{BCS}}(h, T) < 0$. Although it is not difficult to prove that the BCS functional has a minimizer, we refrain from giving a proof here. If we assume that the BCS functional has a minimizer $\Gamma$ then the condition $F^{\mathrm{BCS}}(h, T) < 0$ implies $\alpha = \Gamma_{12} \neq 0$. 

The goal of this paper is to derive an asymptotic formula for $F^{\mathrm{BCS}}(h, T)$ for small $h > 0$. This will allow us to derive Ginzburg--Landau theory and to show how the critical temperature depends on the external electric and magnetic field and on $h$. For our main results to hold, we need the following assumptions.

\begin{asmp}
\label{Assumption_V}
We assume that the interaction potential $V$ is a radial function that satisfies $(1+|\cdot|^2) V\in L^2(\mathbb{R}^3) \cap L^\infty(\Rbb^3)$. Moreover, the electric and the magnetic potentials $W\in W^{1, \infty}(\Rbb^3)$ and $A\in W^{3, \infty}(\Rbb^3; \Rbb^3)$ are $\Lambda_1$-periodic functions, i.e.  $W(x + \lambda) = W(x)$ and $A(x + \lambda) = A(x)$ for $\lambda\in \Lambda_1$ and all $x\in \Rbb^3$. We also assume that $A(0) = 0$.
\end{asmp}

\subsection{The translation-invariant BCS functional}
\label{BCS_functional_TI_Section}

In the absence of external fields we describe the system by translation-invariant states, that is, we assume that the integral kernels of $\gamma$ and $\alpha$ are of the form $\gamma(x-y)$ and $\alpha(x-y)$. The trace per unit volume is in this case defined with respect to a cube with sidelength $1$. We denote the resulting translation-invariant BCS functional by $\mathcal{F}^{\mathrm{BCS}}_{\mathrm{ti},T}$. The translation-invariant BCS functional is studied in detail in \cite{Hainzl2007}, see also the review article \cite{Hainzl2015}. In \cite{Hainzl2007} it has been shown that there is a unique critical temperature $\Tc \geq 0$ such that $\mathcal{F}^{\mathrm{BCS}}_{\mathrm{ti},T}$ has a minimizer with $\alpha \neq 0$ for $T < \Tc$. The normal state in \eqref{Gamma0} with $h=0$ is the unique minimizer if $T\geq \Tc$. Moreover, the critical temperature $\Tc$ can be characterized by a linear criterion: It equals the unique temperature $T$ such that the linear operator
\begin{align*}
K_{T} - V 
\end{align*}
has zero as its lowest eigenvalue. Here $K_T = K_T(-\i \nabla)$ with the symbol 
\begin{align}
	K_T(p) \coloneqq  \frac{p^2 - \mu}{\tanh \frac{p^2-\mu}{2T}}. \label{KT-symbol}
\end{align}
The operator $K_{T} - V $ is understood to act on the space $L_{\mathrm{sym}}^2(\Rbb^3)$ of reflection-symmetric square-integrable functions on $\mathbb{R}^3$. To be precise, the results in \cite{Hainzl2007} have been proven without the assumption $\alpha(-x) = \alpha(x)$ for a.e. $x\in \mathbb{R}^3$. In this case, the operator $K_\Tc - V$ acts on functions in the Hilbert space $L^2(\Rbb^3)$ instead of $L_{\mathrm{sym}}^2(\Rbb^3)$. The results in \cite{Hainzl2007}, however, equally hold in the case of symmetric Cooper pair wave functions. That is, they hold in the same way if $V$ is reflection symmetric and if the translation-invariant BCS functional is minimized over functions $\gamma(x)$ and $\alpha(x)$ that are both assumed to be reflection symmetric.

We note that the function $K_T(p)$ satisfies the inequalities $K_T(p) \geq 2T$ for $\mu \geq 0$, as well as $K_T(p)\geq |\mu|/\tanh(|\mu|/(2T))$ for $\mu < 0$. Our assumptions on $V$ guarantee that the essential spectrum of $K_T-V$ equals that of $K_T$, and hence an eigenvalue below $2T$ for $\mu \geq 0$ or below $|\mu|/\tanh(|\mu|/(2T))$ for $\mu < 0$ is necessarily isolated and of finite multiplicity. This, in particular, applies to an eigenvalue of $K_T - V$ at $0$.

We are interested in the situation, where $\Tc > 0$ and where the translation-invariant BCS functional has a unique minimizer with a radial Cooper pair wave function (s-wave Cooper pairs) for $T$ close to $\Tc$. The following assumptions guarantee that we are in such a situation. Part (b) should be compared to \cite[Theorem~2.8]{DeuchertGeisinger}. 

\begin{asmp}
\label{Assumption_KTc}
We assume that the interaction potential $V$ is such that the following holds:
\begin{enumerate}[(a)]
\item We have $\Tc >0$. 

\item The lowest eigenvalue of $K_{\Tc} - V$ is simple.
\end{enumerate}
\end{asmp}

As has been shown in \cite[Theorem 3]{Hainzl2007}, our first assumption is satisfied if $V \geqslant 0$ does not vanish identically. Throughout this paper we denote by $\alpha_*$ the unique solution to the equation
\begin{align}
K_\Tc \alpha_* = V\alpha_*. \label{alpha_star_ev-equation}
\end{align}
Since $V$ is radial we know that the same is true for $\alpha_*$. Without loss of generality we will assume that $\alpha_*$ is real-valued and satisfies $\Vert \alpha_*\Vert_{L^2(\Rbb^3)} = 1$. If we write the above equation as $\alpha_* = K_{\Tc}^{-1} V\alpha_*$, we see that $V\in L^\infty(\Rbb^3)$ implies $\alpha_*\in H^2(\Rbb^3)$. Moreover, we know from \cite[Proposition 2]{Hainzl2012} that
\begin{align}
\int_{\Rbb^3} \dd x \; \bigl[ |x^\nu \alpha_*(x)|^2 + |x^\nu \nabla \alpha_*(x)|^2 \bigr] < \infty \label{Decay_of_alphastar}
\end{align}
holds for $\nu \in \Nbb_0^3$.


\subsection{The Ginzburg--Landau functional}
\label{Ginzburg-Landau-functional_Section}

We say that a function $\Psi$ on $Q_h$ is \textit{gauge-periodic} if the magnetic translations of the form
\begin{align}
	T_h(\lambda)\Psi(X) &\coloneqq  \e^{\i \Bbold \cdot (\lambda \wedge X)} \; \Psi(X + \lambda ), & \lambda &\in \Lambda_h, \label{Magnetic_Translation_Charge2}
\end{align}
leave $\Psi$ invariant. We highlight that $T(\lambda)$ in \eqref{Magnetic_Translation} equals $T_h(\lambda)$ provided we replace $\Bbold$ by $2\Bbold$. Let $\Lambda_0, \Lambda_2, \Lambda_3 >0$, $\Lambda_1, D \in \Rbb$, and let $\Psi$ be a gauge-periodic function in the case $h=1$. The Ginzburg--Landau functional is defined by
\begin{align}
\EGL(\Psi) &\coloneqq  \frac{1}{|Q_1|} \int_{Q_1} \dd X \; \bigl\{ \Lambda_0 \; |(-\i\nabla + 2\Abold)\Psi(X)|^2 + \Lambda_1 \, W(X)\, |\Psi(X)|^2 \notag \\
&\hspace{150pt} - D \, \Lambda_2\, |\Psi(X)|^2 + \Lambda_3\,|\Psi(X)|^4\bigr\}. \label{Definition_GL-functional}
\end{align}
We highlight the factor $2$ in front of the magnetic vector potential in \eqref{Definition_GL-functional}. Its appearance is due to the fact that $\Psi$ describes the center-of-mass motion of Cooper pairs carrying twice the charge of a single fermion.

The Ginzburg--Landau energy is defined by
\begin{align*}
E^{\mathrm{GL}}(D) \coloneqq  \inf \bigl\{ \EGL(\Psi) : \Psi\in \Hmag^1(Q_1)\bigr\}.
\end{align*}
We also define the critical parameter
\begin{align}
	\Dc &\coloneqq  \frac 1{\Lambda_2} \inf \spec_{\Lmag^2(Q_1)} \bigl(\Lambda_0 \, (-\i \nabla + \Abold)^2 + \Lambda_1 \; W\bigr). \label{Dc_Definition}
\end{align}
As has been shown in \cite[Lemma 2.5]{Hainzl2014}, we have $\EGLGSE < 0$ if $D > \Dc$ and $\EGLGSE =0$ if $D \leq \Dc$.

In our analysis we encounter the Ginzburg--Landau functional in an $h$-dependent version, where $Q_1, \Abold, W$, and $D$ in \eqref{Definition_GL-functional} are replaced by $Q_h, \Abold_h, W_h$, and $h^2 D$ respectively. If we denote this functional by $\EGLh(\Psi)$ we have
\begin{equation*}
	\inf \bigl\{ \EGLh(\Psi) : \Psi\in \Hmag^1(Q_h)\bigr\} = h^4 E^{\mathrm{GL}}(D),
\end{equation*}
which follows by scaling. More precisely, for given $\psi$ the function
\begin{align}
\Psi(X) &\coloneqq  h \; \psi(h \, X), & X\in \Rbb^3, \label{GL-rescaling}
\end{align}
obeys
\begin{align}
\EGLh(\Psi) = h^4 \EGL(\psi). \label{EGL-scaling}
\end{align}


\subsection{Main results}
\label{Main_Result_Section}

Our first main result concerns an asymptotic expansion of the BCS free energy in the small parameter $h>0$. The precise statement is captured in the following theorem.

\begin{bigthm}
\label{Main_Result}
Let Assumptions \ref{Assumption_V} and \ref{Assumption_KTc} hold, let $D \in \Rbb$, and let the coefficients $\Lambda_0, \Lambda_1, \Lambda_2$, and $\Lambda_3$ be given by \eqref{GL-coefficient_1}-\eqref{GL_coefficient_3} below. Then there are constants $C>0$ and $h_0 >0$ such that for all $0 < h \leq h_0$, we have
\begin{align}
F^{\mathrm{BCS}}(h,\, \Tc(1 - Dh^2)) = h^4 \; \bigl( \EGLGSE + R \bigr), \label{ENERGY_ASYMPTOTICS}
\end{align}
with $R$ satisfying the estimate
\begin{align}
Ch \geq R \geq - \Rcal \coloneqq  -C h^{\nicefrac 1{6}}. \label{Rcal_error_Definition}
\end{align}
Moreover, for any approximate minimizer $\Gamma$ of $\FBCS$ at $T = \Tc(1 - Dh^2)$ in the sense that
\begin{align}
\FBCS(\Gamma) - \FBCS(\Gamma_0) \leq h^4 \bigl( \EGLGSE + \rho\bigr) 
\label{BCS_low_energy}
\end{align}
holds for some $\rho \geq 0$, we have the decomposition
\begin{align}
\alpha(X, r ) = \alpha_*(r) \Psi(X)  + \sigma(X,r) \label{Thm1_decomposition}
\end{align}
for the Cooper pair wave function $\alpha = \Gamma_{12}$. Here, $\sigma$ satisfies
\begin{align}
\frac{1}{|Q_h|} \int_{Q_h} \dd X \int_{\Rbb^3} \dd r \; |\sigma(X, r)|^2 &\leq C h^{\nicefrac {11}3}, \label{Thm1_error_bound}
\end{align}
$\alpha_*$ is the normalized zero energy eigenstate of $K_{\Tc}-V$, and the function $\Psi$ obeys
\begin{align}
\EGLh(\Psi) \leq h^4 \left( \EGLGSE + \rho + \Rcal \right). \label{GL-estimate_Psi}
\end{align}
\end{bigthm}

Our second main result is a statement about the dependence of the critical temperature of the BCS functional on $h>0$ and on the external fields.

\begin{bigthm}
\label{Main_Result_Tc} \label{MAIN_RESULT_TC}
Let Assumptions \ref{Assumption_V} and \ref{Assumption_KTc} hold. Then there are constants $C>0$ and $h_0 >0$ such that for all $0 < h \leq h_0$ the following holds:
\begin{enumerate}[(a)]
	\item Let $0 < T_0 < \Tc$. If the temperature $T$ satisfies
	\begin{equation}
		T_0 \leq T \leq \Tc \, ( 1 - h^2 \, ( \Dc + C \, h^{\nicefrac 12}))
		\label{eq:lowertemp}
	\end{equation}
	with $\Dc$ in \eqref{Dc_Definition}, then we have
	\begin{equation*}
		F^{\mathrm{BCS}}(h,T) < 0.
	\end{equation*} 
	\item If the temperature $T$ satisfies
	\begin{equation}
		T \geq \Tc \, ( 1 - h^2 \, ( \Dc - \Rcal ) )
		\label{eq:uppertemp}
	\end{equation}
	with $\Dc$ in \eqref{Dc_Definition} and $\Rcal$ in \eqref{Rcal_error_Definition}, then we have
	\begin{equation*}
		\FBCS(\Gamma) - \FBCS(\Gamma_0) > 0
	\end{equation*}
	unless $\Gamma = \Gamma_0$.
\end{enumerate}
\end{bigthm}

\begin{bems}
\label{Remarks_Main_Result}
\begin{enumerate}[(a)]
\item Theorem~\ref{Main_Result} and Theorem~\ref{Main_Result_Tc} extend similar results in \cite{Hainzl2012} and \cite{Hainzl2014} to the case of general external electric and magnetic fields. In these references the main restriction is that the vector potential is assumed to be periodic, that is, the corresponding magnetic field has vanishing flux through the faces of the unit cell $Q_h$, compare with the discussion in Section~\ref{Magnetically_Periodic_Samples}. Removing this restriction causes major mathematical difficulties already for the constant magnetic field because its vector potential cannot be treated as a perturbation of the Laplacian. More precisely, it was possible in \cite{Hainzl2012,Hainzl2014} to work with a priori bounds for low-energy states that do not involve the external magnetic field. As noticed in the discussion below Remark~6 in \cite{Hainzl2017}, this is not possible if the magnetic field has nonzero flux through the faces of the unit cell. To prove a priori bounds that involve a constant magnetic field one has to deal with the fact that the components of the magnetic momentum operator do not commute, which leads to significant technical difficulties. These difficulties have been overcome in \cite{DeHaSc2021}, which allowed us to extend the results \cite{Hainzl2012,Hainzl2014} to the case of a system in a constant magnetic field. Our proof of Theorem~\ref{Main_Result} and Theorem~\ref{Main_Result_Tc} uses these a priori bounds, and should therefore be interpreted as an extension of the methods in \cite{DeHaSc2021} to the case of general external electric and magnetic fields. The main technical novelty of this article is a further development of the phase approximation method, which has been pioneered in the framework of BCS theory for the case of the constant magnetic field in \cite{Hainzl2017} and \cite{DeHaSc2021}. It allows us to compute the BCS free energy of a class of trial states (Gibbs states) in a controlled way, and is the key new ingredient for our proof of upper and lower bounds for the BCS free energy in the presence of general external fields.

\item When we compare our result in Theorem~\ref{Main_Result} to the main Theorem in \cite{Hainzl2012}, we notice the following differences: (1) We use microscopic coordinates while macroscopic coordinates are used in \cite{Hainzl2012,Hainzl2014}, see the discussion above \cite[Eq.~(1.4)]{Hainzl2012}. (2) Our free energy is normalized by a large volume factor, see \eqref{Trace_per_unit_volume_definition} and \eqref{BCS functional}. This is not the case in \cite{Hainzl2012,Hainzl2014}. Accordingly, the GL energy appears on the order $h^4$ in our setting and on the order $h$ in the setting in \cite{Hainzl2012}. (3) The leading order of the Cooper pair wave function in \cite[Theorem 1]{Hainzl2012} is of the form
\begin{equation}
	\frac{1}{2} \alpha_*(x-y) (\Psi(x) + \Psi(y)).
	\label{DHS1:eq:remarksA1}
\end{equation}
This should be compared to \eqref{Thm1_decomposition}, where relative and center-of-mass coordinates are used. When we use the a priori bound for $\Vert \nabla \Psi \Vert_2$ below Eq.~(5.61) in \cite{Hainzl2012}, we see that this decomposition equals that in \eqref{Thm1_decomposition} to leading order in $h$. 

\item The Ginzburg--Landau energy appears at the order $h^4$. This needs to be compared to the energy of the normal state, which is of order $1$ in $h$. To understand the order of the GL energy we need to realize that each factor of $\Psi$ in $\EGLh$ defined below \eqref{Dc_Definition} carries a factor $h$. This follows from the scaling in \eqref{GL-rescaling} and the fact that the GL energy is normalized by the volume factor $|Q_h|^{-1}$. Moreover, every magnetic momentum operator carries a factor $h$ because $\Psi$ varies on the length scale $h^{-1}$ and the electric potential carries a factor $h^2$. In combination, these considerations explain the size of all terms in the GL functional. It is worth noting that the prefactor $-h^2 D$ in front of the quadratic term without external fields equals $(T-T_{\mathrm{c}})/T_{\mathrm{c}}$. 

\item The size of the remainder in \eqref{Thm1_error_bound} should be compared to the $L^2$-norm per unit volume of the leading order part of the Cooper pair wave function in \eqref{Thm1_decomposition}, which satisfies
\begin{equation*}
	\frac{1}{| Q_h |} \int_{Q_h} \dd X \int_{\mathbb{R}^3} \dd r \; | \alpha_*(r) \Psi(X) |^2 \sim h^2
\end{equation*}
if $D>0$. We highlight that $\alpha_*$ varies on the microscopic length scale $1$ and that $\Psi$ captures the effects of the external fields on the macroscopic length scale $h^{-1}$.

\item Our bounds show that $D$ in Theorem~\ref{Main_Result} can be chosen as a function of $h$ as long as $|D| \leq D_0$ holds for some constant $D_0>0$.

\item The upper bound for the error in \eqref{Rcal_error_Definition} is worse than the corresponding bound in \cite{DeHaSc2021} by the factor $h^{-1}$. It is of the same size as the comparable error term in \cite[Theorem~1]{Hainzl2012}.

\item Theorem~\ref{MAIN_RESULT_TC} gives bounds on the temperature regions where superconductivity is present or absent. The interpretation of the theorem is that the critical temperature of the full model satisfies
\begin{equation*}
	\Tc(h) = \Tc \left( 1 - D_{\mathrm{c}} h^2 \right) + o(h^2),
\end{equation*}
with the critical temperature $\Tc$ of the translation-invariant problem. The coefficient $D_{\mathrm{c}}$ is determined by linearized Ginzburg--Landau theory, see \eqref{Magnetic_Translation_Charge2}. The above equation allows us to compute the upper critical field $B_{\mathrm{c}2}$, above which superconductivity is absent. It also allows to to compute the derivative of $B_{\mathrm{c}2}$ with respect to $T$ at $\Tc$, see \cite[Appendix~A]{Hainzl2017}.

\item We expect that the assumption $0 < T_0 < \Tc$ in part (a) of Theorem~\ref{MAIN_RESULT_TC}, which also appeared in \cite{Hainzl2017,DeHaSc2021}, is only of technical nature. We need it because our trial state analysis breaks down as $T$ approaches zero. We note that there is no such restriction in part (b) of Theorem~\ref{MAIN_RESULT_TC} or in Theorem~\ref{Main_Result}.
\end{enumerate}
\end{bems}


\subsection{Organization of the paper and strategy of proof}

For the convenience of the reader we give here a short summary of the organization of the paper and the proof of our two main results.

In Section~\ref{sec:HeuristicComputation} we provide a brief non-rigorous computation that shows from which terms in the BCS functional the different terms in the GL functional arise. Afterwards we complete in Section~\ref{Preliminaries} the introduction of our mathematical setup. That is, we collect useful properties of the trace per unit volume and introduce the relevant spaces of gauge-periodic functions. 

In Section~\ref{Upper_Bound} we collect the results of our trial state analysis. We introduce a class of Gibbs states with Cooper pair wave functions that admit a product structure of the form $\alpha_*(r) \Psi(X)$ to leading order in $h$. Here, $\alpha_*$ is the ground state wave function in \eqref{alpha_star_ev-equation} and $\Psi$ is a gauge-periodic function. We state and motivate several results concerning the Cooper pair wave function and the BCS energy of these states. Afterwards we use these statements to provide the proof of the upper bound for the BCS free energy in \eqref{ENERGY_ASYMPTOTICS} as well as the proof of Theorem~\ref{Main_Result_Tc} (a). It is important to note that these results are needed again in Section \ref{Lower Bound Part B}, where we give the proof of the lower bound on the BCS free energy in \eqref{ENERGY_ASYMPTOTICS} and the proof of Theorem \ref{Main_Result_Tc} (b). 

In Section~\ref{Proofs} we provide the proofs of the results in Section~\ref{Upper_Bound} concerning the Cooper pair wave function and the BCS free energy of our trial states. It is the main part of our article and contains the main technical novelties. Our approach is based on an application of the phase approximation method for general magnetic fields to our nonlinear setting. The phase approximation is a well-known tool in the physics literature, see, e.g., \cite{Werthamer1966}, and has also been used in the mathematical literature to study spectral properties of Schr\"odinger operators involving a magnetic field, for instance in \cite{NenciuCorn1998,Nenciu2002}. In the case of a constant magnetic field, this method has been pioneered within the framework of linearized BCS theory in \cite{Hainzl2017} and for the full nonlinear model in \cite{DeHaSc2021}. An application of the phase approximation method to the case of a magnetic field with zero flux through the unit cell is contained in unpublished notes by Frank, Geisinger, Hainzl, and Tzaneteas \cite{BdGtoGL}. The main technical novelty in Section~\ref{Proofs} is a further development of the phase approximation method for general external fields in our nonlinear setting. This allows us to compute the BCS free energy of a class of trial states (Gibbs states) in a controlled way, which is the key new ingredient for the proof of upper and lower bounds for the BCS free energy in the presence of general external fields. Our approach should also be compared to the trial state analysis in \cite{Hainzl2012,Hainzl2014}, where a semi-classical expansion is used to treat magnetic fields with zero flux through the unit cell. We highlight that the trial state analysis for general external fields requires considerably more effort than the one for a constant magnetic field in \cite{DeHaSc2021}. This is also reflected in the length of the proofs.

In Section~\ref{Lower Bound Part A}, we prove a priori estimates for BCS states, whose BCS free energy is smaller than or equal to that of the normal state $\Gamma_0$ in \eqref{Gamma0} (low-energy states). More precisely, we show that the Cooper pair wave function of any such state is, to leading order as $h \to 0$, given by $\alpha_*(r) \Psi(X)$ with $\alpha_*$ in \eqref{alpha_star_ev-equation} and with a gauge-periodic function $\Psi$. The proof of the same statement in the case of a constant magnetic field has been the main novelty in \cite{DeHaSc2021}. To treat the case of general external fields, we perturbatively remove the periodic vector potential $A$ and the electric potential $W$. This allows us to reduce the problem to the case of a constant magnetic field treated in \cite{DeHaSc2021}. Similar a priori estimates for the case of magnetic fields with zero flux through the unit cell had been proved for the first time in \cite{Hainzl2012}.

The proofs of the lower bound on \eqref{ENERGY_ASYMPTOTICS} and of Theorem~\ref{Main_Result_Tc}~(b), which go along the same lines as those presented in \cite{Hainzl2012,Hainzl2014,DeHaSc2021}, are given in Section~\ref{Lower Bound Part B}. They complete the proofs of Theorem \ref{Main_Result} and \ref{Main_Result_Tc}. The main idea is to use the a priori estimates in Section \ref{Lower Bound Part A} to replace a general low-energy state in the BCS functional by a Gibbs state, whose Cooper pair wave function has the same leading order behavior for small $h$, in a controlled way. This, in particluar, allows us to estimate the BCS energy of a general low-energy state in terms of that of a Gibbs state, which has been computed in Sections~\ref{Upper_Bound} and \ref{Proofs}. Because of the considerable overlap in content with the related section in \cite{DeHaSc2021}, we shortened the proofs in this section to a minimal length.

Throughout the paper, $c$ and $C$ denote generic positive constants that change from line to line. We allow them to depend on the various fixed quantities like $h_0$, $D_0$, $\mu$, $\Tc$, $V$, $A$, $W$, $\alpha_*$, etc. Further dependencies are highlighted.

\subsection{Heuristic computation of the terms in the Ginzburg--Landau functional}
\label{sec:HeuristicComputation}

In the following we present a brief and non-rigorous computation of the BCS energy of the trial state that we use in the proof of the upper bound for the BCS free energy in Section~\ref{UPPER_BOUND}. The goal is to show from which terms in the BCS functional the different terms in the Ginzburg--Landau (GL) functional arise. A more detailed and more precise discussion of these issues can be found in Section~\ref{UPPER_BOUND}. 

Our trial state (a Gibbs state) is defined by
\begin{equation}
	\begin{pmatrix} \gamma_{\Delta} & \alpha_{\Delta} \\ \overline{\alpha_{\Delta}} & 1 - \overline{\gamma_{\Delta}} \end{pmatrix} =
	\Gamma_{\Delta} = \frac{1}{1+\e^{\beta H_{\Delta}}}.  
	\label{eq:heuristicstrialstate}
\end{equation}
Here,  $\beta^{-1} = T = \Tc(1 - Dh^2)$ and the Hamiltonian is given by
\begin{equation*}
	H_{\Delta} = \begin{pmatrix} (-\mathrm{i} \nabla + \Abold_h )^2 - \mu & \Delta \\ \overline{\Delta} & -\overline{(-\mathrm{i} \nabla + \Abold_h )^2} + \mu \end{pmatrix},
\end{equation*}
where the operator $\Delta$ is defined via its integral kernel
\begin{equation*}
	\Delta(x,y) = -2\, V \alpha_* (x-y)\, \Psi_h \left( \frac{x+y}{2} \right).
	\label{eq:AB1}
\end{equation*}
The function $\Psi_h(X) = h \psi(hX)$ is chosen such that $\psi$ is a minimizer of the Ginzburg--Landau functional in \eqref{Definition_GL-functional}. We therefore have 
\begin{equation*}
	\frac{1}{| Q_h|} \int_{Q_h} \dd X \ | \Psi_h(X) |^2 \sim h^2
\end{equation*}
as well as 
\begin{equation}
	\Tr [ \Delta^* \Delta ] = \frac{4}{| Q_h|} \int_{Q_h} \dd X \; | \Psi_h(X) |^2  \int_{\mathbb{R}^3} \dd r \; | V(r) \alpha_*(r) |^2 \sim h^2.
	\label{eq:heurusticscalingdelta}
\end{equation}
Here $r=x-y$ and $X=(x+y)/2$ denote relative and center-of-mass coordinates. The operator $\Delta$ in \eqref{eq:heuristicstrialstate} is therefore a small perturbation if $0 < h \ll 1$.

The BCS free energy of the trial state $\Gamma_{\Delta}$ is given by
\begin{align*}
	\FBCS(\Gamma_{\Delta}) - \FBCS(\Gamma_0) &= \frac{1}{2} \Tr\left[ H_0 (\Gamma_{\Delta} - \Gamma_0) \right] - T S(\Gamma) + T S(\Gamma_0)  \\
	&\hspace{30pt} - \frac{1}{| Q_h |} \int_{Q_h} \dd X \int_{\mathbb{R}^3} \dd r \; V(r) | \alpha_{\Delta}(X,r) |^2, 
\end{align*}
where $\Gamma_0$ denotes the normal state in \eqref{Gamma0}. Applications of  
\begin{equation*}
	\Tr\left[ H_0 (\Gamma_{\Delta} - \Gamma_0) \right] = \Tr\left[ H_{\Delta} \Gamma_{\Delta} - H_0 \Gamma_0 \right] - \Tr[ (H_{\Delta} - H_0 ) \Gamma_{\Delta} ],
\end{equation*}
and \cite[Eqs.~(4.3-4.5)]{Hainzl2012}, allow us to rewrite this formula as
\begin{align}
	\FBCS(\Gamma_{\Delta}) - \FBCS(\Gamma_0) =& -\frac {1}{2 \beta}\Tr_0 \bigl[ \ln\bigl( 1+\exp\bigr( - \beta H_\Delta \bigr)\bigr) - \ln\bigl( 1+\exp\bigl( -\beta H_0 \bigr)\bigr)\bigr] \nonumber \\
	&+ \frac{\langle \alpha_*, V\alpha_*\rangle_{L^2(\Rbb^3)}}{|Q_h|} \int_{Q_h} \dd X \ |\Psi_h(X)|^2  \nonumber \\   
	&- \frac{1}{|Q_h|} \int_{Q_h} \dd X\int_{\Rbb^3} \dd r\; V(r) \, \bigl|\alpha(X,r) - \alpha_*(r) \Psi_h(X)\bigr|^2. \label{eq:heuristics1}
\end{align}
Here $\Tr_0[A] = \Tr[PAP] + \Tr[QAQ]$ with 
\begin{equation*}
	P = \begin{pmatrix}
		1 & 0 \\ 0 & 0 \end{pmatrix}
\end{equation*} 
and $Q = 1 - P$.

To identify the terms in the GL functional, we need to expand the terms in \eqref{eq:heuristics1} in powers of $h$. To that end, we first expand them up to fourth order in powers of $\Delta$ because the Ginzburg--Landau functional is a fourth order polynomial in $\Psi_h$. Afterwards, we expand the resulting terms in powers of $h$, that is, we use that the external fields $W_h(x) = h^2 W(hx)$ and $\Abold_h(x) = h \Abold(hx)$ as well as the temperature $T = \Tc(1 - Dh^2)$ with $D>0$ depend on $h$. It turns out that the last term on the right side of \eqref{eq:heuristics1} is of the order $o(h^4)$. Since the GL energy appears on the order $h^4$ it does not contribute to it. In our trial state analysis in Sections~\ref{Upper_Bound}~and~\ref{Proofs} we show that there is a linear operator $L_{T,\Abold,W}$ and a cubic map $N_{T,\Abold,W}(\Delta)$ such that
\begin{align}
	-\frac {1}{2\beta}\Tr_0 \bigl[ \ln\bigl( 1+\exp\bigl( - \beta H_\Delta \bigr)\bigr) - \ln\bigl( 1+\exp\bigl( -\beta H_0 \bigr)\bigr)\bigr] & \nonumber \\
	&\hspace{-120pt} = -\frac{1}{4} \langle \Delta, L_{T,\Abold,W} \Delta \rangle + \frac{1}{8} \langle \Delta, N_{T,\Abold,W}(\Delta) \rangle + o(h^4) \label{eq:heuristicquadraticterms}
\end{align}
holds. In combination with the first term in the second line of \eqref{eq:heuristics1}, the quadratic terms in \eqref{eq:heuristicquadraticterms} contain the quadratic terms in the Ginzburg--Landau functional:
\begin{align*}
	&-\frac{1}{4} \langle \Delta, L_{T,\Abold,W} \Delta \rangle + \frac{\langle \alpha_*, V\alpha_*\rangle_{L^2(\Rbb^3)}}{|Q_h|} \int_{Q_h} \dd X \; |\Psi_h(X)|^2 \\
	&= \frac{1}{|Q_h|} \int_{Q_h} \dd X \; \bigl\{ \Lambda_0 \; |(-\i\nabla + 2\Abold_h)\Psi_h(X)|^2 + \Lambda_1 \, W_h(X)\, |\Psi_h(X)|^2 - Dh^2 \, \Lambda_2\, |\Psi_h(X)|^2 \bigr\} \\
	&\hspace{0.5cm} + o(h^4)
\end{align*}
with $\Lambda_0, \Lambda_1,$ and $\Lambda_2$ defined in \eqref{GL-coefficient_1}-\eqref{GL-coefficient_2}. From the quartic term in \eqref{eq:heuristicquadraticterms} we will extract the quartic term in the GL functional:
\begin{equation*}
	\frac{1}{8} \langle \Delta, N_{T,\Abold,W}(\Delta) \rangle = \frac{\Lambda_3}{|Q_h|} \int_{Q_h} \dd X \; |\Psi_h(X)|^4 + o(h^4).
\end{equation*}
The coefficient $\Lambda_3$ is defined in \eqref{GL_coefficient_3}. Accordingly, 
\begin{equation*}
	\FBCS(\Gamma_{\Delta}) - \FBCS(\Gamma_0) = h^4 \left( \EGLGSE + o(1) \right),
\end{equation*}
where we used $\EGLh(\Psi_h) = h^4 \EGLGSE$ in the last step.


\section{Preliminaries}
\label{Preliminaries}


\subsection{Schatten classes}
\label{Schatten_Classes}

The trace per unit volume in \eqref{Trace_per_unit_volume_definition} gives rise to Schatten classes of periodic operators, whose norms play an important role in our proofs. In this section we recall several well-known facts about these norms.

For $1 \leq p < \infty$, the $p$\tho\ local von-Neumann--Schatten class $\Scal^p$ consists of all gauge-periodic operators $A$ having finite $p$-norm, that is, $\Vert A\Vert_p^p \coloneqq  \Tr (|A|^p) <\infty$. The space of bounded gauge-periodic operators $\Scal^\infty$ is equipped with the usual operator norm. We note that the $p$-norm is not monotone decreasing in the index $p$. This should be compared to the usual Schatten norms, where such a property holds, see the discussion below \cite[Eq. (3.9)]{Hainzl2012}.

We recall that the triangle inequality 
\begin{align*}
\Vert A + B\Vert_p \leq \Vert A\Vert_p + \Vert B\Vert_p
\end{align*}
holds for operators in $\Scal^p$ for $1 \leq p \leq \infty$. We also have the generalized version of Hölder's inequality
\begin{align}
\Vert AB\Vert_r \leq \Vert A\Vert_p \Vert B\Vert_q, \label{Schatten-Hoelder}
\end{align}
which holds for $1 \leq p,q,r \leq \infty$ with $\frac{1}{r} = \frac{1}{p} + \frac{1}{q}$. The familiar inequality
\begin{align*}
| \Tr A | \leq \Vert A \Vert_1
\end{align*}
also holds in the case of local Schatten norms.

The above inequalities can be deduced from their versions for the usual Schatten norms, see, e.g., \cite{Simon05}, with the help of the magnetic Bloch--Floquet decomposition. We refer to \cite[Section XIII.16]{Reedsimon4} for an introduction to the Bloch--Floquet transformation and to \cite{Stefan_Peierls} for a treatment of the magnetic case. To be more precise, a gauge-periodic operator $A$ satisfies the unitary equivalence 
\begin{align*}
A \cong \int^{\oplus}_{[0,\sqrt{ 2 \pi }\, h]^3} \mathrm{d} k \;  A_{k},
\end{align*}
which we use to write the trace per unit volume as
\begin{align}
\Tr A = \int_{[0,\sqrt{ 2 \pi }\, h]^3} \frac{\dd k}{(2\pi)^3} \; \Tr_{L^2(Q_h)} A_{k}. \label{eq:ATPUV}
\end{align}
Here, $\Tr_{L^2(Q_h)}$ denotes the usual trace over $L^2(Q_h)$. When we use that $(AB)_k = A_k B_k$ holds for gauge-periodic operators $A$ and $B$, the above mentioned inequalities for the trace per unit volume are implied by their usual versions.


\subsection{Gauge-periodic Sobolev spaces}
\label{Periodic Spaces}

In this section we introduce Banach spaces of gauge-periodic functions, which will be used to describe Cooper pair wave functions of BCS states. 

When working with center-of-mass and relative coordinates $(X,r)$ it is useful to define the magnetic momentum operators
\begin{align}
	\Pi_{\Abold} &\coloneqq  -\i\nabla_X + 2 \Abold(X), & \tilde \pi_{\Abold} &\coloneqq  -\i\nabla_r + \frac 12 \Abold(r). \label{Magnetic_Momenta_full_COM}
\end{align}
We will also use the notation 
\begin{align}
	\Pi &\coloneqq  -\i\nabla_X + 2 \Abold_{\Bbold}(X), & \tilde \pi &\coloneqq  -\i\nabla_r + \frac 12  \Abold_\Bbold(r). \label{Magnetic_Momenta_COM}
\end{align}
If several coordinates appear in an equation we sometimes write $\Pi_X$ and $\tilde \pi_r$ to highlight on which coordinate $\Pi$ and $\tilde \pi$ are acting.

A function $\Psi \in L_\loc^p(\Rbb^3)$ with $1 \leq p \leq \infty$ belongs to the space $L_{\mathrm{mag}}^p(Q_h)$ provided $T_h(\lambda)\Psi = \Psi$ holds for all $\lambda\in\Lambda_h$ (with $T_h(\lambda)$ in \eqref{Magnetic_Translation_Charge2}). We endow $L_{\mathrm{mag}}^p(Q_h)$ with the usual $p$-norm per unit volume
\begin{align}
\Vert \Psi\Vert_{\Lmag^p(Q_h)}^p &\coloneqq  \fint_{Q_h} \dd X \; |\Psi(X)|^p \coloneqq  \frac{1}{|Q_h|} \int_{Q_h} \dd X \; |\Psi(X)|^p \label{Periodic_p_Norm}
\end{align}
if $1 \leq p < \infty$ and with the $L^{\infty}(Q_h)$-norm if $p=\infty$. When it does not lead to confusion we use the abbreviation $\Vert \Psi\Vert_p$.

Analogously, for $m\in \Nbb_0$, we define the Sobolev spaces of gauge-periodic functions corresponding to the constant magnetic field as
\begin{align}
\Hmag^m(Q_h) &\coloneqq  \bigl\{ \Psi\in \Lmag^2(Q_h) :  \Pi^\nu \Psi\in \Lmag^2(Q_h) \quad \forall \nu\in \Nbb_0^3, |\nu|_1\leq m\bigr\}, \label{Periodic_Sobolev_Space}
\end{align}
where $|\nu |_1 \coloneqq  \sum_{i=1}^3 \nu_i$ for $\nu\in \Nbb_0^3$. It is a Hilbert space when endowed with the inner product
\begin{align}
\langle \Phi, \Psi\rangle_{\Hmag^m(Q_h)} &\coloneqq  \sum_{|\nu|_1\leq m} h^{-2 - 2|\nu|_1} \; \langle \Pi^\nu \Phi, \Pi^\nu \Psi\rangle_{\Lmag^2(Q_h)}.  \label{Periodic_Sobolev_Norm}
\end{align}
We note that if $\Psi$ is a gauge-periodic function then so is $\Pi^\nu \Psi$, since the magnetic momentum operator $\Pi$
commutes with the magnetic translations $T_h(\lambda)$ in \eqref{Magnetic_Translation_Charge2}. Furthermore, $\Pi$ is a self-adjoint operator on $\Hmag^1(Q_h)$. 

The norms introduced in \eqref{Periodic_p_Norm} and \eqref{Periodic_Sobolev_Norm} display a scaling behavior with respect to $h$, which is motivated by the Ginzburg--Landau scaling in \eqref{GL-rescaling}. More precisely, whenever $\psi \in \Lmag^p(Q_1)$ and $\Psi(x) = h \psi(hx)$, then
\begin{align}
\Vert \Psi\Vert_{\Lmag^p(Q_h)} = h \, \Vert \psi\Vert_{\Lmag^p(Q_1)} \label{Periodic_p_Norm_scaling}
\end{align}
for every $1\leq p \leq \infty$. That is, $\Psi \sim h$ in any $p$-norm per unit volume.

The inner product in \eqref{Periodic_Sobolev_Norm} is chosen such that
\begin{align*}
\Vert \Psi\Vert_{\Hmag^m(Q_h)} = \Vert \psi\Vert_{\Hmag^m(Q_1)}
\end{align*}
holds. This follows from \eqref{Periodic_p_Norm_scaling} and the fact that $\Vert \Pi^\nu\Psi\Vert_2^2$ scales as $h^{2 + 2|\nu|_1}$ for $\nu\in \Nbb_0^3$. Such scaled norms have also been used in \cite{DeHaSc2021} but not in \cite{Hainzl2012,Hainzl2014}.

For the sake of completeness, let us also mention the following magnetic Sobolev inequality. There is a constant $C>0$ such that for any $h>0$ and any $\Psi\in \Hmag^1(Q_h)$, we have
\begin{align}
\Vert \Psi\Vert_{\Lmag^6(Q_h)}^2 &\leq  C \, h^{-2}\, \Vert \Pi \Psi\Vert_{\Lmag^2(Q_h)}^2. \label{Magnetic_Sobolev}
\end{align}
The proof can be found in \cite{DeHaSc2021} below Eq.~(2.7).


The Cooper pair wave function $\alpha$ of an admissible BCS state $\Gamma$ belongs to the Hilbert--Schmidt class $\Scal^2$ defined in Section \ref{Schatten_Classes}, see the discussion below \eqref{Trace_per_unit_volume_definition}. The symmetry and the gauge-periodicity of the kernel of $\alpha$ in \eqref{alpha_periodicity} can be reformlated as
\begin{align}
\alpha(X,r) &= \e^{\i \Bbold \cdot (\lambda \wedge X)} \; \alpha(X+ \lambda, r), \quad \lambda\in \Lambda_h; & \alpha(X,r) &= \alpha(X, -r) \label{alpha_periodicity_COM}
\end{align}
in terms of center-of-mass and relative coordinates. In other words, $\alpha(X,r)$ is a gauge-periodic function of the center-of-mass coordinate $X \in \mathbb{R}^3$ and a reflection-symmetric function of the relative coordinate $r \in \mathbb{R}^3$. We make use of the unitary equivalence of $\Scal^2$ and the space
\begin{align*}
\Lsymm \coloneqq  \Lmag^2(Q_h) \otimes L_{\mathrm{sym}}^2(\Rbb^3),
\end{align*}
which consists of all square-integrable functions satisfying \eqref{alpha_periodicity_COM}. We also define the norm
\begin{align*}
\Vert \alpha\Vert_{\Lsymm}^2 \coloneqq  \fint_{Q_h} \dd X\int_{\Rbb^3} \dd r \; |\alpha(X, r)|^2 = \frac{1}{|Q_h|} \int_{Q_h} \dd X\int_{\Rbb^3} \dd r \; |\alpha(X, r)|^2.
\end{align*} 
The identity $\Vert \alpha\Vert_2 = \Vert \alpha\Vert_{\Lsymm}$ follows from \eqref{alpha_periodicity_COM}. In the following we therefore identify the scalar products $\langle \cdot, \cdot\rangle$ on $\Lsymm$ and $\Scal^2$ with each other and we do not distinguish between operators in $\Scal^2$ and their kernels as this does not lead to confusion. 

By $H^1(Q_h\times \Rbb_{\mathrm s}^3)$ we denote the Sobolev space of all functions $\alpha\in L^2(Q_h\times \Rbb_{\mathrm s}^3)$, which have finite $H^1$-norm defined by
\begin{align}
\Vert \alpha\Vert_{H^1(Q_h\times \Rbb_{\mathrm s}^3)}^2 &\coloneqq  \Vert \alpha\Vert_2^2 + \Vert \Pi\alpha\Vert_2^2 + \Vert \tilde \pi\alpha\Vert_2^2 \label{H1-norm}
\end{align}
with $\Pi$ and $\tilde \pi$ in \eqref{Magnetic_Momenta_COM}.

We highlight that the norm in \eqref{H1-norm} is equivalent to the two norms 
\begin{align}
\Tr [\alpha\alpha^*] + \Tr [(-\i \nabla + \Abold_\Bbold)\alpha \alpha^* (-\i \nabla + \Abold_\Bbold)]  + \Tr [(-\i \nabla + \Abold_\Bbold) \alpha^* \alpha (-\i \nabla + \Abold_\Bbold)] \label{Norm_equivalence_1}
\end{align}
and
\begin{align}
\Vert \alpha\Vert_2^2 + \Vert (-\i \nabla + \Abold_\Bbold)\alpha\Vert_2^2 + \Vert \alpha (-\i \nabla + \Abold_\Bbold)\Vert_2^2, \label{Norm_equivalence_2}
\end{align}
compare also with the discussion below \eqref{Trace_per_unit_volume_definition}. We also note that the $H^m$-norm in \eqref{Periodic_Sobolev_Norm} and the $H^1$-norm in \eqref{H1-norm} are equivalent to the norms that we obtain when $\Abold_{\mathrm{B}}$ is replaced by $\Abold = \Abold_{\mathrm{B}} + A$ with a periodic vector potential $A \in L^{\infty}(\mathbb{R}^3)$.

\section{Trial States and their BCS Energy}
\label{Upper_Bound} \label{UPPER_BOUND}

In this section we introduce a class of trial states (Gibbs states), state several results concerning their Cooper pair wave function and their BCS free energy, and use these results to prove the upper bound on \eqref{ENERGY_ASYMPTOTICS} as well as Theorem~\ref{Main_Result_Tc}~(a). The trial states $\Gamma_{\Delta}$ are of the form stated in \eqref{eq:heuristicstrialstate}. In Proposition~\ref{Structure_of_alphaDelta} we show that if $\Delta$ is given by $V \alpha_*(r) \Psi(X)$ with a gauge periodic function $\Psi$ that is small in an appropriate sense for small $h$, then $[\Gamma_{\Delta}]_{12} = \alpha_{\Delta} \approx \alpha_*(r) \Psi(X)$ to leading order in $h$. In Proposition~\ref{BCS functional_identity} we prove a representation formula for the BCS functional that allows us to compute the BCS energy of the trial states $\Gamma_\Delta$. Finally, in Theorem~\ref{Calculation_of_the_GL-energy} we extract the terms of the Ginzburg--Landau functional from the BCS free energy of $\Gamma_{\Delta}$. The proofs of these statements are given in Section~\ref{Proofs}. Our trial analysis should be viewed as further development of that in \cite{DeHaSc2021} for the constant magnetic field. The techniques we develop in Sections \ref{Upper_Bound} and \ref{Proofs} are based on gauge-invariant perturbation, which has been pioneered in the framework of linearized BCS theory for a constant external magnetic field in \cite{Hainzl2017}. Our approach should also be compared to the trial state analysis in \cite{Hainzl2012,Hainzl2014}, where a semi-classical expansion is used to treat magnetic fields with zero flux through the unit cell. 


\subsection{The Gibbs states \texorpdfstring{$\Gamma_\Delta$}{GammaDelta}}

For $\Psi\in \Lmag^2(Q_h)$ we define the gap function $\Delta\in \Lsymm$ by
\begin{align}
\Delta(X,r) \coloneqq  \Delta_\Psi(X, r) &\coloneqq  -2 \; V\alpha_*(r) \Psi(X).  \label{Delta_definition}
\end{align}
We also introduce the one-particle Hamiltonian
\begin{align}
\hfrak_{\Abold, W} &\coloneqq  \hfrak_\Abold + W \coloneqq  (-\i \nabla +\Abold_h )^2 + W_h - \mu \label{hfrakAW_definition}
\end{align}
as well as
\begin{align}
H_{\Delta} &\coloneqq  H_0 + \delta \coloneqq  \begin{pmatrix}
\hfrak_{\Abold, W} & 0 \\ 0 & -\ov{\hfrak_{\Abold, W}}
\end{pmatrix} + \begin{pmatrix}
0 & \Delta \\ \ov \Delta & 0
\end{pmatrix} = \begin{pmatrix}
\hfrak_{\Abold, W} & \Delta \\ \ov \Delta & -\ov {\hfrak_{\Abold, W}}
\end{pmatrix}. \label{HDelta_definition}
\end{align}
The Gibbs state at inverse temperature $\beta = T^{-1} >0$ is defined by
\begin{align}
\begin{pmatrix} \gamma_\Delta & \alpha_\Delta \\ \ov{\alpha_\Delta} & 1 - \ov{\gamma_\Delta}\end{pmatrix} = \Gamma_\Delta \coloneqq  \frac{1}{1 + \e^{\beta H_\Delta}}. \label{GammaDelta_definition}
\end{align}
We highlight that the choice $\Delta =0$ yields the normal state $\Gamma_0$ in \eqref{Gamma0}. In our proof of the upper bound for the free energy in \eqref{ENERGY_ASYMPTOTICS} we will choose $\Psi$ as a minimizer of the Ginzburg--Landau functional in \eqref{Definition_GL-functional}, which satisfies the scaling in \eqref{GL-rescaling}. Since the $L^2(\Rbb^3)$-norm of $V\alpha_*$ is of the order $1$, the local Hilbert--Schmidt norm of $\Delta$ is of the order $h$ in this case. In the proof of the lower bound we have less information about the function $\Psi$. The related difficulties are discussed in Remark~\ref{rem:alpha} below. 

\begin{lem}[Admissibility of $\Gamma_\Delta$]
\label{Gamma_Delta_admissible}
Let Assumptions \ref{Assumption_V} and \ref{Assumption_KTc} hold. Then, for any $h>0$, any $T>0$, and any $\Psi\in \Hmag^1(Q_h)$, the state $\Gamma_\Delta$ in \eqref{GammaDelta_definition} is admissible.
\end{lem}

The choice of the states $\Gamma_\Delta$ is motivated by the following observation. Using standard variational arguments one can show that any minimizer $\Gamma$ of the BCS functional satisfies the nonlinear Bogolubov--de Gennes equation
\begin{align}
\Gamma &= \frac 1{1 + \e^{\beta \, \Hbb_{V\alpha}}}, & \Hbb_{V\alpha} = \begin{pmatrix} \hfrak_{\Abold, W} & -2\, V\alpha \\ -2\, \ov{V\alpha} & -\ov{\hfrak_{\Abold, W}}\end{pmatrix}. \label{BdG-equation}
\end{align}
Here, $V\alpha$ is the operator given by the integral kernel $V(r)\alpha(X,r)$. Since we are interested in approximate minimizers of the BCS functional, we choose $\Gamma_{\Delta}$ as an approximate solution to the BdG-equation in \eqref{BdG-equation}. The next result shows that, as far as the leading order behavior of $\alpha_{\Delta}$ is concerned, this is indeed the case. It should be compared to \eqref{Thm1_decomposition}. 


\begin{prop}[Structure of $\alpha_\Delta$]
\label{Structure_of_alphaDelta} \label{STRUCTURE_OF_ALPHADELTA}
Let Assumption \ref{Assumption_V} and \ref{Assumption_KTc} (a) be satisfied and let $T_0>0$ be given. Then, there is a constant $h_0>0$ such that for any $0 < h \leq h_0$, any $T\geq T_0$, and any $\Psi\in \Hmag^2(Q_h)$ the function $\alpha_\Delta$ in \eqref{GammaDelta_definition} with $\Delta \equiv \Delta_\Psi$ as in \eqref{Delta_definition} has the decomposition
\begin{align}
\alpha_\Delta(X,r) &= \Psi(X) \alpha_*(r) - \eta_0(\Delta)(X,r) - \eta_{\perp}(\Delta)(X,r). \label{alphaDelta_decomposition_eq1}
\end{align}
The remainder functions $\eta_0(\Delta)$ and $\eta_\perp(\Delta)$ have the following properties:
\begin{enumerate}[(a)]
\item The function $\eta_0$ satisfies the bound
\begin{align}
\Vert \eta_0\Vert_\Hsymm^2 &\leq  C\; \bigl( h^5 + h^2 \, |T - \Tc|^2\bigr) \; \bigl(  \Vert \Psi\Vert_{\Hmag^1(Q_h)}^6 + \Vert \Psi\Vert_{\Hmag^1(Q_h)}^2\bigr). \label{alphaDelta_decomposition_eq2}
\end{align}

\item The function $\eta_\perp$ satisfies the bound
\begin{align}
\Vert \eta_\perp\Vert_{\Hsymm}^2 + \Vert |r|\eta_\perp\Vert_{\Lsymm}^2 &\leq C \; h^6 \; \Vert \Psi\Vert_{\Hmag^2(Q_h)}^2. \label{alphaDelta_decomposition_eq3}
\end{align}

\item The function $\eta_\perp$ has the explicit form
\begin{align*}
\eta_\perp(X, r) &= \int_{\Rbb^3} \dd Z \int_{\Rbb^3} \dd s \; k_T(Z, r-s) \, V\alpha_*(s) \, \bigl[ \cos(Z\cdot \Pi) - 1\bigr] \Psi(X)
\end{align*}
with $k_T(Z,r)$ defined in Section~\ref{Proofs} below \eqref{MTA_definition}. Moreover, for any radial $f,g\in L^2(\Rbb^3)$ the operator
\begin{align*}
\iiint_{\Rbb^9} \dd Z \dd r \dd s \; f(r) \, k_T(Z, r-s) \, g(s) \, \bigl[ \cos(Z\cdot \Pi) - 1\bigr]
\end{align*}
commutes with $\Pi^2$. In particular, 
if $P$ and $Q$ are two spectral projections of $\Pi^2$ with $P Q = 0$, then $\eta_\perp$ satisfies the orthogonality property
\begin{align}
	\bigl\langle f(r) \, (P \Psi)(X), \, \eta_{\perp}(\Delta_{Q\Psi}) \bigr\rangle = 0.
	\label{alphaDelta_decomposition_eq4}
\end{align}
\end{enumerate}
\end{prop}

\begin{bem}
\label{rem:alpha}
The statement of Proposition \ref{Structure_of_alphaDelta} should be read in two different ways, depending on whether we are interested in proving the upper or the lower bound for the BCS free energy. In the former case, the bound on $\Vert |r|\eta_\perp\Vert_{\Lsymm}$ in part (b) and part (c) are irrelevant. The reason is that the gap function $\Delta\equiv \Delta_\Psi$ is defined with a minimizer $\Psi$ of the GL functional, whose $\Hmag^2(Q_B)$-norm is uniformly bounded. In this case all remainder terms can be estimated using \eqref{alphaDelta_decomposition_eq2} and \eqref{alphaDelta_decomposition_eq3}.

In contrast, in the proof of the lower bound for the BCS free energy in Section~\ref{Lower Bound Part B} we are forced to work with a trial state $\Gamma_{\Delta}$, whose gap function is defined in terms of a function $\Psi$ that is related to a low-energy state of the BCS functional. The properties of such a function are captured in Theorem~\ref{Structure_of_almost_minimizers} below. In this case we only have a bound on the $\Hmag^1(Q_h)$-norm of $\Psi$ at our disposal. To obtain a function in $\Hmag^2(Q_h)$, we introduce a regularized version of $\Psi$ as in \cite[Section~6]{Hainzl2012}, \cite[Section~6]{Hainzl2014}, \cite[Section~7]{Hainzl2017}, and \cite[Section~6]{DeHaSc2021} by $\Psi_\leq \coloneqq  \Idbb_{[0,\varepsilon]}(\Pi^2)\Psi$ for some $h^2 \ll \varepsilon \ll 1$, see Corollary \ref{Structure_of_almost_minimizers_corollary}. The $\Hmag^2(Q_h)$-norm of $\Psi_\leq$ is not uniformly bounded in $h$, see \eqref{Psileq_bounds} below. This causes a certain error term in the proof of the lower bound to be large, a priori. Part (b) and (c) of Proposition~\ref{Structure_of_alphaDelta} are needed to overcome this problem. Since many details of the relevant proof in Section~\ref{Lower Bound Part B} have been omitted because they go along the same lines as those in \cite[Section~6]{DeHaSc2021} we refer to \cite[Remark~3.3]{DeHaSc2021} for more details.
\end{bem}


\subsection{The BCS energy of the states \texorpdfstring{$\Gamma_\Delta$}{GammaDelta}}

In this section we compute the BCS free energy of our trial states $\Gamma_\Delta$. The goal is to show that this energy minus the energy of the normal state $\Gamma_0$ is, to leading order as $h \to 0$, given by the Ginzburg--Landau energy of the function $\Psi$ appearing in the definition of $\Delta$. For a brief heuristic summary of these computations we refer to Section~\ref{sec:HeuristicComputation}. 
We start our discussion by introducing the operators $L_{T, \Abold, W}$ and $N_{T, \Abold, W}$ that naturally appear when the BCS energy of $\Gamma_{\Delta}$ is expanded in powers of the gap function $\Delta$, see \eqref{eq:heuristicquadraticterms}. The Matsubara frequencies are given by
\begin{align}
\omega_n &\coloneqq  \pi (2n+1) T, \qquad n \in \Zbb. \label{Matsubara_frequencies}
\end{align}
For a local Hilbert--Schmidt operator $\Delta$ with integral kernel $\Delta(x,y)$ satisfying \eqref{alpha_periodicity} and $\Delta(x,y) = \Delta(y,x)$ we define the linear map $L_{T, \Abold, W}$ by 
\begin{align}
L_{T, \Abold, W}\Delta &\coloneqq  -\frac 2\beta \sum_{n\in \Zbb} \frac 1{\i \omega_n - \hfrak_{\Abold, W}} \, \Delta \,  \frac 1 {\i \omega_n + \ov{\hfrak_{\Abold, W}}}. \label{LTAW_definition}
\end{align}
The operator $\hfrak_{\Abold, W}$ is defined in \eqref{hfrakAW_definition}.
In the parameter regime we are interested in, we obtain the quadratic terms in the Ginzburg--Landau functional from $\langle \Delta, L_{T, \Abold, W}\Delta\rangle$. The spectral properties of the operator $L_{T, \Abold, W}$ have been studied in great detail for $W=0$ and $\Abold = \Abold_{e_3}$ in \cite{Hainzl2017}. This allows the authors to compute the BCS critical temperature shift caused by a small constant magnetic field within the framework of linearized BCS theory. That this prediction is accurate also if the nonlinear problem is considered has been shown in \cite{DeHaSc2021}.

Moreover, the nonlinear (cubic) map $N_{T, \Abold, W}$ is defined by
\begin{align}
N_{T, \Abold, W}(\Delta) &\coloneqq  \frac 2\beta \sum_{n\in \Zbb} \frac 1{\i \omega_n - \hfrak_{\Abold, W}}\,  \Delta \,  \frac 1{\i\omega_n + \ov{\hfrak_{\Abold, W}}} \, \ov \Delta \,  \frac 1{\i\omega_n - \hfrak_{\Abold, W}}\, \Delta \, \frac 1{\i\omega_n + \ov{\hfrak_{\Abold, W}}}. \label{NTAW_definition}
\end{align}
The expression $\langle \Delta, N_{T, \Abold, W}(\Delta)\rangle$ gives rise to the quartic term in the Ginzburg--Landau functional. The operator $N_{T, \Abold, W}$ also appeared in \cite{BdGtoGL} and in \cite{DeHaSc2021}. 

From the following lemma we know that $L_{T,\Abold,W} \Delta$ and $N_{T, \Abold, W}(\Delta)(\Delta)$ are both in $\Lsymm$ provided $\Delta$ satisfies the symmetry relations in \eqref{alpha_periodicity_COM} and some mild regularity assumptions. 
\begin{lem}
	\label{lem:propsLN}
	The map $L_{T,\Abold,W}$ is a bounded linear operator on $\Lsymm$. Assume that the integral kernel $\Delta \in L^{2}(Q_h \times \mathbb{R}_{\mathrm{s}}^3)$ defines a bounded operator on $L^2(\mathbb{R}^3)$. Then we have $N_{T, \Abold, W}(\Delta) \in \Lsymm$.
\end{lem}



The following representation formula for the BCS functional is the starting point of our proofs of Theorems \ref{Main_Result} and \ref{Main_Result_Tc}. It will be used to prove upper and lower bounds, and is therefore formulated for general BCS states and not only for our trial states. The proof of Proposition~\ref{BCS functional_identity} can be found in \cite[Proposition 3.4]{DeHaSc2021}.

\begin{prop}[Representation formula for the BCS functional]
\label{BCS functional_identity} \label{BCS FUNCTIONAL_IDENTITY}
Let $\Gamma$ be an admissible state. For any $h>0$, let $\Psi\in \Hmag^1(Q_h)$ and let $\Delta \equiv \Delta_\Psi$ be as in \eqref{Delta_definition}. For $T>0$ and if $V\alpha_*\in L^{\nicefrac 65}(\Rbb^3) \cap L^2(\Rbb^3)$, there is an operator $\Rcal_{T, \Abold, W}^{(1)}(\Delta)\in \Scal^1$ such that
\begin{align}
\FBCS(\Gamma) - \FBCS(\Gamma_0)& \notag\\
&\hspace{-70pt}= - \frac 14 \langle \Delta, L_{T, \Abold, W} \Delta\rangle + \frac 18 \langle \Delta, N_{T, \Abold, W} (\Delta)\rangle + \Vert\Psi \Vert_{\Lmag^2(Q_h)} \, \langle \alpha_*, V\alpha_*\rangle_{L^2(\Rbb^3)} \notag \\
&\hspace{-40pt}+ \Tr\bigl[\Rcal_{T, \Abold, W}^{(1)}(\Delta)\bigr] \notag\\
&\hspace{-40pt}+ \frac{T}{2} \Hcal_0(\Gamma, \Gamma_\Delta) - \fint_{Q_h} \dd X \int_{\Rbb^3} \dd r \; V(r) \, \bigl| \alpha(X,r) - \alpha_*(r) \Psi(X)\bigr|^2, \label{BCS functional_identity_eq}
\end{align}
where
\begin{align}
	\Hcal_0(\Gamma, \Gamma_\Delta) \coloneqq  \Tr_0\bigl[ \Gamma(\ln \Gamma - \ln \Gamma_\Delta) + (1 - \Gamma)(\ln(1-\Gamma) - \ln(1 - \Gamma_\Delta))\bigr] \label{Relative_Entropy}
\end{align}
denotes the relative entropy of $\Gamma$ with respect to $\Gamma_\Delta$. Moreover, $\Rcal_{T, \Abold, W}^{(1)}(\Delta)$ obeys the estimate
\begin{align*}
	\Vert\Rcal_{T, \Abold, W}^{(1)}(\Delta) \Vert_1 \leq C \; T^{-5} \; h^6 \; \Vert \Psi\Vert_{\Hmag^1(Q_h)}^6.
\end{align*}
\end{prop}

The definition \eqref{Relative_Entropy} of the relative entropy uses a weaker form of trace called $\Tr_0$, which is defined as follows. We call a gauge-periodic operator $A$ acting on $L^2(\Rbb^3)\oplus L^2(\Rbb^3)$ weakly locally trace class if $P_0AP_0$ and $Q_0AQ_0$ are locally trace class, where
\begin{align}
P_0 = \begin{pmatrix} 1 & 0 \\ 0 & 0 \end{pmatrix} \label{P0}
\end{align}
and $Q_0 = 1-P_0$. For such operators the weak trace per unit volume is defined by
\begin{align}
\Tr_0 (A)\coloneqq  \Tr\bigl( P_0AP_0 + Q_0 AQ_0\bigr). \label{Weak_trace_definition}
\end{align}
If an operator $A$ is locally trace class then it is also weakly locally trace class but the converse statement does not hold in general. The converse is true, however, if $A \geqslant 0$. We highlight that if $A$ is locally trace class then the weak trace per unit volume and the trace per unit volume coincide. Before their appearance in the context of BCS theory in \cite{Hainzl2012,Hainzl2014,DeHaSc2021}, weak traces of the above kind appeared in \cite{HLS05,FLLS11}. 

Let us have a closer look at the right side of \eqref{BCS functional_identity_eq}. From the terms in the first line we will extract the Ginzburg--Landau functional, see Theorem \ref{Calculation_of_the_GL-energy} below. The terms in the second and third line contribute to the remainder. The term in the second line is small in absolute value, but the techniques used to bound the third line differ for upper and lower bounds. This is responsible for the different qualities of the upper and lower bounds in Theorems \ref{Main_Result} and \ref{Main_Result_Tc}, see \eqref{Rcal_error_Definition}. For an upper bound we choose $\Gamma \coloneqq  \Gamma_\Delta$. Hence $\Hcal_0(\Gamma_\Delta, \Gamma_\Delta)=0$ and the last term in \eqref{BCS functional_identity_eq} can be estimated with the help of Proposition~\ref{Structure_of_alphaDelta}. To obtain a lower bound, the third line needs to be bounded from below using the lower bound for the relative entropy in \cite[Lemma~6.1]{DeHaSc2021}, which appeared for the first time in \cite[Lemma~5]{Hainzl2012}.

Before we state the next result, we introduce the function
\begin{align}
\hat{V\alpha_*}(p) \coloneqq  \int_{\Rbb^3} \dx\; \e^{-\i p\cdot x} \, V(x)\alpha_*(x), \label{Gap_function}
\end{align}
which also fixes our convention of the Fourier transform.

\begin{thm}[Calculation of the GL energy]
\label{Calculation_of_the_GL-energy} \label{CALCULATION_OF_THE_GL-ENERGY}
Let Assumptions \ref{Assumption_V} and \ref{Assumption_KTc} (a) hold and let $D\in \Rbb$ be given. Then, there is a constant $h_0>0$ such that for any $0 < h \leq h_0$, any $\Psi\in \Hmag^2(Q_h)$, $\Delta \equiv \Delta_\Psi$ as in \eqref{Delta_definition}, and $T = \Tc(1 - Dh^2)$, we have
\begin{align}
- \frac 14 \langle \Delta, L_{T, \Abold, W} \Delta\rangle + \frac 18 \langle \Delta, N_{T, \Abold, W} (\Delta)\rangle + \Vert \Psi\Vert_{\Lmag^2(Q_h)}^2 \; \langle \alpha_*, V\alpha_*\rangle_{L^2(\Rbb^3)} & \notag\\
&\hspace{-60pt}= \EGLh(\Psi) + R(h). \label{Calculation_of_the_GL-energy_eq}
\end{align}
Here,
\begin{align*}
|R(h)|\leq C \, \bigl[ h^5 \, \Vert \Psi\Vert_{\Hmag^1(Q_h)}^2 + h^6 \, \Vert\Psi\Vert_{\Hmag^2(Q_h)}^2 \bigr] \, \bigl[ 1 + \Vert \Psi\Vert_{\Hmag^1(Q_h)}^2 \bigr]
\end{align*}
and with the functions
%
\begin{align}
g_1(x) &\coloneqq  \frac{\tanh(x/2)}{x^2} - \frac{1}{2x}\frac{1}{\cosh^2(x/2)}, & g_2(x) &\coloneqq  \frac 1{2x} \frac{\tanh(x/2)}{\cosh^2(x/2)}, \label{XiSigma}
\end{align}
the coefficients $\Lambda_0$, $\Lambda_1$, $\Lambda_2$, and $\Lambda_3$ in $\EGLh$ are given by
\begin{align}
\Lambda_0 &\coloneqq  \frac{\beta_c^2}{16} \int_{\Rbb^3} \frac{\dd p}{(2\pi)^3} \; |(-2)\hat{V\alpha_*}(p)|^2 \; \bigl( g_1 (\beta_c(p^2-\mu)) + \frac 23 \beta_c \, p^2\, g_2(\beta_c(p^2-\mu))\bigr), \label{GL-coefficient_1}\\
\Lambda_1 &\coloneqq  \frac{\betac^2}{4} \int_{\Rbb^3} \frac{\dd p}{(2\pi)^3} \; |(-2)\hat{V\alpha_*}(p)|^2 \; g_1(\betac (p^2-\mu)), \label{GL-coefficient_W} \\
\Lambda_2 &\coloneqq  \frac{\beta_c}{8} \int_{\Rbb^3} \frac{\dd p}{(2\pi)^3} \; \frac{|(-2)\hat{V\alpha_*}(p)|^2}{\cosh^2(\frac{\beta_c}{2}(p^2 -\mu))},\label{GL-coefficient_2} \\
\Lambda_3 &\coloneqq  \frac{\beta_c^2}{16} \int_{\Rbb^3} \frac{\dd p}{(2\pi)^3} \; |(-2) \hat{V\alpha_*}(p)|^4 \;  \frac{g_1(\beta_c(p^2-\mu))}{p^2-\mu}.\label{GL_coefficient_3}
\end{align}
%
%
\end{thm}



It has been argued in \cite{Hainzl2012,Hainzl2014} that the coefficients $\Lambda_0$, $\Lambda_2$, and $\Lambda_3$ are positive. The coefficient $\Lambda_1$ can, in principle, have either sign. Its sign is related to the derivative of $T_{\mathrm{c}}$ with respect to $\mu$, see the remark below Eq.~(1.21) in \cite{Hainzl2012}. 


We highlight the small factor $h^5$ in front of the $\Hmag^1$-norm of $\Psi$ in the bound for $|R(h)|$. It is worse than the comparable estimate in \cite[Theorem 3.5]{DeHaSc2021}, which is a consequence of the presence of the periodic vector potential $A$. The error is, however, of the same size as the related error terms in \cite{Hainzl2012,Hainzl2014}. 

Theorem \ref{Calculation_of_the_GL-energy} provides us with a result for the BCS energy for temperatures of the form $T = \Tc(1 - Dh^2)$ with $D \in \mathbb{R}$ fixed. In the proof of Theorem~\ref{Main_Result_Tc}~(a) we also need the information that our system is superconducting for smaller temperatures. The precise statement is captured in the following proposition.

\begin{prop}[A priori bound on Theorem \ref{Main_Result_Tc} (a)]
\label{Lower_Tc_a_priori_bound}
Let Assumptions \ref{Assumption_V} and \ref{Assumption_KTc} (a) hold and let $T_0>0$. Then, there are constants $h_0>0$ and $D_0>0$ such that for all $0 < h \leq h_0$ and all temperatures $T$ obeying
\begin{align*}
 T_0 \leq T < \Tc (1 - D_0 h^2),
\end{align*}
there is a BCS state $\Gamma$ with
\begin{align}
\FBCS(\Gamma) - \FBCS(\Gamma_0) < 0. \label{Lower_critical_shift_2}
\end{align}
\end{prop}


\subsection{The upper bound on \texorpdfstring{(\ref{ENERGY_ASYMPTOTICS})}{(\ref{ENERGY_ASYMPTOTICS})} and proof of Theorem \ref{Main_Result_Tc} (a)}
\label{Upper_Bound_Proof_Section}

The results in the previous section can be used to prove the upper bound on \eqref{ENERGY_ASYMPTOTICS} and Theorem~\ref{Main_Result_Tc} (a). These proof are almost literally the same as in the case of a constant magnetic field, and we therefore refer to \cite[Section 3.3]{DeHaSc2021} for a detailed presentation.

\section{Proofs of the Results in Section \ref{Upper_Bound}}
\label{Proofs}




\subsection{Schatten norm estimates for operators given by product kernels}
\label{Estimates_on_product_wave_functions_Section}

During our trial state analysis, we frequently need Schatten norm estimates for operators defined by integral kernels of the form $\tau(x-y) \Psi((x+y)/2)$. The relevant estimates are provided in the following lemma, whose proof can be found in \cite[Lemma 4.1]{DeHaSc2021}.

\begin{lem}
\label{Schatten_estimate}
Let $h>0$, let $\Psi$ be a gauge-periodic function on $Q_h$ and let $\tau$ be an even and real-valued function on $\Rbb^3$. Moreover, let the operator $\alpha$ be defined via its integral kernel $\alpha(X,r) \coloneqq  \tau(r)\Psi(X)$, i.e., $\alpha$ acts as
\begin{align*}
\alpha f(x) &= \int_{\Rbb^3} \dd y \; \tau(x - y) \Psi\bigl(\frac{x+y}{2}\bigr) f(y), & f &\in L^2(\Rbb^3).
\end{align*}

\begin{enumerate}[(a)]
\item Let $p \in \{2,4,6\}$. If $\Psi\in \Lmag^p(Q_h)$ and $\tau \in L^{\frac {p}{p-1}}(\Rbb^3)$, then $\alpha \in \Scal^p$ and
\begin{align*}
\Vert \alpha\Vert_p \leq C \; \Vert \tau\Vert_{\frac{p}{p-1}} \; \Vert \Psi\Vert_p.
\end{align*}

\item For any $\nu > 3$, there is a $C_\nu >0$, independent of $h$, such that if $(1 +|\cdot|)^\nu \tau\in L^{\nicefrac 65}(\Rbb^3)$ and $\Psi\in \Lmag^6(Q_h)$, then $\alpha \in \Scal^\infty$ and
\begin{align*}
\Vert \alpha\Vert_\infty &\leq C_\nu \, h^{-\nicefrac 12} \; \max\{1 , h^\nu\} \; \Vert (1 + |\cdot|)^\nu \tau\Vert_{\nicefrac 65} \; \Vert \Psi\Vert_6.
\end{align*}
\end{enumerate}
\end{lem}


\subsection{Magnetic resolvent estimates}
\label{DHS2:Phase_approximation_method_Section}

In this section we provide bounds for the resolvent kernel 
\begin{align}
G^z_h(x,y) &\coloneqq  \frac{1}{z - (-\i \nabla + \Abold_h)^2 + \mu}(x,y), & x,y &\in \Rbb^3 \label{Ghz_definition}
\end{align}
of the magnetic Laplacian that will be applied extensively in the proofs of Proposition~\ref{Structure_of_alphaDelta} and Theorem~\ref{Calculation_of_the_GL-energy}. Our analysis is based on gauge-invariant perturbation theory in the spirit of Nenciu, see \cite[Section~V]{Nenciu2002}, and generalizes the analysis for the constant magnetic field in \cite[Section~2]{Hainzl2017} and \cite[Section~4.4.1]{DeHaSc2021}. In case of a bounded magnetic vector potential, versions of some of our results appeared in \cite{BdGtoGL}. 

We introduce the non-integrable phase factor, also called the Wilson line, by
\begin{align}
	\Phi_\Abold(x,y) \coloneqq  -\int_y^x \Abold(u) \cdot \dd u \coloneqq  -\int_0^1 \dd t\; \Abold(y + t(x-y))\cdot (x-y). \label{PhiA_definition}
\end{align}
In case of the constant magnetic field the right side of \eqref{PhiA_definition} reduces to $\frac{\bold{B}}{2}\cdot(x \wedge y)$. We also define the gauge-invariant kernel $g_h^z(x,y)$ via the equation
\begin{align}
	G_h^z(x,y) = &  \, \e^{\i \Phi_{\Abold_h}(x,y)} g_h^z(x,y), & x,y &\in \Rbb^3. \label{ghz_definition}
\end{align}
It should be compared to the translation-invariant (and gauge-invariant) kernel $g_B^z(x-y)$ introduced in \cite[Eq.~(4.27)]{Hainzl2017} for the constant magnetic field. Its gauge-invariance makes it the natural starting point for a perturbative analysis. 

The integral kernel of the operator $(z+\Delta + \mu)^{-1}$ will be denoted by $g_0^z(x-y)$. The main result of this subsection is the following proposition.

\begin{prop}
\label{gh-g_decay}
Assume that $\Abold = \Abold_{e_3} + A$ with $A \in W^{3,\infty}(\mathbb{R}^3,\mathbb{R}^3)$. For $t, \omega\in \Rbb$ let
\begin{align}
	f(t, \omega) \coloneqq  \frac{|\omega| + |t + \mu|}{(|\omega| + (t + \mu)_-)^2}, \label{gh-g_decay_f} 
\end{align}
where $x_- \coloneqq  -\min\{x,0\}$. For any $a\geq 0$, there are constants $\delta_a , C_a > 0$ such that for all $t, \omega\in \Rbb$ and for all $h \geq 0$ with $f(t, \omega) \,  h^2 \leq \delta_a$ there are ($h$-dependent) even $L^1(\Rbb^3)$-functions $\gzfunction^{\i\omega + t}$, $\gzfunction_\nabla^{\i\omega + t}$, $\gzfunctiondiff^{\i\omega + t}$, and $\gzfunctiondiff_\nabla^{\i\omega + t}$ such that 
\begin{align}
|g_h^{\i\omega + t}(x,y)| &\leq \gzfunction^{\i\omega + t} (x-y),  \notag\\
|\nabla_x g_h^{\i\omega + t}(x,y)| &\leq \gzfunction_\nabla^{\i\omega + t} (x-y), \notag \\
|\nabla_y g_h^{\i\omega + t}(x,y)| &\leq \gzfunction_\nabla^{-\i\omega + t} (x-y), \label{gh-g_decay_eq1}
\end{align}
as well as
\begin{align}
|g_h^{\i\omega + t}(x,y) - g_0^{\i\omega + t}(x - y)| &\leq \gzfunctiondiff^{\i\omega + t} (x-y), \notag \\
| \nabla_x g_h^{\i\omega + t}(x,y) - \nabla_x g_0^{\i \omega + t}(x - y)| &\leq \gzfunctiondiff_\nabla^{\i\omega + t} (x-y), \notag \\
|\nabla_y g_h^{\i\omega + t}(x,y) - \nabla_y g_0^{\i \omega + t}(x - y)| &\leq \gzfunctiondiff_\nabla^{-\i\omega + t} (x-y). \label{gh-g_decay_eq2}
\end{align}
Furthermore, we have the estimates
\begin{align}
\Vert \, |\cdot|^a \gzfunction^{\i\omega + t} \Vert_1 &\leq C_a \, f(t, \omega)^{1 + \frac a2}, \notag \\
\Vert \, |\cdot|^a \gzfunction_\nabla^{\i\omega + t} \Vert_1 &\leq C_a \, f(t,\omega)^{\frac 12 + \frac a2} \, \Bigl[ 1 + \frac{|\omega| + |t - \mu|}{|\omega| + (t - \mu)_-} \Bigr], \label{gh-g_decay_eq3}
\end{align}
and
\begin{align}
\Vert \, |\cdot|^a \gzfunctiondiff^{\i\omega + t} \Vert_1 &\leq C_a \, h^3 \, f(t, \omega)^{\frac 52 + \frac a2}, \notag \\
\Vert \, |\cdot|^a \gzfunctiondiff_\nabla^{\i\omega + t} \Vert_1 &\leq C_a \, h^3 \, f(t, \omega)^{2 + \frac a2} \Bigl[ 1 + \frac{|\omega| + |t - \mu|}{|\omega| + (t - \mu)_-} \Bigr]. \label{gh-g_decay_eq4}
\end{align}
\end{prop}

\begin{bem}
The bounds in the above proposition should be compared to those for $g_B^z(x)$ in \cite[Lemma~10]{Hainzl2017} (estimates without gradient) and \cite[Lemma~4.5]{DeHaSc2021} (estimates with gradient). Although the kernel $g_h^z(x,y)$ defined in \eqref{ghz_definition} is not translation-invariant, $|g_h^z(x,y)|$, $|g_h^z(x,y) - g_0^z(x-y)|$, and the same terms with a gradient can be bounded by translation-invariant kernels. Moreover, these translation-invariant kernels satisfy $L^1$-norm bounds that are mostly of the same quality as those obtained for the kernels $|g_B^z(x)|$, $|g_B^z(x) - g_0^z(x)|$, and the same terms with a gradient in \cite{Hainzl2017,DeHaSc2021}. We highlight that, in comparison to \cite[Eq.~(4.34)]{DeHaSc2021}, we lose a power of the small parameter $h$ in the estimate in \eqref{gh-g_decay_eq4}. This is due to the second term in the bracket in \eqref{Thz_definition} below and it is in accordance with comparable bounds in \cite{Hainzl2012} and \cite{Hainzl2014}. The fact that the above kernels can be bounded from above by translation-invariant kernels is an important ingredient for the proofs of Proposition~\ref{Structure_of_alphaDelta} and Theorem~\ref{Calculation_of_the_GL-energy}.
\end{bem}

Before we give the proof of Proposition~\ref{gh-g_decay} we provide two lemmas. The first lemma concerns $L^1$-norm bounds for the kernel $g_0^z$ and its gradient. Its proof can be found in \cite[Lemma 4.4]{DeHaSc2021}. The bound for $g_0^z$ (but not the one for $\nabla g_0^z$) appeared previously in \cite[Lemma~9]{Hainzl2014}.

\begin{lem}
\label{g_decay}
Let $a > -2$. There is a constant $C_a >0$ such that for $t,\omega\in \Rbb$, we have
\begin{align}
\left \Vert \, |\cdot|^a g_0^{\i \omega + t}\right\Vert_1 &\leq C_a \; f(t, \omega)^{1+ \frac a2}
\label{g_decay_eq1}
\end{align}
with $f(t, \omega)$ in \eqref{gh-g_decay_f}. Furthermore, for any $a > -1$, there is a constant $C_a >0$ with
\begin{align}
\left \Vert \, |\cdot|^a \nabla g_0^{\i\omega + t} \right\Vert_1 \leq C_a \; f(t, \omega)^{\frac 12 + \frac a2} \; \Bigl[ 1 + \frac{|\omega| + |t+ \mu|}{|\omega| + (t + \mu)_-}\Bigr]. \label{g_decay_eq2}
\end{align}
\end{lem}

The second lemma provides us with formulas for the gradient of the function $\Phi_\Abold(x,y)$ defined in \eqref{PhiA_definition} with respect to $x$ and $y$. 

\begin{lem}
\label{PhiA_derivative}
Assume that $\Abold = \Abold_{e_3} + A$ with $A \in W^{2,\infty}(\mathbb{R}^3,\mathbb{R}^3)$. Then we have
\begin{align}
\nabla_x \Phi_\Abold(x,y) &= -\Abold(x) + \tilde \Abold(x,y), &  \nabla_y \Phi_\Abold(x,y) &= \Abold(y) - \tilde \Abold(y,x), \label{PhiA_derivative_eq1}
\end{align}
where
\begin{align}
\tilde \Abold (x,y) \coloneqq  \int_0^1 \dt \; t \curl \Abold(y + t(x - y)) \wedge (x-y) \label{Atilde_definition}
\end{align}
is the transversal Poincar\'e gauge relative to $y$.
\end{lem}

\begin{bem}
	The function $\Phi_\Abold(x,y)$ is a gauge transformation that relates $\Abold(x)$ and $\tilde \Abold(x, y)$.
\end{bem}

\begin{proof}[Proof of Lemma~\ref{PhiA_derivative}]
From Morrey's inequality we know that $\curl \Abold$ is Lipschitz continuous, and hence the line integral in \eqref{Atilde_definition} is well defined. For two vector fields $v$ and $w$ we have
\begin{align}
\nabla (v \cdot w) = (v\cdot \nabla)w + (w\cdot \nabla) v + v\wedge \curl w + w\wedge \curl v. \label{PhiA_derivative_4}
\end{align}
We apply this equality for fixed $y\in \Rbb^3$ to
\begin{align*}
v(x) &= \int_0^1\, \dt \; \Abold(y + t(x-y)), & w(x) &= x-y.
\end{align*}
Our definition implies $\curl w = 0$ and we find that
\begin{align}
-\nabla_x \Phi_\Abold(x,y) &= \Bigl( \int_0^1 \dt \; \Abold(y + t(x-y)) \cdot \nabla_x\Bigr) (x-y) \notag \\
&\hspace{-60pt} + \lk (x-y) \cdot \nabla_x\rk \int_0^1 \dt \; \Abold(y + t(x-y)) + (x-y) \wedge \int_0^1 \dt \; t \, \curl \Abold(y + t(x-y)). \label{PhiA_derivative_1}
\end{align}
The first term on the right side equals
\begin{align}
\sum_{i=1}^3 \int_0^1\dt \; \Abold_i(y + t(x-y)) \, \partial_i (x-y) &= \int_0^1 \dt \; \Abold(y + t(x-y)). \label{PhiA_derivative_2}
\end{align}
To rewrite the second term on the right side of \eqref{PhiA_derivative_1}, we use integration by parts and find
\begin{align}
\left( (x-y)\cdot \nabla_x\right) \int_0^1 \dt \; \Abold(y + t(x-y)) &= \int_0^1 \dt \; t \, \frac{\dd}{\dt} \Abold(y + t(x-y)) \notag \\
&\hspace{-50pt} = t \, \Abold(y + t(x-y))\Big|_0^1 - \int_0^1 \dt \; \Abold(y + t(x-y)). \label{PhiA_derivative_3}
\end{align}
Therefore, the sum of the terms in \eqref{PhiA_derivative_2} and \eqref{PhiA_derivative_3} equals $\Abold(x)$. Since the last term on the right side of \eqref{PhiA_derivative_1} equals $-\tilde \Abold(x,y)$, this proves the first equation in \eqref{PhiA_derivative_eq1}. The second equation follows from $\Phi_\Abold(x,y) = -\Phi_\Abold(y,x)$.
\end{proof}

\begin{proof}[Proof of Proposition~\ref{gh-g_decay}]
We use the abbreviation $z = \i \omega + t$ throughout the proof. In the first step we express $G_h^z(x,y)$ in \eqref{Ghz_definition} in terms of the kernel
\begin{align}
\tilde G_h^z(x,y) \coloneqq  \e^{\i \Phi_{\Abold_h}(x,y)} \, g_0^z(x-y). \label{Gtildehz_definition}
\end{align}
From \eqref{PhiA_derivative_eq1} we know that
\begin{align}
(-\i \nabla_x + \Abold_h(x)) \; \e^{\i \Phi_{\Abold_h}(x,y)} = \e^{\i \Phi_{\Abold_h}(x,y)}\; (-\i\nabla_x + \tilde \Abold_h(x,y)), \label{PhiA_Magnetic_Momentum_Action}
\end{align}
where $\tilde \Abold(x,y)$ in \eqref{Atilde_definition} is our vector potential in Poincar\'e gauge. Furthermore, a short computation shows that \eqref{PhiA_Magnetic_Momentum_Action} implies the operator equation
\begin{align}
(z - (-\i \nabla + \Abold_h)^2 + \mu) \tilde G_h^z = \Idbb - T_h^z,  \label{Ghz_Thz_relation}
\end{align}
where $T_h^z$ is the operator defined by the integral kernel
\begin{align}
T_h^z (x,y) \coloneqq  \e^{\i \Phi_{\Abold_h}(x,y)} \bigl( 2 \, \tilde \Abold_h(x,y) (-\i \nabla_x) -\i \divv_x \tilde \Abold_h(x,y) + |\tilde \Abold_h(x,y)|^2 \bigr) g_0^z(x-y). \label{Thz_definition}
\end{align}
Since $g_0^z$ is a radial function and the vector $\tilde \Abold(x,y)$ is perpendicular to $x-y$ the first term on the right side vanishes. The operator $T_h^z$ also appears in \cite{Nenciu2002}.

We claim that
\begin{align}
|T_h^z(x,y) | &\leq M_\Abold \, h^3 \; \eta^z(x-y) \label{Thz_boundedness_1}
\end{align}
holds, where $\eta^z$ is the ($h$-dependent) function
\begin{align}
\eta^z(x) &\coloneqq  \bigl( |x| + \Vert \curl \Abold\Vert_\infty^{\nicefrac 12} \, h \, |x|^2\bigr) \, |g_0^z(x)| \label{hz_definition}
\end{align}
and
\begin{align}
M_\Abold \coloneqq  \max\bigl\{ \Vert \curl(\curl \Abold) \Vert_\infty \, , \, \Vert \curl \Abold \Vert_\infty^{\nicefrac 32} \bigr\}. \label{MA_definition}
\end{align}
We note that the bound in \eqref{Thz_boundedness_1} holds as an equality with a similar function on the right side in the case of the constant magnetic field, see the proof of Lemma~10 in \cite{Hainzl2017}.

To prove \eqref{Thz_boundedness_1}, we first derive a bound for $|\divv_x \tilde \Abold(x,y)|$. For two vector fields $v$ and $w$ we have 
\begin{align*}
	\divv (v \wedge w) = \curl (v) \cdot w - \curl (w) \cdot v.
\end{align*}
We apply the above equality with the choice $v(x) = \curl \Abold(y + t(x-y))$, $w(x) = x-y$ and use $\curl w =0$ to write
\begin{align*}
	\divv_x \bigl(\curl \Abold(y+t(x-y)) \wedge (x-y)\bigr) &= t \; \curl(\curl \Abold) (y + t (x-y)) \cdot (x-y).
\end{align*}
Since $\curl(\curl \Abold)$ is Lipschitz continuous, which follows from $A \in W^{3,\infty}(\mathbb{R}^3,\mathbb{R}^3)$, we conclude that
\begin{align}
	|\divv_x \tilde \Abold(x,y)| &\leq \int_0^1\dt \; t^2\, | \curl (\curl \Abold)(y + t(x-y))\cdot (x-y)| \notag\\
	&\leq \Vert \curl (\curl \Abold) \Vert_\infty \; |x-y|. \label{Thz_boundedness_2}
\end{align}
We also have 
\begin{align}
	|\tilde \Abold(x,y)| &\leqslant \int_0^1 \dt \; t \; |\curl \Abold(y + t(x-y)) \wedge (x-y)| \notag \\
	&\leq \Vert \curl \Abold\Vert_\infty \, |x-y|. \label{Thz_boundedness_3}
\end{align}
In combination with $\Vert \curl(\curl \Abold_h)\Vert_\infty \leq M_\Abold h^3$ and $\Vert \curl \Abold_h\Vert_\infty^2 \leq M_\Abold \, \Vert \curl A\Vert_\infty^{\nicefrac 12} h^4$, this proves \eqref{Thz_boundedness_1}. 

Next, we have a closer look at the function $\eta^z$. Lemma~\ref{g_decay} and the assumption $f(t, \omega) \,  h^2 \leq \delta_a$ imply the bound
\begin{align}
\Vert \, |\cdot|^a \eta^z\Vert_1 \leq C_a \; f(t, \omega)^{\frac 32 + \frac a2} \bigl[ 1 + h \, f(t,\omega)^{\nicefrac 12}\bigr] \leq C_a \, f(t, \omega)^{\frac 32 + \frac a2}.
\label{eq:boundL1hz}
\end{align}
In particular, 
\begin{align}
M_\Abold \, h^3 \, \Vert \eta^z\Vert_1 \leq C \, M_\Abold \, h^3 \, f(t,\omega)^{\frac 32} \leq \frac 12 \label{gh-g_decay_1}
\end{align}
for all allowed $t, \omega$ and $h$ provided $\delta_a$ is chosen small enough. The bound in \eqref{Thz_boundedness_1} and an application of Young's inequality therefore show that the operator norm of $T_h^z$ satisfies
\begin{align}
\Vert T_h^z\Vert_\infty \leq M_\Abold \, h^3 \; \Vert \eta^z \Vert_1 \leqslant \frac{1}{2}. \label{Thz_boundedness_eq2}
\end{align}
We use \eqref{Ghz_Thz_relation} and this bound to write the resolvent of the magnetic Laplacian as 
\begin{align}
\frac 1{z - (-\i \nabla + \Abold_h)^2 + \mu} = \tilde G_h^z\; \frac{1}{1-T_h^z} = \tilde G_h^z\; \sum_{j=0}^\infty \bigl(T_h^z\bigr)^j. \label{Neumann-series}
\end{align}
This finishes the first step of our proof. In the second step we use \eqref{Neumann-series} to prove the claimed bounds for the integral kernel $g_h^z(x,y)$. Our first goal is to prove the bounds for $g_h^z(x,y)$ without a gradient.

In the following we use the notation $\mathcal{S}_h^z = \sum_{j=1}^\infty ( T_h^z )^j$. Eq.~\eqref{Neumann-series} allows us to write the kernel $g_h^z(x,y)$ as
\begin{align}
	g_h^z(x,y) = g_0^z(x-y) + \e^{-\i \Phi_{\Abold_h}(x,y)} \int_{\Rbb^3} \dd u \; \e^{\i \Phi_{\Abold_h}(x, u)} g_0^z(x-u) \; \Scal_h^z(u,y). \label{GAz-GtildeAz_decay_3}
\end{align}
We use \eqref{Thz_boundedness_1} to bound the integral kernel of the operator $\mathcal{S}_h^z$ by
\begin{equation}
	| \mathcal{S}_h^z (x,y) | \leqslant \sum_{j=1}^{\infty} (M_{\Abold} h^3)^{j} (\eta^z)^{*j}(x-y) \eqqcolon s^z(x-y),
	\label{tau_definition}
\end{equation}
where $(\eta^z)^{*j}$ denotes the $j$-fold convolution of $\eta^z$ with itself. An application of the inequality
\begin{align}
	|x_1 + \cdots + x_j|^a \leq j^{(a-1)_+} \bigl(|x_1|^a + \cdots + |x_j|^a \bigr) \label{DHS2:convexity}
\end{align}
with $a \geq 0$ and $x_+ = \max\{ 0,x \}$ allows us to see that the function $s^z$ satisfies the pointwise bound
\begin{align*} 
|x|^a \, s^z(x) \leq \sum_{j=1}^\infty (M_\Abold h^3)^j \sum_{m=1}^j j^{(a-1)_+} \, \eta^z * \cdots * \bigl( |\cdot|^a \eta^z\bigr) * \cdots * \eta^z(x).
\end{align*}
Here, $|\cdot|^a \eta^z$ appears in the $m$\tho\ slot. An application of \eqref{eq:boundL1hz}, \eqref{gh-g_decay_1}, and Young's inequality therefore implies
\begin{align}
\int_{\mathbb{R}^3} \dd x \ |x|^a s^z(x) &\leq M_{\Abold} h^3 \, \Vert \, |\cdot|^a \eta^z\Vert_1 \sum_{j=1}^\infty \frac{j^{1 + (a-1)_+}}{2^{j-1}} \leq C_a \, h^3 \, f(t, \omega)^{\frac 32 + \frac a2}. \label{gh-g_decay_2} 
\end{align}
Let us also define the functions 
\begin{align*}
	\rho^z(x) &\coloneqq |g_0^z(x)| + |g_0^z| \ast s^z(x), & \tau^z(x) &\coloneqq |g_0^z| \ast s^z(x).
\end{align*}
From \eqref{GAz-GtildeAz_decay_3} we know that 
\begin{align*}
	|g_h^z(x,y)| &\leqslant \rho^z(x-y)   & |g_h^z(x,y) - g_0^z(x-y)| &\leqslant \tau^z(x-y).
\end{align*}
The claimed bounds for the $L^1(\mathbb{R}^3)$-norms of $\rho^z$ and $\tau^z$ in \eqref{gh-g_decay_eq3} and \eqref{gh-g_decay_eq4} follow from Lemma~\ref{g_decay}, \eqref{tau_definition}, and \eqref{gh-g_decay_2}. It remains to prove the bounds involving a gradient.

An application of Lemma~\ref{PhiA_derivative} and \eqref{Thz_boundedness_3} show
\begin{align*}
|\nabla_x \e^{-\i \Phi_\Abold(x,y)} \e^{\i \Phi_\Abold(x,u)}| \leq |\tilde \Abold(x,y)| + |\tilde \Abold(x,u)| \leqslant C h^2 \left( |x-y| + |x -u| \right).
 \end{align*}
In combination with \eqref{GAz-GtildeAz_decay_3} and \eqref{tau_definition}, this implies 
\begin{equation*}
	|\nabla_x g_h^z(x,y)| \leq |\nabla g_0^z(x-y)| + C h^2 \int_{\mathbb{R}^3} \dd u \left( |x-u| + |u-y| \right) |g_0^z(x-u)| s^z(u-y). 
\end{equation*}
We denote the right side of the above equation by $\rho_{\nabla}^z(x-y)$. The claimed bound for $\rho_{\nabla}^z$ follows immediately from those for $g_0^z$ and $s^z$, see Lemma~\ref{g_decay} and \eqref{gh-g_decay_2}, and the assumption $f(t, \omega) \,  h^2 \leq \delta_a$. A bound for $|\nabla_x g_h^z(x,y) - \nabla g_0^z(x-y)|$ can be obtained similarly. The bounds for $|\nabla_y g_h^z(x,y)|$ and $|\nabla_y g_h^z(x,y) - \nabla g_0^z(x-y)|$ can be obtained when we use the identity $G_h^z(x,y) = \ov{G_h^{\ov z}(y,x)}$. This proves Proposition~\ref{gh-g_decay}.
\end{proof}

\subsection{Proof of Lemma \ref{Gamma_Delta_admissible}}
\label{sec:proofofadmissibilitya}

Let us recall the definition of $\Gamma_\Delta$ in \eqref{GammaDelta_definition}. From its definition we infer that it is a gauge-periodic generalized fermionic one-particle density matrix. Consequently, we only need to verify the trace class condition in \eqref{Gamma_admissible}.

We use the identity $(\exp(x)+1)^{-1} = (1-\tanh(x/2))/2$ to write $\Gamma_\Delta$ as 
\begin{align}
	\Gamma_\Delta &= \frac 12 - \frac 12 \tanh\bigl( \frac \beta 2 H_\Delta\bigr). \label{GammaDelta_tanh_relation1}
\end{align}
Let us also recall the Mittag--Leffler series expansion
\begin{align}
	\tanh\bigl( \frac \beta 2 z\bigr) &= -\frac{2}{\beta} \sum_{n\in \Zbb} \frac{1}{\i\omega_n - z}, \label{tanh_Matsubara}
\end{align}
see e.g. \cite[Eq. (3.12)]{DeHaSc2021}. Its convergence becomes manifest by combining the $+n$ and $-n$ terms. When we use \eqref{GammaDelta_tanh_relation1}, and the resolvent identity
\begin{align}
	(z- T)^{-1} = (z-S)^{-1} + (z-T)^{-1} \; (S - T)\; (z- S)^{-1} \label{Resolvent_Equation}
\end{align}
for two operators $S$ and $T$, we find 
\begin{align}
	\Gamma_\Delta = \frac 12 - \frac 12 \tanh\bigl( \frac \beta 2 H_\Delta\bigr) = \frac 12 + \frac{1}{\beta} \sum_{n\in \Zbb} \frac{1}{\i \omega_n - H_\Delta} = \Gamma_0 + \Ocal + \Qcal_{T,\Abold, W}(\Delta). \label{alphaDelta_decomposition_1}
\end{align}
Here $\Gamma_0$ denotes the normal state in \eqref{Gamma0} and
\begin{align}
	\Ocal &\coloneqq  \frac 1\beta \sum_{n\in \Zbb} \frac{1}{ \i \omega_n - H_0} \delta \frac{1}{ \i \omega_n - H_0}, \notag \\
	\Qcal_{T,\Abold, W}(\Delta) &\coloneqq  \frac 1\beta \sum_{n\in \Zbb} \frac{1}{ \i \omega_n - H_0} \delta\frac{1}{ \i \omega_n - H_0} \delta \frac{1}{ \i \omega_n - H_\Delta} \label{alphaDelta_decomposition_2}
\end{align}
with $\delta$ in \eqref{HDelta_definition}. 

Since the diagonal components of the operator $\Ocal$ equal zero, this term does not contribute to the $11$-component $\gamma_{\Delta}$ of $\Gamma_{\Delta}$. In the following, we use the notation $\pi = - \i \nabla + \Abold_{\Bbold}$. To see that $(1 + \pi^2) [\Qcal_{T, \Abold, W}(\Delta)]_{11}$ is locally trace class, we use
\begin{align*}
	\frac{1}{\i \omega_n \pm H_0}\,  \delta\,  \frac{1}{\i \omega_n \pm H_0} \, \delta 
	&= \begin{pmatrix}
		\frac{1}{\i \omega_n \pm \hfrak_{\Abold, W}} \, \Delta \frac{1}{\i \omega_n \mp \ov{\hfrak_{\Abold, W}}} \, \ov \Delta & 0 \\ 0 & \frac{1}{\i \omega_n \mp \ov{\hfrak_{\Abold, W}}}\,  \ov \Delta \frac{1}{\i \omega_n \pm \hfrak_{\Abold, W}} \, \Delta
	\end{pmatrix} 
\end{align*}
to write it as
\begin{align*}
	\bigl[ \Qcal_{T, \Abold, W}(\Delta)\bigr]_{11} = \frac 1\beta \sum_{n\in \Zbb} \frac{1}{\i\omega_n - \hfrak_{\Abold, W}} \, \Delta \, \frac{1}{\i\omega_n + \ov{\hfrak_{\Abold, W}}} \, \ov \Delta \, \Bigl[ \frac{1}{\i\omega_n - H_\Delta}\Bigr]_{11}.
\end{align*}
The operator $\hfrak_{\Abold, W}$ is defined in \eqref{hfrakAW_definition}. An application of Hölder's inequality in \eqref{Schatten-Hoelder} shows that the local trace norm of the term inside the sum is bounded by
\begin{align*}
	\Bigl\Vert (1 + \pi^2) \frac{1}{\i \omega_n - \hfrak_{\Abold, W}}\Bigr\Vert_\infty  \; \frac{1}{|\omega_n|^2} \; \Vert \Delta\Vert_2^2.
\end{align*}
Using Cauchy-Schwarz, we see that $\hfrak_{\Abold, W} \leqslant C (1+\pi^2)$, which implies that the operator norm in the above equation is bounded uniformly for $n \in \mathbb{Z}$. Since $|\omega_n|^{-2}$ is summable in $n$ and $\Delta$ is locally Hilbert-Schmidt these considerations show that $(1+\pi^2)[\Qcal_{T,\Abold, W}(\Delta)]_{11}$ is locally trace class. It remains to show that $(1 + \pi^2) \gamma_0$ with $\gamma_0$ in \eqref{Gamma0} is locally trace class. 

To that end, we first note that
\begin{equation*}
	\left\Vert (1+\pi^2) \gamma_0 \right\Vert_1 \leqslant \left\Vert (1+\pi^2) \frac{1}{1 + \hfrak_{\Abold, W} + \mu} \right\Vert_{\infty} \ \left\Vert (1+\hfrak_{\Abold, W} + \mu) \gamma_0 \right\Vert_{1}.
\end{equation*}
We argue as above to see that the first norm on the right side is finite. To obtain a bound for the second norm, we first note that there is a constant $C>0$ such that $(1+x)(\exp(\beta (x-\mu))+1)^{-1} \leqslant C \exp(-\beta x/2)$ holds for $x > a > -\infty $. The constant $C$ depends on $\beta,\mu$, and $a$. Accordingly, 
\begin{equation*}
	\left\Vert (1+\hfrak_{\Abold, W} + \mu) \gamma_0 \right\Vert_{1} \leqslant C \Tr \exp(-\beta \hfrak_{\Abold, W}/2).
\end{equation*}
From Corollary~A.1.2 and Corollary~B.13.3 in \cite{Simon82} we know that for any $t>0$ the operator $\exp(-t \hfrak_{\Abold, W})$ has an integral kernel $k_{t}(x,y)$ that satisfies
\begin{equation*}
	\left\Vert k_t \right\Vert_{2,\infty}^2 = \esssup_{x \in \mathbb{R}^3} \int_{\mathbb{R}^3} \dd y \ | k_{t}(x,y)|^2 < \infty.
\end{equation*}
Accordingly,
\begin{equation*}
	\Tr \exp(-\beta \hfrak_{\Abold, W}/2) = \left\Vert \exp(-\beta \hfrak_{\Abold, W}/4) \right\Vert_2^2 = \fint_{Q_h} \dd x \int_{\mathbb{R}^3} \dd y \ |k_{\beta/4}(x,y)|^2 \leqslant \Vert k_{\beta/4} \Vert_{2,\infty}^2.
\end{equation*}
We conclude that $(1+\pi^2) \gamma_0$ is locally trace class. This ends the proof of Lemma~\ref{Gamma_Delta_admissible}.

\subsection{Proof of Lemma~\ref{lem:propsLN}}
We start by proving that $L_{T,\Abold,W}$ is a bounded linear operator on $\Lsymm$. To that end, we first check that for $\Delta \in \Lsymm$, we have $L_{T,\Abold,W} \Delta \in \Scal^2$. Using H\"older's inequality for the trace per unit volume in \eqref{Schatten-Hoelder} and $\omega_n = \pi (2n+1) T $, we see that
\begin{equation}
	\sum_{n \in \mathbb{Z}} \left\Vert  \frac 1{\i \omega_n - \hfrak_{\Abold, W}} \, \Delta \,  \frac 1 {\i \omega_n + \ov{\hfrak_{\Abold, W}}}  \right\Vert_2 \leqslant \Vert \Delta \Vert_2 \sum_{n \in \mathbb{Z}} \frac{1}{\omega_n^2} \leqslant C  \Vert \Delta \Vert_2 \sum_{n \in \mathbb{Z}} \frac{1}{(2n+1)^2},
	\label{eqA:1B}
\end{equation} 
which proves the claim. 

Next, we show that $L_{T, \Abold, W}\Delta$ satisfies \eqref{alpha_periodicity_COM}. To that end, we need the identity
\begin{align}
	\frac{1}{\i\omega_n + \ov{\hfrak_{\Abold, W}} - \mu} (x,y) = -\frac{1}{-\i \omega_n - \hfrak_{\Abold, W} + \mu}(y,x).
	\label{GAz_Kernel_of_complex_conjugate}
\end{align}
It follows from $\ov{ \Rcal^* (x,y) } = \Rcal(y,x)$ for a general operator $\Rcal$ with kernel $\Rcal(x,y)$ and
\begin{align*}
	\frac{1}{z - \ov{\hfrak_{\Abold, W}} + \mu} = \ov{\Bigl(\frac{1}{z - \hfrak_{\Abold, W} + \mu} \Bigr)^*}.
\end{align*}
Using the coordinate transformation $(w_1,w_2) \mapsto (w_1+v,w_2+v)$ and the above identities for the resolvent kernel, we see that
\begin{align}
	L_{T,\Abold,W} \Delta(X+v,r) &= \frac{2}{\beta} \sum_{n \in \mathbb{Z}} \iint_{\Rbb^3\times \Rbb^3} \dd w_1 \dd w_2 \; \Delta \bigl( \frac{w_1+w_2}{2} + v, w_1 - w_2\bigr) \notag \\
	&\hspace{-60pt} \times  \frac{1}{\i\omega_n - \hfrak_{\Abold, W}+ \mu}\bigl( X + v + \frac r2, w_1\bigr) \frac{1}{-\i\omega_n - \hfrak_{\Abold, W}+ \mu}\bigl( X + v - \frac r2, w_2\bigr) 
	%
%
%
	\label{eqA:2B}
\end{align}
holds. We highlight that we wrote $\Delta$ in terms of relative and center-of-mass coordinates in \eqref{eqA:2B}. We have $T(v) \hfrak_{\Abold, W} T(v)^* = \hfrak_{\Abold, W}$, $v \in \Lambda_h$, where $T(v)$ is the magnetic translation in \eqref{Magnetic_Translation}, and hence the same identity with $\hfrak_{\Abold, W}$ replaced by its resolvent. In combination with
\begin{align*}
	&\iint_{\Rbb^3\times \Rbb^3} \dd x \dd y \; \overline{\varphi(x)} \, [T(v)(z - \hfrak_{\Abold, W})^{-1}T(v)^*](x,y) \, \varphi(y) \\
	&\hspace{3cm}= \int_{\mathbb{R}^6} \dd x \dd y \; \overline{\varphi(x)} \, \e^{\i \frac{\Bbold}{2} \cdot (v \wedge (x-y))}  \, (z - \hfrak_{\Abold, W})^{-1}(x+v,y+v) \, \varphi(y), 
\end{align*}
which holds for $z \in \rho(\hfrak_{\Abold, W})$, this proves that the resolvent kernel of $\hfrak_{\Abold, W}$ obeys the first relation in \eqref{alpha_periodicity}. When we combine this and the fact that $\Delta$ satisfies the second relation in \eqref{alpha_periodicity}, we see that we pick up the total phase
\begin{align*}
	\e^{\i \frac{\Bbold}{2} \cdot (v\wedge (X + \frac r2 - w_1))} \, \e^{\i \frac{\Bbold}{2} \cdot (v\wedge (X - \frac r2 - w_2))} \, \e^{\i \frac{\Bbold}{2} \cdot (v\wedge (w_1 + w_2))} = \e^{\i \Bbold\cdot (v\wedge X)}
\end{align*}
in \eqref{eqA:2B}. This proves the first equation in \eqref{alpha_periodicity_COM} for $L_{T, \Abold, W} \Delta(X,r)$. 

To see that the second equation in \eqref{alpha_periodicity_COM} is true, we further observe that the Matsubara frequencies obey $-\omega_n = \omega_{-(n+1)}$. Hence, the index shift $n \mapsto -n-1$ and \eqref{GAz_Kernel_of_complex_conjugate} imply the desired symmetry. This proves that $L_{T,\Abold,W}$ is a bounded linear operator on $\Lsymm$. 

To show that $N_{T,\Abold,W}(\Delta) \in \mathcal{S}^2$, we use the bound
\begin{align*}
	&\sum_{n \in \mathbb{Z}} \left\Vert \frac 1{\i \omega_n - \hfrak_{\Abold, W}}\,  \Delta \,  \frac 1{\i\omega_n + \ov{\hfrak_{\Abold, W}}} \, \ov \Delta \,  \frac 1{\i\omega_n - \hfrak_{\Abold, W}}\, \Delta \, \frac 1{\i\omega_n + \ov{\hfrak_{\Abold, W}}} \right\Vert_2 \\
	&\hspace{8cm} \leq C \, \Vert \Delta \Vert_2 \ \Vert \Delta \Vert_{\infty}^2 \ \sum_{n \in \mathbb{Z}} \frac{1}{(2n+1)^4}.
\end{align*} 
The proof of \eqref{alpha_periodicity_COM} for $N_{T,\Abold,W}(\Delta)$ goes along the same lines as that for $L_{T,\Abold,W} \Delta$. This proves Lemma~\ref{lem:propsLN}.

\subsection{Proof of Theorem \ref{Calculation_of_the_GL-energy}}
\label{Calculation_of_the_GL-energy_proof_Section}

Before we start with the proof of Theorem~\ref{Calculation_of_the_GL-energy}, we briefly mention the two main steps. In the first step we compute $\langle \Delta, L_{T, \Abold, W} \Delta\rangle$. To that end, we decompose the operator $L_{T, \Abold, W}$ into several increasingly simpler parts, which allows us to extract the quadratic terms in the GL functional. The related analysis can be found in Sections~\ref{sec:Q1}--\ref{Summary_quadratic_terms_Section}. The quartic term in the GL functional emerges from $\langle \Delta, N_{T, \Abold, W}(\Delta) \rangle$. In the second step we study the nonlinear operator $N_{T, \Abold, W}$ and introduce comparable steps of simplification as for $L_{T, \Abold, W}$. The related analysis starts in Section~\ref{sec:NT}. The relation to the existing literature will be discussed mostly in Sections~\ref{sec:repLT}, \ref{Approximation_of_LTA^W_Section}, and \ref{sec:NT} after the relevant mathematical objects have been introduced. 



\subsubsection{Decomposition of \texorpdfstring{$L_{T, \Abold, W}$}{LTAW} --- separation of \texorpdfstring{$W$}{W}}
\label{sec:Q1}

We use the resolvent equation in \eqref{Resolvent_Equation} to decompose the operator $L_{T, \Abold, W}$ in \eqref{LTAW_definition} as
\begin{align}
L_{T, \Abold, W} = L_{T, \Abold} + \Woperator_{T, \Abold} + \Rcal_{T, \Abold,W}^{(2)}, \label{LTAW_decomposition}
\end{align}
where $L_{T, \Abold} = L_{T, \Abold, 0}$, 
\begin{align}
\Woperator_{T, \Abold} \Delta &\coloneqq -\frac 2 \beta \sum_{n\in \Zbb} \Bigl[ \frac 1{\i \omega_n - \hfrak_\Abold} \, W_h \, \frac 1{\i \omega_n - \hfrak_\Abold} \, \Delta \, \frac{1}{\i \omega_n + \ov{\hfrak_\Abold}} \notag \\
&\hspace{100pt} - \frac{1}{\i\omega_n - \hfrak_\Abold} \, \Delta \, \frac{1}{\i \omega_n + \ov{\hfrak_\Abold}} \, W_h \, \frac{1}{\i\omega_n + \ov{\hfrak_\Abold}}\Bigr], \label{LTA^W_definition} 
\end{align}
and
\begin{align}
\Rcal_{T, \Abold, W}^{(2)} \Delta &\coloneqq -\frac 2\beta \sum_{n\in \Zbb} \Bigl[ \frac{1}{\i\omega_n - \hfrak_\Abold} \, W_h \, \frac{1}{\i \omega_n - \hfrak_\Abold} \, W_h \, \frac{1}{\i\omega_n - \hfrak_{\Abold, W}} \, \Delta \, \frac{1}{\i \omega_n + \ov{\hfrak_\Abold}} \notag \\
&\hspace{60pt} + \frac{1}{\i\omega_n - \hfrak_\Abold} \, \Delta \, \frac{1}{\i\omega_n + \ov{\hfrak_\Abold}} \, W_h \, \frac{1}{\i\omega_n + \ov{\hfrak_{\Abold, W}}} \, W_h \, \frac{1}{\i \omega_n + \ov{\hfrak_\Abold}} \notag \\
&\hspace{60pt} -\frac{1}{\i \omega_n - \hfrak_\Abold} \, W_h \, \frac{1}{\i\omega_n - \hfrak_{\Abold, W}} \, \Delta \, \frac{1}{\i\omega_n + \ov{\hfrak_{\Abold, W}}} \, W_h \, \frac 1{\i\omega_n + \ov{\hfrak_\Abold}} \Bigr]. \label{RTAW2_definition}
\end{align}
In the special case of a constant magnetic field, the operator $L_{T, \Abold}$ appeared for the first time in \cite{Hainzl2017}. The operator $\Woperator_{T, 0}$ (case of no external magnetic fields) was studied in \cite{ProceedingsSpohn}. In the sections \ref{sec:repLT}--\ref{Analysis_of_MTA_Section} we analyze $\langle \Delta, L_{T, \Abold} \Delta \rangle$ and extract the first and the third term in the GL functional from it. Afterwards, we study in Sections~\ref{LTAW_action_Section} and \ref{Approximation_of_LTA^W_Section} the quadratic form $\langle \Delta, \Woperator_{T, \Abold} \Delta \rangle$, which contributes the second term of the GL functional. Finally, in Section~\ref{Summary_quadratic_terms_Section}, we collect the results of the previous sections and provide a bound for $\langle \Delta, \Rcal_{T, \Abold, W}^{(2)}\Delta \rangle$ showing that this term does not contribute to the GL functional.



\subsubsection{A representation formula for \texorpdfstring{$L_{T,\Abold}$}{LTA} and an outlook on the quadratic terms}
\label{sec:repLT}

In the following subsections we compute the contribution from $\langle \Delta, L_{T, \Abold} \Delta \rangle$ to the Ginzburg--Landau energy. Our analysis is based on the following representation formula for the operator $L_{T, \Abold}$, which characterizes its action solely in terms of relative and center-of-mass coordinates.


\begin{lem}
\label{LTA_action}
The operator $L_{T,\Abold} \colon \Lsymm \ra \Lsymm$ acts as
\begin{align*}
(L_{T,\Abold}\alpha) (X,r) &= \iint_{\Rbb^3\times \Rbb^3} \dd Z \dd s \; k_{T, \Abold}(X, Z, r, s) \; (\e^{\i Z \cdot (-\i \nabla_X)}\alpha) (X,s)
\end{align*}
with
\begin{align}
k_{T,\Abold}(X, Z, r, s) \coloneqq  \frac 2\beta \sum_{n\in \Zbb} k_{T, \Abold}^n(X, Z, r, s) \; \e^{\i \tilde \Phi_{\Abold_h}(X, Z, r, s)}, \label{kTA_definition}
\end{align}
where
\begin{align}
k_{T, \Abold}^n(X, Z, r, s) &\coloneqq  g_h^{\i\omega_n} \bigl(X + \frac r2, X + Z + \frac s2\bigr) \; g_h^{-\i\omega_n}\bigl(X - \frac r2, X + Z - \frac s2\bigr) \label{kTAn_definition}
\end{align}
with $g_h^z$ in \eqref{ghz_definition} and
\begin{align}
\tilde \Phi_\Abold (X, Z, r, s) &\coloneqq  \Phi_\Abold\bigl(X + \frac r2, X + Z + \frac s2\bigr) + \Phi_\Abold\bigl(X - \frac r2, X + Z - \frac s2\bigr) \label{LTA_PhitildeA_definition}
\end{align}
with $\Phi_\Abold$ in \eqref{PhiA_definition}.
\end{lem}
\begin{bem}
	Lemma~\ref{LTA_action} should be compared to the representation formula in \cite[Lemma~11]{Hainzl2017} for the operator $L_{T, B}$ in \cite[Eq.~(8)]{Hainzl2017}. The differences between the two representation formulas are related to the fact that our magnetic field is non-constant. Accordingly, the kernel $g_h^z(x,y)$ is not translation-invariant and $\Phi_{\Abold_h}(x,y)$ does not simply equal $\frac{\bold{B}}{2}  \cdot ( x \wedge y)$. This results in the dependence of the function $k_{T, \Abold}^n$ on the coordinate $X$ and in the fact that the representation formula in Lemma~\ref{LTA_action} is not symmetric under the transformation $Z \mapsto -Z$. As a consequence, the operator $\cos(Z\cdot \Pi)$ in \cite[Lemma~11]{Hainzl2017} is replaced by $\exp(\i Z\cdot (-\i \nabla_X))$. Both the cosine function and the full magnetic momentum operator in its argument, will be recovered at a later stage, see \eqref{MtildeTA_definition} below. The main guiding principle behind the definition of the above representation formula and its subsequent analysis is that we should think of $\Phi_{\Abold_h}(x,y)$ as a generalization of $\frac{\bold{B}}{2}  \cdot ( x \wedge y)$. The latter has more convenient algebraic properties, which is responsible for much of the simplicity of the analysis in \cite{Hainzl2017,DeHaSc2021} in comparison to the present work. In case of $\Phi_{\Abold_h}(x,y)$ we do computations as if similar relations were satisfied and afterwards carefully bound the emergent remainder terms.
\end{bem}

\begin{proof}[Proof of Lemma \ref{LTA_action}]
With \eqref{GAz_Kernel_of_complex_conjugate} applied to $W =0$ the integral kernel of $L_{T, \Abold}$ can be written as
\begin{align*}
L_{T,\Abold} \alpha(x, y) &= \frac 2\beta \sum_{n\in \Zbb} \iint_{\Rbb^3\times \Rbb^3} \dd u\dd v\; G_h^{\i\omega_n} (x, u) \, G_h^{-\i\omega_n}(y, v) \, \alpha(u,v).
\end{align*}
We note that $\alpha$ and $L_{T,\Abold} \alpha$ are not yet written in terms of relative and center-of-mass coordinates, which will be done in the next step. To that end, we define the coordinates $X =\frac{x+y}{2}$, $r=x-y$,
\begin{align}
u &= X + Z + \frac s2, & v = X + Z - \frac s2,
\end{align}
and introduce the notation $\zeta_X^r \coloneqq  X + \frac r2$. This allows us to write the above equation as
\begin{align*}
	L_{T,\Abold} \alpha(X, r) &= \frac 2\beta \sum_{n\in \Zbb} \iint_{\Rbb^3\times \Rbb^3} \dd Z\dd s\; G_h^{\i\omega_n} (\zeta_X^r, \zeta_{X+Z}^s) \, G_h^{-\i\omega_n}(\zeta_X^{-r}, \zeta_{X+Z}^{-s}) \, \alpha(X + Z, s).
\end{align*} 
We highlight that, by a slight abuse of notation, we denoted the kernels of $\alpha$ and $L_{T,\Abold} \alpha$ when expressed in terms of relative and center-of-mass coordinates still by the same symbols.  When we use the relation $G^z_h(x,y) = \exp( \i \Phi_{\Abold}(x,y) ) g_h^z(x,y)$ and
\begin{align*}
\alpha(X+Z,s) = \e^{\i Z\cdot (-\i \nabla_X)} \alpha(X, s),
\end{align*}
we see that the above identity implies the claimed formula.
\end{proof}

We analyze the operator $L_{T,\Abold}$ in four steps. In the first three steps, we introduce three operators of increasing simplicity in their dependence on $\Abold_h$. More precisely, we write it as 
\begin{align}
L_{T,\Abold} = (L_{T,\Abold} - \tilde L_{T,\Abold}) + (\tilde L_{T,\Abold} - \tilde M_{T,\Abold}) + (\tilde M_{T,\Abold} - M_{T,\Abold}) + M_{T, \Abold} \label{LTA_decomposition}
\end{align}
with the operators $\tilde L_{T,\Abold}$, $\tilde M_{T, \Abold}$, and $M_{T,\Abold}$ defined below in \eqref{LtildeTA_definition}, \eqref{MtildeTA_definition}, and \eqref{MTA_definition}, respectively. As we will show, the operators in brackets in \eqref{LTA_decomposition} do not contribute to the GL functional. The operator $\tilde L_{T,\Abold}$ is obtained from $L_{T,\Abold}$ when we replace the kernels $g_h^{z}(x,y)$ in \eqref{kTAn_definition} by $g_0^z(x-y)$. To obtain the operator $\tilde M_{T,\Abold}$ from $\tilde L_{T,\Abold}$, we need to replace the phase factor in the definition of $k_{T,\Abold}^n(X,Z,r,s)$ by $\exp( -\i (r-s) \cdot D \Abold_h(X) (r+s)/4 )$, where $D \Abold_h$ denotes the Jacobi matrix of $\Abold_h$, and $\exp(\i Z \cdot (-\i \nabla_X))$ by $\cos( Z \cdot \Pi_{\Abold_h})$. Finally, $M_{T,\Abold}$ emerges when we replace $\exp( -\i (r-s) \cdot D \Abold_h(X) (r+s)/4 )$ in the definition of $\tilde M_{T, \Abold}$ by $1$. In the fourth and final step we extract the quadratic terms in the GL functional (except the one proportional to $W$) as well as a term that cancels the last term on the left side of \eqref{Calculation_of_the_GL-energy_eq} from $\langle \Delta, M_{T,\Abold} \Delta \rangle$. To that end, we expand the operator $\cos( Z \cdot \Pi_{\Abold_h})$ up to second order in powers $Z \cdot \Pi_{\Abold_h}$ and use $T = T_{\mathrm{c}}(1 - D h^2)$.

In the case of a constant magnetic field a similar decomposition of $L_{T,\Abold}$ has been introduced for the first time in \cite{Hainzl2017}. In this reference the operator $\tilde L_{T,\Abold}$ is called $M_{T,\Abold}$ and our $M_{T,\Abold}$ is called $N_{T,\Bbold}$. We did not follow the notation in \cite{Hainzl2017} because the symbol $N_{T,\Abold,W}$ is reserved for the nonlinear term in our paper. The decomposition of $L_{T,\Abold}$ in \cite{Hainzl2017} has also been used in \cite{DeHaSc2021}. In comparison to these two references we have an additional term (the operator $\tilde M_{T,\Abold}$) in our decomposition of $L_{T,\Abold}$, which is a consequence of the fact that we are dealing with a general magnetic field. Generally speaking, the magnetic vector potential $\Abold_h$ is more difficult to treat than $\Abold_{\Bbold}$ because several algebraic relations that hold for the latter do not hold for the former. Our main contribution in this section is that we overcome the related mathematical difficulties in the computation of the above terms. Another main difference between our work and \cite{Hainzl2017} is that we additionally need $\Hmag^1(Q_h)$-norm estimates. In the special case of a constant magnetic field such bounds have been proved in \cite{DeHaSc2021}. It should also be noted that $L_{T,\Abold}$ acts on $L^2(\mathbb{R}^6)$ in \cite{Hainzl2017}, while it acts on $\Lsymm$ in our case and in \cite{DeHaSc2021}.

\subsubsection{Approximation of \texorpdfstring{$L_{T, \Abold}$}{LTA}}
\label{Approximation_of_LTAW_Section}

\paragraph{The operator $\tilde L_{T, \Abold}$.}
\label{para:approx1}

We define the operator $\tilde L_{T, \Abold}$ by
\begin{align}
\tilde L_{T, \Abold}\alpha(X,r) &\coloneqq  \iint_{\Rbb^3 \times \Rbb^3} \dd Z \dd s \; \tilde k_{T, \Abold}(X, Z, r,s) \; ( \e^{\i Z \cdot (-\i \nabla_X)} \alpha)(X,s) \label{LtildeTA_definition}
\end{align}
with
\begin{align}
\tilde k_{T, \Abold} (X, Z, r,s) \coloneqq  \frac 2\beta \sum_{n\in \Zbb} k_T^n(Z, r-s) \; \e^{\i \tilde \Phi_{\Abold_h}(X, Z, r, s)}, \label{ktildeTA_definition}
\end{align}
where $\tilde \Phi_\Abold$ is defined in \eqref{LTA_PhitildeA_definition} and
\begin{align}
k_T^n(Z, r) \coloneqq  k_{T, 0}^n(0, Z, r, 0) = g_0^{\i\omega_n}\bigl(Z - \frac{r}{2} \bigr) \, g_0^{-\i \omega_n}\bigl( Z + \frac{r}{2} \bigr). \label{kTn_definition}
\end{align}

The following proposition allows us to replace the operator $L_{T, \Abold}$ by $\tilde L_{T, \Abold}$ in the computation of the BCS energy of our trial state.

\begin{prop}
\label{LTA-LtildeTA}
Assume that $\Abold = \Abold_{e_3} + A$ with $A \in W^{2,\infty}(\mathbb{R}^3,\mathbb{R}^3)$ periodic, let $|\cdot|^k V\alpha_*\in L^2(\Rbb^3)$ for $k \in \{ 0,1 \}$, $\Psi\in \Hmag^1(Q_h)$, and denote $\Delta \equiv \Delta_\Psi$ as in \eqref{Delta_definition}. For any $T_0 > 0$ there is $h_0>0$ such that for any $0 < h \leq h_0$ and any $T\geq T_0$ we have
\begin{align*}
\Vert L_{T, \Abold} \Delta - \tilde L_{T, \Abold} \Delta\Vert_\Hsymm^2 \leq C \; h^8 \; \left( \Vert V\alpha_*\Vert_2^2 + \Vert \ |\cdot| V\alpha_*\Vert_2^2 \right) \; \Vert \Psi\Vert_{\Hmag^1(Q_h)}^2.
\end{align*}
\end{prop}

\begin{bem}
\label{rem:A1}
In order to prove Theorem~\ref{Calculation_of_the_GL-energy}, we only need the bound
\begin{equation*}
	| \langle \Delta, (L_{T, \Abold} - \tilde L_{T, \Abold})\Delta\rangle | \leqslant C \; h^5 \; \left( \Vert V\alpha_*\Vert_2^2 + \Vert \ |\cdot| V\alpha_*\Vert_2^2 \right) \; \Vert \Psi\Vert_{\Hmag^1(Q_h)}^2,
\end{equation*}
which is a direct consequence of Cauchy--Schwarz, Proposition~\ref{LTA-LtildeTA}, Lemma~\ref{Schatten_estimate}, and \eqref{Periodic_Sobolev_Norm}. We prove the more general statement in Proposition~\ref{LTA-LtildeTA} here because we need it in the proof of Proposition~\ref{Structure_of_alphaDelta} in Section~\ref{sec:proofofadmissibility} below. 
\end{bem}

Let us define the functions
\begin{align}
F_{T,h}^{a} &\coloneqq  \frac 2\beta \sum_{m=0}^a \sum_{n\in \Zbb}    \bigl(|\cdot|^m \, \gzfunctiondiff^{\i \omega_n} \bigr) * \gzfunction^{-\i\omega_n} + \gzfunctiondiff^{\i\omega_n} * \bigl(|\cdot|^m \, \gzfunction^{-\i\omega_n} \bigr) \notag \\
&\hspace{80pt}+ \bigl(|\cdot|^a \, |g_0^{\i\omega_n}|\bigr) * \gzfunctiondiff^{-\i\omega_n} + |g_0^{\i\omega_n}| * \bigl(|\cdot|^a \, \gzfunctiondiff^{-\i\omega_n} \bigr) \label{LTA-LtildeTA_FTA_definition}
\end{align}
with $a \geq 0$ and
\begin{align}
G_{T, h} &\coloneqq  \frac 2\beta \sum_{\# \in \{ \pm \}} \sum_{n\in \Zbb} \gzfunctiondiff_\nabla^{\# \i\omega_n} * \gzfunction^{-\i\omega_n} + \gzfunctiondiff^{\i\omega_n} * \gzfunction_\nabla^{-\# \i \omega_n}  + |\nabla g_0^{\# \i\omega_n}| * \gzfunctiondiff^{-\i\omega_n} + |g_0^{\i\omega_n}| * \gzfunctiondiff_\nabla^{-\# \i\omega_n},  \label{LTA-LtildeTA_GTA_definition}
\end{align}
which play a prominent role in the proof of Proposition~\ref{LTA-LtildeTA}. We also recall the definition of the Matsubara frequencies $\omega_n$ in \eqref{Matsubara_frequencies}, of $g_0^z$ in \eqref{ghz_definition}, and of the functions $\rho^z$, $\tau^z$, $\rho_{\nabla}^z$, and $\tau_{\nabla}^z$ in Proposition~\ref{gh-g_decay}. 

We claim that for any $T_0 > 0$ there is $h_0>0$ such that for any $0 < h \leq h_0$ and any $T\geq T_0$ we have 
\begin{align}
\Vert F_{T, h}^a \Vert_1 + \Vert G_{T, h} \Vert_1 \leq C_a \; h^3. \label{LTA-LtildeTA_FThGTh}
\end{align}
To prove this claim, we apply Young's inequality, Proposition~\ref{gh-g_decay},  and Lemma~\ref{g_decay}, and note that the function $f(t, \omega)$ in \eqref{gh-g_decay_f} obeys the estimate
\begin{align}
f(0, \omega_n) &\leq C \; |2n+1|^{-1}. \label{g0_decay_f_estimate1}
\end{align}
Moreover,
\begin{align}
1+\frac{|\omega_n| + |\mu|}{|\omega_n| + \mu_-} \leq C. \label{g0_decay_f_estimate2}
\end{align}
In combination, these considerations prove our claim. We are now prepared to give the proof of the above proposition.

\begin{proof}[Proof of Proposition \ref{LTA-LtildeTA}]
We have
\begin{align}
\Vert L_{T, \Abold}\Delta - \tilde L_{T, \Abold} \Delta\Vert_\Hsymm^2 &= \Vert L_{T, \Abold}\Delta - \tilde L_{T, \Abold} \Delta\Vert_2^2 \notag \\
&\hspace{-50pt} + \Vert \Pi (L_{T, \Abold}\Delta - \tilde L_{T, \Abold} \Delta)\Vert_2^2 + \Vert \tilde \pi(L_{T, \Abold}\Delta - \tilde L_{T, \Abold} \Delta)\Vert_2^2 \label{LTA-LtildeTBA_8}
\end{align}
and we claim that the first term on the right side satisfies
\begin{align}
\Vert L_{T, \Abold}\Delta - \tilde L_{T, \Abold} \Delta\Vert_2^2 &\leq 4 \; \Vert \Psi\Vert_2^2 \; \Vert F_{T, h}^0 * |V\alpha_*| \, \Vert_2^2. \label{LTA-LtildeTBA_1}
\end{align}
Using Young's inequality, \eqref{Periodic_Sobolev_Norm}, and \eqref{LTA-LtildeTA_FThGTh}, we see that the right side of \eqref{LTA-LtildeTBA_1} is bounded by a constant times $h^8 \Vert V\alpha_*\Vert_2^2  \Vert \Psi\Vert_{\Hmag^1(Q_h)}^2$. If \eqref{LTA-LtildeTBA_1} holds, this therefore proves the claimed bound for this term.

To prove \eqref{LTA-LtildeTBA_1}, we start by noting that
\begin{align}
\Vert L_{T, \Abold}\Delta - \tilde L_{T, \Abold}\Delta\Vert_2^2 & \leq 4 \int_{\Rbb^3} \dd r \iint_{\Rbb^3 \times \Rbb^3} \dd Z\dd Z' \iint_{\Rbb^3 \times \Rbb^3} \dd s\dd s' \; |V\alpha_*(s)| \; |V\alpha_*(s')|  \notag\\
&\hspace{60pt} \times \esssup_{X\in \Rbb^3} | (k_{T, \Abold} - \tilde k_{T, \Abold}) (X, Z, r, s)| \notag\\
&\hspace{60pt} \times \esssup_{X\in \Rbb^3} |(k_{T, \Abold} - \tilde k_{T, \Abold}) (X, Z', r, s')|  \notag\\
&\hspace{30pt}\times \fint_{Q_h} \dd X \; |\e^{\i Z \cdot (-\i \nabla_X)} \Psi(X)| \; | \e^{\i Z'\cdot (-\i \nabla_X)}\Psi(X)|. \label{Expanding_the_square}
\end{align}
Since the norm of the operator $\e^{\i Z\cdot (-\i \nabla_X)}$ equals $1$, we have
\begin{align}
\fint_{Q_h} \dd X \; |\e^{\i Z \cdot (-\i \nabla_X)} \Psi(X)| \; | \e^{\i Z'\cdot (-\i \nabla_X)}\Psi(X)| &\leq \Vert \Psi\Vert_2^2. \label{LTA-LtildeTBA_3}
\end{align}
Consequently, \eqref{Expanding_the_square} yields
\begin{align}
\Vert L_{T, \Abold}\Delta - \tilde L_{T, \Abold}\Delta\Vert_2^2 & \leq 4\; \Vert \Psi\Vert_2^2 \notag\\
&\hspace{-80pt} \times \int_{\Rbb^3 }\dd r\, \Bigl| \iint_{\Rbb^3 \times \Rbb^3} \dd Z\dd s \; \esssup_{X\in \Rbb^3} |(k_{T, \Abold} - \tilde k_{T, \Abold}) (X, Z, r, s)|\; |V\alpha_*(s)|\Bigr|^2. \label{LTA-LtildeTBA_2}
\end{align}
We note that
\begin{align}
k_{T, \Abold}^n(X, Z, r, s) - k_T^n(Z, r - s) & \notag \\
&\hspace{-120pt} = \bigl( g_h^{\i \omega_n} - g_0^{\i \omega_n}\bigr) \bigl( X+ \frac r2, X + Z+ \frac s2\bigr) \, g_h^{-\i \omega_n} \bigl( X - \frac r2, X + Z - \frac s2\bigr) \notag \\
&\hspace{-100pt} + g_0^{\i \omega_n} \bigl(Z - \frac {r-s}2\bigr) \, \bigl( g_h^{-\i \omega_n} - g_0^{-\i \omega_n} \bigr) \bigl( X - \frac r2 , X + Z - \frac s2\bigr) \label{LTA-LtildeTBA_15}
\end{align}
and hence, by Proposition~\ref{gh-g_decay}, the integrand in \eqref{LTA-LtildeTBA_2} is bounded by
\begin{align}
\bigl|(k_{T, \Abold} - \tilde k_{T, \Abold}) (X, Z, r, s)\bigr| &\leq  \frac 2\beta\,  \smash{\sum_{n\in \Zbb}} \, \bigl[ \gzfunctiondiff^{\i\omega_n} \bigl( Z - \frac {r-s}2\bigr) \; \gzfunction^{-\i\omega_n} \bigl( Z + \frac {r-s}2\bigr) \notag\\
&\hspace{50pt}+ |g_0^{\i\omega_n}|\bigl( Z -\frac {r-s}2\bigr)\;  \gzfunctiondiff^{-\i\omega_n} \bigl( Z +\frac {r-s}2\bigr)\bigr]. \label{LTA-LtildeTBA_11}
\end{align}
We combine \eqref{LTA-LtildeTBA_11} and the fact that the functions in \eqref{LTA-LtildeTBA_11} are even (see Proposition~\ref{gh-g_decay}), to show
\begin{align}
\int_{\Rbb^3} \dd Z \; \esssup_{X\in \Rbb^3} |(k_{T, \Abold} - \tilde k_{T, \Abold})(X, Z, r, s)| \leq F_{T, h}^0(r-s), \label{LTA-LtildeTBA_5}
\end{align}
where $F_{T, \Abold}^0$ is the function in \eqref{LTA-LtildeTA_FTA_definition}. When we apply \eqref{LTA-LtildeTBA_5} to \eqref{LTA-LtildeTBA_2}, we obtain \eqref{LTA-LtildeTBA_1}. 

Let us pause for a moment and highlight the main idea behind the above bound because it will reappear frequently in the subsequent analysis. The kernels in \eqref{LTA-LtildeTBA_15} are not translation-invariant. From Proposition~\ref{gh-g_decay} we know, however, that they can by bounded by translation-invariant kernels. When we do this, we see that we obtain convolutions of translation-invariant kernels after the integration over $Z$ has been carried out. These convolutions can now be estimated with the $L^1$-norm bounds in Proposition~\ref{gh-g_decay}. We highlight that this emergent simplicity is difficult to find when working in the operator picture. Our analysis is inspired by the analysis for the constant magnetic field in \cite{Hainzl2017}, where much of the above structure is more apparent.

For the second term on the right side of \eqref{LTA-LtildeTBA_8}, we claim the bound
\begin{align}
\Vert \Pi(L_{T, \Abold}\Delta - \tilde L_{T, \Abold}\Delta)\Vert_2^2 &\leq C \, h^2 \; \Vert (F_{T, h}^1 + G_{T, h} ) * |V\alpha_*| \,\Vert_2^2 \;  \Vert \Psi\Vert_{\Hmag^1(Q_h)}^2 \notag \\
&\leq C \, h^8 \; \Vert V\alpha_*\Vert_2^2  \; \Vert \Psi\Vert_{\Hmag^1(Q_h)}^2. \label{LTA-LtildeTBA_6}
\end{align}
The second inequality follows from Young's inequality and \eqref{LTA-LtildeTA_FThGTh}. To prove the first inequality in \eqref{LTA-LtildeTBA_6}, we note that the gradient can act either on $\Psi$ or on $k_{T,\Abold}-\widetilde{k}_{T,\Abold}$ and we start by considering the term, where it acts on $\Psi$. We therefore apply \eqref{Expanding_the_square} with $\exp(\i Z \cdot (-\i \nabla_X))$ replaced by $\Pi_X \, \exp(\i Z \cdot (-\i \nabla_X))$ (we recall that $\Pi_X = -\i \nabla_X + 2 \Abold_{\Bbold}(X)$) and replace \eqref{LTA-LtildeTBA_3} by
\begin{align*}
&\fint_{Q_h} \dd X \; |\Pi_X \, \e^{\i Z \cdot (-\i \nabla_X)}\Psi(X)| \; |\Pi_X \, \e^{\i Z' \cdot (-\i \nabla_X)}\Psi(X)| \\
&\hspace{6cm}\leq \Vert \Pi_X \, \e^{\i Z \cdot (-\i \nabla_X)}\Psi\Vert_2\; \Vert \Pi_X \, \e^{\i Z' \cdot (-\i \nabla_X)}\Psi\Vert_2.
\end{align*}
From 
a direct computation, we know that
\begin{align}
\Pi_X\, \e^{\i Z\cdot (-\i \nabla_X)} = \e^{\i Z \cdot (-\i \nabla_X)} \bigl[\Pi_X - \Bbold \wedge Z\bigr]. \label{PiU_intertwining}
\end{align}
Using this and \eqref{Periodic_Sobolev_Norm}, we find
\begin{align}
\Vert \Pi_X \, \e^{\i Z\cdot (-\i \nabla_X)} \Psi\Vert_2 &\leq \Vert \Pi\Psi\Vert_2 +  |\Bbold| \, |Z| \; \Vert \Psi\Vert_2 \leq C\, h^2 \; \Vert \Psi\Vert_{\Hmag^1(Q_h)} \; (1 + |Z|), \label{PiXcos_estimate}
\end{align}
which subsequently proves
\begin{align}
\Vert \Pi(L_{T, \Abold}\Delta - \tilde L_{T, \Abold}\Delta)\Vert_2^2 &\leq C \, h^2 \; \Vert \Psi\Vert_{\Hmag^1(Q_h)}^2  \notag\\
&\hspace{-110pt} \times  \int_{\Rbb^3} \dd r \Bigl( \; \Bigl| \iint_{\Rbb^3 \times \Rbb^3} \dd Z\dd s \; h \, (1+ |Z|) \; \esssup_{X\in \Rbb^3} |(k_{T, \Abold} - \tilde k_{T, \Abold})(X, Z, r,s)| |V\alpha_*(s)|\Bigr|^2  \notag \\
&\hspace{-2.5cm} + \Bigl| \iint_{\Rbb^3 \times \Rbb^3} \dd Z\dd s \; \esssup_{X\in \Rbb^3} |(\nabla_X k_{T, \Abold} - \nabla_X \tilde k_{T, \Abold})(X, Z, r, s)| \, |V\alpha_*(s)|\Bigr|^2 \; \Bigr). \label{LTA-LtildeTBA_9}
\end{align}
We claim that 
\begin{align}
|\nabla_X \tilde\Phi_\Abold(X, Z, r, s) | \leq C \, \Vert D\Abold\Vert_\infty  \Bigl( \bigl| Z + \frac{r-s}{2}\bigr| + \bigl| Z - \frac{r-s}{2}\bigr|\Bigr) \label{LTA-LtildeTBA_16}
\end{align}
holds, where $D\Abold$ denotes the Jacobi matrix of $\Abold$. To see this, we use Lemma~\ref{PhiA_derivative} and compute
\begin{align*}
\nabla_X \tilde \Phi_\Abold(X, Z, r, s) &= \Abold\bigl( X + Z + \frac s2\bigr) - \Abold\bigl( X + \frac r2\bigr) + \Abold\bigl(X + Z - \frac s2\bigr) - \Abold\bigl( X - \frac r2\bigr) \\
&\hspace{40pt} + \tilde \Abold\bigl( X + \frac r2, X + Z + \frac s2\bigr) - \tilde \Abold \bigl( X + Z + \frac s2, X +\frac r2\bigr) \\
&\hspace{40pt} + \tilde \Abold\bigl( X - \frac r2, X + Z - \frac s2\bigr) - \tilde \Abold \bigl( X + Z - \frac s2, X - \frac r2\bigr).
\end{align*}
The claim is a direct computation of this equality and a first order Taylor approximation. In combination, Proposition~\ref{gh-g_decay}, \eqref{LTA-LtildeTBA_15}, and \eqref{LTA-LtildeTBA_16} imply that, for $h$ small enough,
\begin{align*}
\int_{\Rbb^3} \dd Z \; \esssup_{X\in \Rbb^3} |(\nabla_X k_{T, \Abold} - \nabla_X \tilde k_{T, \Abold})(X, Z, r, s)| &\leq \bigl( G_{T, h} + F_{T, h}^1\bigr) (r-s).
\end{align*}
The functions $F_{T, h}^1$ and $G_{T, h}$ are defined in \eqref{LTA-LtildeTA_FTA_definition} and \eqref{LTA-LtildeTA_FTA_definition}, respectively. A similar argument that uses $|Z| \leq
| Z + \frac{r}{2}| + | Z - \frac{r}{2}|$ shows
\begin{equation*}
	\int_{\Rbb^3 } \dd Z \; (1+ |Z|) \; \esssup_{X\in \Rbb^3} |(k_{T, \Abold} - \tilde k_{T, \Abold})(X, Z, r,s)| \leqslant F_{T, h}^1(r-s).  
\end{equation*}
When we insert these two bounds into \eqref{LTA-LtildeTBA_9}, this proves \eqref{LTA-LtildeTBA_6}. It remains to consider the third term on the right side of \eqref{LTA-LtildeTBA_8}.

We claim that it satisfies
\begin{align}
\Vert \tilde \pi (L_{T, \Abold}\Delta - \tilde L_{T, \Abold} \Delta)\Vert_2^2 &\leq C \; \Vert \Psi\Vert_2^2 \; \Vert (F_{T, h}^1 + G_{T, h}) * | \ |\cdot| V\alpha_*| \, \Vert_2^2 \notag \\
&\leq C \, h^8\; \Vert \ |\cdot| \ V\alpha_*\Vert_2^2  \; \Vert \Psi\Vert_{\Hmag^1(Q_h)}^2. \label{LTA-LtildeTBA_4}
\end{align}
The second inequality follows from Young's inequality, \eqref{Periodic_Sobolev_Norm}, and \eqref{LTA-LtildeTA_FThGTh}. To see that the first inequality holds, we first estimate
\begin{align}
\Vert \tilde \pi (L_{T, \Abold}\Delta - \tilde L_{T, \Abold}\Delta)\Vert_2^2 & \leq 4 \;\Vert \Psi\Vert_2^2 \notag \\
&\hspace{-100pt} \times  \int_{\Rbb^3} \dd r \, \Bigl| \iint_{\Rbb^3 \times \Rbb^3} \dd Z\dd s \; \esssup_{X\in \Rbb^3} |(\tilde \pi k_{T, \Abold} - \tilde \pi\tilde k_{T, \Abold} )(X, Z, r, s)| \;|V\alpha_*(s)| \Bigr|^2.  \label{LTA-LtildeTBA_10}
\end{align}
We start by noting that
\begin{align*}
&\tilde \pi k_{T, \Abold}(X, Z, r, s) - \tilde \pi \tilde k_{T, \Abold}(X, Z, r, s) \\
&= \e^{\mathrm{i} \widetilde{\Phi}_{\Abold}(X,Z,r,s)} \bigl(-\i \nabla_r + \frac 14 \Bbold \wedge r + \nabla_r \widetilde{\Phi}_{\Abold}(X,Z,r,s) \bigr) \frac{2}{\beta} \sum_{n\in \Zbb} ( k_{T, \Abold}^n(X, Z, r, s) - \tilde k_T^n(Z, r-s) )
\end{align*}
and
\begin{align*}
	\nabla_r \tilde \Phi_{\Abold}(X, Z, r, s) &= -\frac 12 \bigl[ \Abold\bigl( X + \frac r2\bigr) - \Abold\bigl( X - \frac r2\bigr) \bigr] \\
	&\hspace{30pt} + \frac 12 \bigl[ \tilde \Abold\bigl( X + \frac r2, X + Z + \frac s2\bigr) - \tilde \Abold \bigl( X - \frac r2, X + Z - \frac s2\bigr)\bigr],
\end{align*}
which follows from Lemma~\ref{PhiA_derivative}. We combine these two identities and estimate
\begin{align}
|\tilde \pi k_{T, \Abold} - \tilde \pi \tilde k_{T, \Abold}| \leq& \frac 2\beta \sum_{n\in \Zbb} |\nabla_r k_{T, \Abold}^n - \nabla_r k_T^n| + |k_{T, \Abold} - \tilde k_{T, \Abold}| \left( \frac{| \Bbold |}{4} + h^2 \Vert D \Abold \Vert_{\infty} \right) \nonumber \\
&\hspace{2.5cm}\times \left( | r-s| + |s| + \left| Z + \frac{r-s}{2} \right| + \left| Z - \frac{r-s}{2} \right| \right). \label{LTA-LtildeTBA_13}
\end{align}
We apply 
\begin{align}
|r-s|^a &= \bigl| \frac{r-s}{2} + Z + \frac{r-s}{2} - Z\bigr|^a \leq 2^{(a-1)_+} \Bigl( \bigl| Z - \frac{r-s}{2}\bigr|^a + \bigl| Z + \frac{r-s}{2}\bigr|^a\Bigr), \label{r-s_estimate}
\end{align}
which holds for $a \geq 0$, with $a=1$ to bound $|r-s|$ in \eqref{LTA-LtildeTBA_13}. When we put \eqref{LTA-LtildeTBA_13} and \eqref{r-s_estimate} together, and additionally use  Lemma~\ref{gh-g_decay}, \eqref{LTA-LtildeTBA_15} and \eqref{LTA-LtildeTBA_13}, we find 
\begin{align}
\int_{\Rbb^3} \dd Z \; \esssup_{X\in \Rbb^3} |(\tilde \pi k_{T, \Abold} - \tilde \pi\tilde k_{T, \Abold}) (X, Z, r, s)| \leq C \bigl(F_{T, h}^1 + G_{T, h}\bigr)(r-s) (1+|s|). \label{LTA-LtildeTBA_7}
\end{align}
Finally, \eqref{LTA-LtildeTBA_7} and \eqref{LTA-LtildeTBA_10} imply \eqref{LTA-LtildeTBA_4}, which finishes the proof of Proposition~\ref{LTA-LtildeTA}.
\end{proof}

\paragraph{The operator $\tilde M_{T,\Abold}$.} 

We define $\tilde M_{T,\Abold}$ by
\begin{align}
\tilde M_{T,\Abold}\alpha(X,r) \coloneqq  \iint_{\Rbb^3 \times \Rbb^3} \dd Z\dd s \; k_T(Z , r-s) \, \e^{-\i \frac{r-s}{4} \cdot D\Abold_h (X)(r+s)} \, (\cos(Z \cdot \Pi_{\Abold_h}) \alpha) (X,s), \label{MtildeTA_definition}
\end{align}
where 
\begin{equation}
	k_T(Z, r) \coloneqq  \frac{2}{\beta} \sum_{n\in \Zbb}  g_0^{\i\omega_n}\bigl(Z - \frac{r}{2} \bigr) \, g_0^{-\i \omega_n}\bigl( Z + \frac{r}{2} \bigr).
	\label{eq:AKT0}
\end{equation}
By $(D\Abold)_{ij} \coloneqq  \partial_j \Abold_i$ we denote the Jacobi matrix of $\Abold$. Here and in the following we use the notation that $D\Abold_h (X)(r+s)$ denotes the matrix $D\Abold_h (X)$ applied to $r+s$. Accordingly, $\frac{r-s}{4} \cdot D\Abold_h (X)(r+s)$ is the inner product of this vector with $\frac{r-s}{4}$.

The following proposition allows us to replace $\tilde L_{T, \Abold}$ by $\tilde M_{T, \Abold}$ in our computations.


\begin{prop}
\label{LtildeTA-MtildeTA}
Assume that $\Abold = \Abold_{e_3} + A$ with a periodic vector potential $A \in W^{2,\infty}(\mathbb{R}^3,\mathbb{R}^3)$, let $|\cdot|^k V\alpha_*\in L^2(\Rbb^3)$ for $k \in \{ 0,1,2 \}$, $\Psi\in \Hmag^1(Q_h)$, and denote $\Delta \equiv \Delta_\Psi$ as in \eqref{Delta_definition}. For any $T_0 > 0$ there is $h_0>0$ such that for any $0 < h \leq h_0$ and any $T\geq T_0$ we have
\begin{align}
\Vert \tilde L_{T, \Abold}\Delta - \tilde M_{T,\Abold}\Delta \Vert_\Hsymm^2 &\leq C\; h^8 \; \max_{k =0,1,2} \Vert \, |\cdot|^k V\alpha_*\Vert_2^2 \;\Vert \Psi\Vert_{\Hmag^1(Q_h)}^2. \label{LtildeTA-MtildeTA_eq1}
\end{align}
\end{prop}

Before we give the proof of Proposition~\ref{LtildeTA-MtildeTA}, we state and prove two preparatory lemmas. The first lemma allows us to extract the term $\Phi_{2A_h}(X, X + Z)$ and to replace $\e^{\i Z \cdot (-\i \nabla)}$ by $\e^{\i Z \cdot \Pi_{\Abold_h}}$ in the definition of $\tilde L_{T, \Abold}$. In \cite{Hainzl2017,DeHaSc2021}, where only the constant magnetic field is present, this approximation holds as an algebraic identity. A version of the first bound in \eqref{Phi_Approximation_eq1} has, in case of a periodic vector potential, been proved in \cite{BdGtoGL}. In this reference $Z$ does not appear on the right side, which is why this bound appears to be wrong.
\begin{lem}
\label{Phi_Approximation}
Assume that $\Asolo \in \Wvec{2}$. We then have
\begin{align}
\sup_{X\in \Rbb^3} \bigl| \tilde \Phi_\Asolo(X, Z, r, s) - \Phi_{2\Asolo}(X,X+Z) + \frac 14 (r-s)\cdot D \Asolo(X)(r+s)\bigr| & \notag\\
&\hspace{-200pt} \leq C \; \Vert D^2\Asolo\Vert_\infty \; \bigl( |Z| +  |r-s|\bigr)  \bigl( |s|^2 + |r-s|^2\bigr), \label{Phi_Approximation_eq1}
\end{align}
where $\Phi_\Asolo$ is defined in \eqref{PhiA_definition} and $\tilde \Phi_\Asolo$ is defined in \eqref{LTA_PhitildeA_definition}. We also have
\begin{align}
\sup_{X\in \Rbb^3} \bigl| \nabla_X \tilde \Phi_\Asolo(X, Z, r, s) - \nabla_X \Phi_{2\Asolo} (X, X + Z) \bigr| & \notag\\
&\hspace{-100pt} \leq C \; \Vert D^2\Asolo \Vert_\infty\; \bigl( |Z| + |r-s|\bigr)  \bigl(|s| + |r-s|\bigr) \label{Phi_Approximation_eq3}
\end{align}
as well as
\begin{align}
\sup_{X\in \Rbb^3} \bigl| \nabla_X (r-s) \cdot D\Asolo(X) \cdot (r+s) \bigr| \leq C \, \Vert D^2\Asolo\Vert_\infty \, |r-s| \, \bigl( |s| + |r-s|\bigr) \label{Phi_Approximation_eq4}
\end{align}
and
\begin{align}
\sup_{X\in \Rbb^3} \bigl| \nabla_r \tilde \Phi_\Asolo(X, Z, r, s) + \frac 14 \nabla_r (r-s)\cdot D \Asolo(X)(r+s) \bigr| &\notag\\
&\hspace{-100pt}\leq C\, \Vert D^2\Asolo\Vert_\infty \, \bigl( |s|^2 + |r-s|^2 + |Z|^2\bigr). \label{Phi_Approximation_eq6}
\end{align}
The same bounds hold if $\Asolo$ is replaced by $\Abold_{e_3} + \Asolo$.
\end{lem}


\begin{proof}[Proof of Lemma \ref{Phi_Approximation}]
We use the notation $\zeta_X^r \coloneqq  X + \frac r2$ and start by writing
\begin{align}
\tilde \Phi_\Asolo(X, Z, r, s) &= \int_0^1 \dt\; \bigl[\Asolo\bigl( \zeta_{X + Z - tZ}^{s + t(r-s)} \bigr) + \Asolo\bigl(\zeta_{X + Z - tZ}^{-s - t(r-s)}\bigr) \bigr] \cdot Z \notag \\
& \hspace{50pt} -  \int_0^1 \dt\; \bigl[\Asolo\bigl(\zeta_{X + Z - tZ}^{s + t(r-s)}\bigr) - \Asolo\bigl( \zeta_{X + Z - tZ}^{-s - t(r-s)} \bigr) \bigr] \cdot \frac{r-s}{2}. \label{Phi_Approximation_1}
\end{align}
A second order Taylor expansion in the variable $\pm \frac 12 (s + t(r-s))$ allows us to show that
\begin{align}
\Bigl| \Asolo\bigl(X + Z- tZ \pm \frac{s + t(r-s)}{2}\bigr) - \Asolo(X + Z-tZ) \mp \frac 12 D\Asolo(X + Z-tZ) (s + t(r-s)) \Bigr|   & \notag\\
&\hspace{-220pt} \leq C\; \Vert D^2\Asolo\Vert_\infty \; \bigl(|s|^2 +|r-s|^2\bigr). \label{Phi_Approximation_2}
\end{align}
For the first term on the right side of \eqref{Phi_Approximation_1}, this implies
\begin{align}
\Bigl| \int_0^1 \dt\; \bigl[\Asolo\bigl( \zeta_{X + Z - tZ}^{s + t(r-s)} \bigr) + \Asolo\bigl( \zeta_{X + Z - tZ}^{-s - t(r-s)} \bigr) \bigr] \cdot Z  -\Phi_{2\Asolo}(X,X+Z) \Bigr| & \notag\\
&\hspace{-100pt}  \leq C \, \Vert D^2\Asolo\Vert_\infty \, |Z| \, \bigl(|s|^2 + |r-s|^2\bigr), \label{Phi_Approximation_4}
\end{align}
and we find
\begin{align}
\Bigl| - \int_0^1 \dt\; \bigl[\Asolo\bigl( \zeta_{X + Z - tZ}^{s + t(r-s)} \bigr) - \Asolo\bigl(\zeta_{X + Z - tZ}^{-s - t(r-s)}\bigr) \bigr] \cdot \frac{r-s}{2}+ \frac 14 (r-s) \cdot D\Asolo(X) (r+s) \Bigr| & \notag\\
&\hspace{-350pt} \leq \Bigl| -\frac{r-s}2 \cdot \int_0^1 \dd t \; \bigl[ D\Asolo(X + Z - tZ) - D\Asolo(X) \bigr](s + t(r-s)) \Bigr| \notag \\
&\hspace{-150pt} + C\, \Vert D^2\Asolo\Vert_\infty \, \bigl( |s|^2 + |r-s|^2\bigr) \, |r-s| \notag \\
&\hspace{-350pt} \leq C\, \Vert D^2\Asolo \Vert_\infty \, |r-s| \, \bigl[|s|^2 + |r-s|^2 + \bigl(|s| + |r-s|\bigr) |Z| \bigr] \label{Phi_Approximation_3}
\end{align}
for the second term. Adding up \eqref{Phi_Approximation_4} and \eqref{Phi_Approximation_3} proves \eqref{Phi_Approximation_eq1}. The proof of \eqref{Phi_Approximation_eq3} results from a first order Taylor expansion and that of \eqref{Phi_Approximation_eq4} is a straightforward computation. It remains to prove \eqref{Phi_Approximation_eq6}.


When we differentiate \eqref{Phi_Approximation_1} with respect to $r$ this yields
\begin{align}
\nabla_r \tilde \Phi_\Asolo(X, Z, r, s) &= \int_0^1 \dt\; \frac t2 \ Z \cdot \bigl[D\Asolo\bigl( \zeta_{X + Z - tZ}^{s + t(r-s)} \bigr) - D\Asolo\bigl(\zeta_{X + Z - tZ}^{-s - t(r-s)}\bigr) \bigr] \notag \\
& \hspace{50pt} -  \int_0^1 \dt\; \frac t2 \ \frac{r-s}{2} \cdot \bigl[D\Asolo\bigl(\zeta_{X + Z - tZ}^{s + t(r-s)} \bigr) + D\Asolo\bigl(\zeta_{X + Z - tZ}^{-s - t(r-s)} \bigr) \bigr]  \notag \\
& \hspace{50pt} - \frac 12 \int_0^1 \dt\; \bigl[\Asolo\bigl(\zeta_{X + Z - tZ}^{s + t(r-s)} \bigr) - \Asolo\bigl(\zeta_{X + Z - tZ}^{-s - t(r-s)} \bigr) \bigr]. \label{Phi_Approximation_7}
\end{align}
A first order Taylor expansion shows that the absolute value of the first term on the right side of \eqref{Phi_Approximation_7} is bounded by $C\Vert D^2\Asolo\Vert_\infty (|s| + |r-s|) |Z|$. We also note that
\begin{align}
\frac 14 \nabla_r (r - s) \cdot D\Asolo(X) (r + s) = \frac 14 (r-s) \cdot D\Asolo(X) + \frac 14 D\Asolo(X) (r+s). \label{Phi_Approximation_9}
\end{align}
The second term on the right side of \eqref{Phi_Approximation_7} obeys
\begin{align}
\Bigl| \int_0^1 \dt\; t \; \frac{r-s}{4} \cdot \bigl[D\Asolo\bigl(\zeta_{X + Z - tZ}^{s + t(r-s)} \bigr) + D\Asolo\bigl(\zeta_{X + Z - tZ}^{-s - t(r-s)} \bigr) \bigr] - \frac 14 (r-s) \cdot D\Asolo(X) \bigr| & \notag \\
&\hspace{-360pt} \leq \Bigl| \int_0^1 \dd t \; \frac{t}{2} \; (r-s) \cdot \bigl[ D\Asolo(X + Z - tZ) - D\Asolo(X) \bigr] \Bigr| + C \, \Vert D^2\Asolo\Vert_\infty \bigl( |s| + |r-s|\bigr) |r-s| \notag\\
&\hspace{-360pt} \leq C \, \Vert D^2\Asolo\Vert_\infty \, \bigl[ |r-s| \, |Z| + \bigl( |s| + |r-s|\bigr) |r-s|\bigr] . \label{Phi_Approximation_8}
\end{align}
For the third term on the right side of \eqref{Phi_Approximation_7}, we use the bound
\begin{align*}
\Bigl| \frac 12 \int_0^1 \dt\; \bigl[\Asolo\bigl(\zeta_{X + Z - tZ}^{s + t(r-s)} \bigr) - \Asolo\bigl(\zeta_{X + Z - tZ}^{-s - t(r-s)} \bigr) \bigr] - \frac 14  D\Asolo(X) (r+ s)\Bigr| \leq \Tcal_+ + \Tcal_- + \Tcal,
\end{align*}
where
\begin{align*}
\Tcal_\pm &\coloneqq  \Bigl| \frac 12 \int_0^1 \dd t \; \bigl[ \Asolo\bigl( \zeta_{X + Z - tZ}^{\pm s \pm t(r-s)}\bigr) \mp \Asolo(X + Z - tZ) - \frac 12 D\Asolo(X + Z- tZ) (s+ t(r-s)) \bigr]\Bigr|
\end{align*}
and
\begin{align*}
\Tcal &\coloneqq  \Bigl| \frac 12 \int_0^1 \dd t \; D\Asolo(X + Z- tZ) (s + t(r-s)) - \frac 14 D\Asolo(X) (r+s)\Bigr|.
\end{align*}
By \eqref{Phi_Approximation_2}, we have
\begin{align*}
\Tcal_\pm &\leq C \, \Vert D^2 \Asolo\Vert_\infty \, \bigl( |s|^2 + |r-s|^2\bigr), & \Tcal &\leq C\, \Vert D^2\Asolo\Vert_\infty \, \bigl( |s| + |r-s|\bigr ) \, |Z|.
\end{align*}
In combination with \eqref{Phi_Approximation_8}, these considerations imply \eqref{Phi_Approximation_eq6}. The proof for $\Abold_{e_3} + \Asolo$ is literally the same. 
\end{proof}

The next lemma is a substitute for the identity 
\begin{align}
	\e^{\i \Bbold \cdot (X \wedge Z)} \e^{\i Z \cdot P_X} = \e^{\i Z \cdot \Pi_X }
	\label{Magnetic_Translation_constant_Decomposition}
\end{align}
in the case of a general magnetic field. It holds because $\Bbold \cdot (X\wedge Z) = Z \cdot (\Bbold \wedge X)$ and the latter commutes with $Z \cdot \Pi_X$. Here, we used the notations $P = -\i \nabla$ for the momentum operator and $\Pi = -\i \nabla + 2 \Abold_{\Bbold}$ for its magnetic counterpart. Sometimes when several variables appear in an equation we write, e.g., $P_X$, $\Pi_X$, etc.\ to indicate on which variable $P$, $ \Pi$, etc.\ is acting. 

\begin{lem}
\label{Magnetic_Translation_Representation}
Assume that $\Asolo \in L^{\infty}(\mathbb{R}^3,\mathbb{R}^3)$ is a periodic function. Then
\begin{align}
\e^{\i \Phi_{2\Asolo_h}(X , X +Z)} \, \e^{ \i Z\cdot \Pi_X } = \e^{ \i Z\cdot \Pi_{\Abold_h} },
\end{align}
where $\Pi_{\Abold_h} = P_X + 2 \Abold_h(X)$ is understood to act on the $X$ coordinate.
\end{lem}

The above lemma is a consequence of the following more abstract proposition, whose proof can be found in \cite[p. 290]{Werthamer1966}. For the sake of completeness we repeat it here.

\begin{prop}
\label{Operator_Equality_Abstract}
Let $\Hcal$ be a separable Hilbert space, let $P\colon \Dcal(P)\ra \Hcal$ be a densely defined self-adjoint operator, and assume that $Q$ is bounded and self-adjoint. Assume further that $[\e^{\i tP} \, Q \, \e^{-\i tP} , \e^{\i sP} \, Q \, \e^{-\i sP} ] =0$ for every $t,s\in [0,1]$. Then, we have
\begin{align*}
\exp\Bigl(\i \int_0^1\dt \; \e^{\i tP} \, Q\, \e^{-\i tP}\Bigr) \, \e^{\i P} = \e^{\i(P + Q)}.
\end{align*}
\end{prop}

\begin{proof}
For $s\in \Rbb$ we define $Q(s) \coloneqq  \e^{\i sP} \, Q \, \e^{-\i sP}$ and $W(s) \coloneqq  \e^{\i s(P+Q)} \, \e^{-\i sP}$. On $\Dcal(P)$, we may differentiate $W$ to get
\begin{align*}
-\i W'(s) = \e^{\i s(P+Q)} (P+Q) \, \e^{-\i sP} - \e^{\i s(P+Q)} \, P \, \e^{-\i sP} 
=  \e^{\i s(P+Q)} \, Q \, \e^{-\i sP} = W(s) \, Q(s).
\end{align*}
Using that $Q$ is bounded we conclude that this identity also holds on $\mathcal{H}$. Hence, $W$ satisfies the linear differential equation $W'(s) = \i W(s) \, Q(s)$. Since by assumption $[Q(s),Q(t)] = 0$ holds for all $s,t \in [0,1]$, we conclude that the unique solution to this equation can be written as  
\begin{align*}
\tilde W(s) \coloneqq  \exp\Bigl(\i \int_0^s \dt\; Q(t)\Bigr),
\end{align*}
and hence
\begin{align*}
\exp\Bigl(\i \int_0^s \dt\; Q(t)\Bigr) = \e^{\i s(P+Q)}\e^{-\i sP}.
\end{align*}
With the choice $s =1$ this equation proves the claim.
\end{proof}

\begin{proof}[Proof of Lemma \ref{Magnetic_Translation_Representation}]
We first show that
\begin{equation}
	\e^{ \i \Phi_{2\Asolo_h}(X, X+Z) } \, \e^{ \i Z \cdot P_X } = \e^{ \i Z \cdot (P_X + 2\Asolo_h(X)) }
	\label{eq:Afirststep}
\end{equation}	
holds.	To that end, we apply Proposition~\ref{Operator_Equality_Abstract} with the choices $P = Z\cdot P_X$, where $P_X = -\i \nabla_X$, and $Q = 2 Z\cdot \Asolo_h(X)$ and find
\begin{align*}
	\exp\Bigl(\i \int_0^1\dt \; \e^{\i t Z\cdot P_X} \, 2 Z\cdot \Asolo_h(X) \, \e^{-\i t Z \cdot P_X}\Bigr) \, \e^{\i Z\cdot P_X} = \e^{\i(Z \cdot (P_X + 2 \Asolo(X)))}.
\end{align*}
It remains to compute the integral in the exponential, which reads	
\begin{equation*}
	\int_0^1\dt \; \e^{\i t Z\cdot P_X} \, 2 Z\cdot \Asolo_h(X) \, \e^{-\i t Z \cdot P_X} = 2 \int_0^1\dt \; Z\cdot \Asolo_h(X+tZ) = \Phi_{2\Asolo_h}(X,X+Z).
\end{equation*}
To obtain the last equality we applied the coordinate transformation $t\mapsto 1 -t$. In combination, these considerations prove \eqref{eq:Afirststep}. 

Next, we use \eqref{eq:Afirststep} and \eqref{Magnetic_Translation_constant_Decomposition} to see that
\begin{align*}
	\e^{\i \Phi_{2\Asolo_h}(X , X +Z)} \, \e^{ \i Z\cdot \Pi_X } &= \e^{ \i \Phi_{2\Asolo_h}(X, X+Z) } \, \e^{ \i Z \cdot P_X } \; \e^{\i \Bbold \cdot(X \wedge Z)} = \e^{ \i Z \cdot (P_X + 2\Asolo_h(X)) } \; \e^{\i \Bbold \cdot(X \wedge Z)} \\
	&= \e^{ \i Z\cdot \Pi_{\Abold_h} }
\end{align*}
holds. This proves Lemma~\ref{Magnetic_Translation_Representation}.
\end{proof}

For $a\in \Nbb_0$ we define the functions
\begin{align}
F_T^{a} \coloneqq  \frac 2\beta \sum_{n\in \Zbb} \sum_{m=0}^a \sum_{b = 0}^m \binom mb \; \bigl(|\cdot|^{b}\,  |g_0^{\i\omega_n}|\bigr) * \bigl(|\cdot|^{m-b} \, |g_0^{-\i\omega_n}| \bigr) \label{LtildeTA-MtildeTA_FT_definition}
\end{align}
and
\begin{align}
G_T^a &\coloneqq  \smash{\frac 2\beta \sum_{n\in \Zbb} \sum_{m=0}^a \sum_{b=0}^m \binom mb} \; \bigl( |\cdot|^b \, |\nabla g_0^{\i \omega_n}| \bigr) * \bigl( |\cdot|^{m-b} \, |g_0^{-\i\omega_n}|\bigr) \notag \\
&\hspace{150pt} + \bigl( |\cdot|^b \, |g_0^{\i\omega_n}| \bigr) * \bigl( |\cdot|^{m-b} \, |\nabla g_0^{-\i\omega_n}| \bigr), \label{LtildeTA-MtildeTA_GT_definition}
\end{align}
which play an prominent role in the proof of Proposition~\ref{LtildeTA-MtildeTA}. An application of Lemma~\ref{g_decay},  \eqref{g0_decay_f_estimate1}, and \eqref{g0_decay_f_estimate2} shows that for $T \geq T_0 > 0$ and $a\in \Nbb_0$, we have
\begin{align}
\Vert F_{T}^a\Vert_1 + \Vert G_T^a \Vert_1 &\leq C_{a}. \label{LtildeTA-MtildeTA_FTGT}
\end{align}
We are now prepared to give the proof of Proposition~\ref{LtildeTA-MtildeTA}.
\begin{proof}[Proof of Proposition \ref{LtildeTA-MtildeTA}]
We use \eqref{Magnetic_Translation_constant_Decomposition} and $\Phi_{\Abold_{\Bbold}}(x,y)= \frac{\bold{B}}{2} \cdot ( x \wedge y)$ to write the operator $\tilde L_{T, \Abold}$ as
\begin{align*}
\tilde L_{T, \Abold} \alpha(X, r) &= \iint_{\Rbb^3 \times \Rbb^3} \dd Z \dd s \; \e^{\i \frac{\Bbold}{4} \cdot (r\wedge s)} k_T(Z, r - s) \, \e^{\i \tilde \Phi_{\Asolo_h}(X, Z, r, s)} \, (\e^{\i Z \cdot \Pi} \alpha)(X, s),
\end{align*}
where $\Pi = -\i \nabla + 2 \Abold_{\Bbold}$. With the identities
\begin{align*}
- \frac{r-s}{4} \cdot D \Abold_{\Bbold} (r + s) &= \frac{\Bbold}{4} \cdot (r\wedge s), & \Phi_{2\Abold_{\Bbold}}(X, X + Z) &= Z \cdot (\Bbold \wedge X),
\end{align*}
and \eqref{LTA_PhitildeA_definition} we check that
\begin{align*}
\tilde \Phi_{\Abold_{\Bbold}} (X, Z, r, s) &= - \frac{r-s}{4} \cdot D\Abold_{\Bbold}(X)  (r+s) + \Phi_{2\Abold_{\Bbold}}(X, X + Z).
\end{align*}
In combination with Lemma~\ref{Magnetic_Translation_Representation} and the fact that the integrand in the definition of $\tilde M_{T, \Abold}$ is an even function of $Z$, this allows us to write $\tilde M_{T, \Abold}$ as
\begin{align*}
\tilde M_{T, \Abold} \alpha(X, r) &= \smash{\iint_{\Rbb^3 \times \Rbb^3}} \dd Z \dd s \; \e^{\i \frac \Bbold 4 \cdot (r\wedge s)} k_T(Z, r-s) \, \e^{\i \Phi_{2\Asolo_h} (X, X+Z)} \, \e^{-\i \frac{r-s}{4} \cdot D\Asolo_h(X) (r + s)}  \\
&\hspace{230pt} \times (\e^{\i Z \cdot \Pi} \alpha)(X, s),
\end{align*}
and consequently, 
\begin{align}
\bigl(\tilde L_{T, \Abold} \Delta - \tilde M_{T, \Abold} \Delta\bigr) (X, r) &= -2 \iint_{\Rbb^3 \times \Rbb^3} \dd Z \dd s \; \e^{\i \frac \Bbold 4 \cdot (r\wedge s)} k_T(Z , r-s) \, V\alpha_*(s) \, (\e^{\i Z \cdot \Pi} \Psi)(X)\notag \\
&\hspace{-20pt} \times \bigl[ \e^{\i \tilde \Phi_{\Asolo_h}(X, Z, r, s)} - \e^{\i \Phi_{2\Asolo_h}(X, X + Z)}\e^{-\i \frac{r-s}{4} \cdot D\Asolo_h(X)(r + s)} \bigr]. \label{LtildeTA-MtildeTA_1}
\end{align}

We claim that
\begin{align}
\Vert \tilde L_{T, \Abold} \Delta - \tilde M_{T, \Abold} \Delta \Vert_2^2 &\leq C \, \Vert \Psi\Vert_2^2 \; \Vert D^2A_h\Vert_\infty^2 \; \Vert F_T^3 * |V\alpha_*| + F_T^1 * |\cdot|^2\, |V\alpha_*|  \, \Vert_2^2 \notag \\
& \leqslant C\; h^8 \; \max_{k =0,1,2} \Vert \, |\cdot|^k V\alpha_*\Vert_2^2 \;\Vert \Psi\Vert_{\Hmag^1(Q_h)}^2.
\label{LtildeTA-MtildeTA_6}
\end{align}
The second bound in the above equation follows from Young's inequality, \eqref{Periodic_Sobolev_Norm}, and \eqref{LtildeTA-MtildeTA_FTGT}. To prove the first bound in \eqref{LtildeTA-MtildeTA_6}, we use \eqref{LTA-LtildeTBA_3} and find
\begin{align}
\Vert \tilde L_{T, \Abold} \Delta - \tilde M_{T, \Abold} \Delta \Vert_2^2 &\leq 4 \, \Vert \Psi\Vert_2^2 \int_{\Rbb^3} \dd r \, \Bigl| \iint_{\Rbb^3 \times \Rbb^3} \dd Z \dd s \; |k_T(Z, r-s)| \, |V\alpha_*(s)| \notag \\
&\hspace{-30pt} \times \sup_{X\in \Rbb^3} \bigl| \e^{\i \tilde \Phi_{\Asolo_h}(X, Z, r, s) - \i \Phi_{2\Asolo_h}(X, X+ Z) + \i \frac{r-s}{4} \cdot D\Asolo_h(X) (r + s)} -1\bigr| \Bigr|^2. \label{LtildeTA-MtildeTA_7}
\end{align}
Furthermore, an application of Lemma~\ref{Phi_Approximation} implies
\begin{align}
\int_{\Rbb^3} \dd Z \; |Z|^a \; |k_T(Z, r-s)| \,  \sup_{X\in \Rbb^3} \bigl| \e^{\i \tilde \Phi_{\Asolo}(X, Z, r, s) - \i \Phi_{2\Asolo}(X, X+ Z) + \i \frac{r-s}{4} \cdot D\Asolo(X) (r + s)} -1\bigr| & \notag \\
&\hspace{-280pt} \leq C \; \Vert D^2\Asolo\Vert_\infty \bigl[  F_T^{3+a}(r-s) + F_T^{1+a}(r-s)\; |s|^2\bigr] \label{LtildeTA-MtildeTA_2}
\end{align}
with $F_T^a$ in \eqref{LtildeTA-MtildeTA_FT_definition}. We need this bound here only for the case $a=0$ but state it for general $a\geq0$ for later reference. In combination with \eqref{LtildeTA-MtildeTA_7}, this proves \eqref{LtildeTA-MtildeTA_6}.

We claim that the term involving $\Pi = -\i \nabla + 2 \Abold_{\Bbold}$, which is understood to act on the center-of-mass coordinate, is bounded by
\begin{align}
\Vert \Pi (\tilde L_{T, \Abold}\Delta - \tilde M_{T, \Abold}\Delta)\Vert_2^2 &\leq C\,  h^2 \, \Vert \Psi\Vert_{\Hmag^1(Q_h)}^2 \, \Vert D^2 \Asolo_h\Vert_\infty^2 \, \bigl( 1 +  \Vert D\Asolo_h\Vert_\infty^2\bigr) \notag \\
&\hspace{-1cm} \times \bigl[ \Vert F_T^4 * |V\alpha_*|\, \Vert_2^2 + \Vert F_T^1 * |\cdot| \, |V\alpha_*| \, \Vert_2^2 + \Vert F_T^2* |\cdot|^2|V\alpha_*| \, \Vert_2^2\bigr]. \label{LtildeTA-MtildeTA_3}
\end{align}
If this holds, the desired bound for this term follows from Young's inequality and \eqref{LtildeTA-MtildeTA_FTGT}. To prove \eqref{LtildeTA-MtildeTA_3}, we use \eqref{LtildeTA-MtildeTA_1} and argue as in the proof of \eqref{LTA-LtildeTBA_9} to see that
\begin{align}
&\Vert \Pi (\tilde L_{T, \Abold}\Delta - \tilde M_{T, \Abold}\Delta)\Vert_2^2 \leq C \, h^2 \, \Vert \Psi\Vert_{\Hmag^1(Q_h)}^2 \int_{\Rbb^3} \dd r\, \Big( \; \Bigl| \iint_{\Rbb^3 \times \Rbb^3} \dd Z \dd s \;  \, |V\alpha_*(s)| \notag \\
&\hspace{0.1cm} \times |k_T(Z, r-s)| \Bigl[ h\,  (1 + |Z|)\,   \sup_{X\in\Rbb^3} \bigl| \e^{\i \tilde \Phi_{\Asolo_h}(X, Z, r, s) - \i \Phi_{2\Asolo_h}(X, X+ Z) + \i \frac{r-s}{4} \cdot D\Asolo_h(X) (r + s)} -1\bigr| \notag \\
&\hspace{1.6cm} + \sup_{X\in \Rbb^3} \bigl| \nabla_X \e^{\i \tilde \Phi_{\Asolo_h}(X, Z, r, s)} - \nabla_X \e^{\i \Phi_{2\Asolo_h}(X, X + Z)} \e^{-\i \frac{r-s}{4} \cdot D\Asolo_h(X)(r + s)} \bigr| \Bigr] \Bigr|^2 \Big). \label{LtildeTA-MtildeTA_4}
\end{align}
The difference of the phases involving a gradient can be estimated as
\begin{align*}
&\bigl| \nabla_X \e^{\i \tilde \Phi_{\Asolo}(X, Z, r, s)} - \nabla_X \e^{\i \Phi_{2\Asolo}(X, X + Z)} \e^{-\i \frac{r-s}{4} \cdot D\Asolo(X)(r + s)} \bigr| \notag \\
&\hspace{1cm} \leq \bigl| \e^{\i \tilde \Phi_{\Asolo}(X, Z, r, s) - \i \Phi_{2\Asolo}(X, X+Z) + \i \frac{r-s}{4} \cdot D\Asolo(X) (r + s)} - 1\bigr| |\nabla_X \tilde \Phi_{\Asolo}(X, Z, r, s)| \notag\\
&\hspace{1.5cm} + \bigl| \nabla_X \tilde \Phi_{\Asolo}(X, Z, r, s) - \nabla_X \Phi_{2\Asolo}(X, X + Z) \bigr| + \bigl|\nabla_X (r-s) \cdot D\Asolo(X) (r+s)\bigr|, 
\end{align*}
which, by \eqref{LTA-LtildeTBA_16} and Lemma~\ref{Phi_Approximation}, is bounded by
\begin{align*}
%
& C \, \Vert D^2\Asolo \Vert_\infty \, \bigl(1 + \Vert D\Asolo\Vert_\infty\bigr) \\
&\hspace{50pt} \times \bigl[ \bigl( |s|^2 + |r-s|^2 \bigr) \bigl( 1+ |Z|^2 + |r-s|^2\bigr) + \bigl( |s| + |r-s|\bigr) \bigl( |Z| + |r-s|\bigr)\bigr].
\end{align*}
Accordingly,
\begin{align*}
	&\int_{\Rbb^3} \dd Z \; |k_T(Z, r-s)| \, \sup_{X\in \Rbb^3} \bigl| \nabla_X \e^{\i \tilde \Phi_{\Asolo}(X, Z, r, s)} - \nabla_X \e^{\i \Phi_{2\Asolo}(X, X + Z)} \e^{-\i \frac{r-s}{4} \cdot D\Asolo(X)(r + s)} \bigr| \\
	&\hspace{40pt} \leq C \, \Vert D^2\Asolo\Vert_\infty \bigl( 1 + \Vert D\Asolo\Vert_\infty\bigr) \\
	&\hspace{60pt} \times \bigl[ F_T^4(r-s) + F_T^1(r-s) \, |s| + F_T^2(r-s) \, |s|^2 \bigr].
\end{align*}
Using \eqref{LtildeTA-MtildeTA_2}, we also find
\begin{align*}
	&\int_{\Rbb^3} \dd Z \; |k_T(Z, r-s)| \, h\,  (1 + |Z|)  \sup_{X\in\Rbb^3} \bigl| \e^{\i \tilde \Phi_{\Asolo}(X, Z, r, s) - \i \Phi_{2\Asolo}(X, X+ Z) + \i \frac{r-s}{4} \cdot D\Asolo(X) (r + s)} -1\bigr| \\
	&\hspace{40pt} \leq C \, \Vert D^2\Asolo\Vert_\infty  \; \bigl[ F_T^4(r-s) + F_T^2(r-s) \, |s|^2 \bigr].
\end{align*}
In combination with \eqref{LtildeTA-MtildeTA_4}, this proves \eqref{LtildeTA-MtildeTA_3}.

We claim that the term involving $\tilde \pi = -\i \nabla + \frac{1}{2} \Abold_{\Bbold}$, which is understood to act on the relative coordinate, is bounded by
\begin{align}
\Vert \tilde \pi (\tilde L_{T, \Abold} \Delta - \tilde M_{T, \Abold}\Delta)\Vert_2^2 &\leq C \, h^2 \, \Vert \Psi\Vert_{\Hmag^1(Q_h)}^2 \, \Vert D^2\Asolo_h\Vert_\infty^2 \, \bigl( 1 + \Vert \Asolo_h\Vert_\infty^2 + \Vert D\Asolo_h\Vert_\infty^2\bigr) \notag\\
&\hspace{1cm}\times \bigl \Vert (F_T^4+ G_T^2) * ( 1 + |\cdot|^2 ) |V\alpha_*|  \bigr\Vert_2^2 \notag \\
& \leqslant C \; h^8 \;  \Vert \Psi\Vert_{\Hmag^1(Q_h)}^2 \; \Vert (1+|\cdot|^2) V \alpha_* \Vert_2^2 . \label{LtildeTA-MtildeTA_5} 
\end{align}
The second inequality is a consequence of Young's inequality and \eqref{LtildeTA-MtildeTA_FTGT}. To prove the first inequality in \eqref{LtildeTA-MtildeTA_5}, we first use \eqref{LtildeTA-MtildeTA_1} and argue as in the proof of \eqref{LTA-LtildeTBA_9} to see that
\begin{align*}
&\Vert \tilde \pi (\tilde L_{T, \Abold}\Delta - \tilde M_{T, \Abold}\Delta)\Vert_2^2 \leq C \, h^2 \, \Vert \Psi\Vert_{\Hmag^1(Q_h)}^2 \, \int_{\Rbb^3} \dd r \, \Big( \; \Bigl| \iint_{\Rbb^3 \times \Rbb^3} \dd Z \dd s \; |V\alpha_*(s)| \\
&\hspace{1cm} \times \bigl|\tilde \pi k_T(Z, r-s) \e^{\i \frac{\Bbold}{4} \cdot (r\wedge s)}\bigr| \sup_{X\in \Rbb^3} \bigl| \e^{\i \tilde \Phi_{\Asolo_h}(X, Z, r, s) - \i \Phi_{2\Asolo_h}(X, X+ Z) + \i \frac{r-s}{4} \cdot D\Asolo_h(X) (r + s)} -1\bigr| \\
&\hspace{0.5cm} + |k_T(Z, r-s)| \, \sup_{X\in \Rbb^3} \bigl| \nabla_r \e^{\i \tilde \Phi_{\Asolo_h}(X, Z, r, s)} - \nabla_r \e^{\i \Phi_{2\Asolo_h}(X, X+ Z)} \e^{-\i \frac{r-s}{4} \cdot D\Asolo_h(X) (r + s)}\bigr| \Bigr|^2 \Big).
\end{align*}
A brief computation shows that the operator $\tilde \pi$ obeys the following intertwining relation with respect to $\e^{\i \frac \Bbold 4 \cdot (r\wedge s)}$:
\begin{align}
	\tilde \pi_r \, \e^ {\frac \i 4 \Bbold \cdot (r \wedge s)} = \e^{\frac \i 4 \Bbold \cdot (r\wedge s)} \bigl( -\i \nabla_r + \frac 14 \Bbold \wedge (r-s)\bigr). \label{tildepi_magnetic_phase}
\end{align}
The notation $\tilde \pi_r$ in the above equation highlights on which of the two variables the operator $\tilde \pi$ is acting. An application of this identity shows
\begin{align}
|\tilde \pi k_T(Z,r-s) \e^{\i \frac{\Bbold}{4} (r\wedge s)}| &\leq |\nabla_r k_T(Z, r-s)| + \frac{|\Bbold|}{4} \, |r-s| \, |k_T(Z,r-s)|. \label{eq:A17}
\end{align}
Hence, a computation similar to that leading to \eqref{LtildeTA-MtildeTA_2} shows that 
\begin{align}
\int_{\Rbb^3} \dd Z \; |\tilde \pi k_T(Z, r-s) \e^{\i \frac{\Bbold}{4} \cdot (r\wedge s)}| & \notag \\
&\hspace{-110pt} \times \sup_{X\in \Rbb^3} \bigl| \e^{\i \tilde \Phi_{\Asolo}(X, Z, r, s) - \i \Phi_{2\Asolo}(X, X+ Z) + \i \frac{r-s}{4} \cdot D\Asolo(X) (r + s)} -1\bigr| & \notag \\
&\hspace{-90pt} \leq C \; \Vert D^2\Asolo \Vert_\infty \; \bigl[ (F_T^4 + G_T^3)(r-s) + (F_T^2 + G_T^1)(r-s)\; |s|^2\bigr]  \label{LtildeTA-MtildeTA_9}
\end{align}
with the function $F_T^a$ in \eqref{LtildeTA-MtildeTA_FT_definition} and $G_T^a$ in \eqref{LtildeTA-MtildeTA_GT_definition}. Let us also note that we estimate the factor $|\Bbold|$ coming from the second term in \eqref{eq:A17} by $1$.

We also have 
\begin{align}
&\bigl| \nabla_r \e^{\i \tilde \Phi_{\Asolo}(X, Z, r, s)} - \nabla_r \e^{\i \Phi_{2\Asolo}(X, X+ Z) - \i \frac{r-s}{4} \cdot D\Asolo(X) (r + s)}\bigr| \notag \\
&\hspace{3cm} \leq \bigl| \e^{\i \tilde \Phi_{\Asolo} (X, Z, r, s) - \i \Phi_{2\Asolo}(X, X + Z)+ \i \frac{r-s}{4} \cdot D\Asolo(X) (r + s)} - 1\bigr| |\nabla_r \tilde \Phi_{\Asolo}(X, Z, r, s)| \notag \\
&\hspace{3.5cm} + \bigl| \nabla_r \tilde \Phi_{\Asolo}(X, Z,r, s) + \nabla_r \frac{r-s}{4} \cdot D\Asolo(X) (r+s)\bigr|. \label{eq:A16}
\end{align}
We use 
\begin{align}
	|\nabla_r \tilde \Phi_{\Asolo}(X, Z, r, s)| &\leq C \; \bigl( \Vert \Asolo\Vert_\infty +  \Vert D\Asolo\Vert_\infty \bigr) \, \bigl( \bigl| Z + \frac{r-s}{2} \bigr| + \bigl| Z - \frac{r-s}{2}\bigr| \bigr) \label{LTA-LtildeTBA_17}
\end{align}
and Lemma~\ref{Phi_Approximation} to see that the right side of \eqref{eq:A16} is bounded by
\begin{align*}	
%
&C \, \Vert D^2\Asolo\Vert_\infty \bigl( 1+ \Vert \Asolo\Vert_\infty + \Vert D \Asolo \Vert_\infty \bigr) \; \bigl[  |s|^2 + |r-s|^2 + |Z|^2  \\
&\hspace{40pt} + \bigl( |s| + |r-s|\bigr) \bigl( |Z|+ |r-s|\bigr) \bigl( \bigl| Z+\frac{r-s}{2} \bigr| + \bigl| Z-\frac{r-s}{2} \bigr| \bigr)  \bigr].
\end{align*}
In combination with \eqref{r-s_estimate} these considerations imply
\begin{align*}
\int_{\Rbb^3} \dd Z \; |k_T(Z,r-s)| \, \sup_{X\in \Rbb^3} \bigl| \nabla_r \e^{\i \tilde \Phi_{\Asolo}(X, Z, r, s)} - \nabla_r \e^{\i \Phi_{2\Asolo}(X, X+ Z)} \e^{-\i \frac{r-s}{4} \cdot D\Asolo(X) (r + s)}\bigr| & \\
&\hspace{-350pt} \leq C\, \Vert D^2\Asolo\Vert_\infty \bigl( 1 + \Vert \Asolo \Vert_\infty + \Vert D\Asolo \Vert_\infty \bigr) \\
&\hspace{-300pt} \times \bigl[ F_T^3 (r-s) + F_T^2 (r-s) \, |s| + F_T^0 (r-s) \, |s|^2\bigr].
\end{align*}
When we combine this with \eqref{LtildeTA-MtildeTA_9}, we obtain \eqref{LtildeTA-MtildeTA_5}.
\end{proof}

\paragraph{The operator $M_{T,\Abold}$.} We define the operator $M_{T,\Abold}$ by
\begin{align}
M_{T,\Abold} \alpha(X,r) \coloneqq  \iint_{\Rbb^3 \times \Rbb^3} \dd Z \dd s \; k_T(Z, r-s) \;(\cos(Z\cdot \Pi_{\Abold_h})\alpha)(X,s), \label{MTA_definition}
\end{align}
where $k_T$ is defined below \eqref{MtildeTA_definition}. In our calculation, we may replace $\tilde M_{T, \Abold}$ by $M_{T,\Abold}$ due to the following error bound.

\begin{prop}
\label{MtildeTA-MTA}
Assume that $\Abold = \Abold_{e_3} + A$ with a $A \in W^{2,\infty}(\mathbb{R}^3,\mathbb{R}^3)$ periodic, let $|\cdot|^k V\alpha_*\in L^2(\Rbb^3)$ for $k \in \{ 0,1,2 \}$, $\Psi\in \Hmag^1(Q_h)$, and denote $\Delta \equiv \Delta_\Psi$ as in \eqref{Delta_definition}. For any $T_0 > 0$ there is $h_0>0$ such that for any $0 < h \leq h_0$ and any $T\geq T_0$ we have
\begin{align}
	\Vert \tilde M_{T,\Abold}\Delta - M_{T,\Abold}\Delta \Vert_\Hsymm^2 &\leq C\;h^6 \; \bigl( \max_{k\in \{0,1,2\}} \Vert \, |\cdot|^k V\alpha_*\Vert_2^2\bigr) \;\Vert \Psi\Vert_{\Hmag^1(Q_h)}^2. \label{MtildeTA-MTA_eq1}
\end{align}
Furthermore,
\begin{align}
	|\langle \Delta, \tilde M_{T,\Abold}\Delta - M_{T,\Abold}\Delta\rangle| &\leq C \;h^6 \; \bigl( \Vert V\alpha_*\Vert_2^2 + \Vert\,  |\cdot|^2 V\alpha_*\Vert_2^2 \bigr) \;\Vert \Psi\Vert_{\Hmag^1(Q_h)}^2. \label{MtildeTA-MTA_eq2}
\end{align}
\end{prop}

\begin{bem}
The two bounds in \eqref{MtildeTA-MTA_eq1} and \eqref{MtildeTA-MTA_eq2} are needed for the proof of Proposition~\ref{Structure_of_alphaDelta} and Theorem~\ref{Calculation_of_the_GL-energy}, respectively. We highlight that the bound in \eqref{MtildeTA-MTA_eq1} is not strong enough to be useful in the proof of Theorem~\ref{Calculation_of_the_GL-energy}. More precisely, if we apply Cauchy--Schwarz and use Lemma~\ref{Schatten_estimate} as well as \eqref{MtildeTA-MTA_eq1} to estimate $\Vert \Delta \Vert_2$, we obtain a bound that is only of the order $h^4$. This is not good enough because $h^4$ is the order of the Ginzburg--Landau energy. To obtain \eqref{MtildeTA-MTA_eq2}, we exploit the fact that $V\alpha_*$ is real-valued, which allows us to replace $\exp(-\i \frac{r-s}{4} \cdot D\Abold_h(X)(r+s))$ in the definition of $\tilde M_{T, \Abold}$ by $\cos(\frac{r-s}{4} \cdot D\Abold_h(X)(r+s))$ and to win an additional factor $h^2$. 
%
%
\end{bem}

\begin{proof}[Proof of Proposition \ref{MtildeTA-MTA}]
The proof is similar to that of Proposition \ref{LTA-LtildeTA}. We begin by proving \eqref{MtildeTA-MTA_eq1} and claim that
\begin{align}
\Vert \tilde M_{T, \Abold}\Delta - M_{T, \Abold}\Delta\Vert_2^2 &\leq C \, \Vert \Psi\Vert_2^2 \, \Vert D\Abold_h\Vert_\infty^2 \, \Vert F_T^2 * |V\alpha_*| + F_T^1 * |\cdot|\, |V\alpha_*|\, \Vert_2^2 \label{MtildeTA-MTA_6}
\end{align}
with the function $F_T^a$ in \eqref{LtildeTA-MtildeTA_FT_definition}. If this holds, the desired bound for this term follows from Young's inequality, \eqref{Periodic_Sobolev_Norm}, and the $L^1$-norm estimate on $F_T^a$ in \eqref{LtildeTA-MtildeTA_FTGT}. To prove \eqref{MtildeTA-MTA_6}, we argue as in \eqref{Expanding_the_square}--\eqref{LTA-LtildeTBA_2} and obtain
\begin{align}
\Vert \tilde M_{T, \Abold}\Delta - M_{T, \Abold}\Delta\Vert_2^2 & \leq 4 \, \Vert \Psi\Vert_2^2 \notag \\
&\hspace{-90pt} \times \int_{\Rbb^3} \dd r\,  \Bigl| \iint_{\Rbb^3 \times \Rbb^3}\dd Z \dd s \; |k_T (Z, r-s)| \, \sup_{X\in \Rbb^3} \bigl| \e^{-\i \frac{r-s}{4} \cdot D\Abold_h(X) (r+s)}  - 1\bigr| \;|V\alpha_*(s)|\Bigr|^2. \label{MtildeTA-MTA_4}
\end{align}
When we combine the bound
\begin{align}
\bigl| (r-s) \cdot D\Abold(X) (r+s)\bigr| &\leq \Vert D\Abold \Vert_\infty \, |r-s| \, \bigl( 2 |s| + |r-s|\bigr) \label{MtildeTA-MTA_1}
\end{align}
and the estimate for $|r-s|$ in \eqref{r-s_estimate}, we obtain
\begin{align}
\int_{\Rbb^3} \dd Z \; |k_T (Z, r-s)| \, \sup_{X\in \Rbb^3} \bigl| \e^{-\i \frac{r-s}{4} \cdot D\Abold(X) (r+s)}  - 1\bigr| & \notag \\
&\hspace{-100pt} \leq C\, \Vert D\Abold\Vert_\infty \bigl[ F_T^2(r-s) + F_T^1 (r - s)\;|s|\bigr]. \label{MtildeTA-MTA_5}
\end{align}
In combination with \eqref{MtildeTA-MTA_4}, this proves \eqref{MtildeTA-MTA_6}.

We also claim that the term involving $\Pi$ is bounded by
\begin{align}
\Vert \Pi (\tilde M_{T, \Abold}\Delta - M_{T, \Abold}\Delta)\Vert_2^2 &\leq C\, h^2 \; \Vert \Psi\Vert_{\Hmag^1(Q_h)}^2 \notag \\
&\hspace{-70pt} \times \bigl[ ( |\Bbold|^2 + \Vert D^2\Asolo_h\Vert_\infty^2 )(1+ |\Bbold|^2 + \Vert D^2\Asolo_h\Vert_\infty^2 ) \, \Vert F_T^3 *|V\alpha_*| + F_T^2 * |\cdot| \, |V\alpha_*| \, \Vert_2^2 \notag \\
&\hspace{-50pt} + \Vert D^2\Abold_h\Vert_\infty^2\, \Vert F_T^2 * |V\alpha_*| + F_T^1 * |\cdot|\, |V\alpha_*|\, \Vert_2^2 \bigr]. \label{MtildeTA-MTA_2}
\end{align}
If this is correct, an application Young's inequality and \eqref{LtildeTA-MtildeTA_FTGT} shows the desired bound for this term. To prove \eqref{MtildeTA-MTA_2}, we first 
show that 
\begin{align}
\Vert \Pi \cos(Z\cdot \Pi_{\Abold_h}) \Psi\Vert_2 \leq C \, h^2 \, (1 + |Z|) \, \Vert \Psi\Vert_{\Hmag^1(Q_h)} \label{MtildeTA-MTA_Lemma}
\end{align}
holds. From Lemma~\ref{Magnetic_Translation_Representation} we know that
\begin{align*}
	\e^{\pm \i Z \cdot \Pi_{\Abold_h}} = \e^{\pm \i Z \cdot \Pi} \e^{-\i \Phi_{2\Asolo_h}(X, X \mp Z)}
\end{align*}
holds. An application of the intertwining relation in \eqref{PiU_intertwining} for $\Pi_X$ and $\e^{\i Z\cdot \Pi_X}$ therefore shows
\begin{align*}
	[\Pi, \e^{\pm \i Z\cdot \Pi_{\Abold_h}}] = \e^{\pm \i Z\cdot \Pi_{\Abold_h}} \bigl[ \mp 2\, \Bbold\wedge Z - \nabla_X \Phi_{2\Asolo_h}(X, X \mp Z)\bigr].
\end{align*}
We highlight that $\Pi$ and $\Pi_{\Abold_h}$ in the two equations above act on the coordinate $X$. Using Lemma~\ref{PhiA_derivative}, we check that
\begin{align*}
	\nabla_X \Phi_{2\Asolo}(X, X \mp Z) = 2\Asolo(X\mp Z) - 2\Asolo(X) + 2\tilde \Asolo(X, X\mp Z) - 2\tilde \Asolo (X\mp Z, X)
\end{align*}
holds. Accordingly, $|\nabla_X\Phi_{2\Asolo}(X, X \mp Z)|\leq C\, \Vert D\Asolo\Vert_\infty |Z|$, which implies
\begin{align}
	\Vert [\Pi, \cos(Z\cdot \Pi_{\Abold_h})]\Psi\Vert_2 \leq C \, ( | \Bbold | + \Vert D\Asolo_h\Vert_\infty ) \, |Z| \, \Vert \Psi\Vert_2 \label{PiXcosPiA_estimate}
\end{align}
as well as \eqref{MtildeTA-MTA_Lemma}. Finally, a computation similar to the one leading to \eqref{LTA-LtildeTBA_9}, Lemma~\ref{Phi_Approximation}, \eqref{MtildeTA-MTA_5}, \eqref{MtildeTA-MTA_Lemma}, and the above considerations prove  \eqref{MtildeTA-MTA_2}. It remains to consider the term proportional to $\tilde \pi$.
%
%

With an argument that is similar to the one leading to \eqref{LTA-LtildeTBA_10}, we see that
\begin{align*}
\Vert \tilde \pi (\tilde M_{T, \Abold}\Delta - M_{T, \Abold}\Delta)\Vert_2^2 & \leq 4\, \Vert \Psi\Vert_2^2 \\
&\hspace{-100pt} \times \int_{\Rbb^3} \dd r \, \Bigl| \iint_{\Rbb^3 \times \Rbb^3}\dd Z \dd s \; \bigl| \tilde \pi k_T (Z, r-s) \bigl[ \e^{-\i \frac{r-s}{4} \cdot D\Abold_h(X) \cdot (r + s)} - 1\bigr] \bigr| \; |V\alpha_*(s)|\Bigr|^2.
\end{align*}
We also note that
\begin{align*}
\bigl|\nabla_r (r - s) D\Abold(X) (r+s) \bigr| \leq C \, \Vert D\Abold\Vert_\infty \bigl( |s| + |r-s|\bigr).
\end{align*}
In combination with \eqref{MtildeTA-MTA_1} and $|\Abold_{\Bbold}(r)| \leqslant | \Bbold | \ (|r-s| + |s|)$, this implies
\begin{align*}
&\int_{\Rbb^3} \dd Z\; \bigl| \tilde \pi k_T (Z, r-s) \bigl[ \e^{-\i \frac{r-s}{4} \cdot D\Abold(X) \cdot (r+s)} - 1\bigr] \bigr| \\
&\hspace{2cm}\leq C\, ( |\Bbold| + \Vert D\Asolo\Vert_\infty ) \; \bigl( (F_T^2 + G_T^2)(r-s) + (F_T^0 + G_T^1)(r-s) \; |s|  \bigr)
\end{align*}
with the function $F_T^a$ in \eqref{LtildeTA-MtildeTA_FT_definition} and $G_T^a$ in \eqref{LtildeTA-MtildeTA_GT_definition}. When we apply Young's inequality and \eqref{LtildeTA-MtildeTA_FTGT}, we obtain \eqref{MtildeTA-MTA_eq1}. It remains to prove \eqref{MtildeTA-MTA_eq2}. 

To that end, we need to consider 
\begin{align}
\langle \Delta, \tilde M_{T, \Abold} \Delta - M_{T, \Abold} \Delta \rangle &= 4 \iint_{\Rbb^3 \times \Rbb^3} \dd r \dd s \; \bigl( \e^{-\i \frac{r-s}{4} \cdot D\Abold_h(X) \cdot (r+s)} - 1 \bigr) V\alpha_*(r) V\alpha_*(s)  \notag\\
&\hspace{-20pt} \times \int_{\Rbb^3} \dd Z \; k_T(Z, r-s)  \fint_{Q_h} \dd X \; \ov{\Psi(X)} \cos(Z\cdot \Pi_{\Abold_h})\Psi(X). \label{MtildeTA-MTA_7}
\end{align}
The left side of this equation is real-valued. It therefore equals $1/2$ times the right side plus $1/2$ times the complex conjugate of the right side. When we use that $V \alpha_*$ is real-valued, that the Matsubara frequencies in \eqref{Matsubara_frequencies} satisfy $-\omega_n = \omega_{-(n+1)}$, and the transformation $n \mapsto -n-1$ in the sum in the definition of $k_T(Z,r-s)$, we see that the complex conjugate of the right side equals the same expression with $\exp(-\i \frac{r-s}{4} \cdot D\Abold_h(X) (r+s))$ replaced by its complex conjugate. Using this and the identity $\cos(x) -1 =- 2\sin^2(\frac x2)$ we find
%
\begin{align}
\langle \Delta, \tilde M_{T, \Abold} \Delta - M_{T, \Abold} \Delta \rangle & = -8 \iint_{\Rbb^3 \times \Rbb^3} \dd r \dd s \; \sin^2\left( \frac {(r-s)} {8} \cdot D\Abold_h(X) \cdot (r+s)\right)  \notag \\
&\hspace{-3cm}\times V\alpha_*(r) V\alpha_*(s) \int_{\Rbb^3} \dd Z \; k_T(Z, r-s)  \fint_{Q_h} \dd X \; \ov{\Psi(X)} \cos(Z\cdot \Pi_{\Abold_h})\Psi(X). \label{MtildeTA-MTA_8}
\end{align}
Furthermore, \eqref{MtildeTA-MTA_1} implies
\begin{align*}
\sin^2\bigl(\frac 18 (r-s) \cdot D\Abold_h(X)(r+s) \bigr) \leq C \, \Vert D\Abold_h\Vert_\infty^2\,  |r-s|^2 \, \bigl( |s|^2+ |r-s|^2\bigr).
\end{align*}
We use this bound, \eqref{MtildeTA-MTA_8}, and $\Vert \cos(Z\cdot\Pi_{\Abold_h}) \Vert_{\infty} \leqslant 1$ to see that 
\begin{align*}
|\langle \Delta, \tilde M_{T, \Abold}\Delta - M_{T, \Abold}\Delta\rangle | &\leq \Vert D\Abold_h\Vert_\infty^2 \; \Vert \Psi\Vert_2^2 \; \bigl\Vert |V\alpha_*| \; \bigl( F_T^4 * |V\alpha_*| + F_T^2 * |\cdot |^2 |V\alpha_*| \bigr) \bigr\Vert_1.
\end{align*}
The desired bound in \eqref{MtildeTA-MTA_eq2} follows when we apply Young's inequality and use \eqref{Periodic_Sobolev_Norm} as well as the $L^1$-norm estimate for $F_T^a$ in \eqref{LtildeTA-MtildeTA_FTGT}. This completes the proof of Proposition~\ref{MtildeTA-MTA}.
\end{proof}

%


\subsubsection{Analysis of \texorpdfstring{$M_{T, \Abold}$}{MTA} and calculation of two quadratic terms  in the Ginz\-burg--Landau functional}
\label{Analysis_of_MTA_Section}

We decompose as $M_{T, \Abold} = M_T^{(1)} + M_{T, \Abold}^{(2)} + M_{T, \Abold}^{(3)}$, where
\begin{align}
M_T^{(1)} \alpha(X,r) &\coloneqq  \iint_{\Rbb^3\times \Rbb^3} \dd Z \dd s \; k_T(Z, r-s) \; \alpha(X,s), \label{MT1_definition}\\
M_{T, \Abold}^{(2)} \alpha(X, r) &\coloneqq   -\frac 12 \iint_{\Rbb^3\times \Rbb^3} \dd Z \dd s\; k_T(Z, r-s) \;  (Z\cdot \Pi_{\Abold_h})^2 \; \alpha(X, s), \label{MTB2_definition}\\
M_{T, \Abold}^{(3)} \alpha(X,r) &\coloneqq  \iint_{\Rbb^3\times \Rbb^3} \dd Z \dd s\; k_T(Z, r-s) \; \Rcal(Z\cdot \Pi_{\Abold_h}) \; \alpha(X, s), \label{MTA3_definition}
\end{align}
and $\Rcal(x) = \cos(x) - 1 + \frac 12 x^2$.

\paragraph{The operator $M_T^{(1)}$.} From the quadratic form $\langle \Delta, M_T^{(1)} \Delta \rangle$ we extract the quadratic term without external fields or a gradient in the Ginzburg--Landau functional in \eqref{Definition_GL-functional}. We also obtain a term that cancels the last term on the left side of \eqref{Calculation_of_the_GL-energy}. The relevant computation can be found in \cite[Proposition 4.11]{DeHaSc2021}. For the sake of completeness, we state the result also here. We recall that $\Delta \equiv \Delta_\Psi = -2 V\alpha_* \Psi$ as in \eqref{Delta_definition}.

\begin{prop}
\label{MT1}
Let $V\alpha_*\in L^2(\Rbb^3)$, $\Psi \in \Lmag^2(Q_h)$, and $\Delta \equiv \Delta_\Psi$ as in \eqref{Delta_definition}. Then the following statements hold.
\begin{enumerate}[(a)]
\item We have $M_{\Tc}^{(1)} \Delta (X, r) = -2\, \alpha_* (r) \Psi(X)$.

\item For any $T_0>0$ there is a constant $c>0$ such that for $T_0 \leq T \leq \Tc$ we have
\begin{align*}
\langle \Delta, M_T^{(1)} \Delta - M_{\Tc}^{(1)} \Delta \rangle \geq  c \, \frac{\Tc - T}{\Tc} \; \Vert \Psi\Vert_2^2.
\end{align*}

\item Given $D\in \Rbb$ there is $h_0>0$ such that for $0< h\leq h_0$, and $T = \Tc (1 - Dh^2)$ we have
\begin{align*}
	\langle \Delta, M_T^{(1)} \Delta -  M_{\Tc}^{(1)} \Delta\rangle  = 4\; Dh^2 \; \Lambda_2 \; \Vert \Psi\Vert_2^2 + R(\Delta)
\end{align*}
with the coefficient $\Lambda_2$ in \eqref{GL-coefficient_2}, and
\begin{align*}
	|R(\Delta)| &\leq C \; h^6 \; \Vert V\alpha_*\Vert_2^2\; \Vert \Psi\Vert_{\Hmag^1(Q_h)}^2.
\end{align*}

\item Assume additionally that $| \cdot | V\alpha_*\in L^2(\Rbb^3)$. Then, there is $h_0>0$ such that for any $0< h \leq h_0$, any $\Psi\in \Hmag^1(Q_h)$, and any $T \geq T_0 > 0$ we have 
\begin{align*}
	\Vert M_T^{(1)}\Delta - M_{\Tc}^{(1)}\Delta\Vert_{\Hsymm}^2 &\leq C \, h^2 \, | T - \Tc |^2 \,  \bigl( \Vert V\alpha_*\Vert_2^2 + \Vert \, |\cdot| V\alpha_*\Vert_2^2\bigr) \Vert\Psi\Vert_{\Hmag^1(Q_h)}^2.
\end{align*}
\end{enumerate}
\end{prop}
\begin{bem}
	Parts (a) and (c) of the above proposition are needed for the proof of Theorem~\ref{Calculation_of_the_GL-energy}, part (b) is needed for the proof of Proposition~\ref{Lower_Tc_a_priori_bound}, and part (d) for the proof of Proposition~\ref{Structure_of_alphaDelta}. 
\end{bem}

\paragraph{The operator $M_{T,\Abold}^{(2)}$.} The kinetic term in the Ginzburg--Landau functional in \eqref{Definition_GL-functional} is contained in $\langle \Delta, M_{T,\Abold}^{(2)} \Delta \rangle$ with $M_{T,\Abold}^{(2)}$ defined in \eqref{MTB2_definition}. The following proposition allows us to extract this term.

\begin{prop}
\label{MTB2}
Let $V\alpha_* \in L^2(\Rbb^3)$ be a radial function, let $A\in W^{1,\infty}(\mathbb{R}^3,\mathbb{R}^3)$ be periodic, assume $\Psi\in \Hmag^1(Q_h)$, and denote $\Delta \equiv \Delta_\Psi$ as in \eqref{Delta_definition}. We then have
\begin{align}
\langle \Delta, M_{\Tc,\Abold}^{(2)} \Delta\rangle = - 4\; \Lambda_0 \; \Vert \Pi_{\Abold_h} \Psi\Vert_2^2 \label{MTA2_1}
\end{align}
with $\Lambda_0$ in \eqref{GL-coefficient_1}. Moreover, for any $T \geq T_0 > 0$ we have
\begin{align}
	|\langle \Delta,  M_{T,\Abold}^{(2)} \Delta - M_{\Tc,\Abold}^{(2)}\Delta\rangle| \leq C\; h^4 \; |T - \Tc| \; \Vert V\alpha_*\Vert_2^2 \; \Vert \Psi\Vert_{\Hmag^1(Q_h)}^2. \label{MTA2_2}
\end{align}
\end{prop}

\begin{proof}
The proof is analogous to the proof of \cite[Proposition 4.13]{DeHaSc2021} with the obvious replacements, and is therefore omitted.
\end{proof}

\paragraph{The operator $M_{T,\Abold}^{(3)}$.} The term $\langle \Delta, M_{T,\Abold}^{(3)} \Delta \rangle$ with $M_{T,\Abold}^{(3)}$ in \eqref{MTA3_definition} does not contribute to the Ginzburg--Landau energy. To obtain a bound for it, we need, as in \cite{DeHaSc2021}, the $\Hmag^2(Q_h)$-norm of $\Psi$. 

\begin{prop}
\label{MTB3}
Let $V\alpha_* \in L^2(\Rbb^3)$, let $A\in W^{1,\infty}(\mathbb{R}^3,\mathbb{R}^3)$ be periodic, assume that $\Psi\in \Hmag^1(Q_h)$, and denote $\Delta \equiv \Delta_\Psi$ as in \eqref{Delta_definition}. For any $T_0>0$ there is $h_0>0$ such that for any $T\geq T_0$ and any $0 < h \leq h_0$ we have
\begin{align*}
|\langle \Delta,  M_{T,\Abold}^{(3)} \Delta\rangle| &\leq C \; h^6 \; \Vert V\alpha_*\Vert_2^2 \; \Vert \Psi\Vert_{\Hmag^2(Q_h)}^2.
\end{align*}
\end{prop}

Before we give the proof of Proposition~\ref{MTB3}, we state and prove the following lemma.

\begin{lem}
\label{ZPiX_inequality_quartic}
Assume that $\Abold = \Abold_{e_3} + \Asolo$ with a periodic function $\Asolo \in W^{2,\infty}(\mathbb{R}^3,\mathbb{R}^3)$. 
\begin{enumerate}[(a)]
\item For $\varepsilon >0$ we have
\begin{align}
|Z\cdot \Pi_\Abold|^4 &\leq C\; |Z|^4 \; \bigl[ \Pi_\Abold^4 + \varepsilon \, \Pi_\Abold^2 + \, |\curl \Abold|^2 + \varepsilon^{-1} \, \bigl( |\curl(\curl \Abold)|^2  + |\nabla (\curl \Abold)|^2 \bigr) \bigr]. \label{ZPiX_inequality_quartic_eq}
\end{align}
The gradient in the last term is understood to act on each component of $\curl A$ separately. The result is a vector field with nine components.
\item Assume that $\Psi\in \Hmag^2(Q_h)$. There is a constant $h_0>0$ such that for any $0 < h \leq h_0$, we have
\begin{align*}
\langle \Psi, \, |Z\cdot \Pi_{\Abold_h}|^4 \, \Psi\rangle &\leq C\; h^6 \; |Z|^4 \; \Vert \Psi\Vert_{\Hmag^2(Q_h)}^2.
\end{align*}
\end{enumerate}
\end{lem}

\begin{proof}
We first give the proof of part (a) and start by noting that
\begin{align}
\bigl[\Pi_\Abold^{(i)}, \Pi_\Abold^{(j)}\bigr] = -\i \sum_{k=1}^3 \varepsilon_{ijk} \; (\curl \Abold)_k \label{ZPiX_inequality_quartic_3}
\end{align}
with the Levi--Civita symbol $\varepsilon_{ijk}$, which is defined as $1$ if $(i, j, k)$ is a cyclic permutation of $\{1, 2, 3\}$, as $-1$ if it is an anticyclic permutation, and zero if at least two indices coincide. We claim that
\begin{align}
\Pi_\Abold \; \Pi_\Abold^2\; \Pi_\Abold = \Pi_\Abold^4 + 2 \, |\curl \Abold|^2 - \curl (\curl \Abold) \cdot \Pi_\Abold. \label{PiPi2Pi_equality}
\end{align}
If this holds, then we can use the fact that all terms in the above equation except for the last are self-adjoint, to see that
\begin{align}
\bigl[ \curl (\curl \Abold) , \Pi_\Abold \bigr] = 0. \label{ZPiX_inequality_quartic_4}
\end{align}
To prove \eqref{PiPi2Pi_equality}, we first note that
\begin{align}
\Pi_\Abold \; \Pi_\Abold^2 &= \Pi_\Abold^2 \; \Pi_\Abold +  2 \sum_{i=1}^3 [\Pi_\Abold, \Pi_\Abold^{(i)}] \, \Pi_\Abold^{(i)} + \sum_{i=1}^3 \bigl[\Pi_\Abold^{(i)} , [\Pi_\Abold, \Pi_\Abold^{(i)}] \bigr].
%
\label{ZPiX_inequality_quartic_6}
\end{align}
Moreover, an application of \eqref{ZPiX_inequality_quartic_3} shows that
\begin{align}
%
\sum_{i=1}^3 [\Pi_\Abold, \Pi_\Abold^{(i)}] \, \Pi_\Abold^{(i)} = \i (\curl \Abold) \wedge \Pi_\Abold \label{ZPiX_inequality_quartic_7}
\end{align}
and
\begin{align}
%
\sum_{i=1}^3 \bigl[\Pi_\Abold^{(i)} , [\Pi_\Abold, \Pi_\Abold^{(i)}] \bigr] = - \curl(\curl \Abold). \label{ZPiX_inequality_quartic_8}
\end{align}
We combine \eqref{ZPiX_inequality_quartic_6}-\eqref{ZPiX_inequality_quartic_8} and find
\begin{align*}
\Pi_\Abold \; \Pi_\Abold^2 \; \Pi_\Abold = \Pi_\Abold^4 + 2\i \bigl( (\curl \Abold) \wedge \Pi_\Abold\bigr) \cdot \Pi_\Abold - \curl (\curl \Abold) \cdot \Pi_\Abold
\end{align*}
Using additionally the identity $\i  ( (\curl \Abold) \wedge \Pi_\Abold) \cdot \Pi_\Abold = |\curl \Abold|^2$, we conclude that \eqref{PiPi2Pi_equality} holds.

Our next goal is to prove the formula
\begin{align}
(Z\cdot \Pi_\Abold) \, \Pi_\Abold^2\, (Z\cdot \Pi_\Abold) &= \Pi_\Abold \, (Z\cdot \Pi_\Abold)^2 \, \Pi_\Abold + (Z \cdot \curl(\curl \Abold)) \; (Z \cdot \Pi_\Abold) \notag \\
&\hspace{100pt} + (Z \wedge \Pi_\Abold) \cdot \bigl( (Z\cdot \nabla) (\curl \Abold)\bigr). \label{ZPiX_inequality_quartic_1}
\end{align}
We note that
\begin{align}
(Z\cdot \Pi_\Abold) \, \Pi_\Abold^2\, (Z\cdot \Pi_\Abold) = \sum_{i,j =1}^3 Z_iZ_j \; \Pi_\Abold^{(i)} \, \Pi_\Abold^2 \, \Pi_\Abold^{(j)}  \label{ZPiX_inequality_quartic_2}
\end{align}
and
\begin{align*}
\Pi_\Abold^{(i)} \, \Pi_\Abold^2 \, \Pi_\Abold^{(j)} = \sum_{m = 1}^3 \bigl( \Pi_\Abold^{(m)} \, \Pi_\Abold^{(i)} \, \Pi_\Abold^{(j)} \, \Pi_\Abold^{(m)} + \Pi_\Abold^{(m)} \, \Pi_\Abold^{(i)} \, [ \Pi_\Abold^{(m)}, \Pi_\Abold^{(j)}] + [\Pi_\Abold^{(i)}, \Pi_\Abold^{(m)}] \, \Pi_\Abold^{(m)} \, \Pi_\Abold^{(j)} \bigr).
\end{align*}
The sum in \eqref{ZPiX_inequality_quartic_2} is left unchanged when we exchange the indices $i$ and $j$. Motivated by this, we combine $\frac 12$ times the original term and $\frac 12$ times the term with $i$ and $j$ interchanged and find
\begin{align}
\Pi_\Abold^{(i)} \, \Pi_\Abold^2 \, \Pi_\Abold^{(j)} + \Pi_\Abold^{(j)} \, \Pi_\Abold^2 \, \Pi_\Abold^{(i)} &= \sum_{m=1}^3 \bigl( \Pi_\Abold^{(m)} \, \Pi_\Abold^{(i)} \, \Pi_\Abold^{(j)} \, \Pi_\Abold^{(m)} + \Pi_\Abold^{(m)} \, \Pi_\Abold^{(j)} \, \Pi_\Abold^{(i)} \, \Pi_\Abold^{(m)} \notag \\
&\hspace{-30pt} + \bigl[ [\Pi_\Abold^{(i)}, \Pi_\Abold^{(m)}] , \Pi_\Abold^{(m)} \, \Pi_\Abold^{(j)}\bigr] + \bigl[ [\Pi_\Abold^{(j)}, \Pi_\Abold^{(m)}] , \Pi_\Abold^{(m)} \, \Pi_\Abold^{(i)}\bigr] \bigr). \label{ZPiX_inequality_quartic_9}
\end{align}
Using the commutator identity $[A, BC] = B\, [A, C] + [A, B] \, C$ we write the third term as
\begin{align}
\bigl[ [\Pi_\Abold^{(i)}, \Pi_\Abold^{(m)}] , \Pi_\Abold^{(m)} \, \Pi_\Abold^{(j)}\bigr] &= \Pi_\Abold^{(m)} \bigl[ [\Pi_\Abold^{(i)}, \Pi_\Abold^{(m)}] , \Pi_\Abold^{(j)}\bigr] + \bigl[ [\Pi_\Abold^{(i)}, \Pi_\Abold^{(m)}] , \Pi_\Abold^{(m)} \bigr] \, \Pi_\Abold^{(j)} \label{ZPiX_inequality_quartic_10}
\end{align}
and likewise for the term with $i$ and $j$ interchanged. 
We also use \eqref{ZPiX_inequality_quartic_8} to see that
\begin{align*}
\frac 12 \sum_{i,j =1}^3 Z_iZ_j \Bigl( \bigl[ [\Pi_\Abold^{(j)} , \Pi_\Abold] , \Pi_\Abold\bigr] \Pi_\Abold^{(i)} +  \bigl[ [\Pi_\Abold^{(i)} , \Pi_\Abold] , \Pi_\Abold\bigr] \Pi_\Abold^{(j)} \Bigr) = (Z \cdot \curl(\curl \Abold) )\; (Z\cdot \Pi_\Abold)
\end{align*}
holds. Concerning the first term on the right side of \eqref{ZPiX_inequality_quartic_10}, \eqref{ZPiX_inequality_quartic_3} can be used to show
\begin{align*}
\bigl[ [\Pi_\Abold^{(i)}, \Pi_\Abold^{(m)}] , \Pi_\Abold^{(j)}\bigr] = \sum_{k=1}^3 \varepsilon_{imk} \, \partial_j (\curl \Abold)_k),
\end{align*}
which implies
\begin{align}
&\frac 12 \sum_{i,j=1}^3 Z_i Z_j \bigl( \Pi_{\Abold} \bigl[ [ \Pi_{\Abold}^{(i)}, \Pi_{\Abold} ], \Pi_{\Abold}^{(j)} \bigr] + \Pi_{\Abold} \bigl[ [ \Pi_{\Abold}^{(j)}, \Pi_{\Abold} ], \Pi_{\Abold}^{(i)} \bigr] \bigr) = (Z \wedge \Pi_\Abold) \cdot \bigl( (Z\cdot \nabla) (\curl \Abold)\bigr). \label{ZPiX_inequality_quartic_11}
\end{align}
When we combine \eqref{ZPiX_inequality_quartic_2}-\eqref{ZPiX_inequality_quartic_11}, this proves proves \eqref{ZPiX_inequality_quartic_1}. We are now prepared to give the proof of \eqref{ZPiX_inequality_quartic_eq}.

%

We start by noting that $|A + B + C|^2 \leq 3(|A|^2 + |B|^2 + |C|^2)$ holds for three linear operators $A, B, C$, which implies
\begin{align}
(Z\cdot \Pi_\Abold)^2 &\leq 3 \; \bigl( Z_1^2 \; (\Pi_\Abold^{(1)})^2 + Z_2^2 \; (\Pi_\Abold^{(2)})^2 + Z_3^2 \; (\Pi_\Abold^{(3)})^2\bigr) \leq 3 \; |Z|^2 \; \Pi_\Abold^2. \label{ZPiX_inequality}
\end{align}
We use \eqref{PiPi2Pi_equality}, \eqref{ZPiX_inequality_quartic_1}, and \eqref{ZPiX_inequality} to show
\begin{align*}
(Z\cdot \Pi_\Abold) \, \Pi_\Abold^2\, (Z\cdot \Pi_\Abold) &\leq 3\, |Z|^2 \, \bigl(\Pi_\Abold^4+ 2\, |\curl \Abold|^2 - \curl(\curl \Abold) \cdot \Pi_\Abold\bigr) \\
&\hspace{10pt} + (Z \cdot \curl(\curl \Abold)) \, (Z\cdot \Pi_\Abold) + (Z \wedge \Pi_\Abold) \cdot \bigl( (Z\cdot \nabla) (\curl \Abold)\bigr). 
\end{align*}
Next, we write $|Z\cdot \Pi_\Abold|^4 = (Z\cdot \Pi_\Abold)  (Z\cdot \Pi_\Abold)^2 (Z\cdot \Pi_\Abold)$, apply \eqref{ZPiX_inequality} to the term in the middle, and find 
\begin{align*}
|Z\cdot \Pi_\Abold|^4 &\leq 9 \, |Z|^4 \bigl( \Pi_\Abold^4 + 2\, |\curl \Abold|^2 - \curl (\curl \Abold) \cdot \Pi_\Abold \bigr) \\
&\hspace{50pt} + 3 \, |Z|^2 \bigl[ (Z \cdot \curl(\curl \Abold)) \, (Z\cdot \Pi_\Abold) + (Z \wedge \Pi_\Abold) \cdot \bigl( (Z\cdot \nabla) (\curl \Abold)\bigr)\bigr].
\end{align*}
Moreover, $AB + BA \leq  \varepsilon \, A^2 + \frac{1}{4 \varepsilon} \, B^2$ for $\varepsilon>0$ and self-adjoint operators $A$ and $B$, \eqref{ZPiX_inequality}, and $|Z\wedge \Pi_\Abold|^2 \leqslant 3 |Z|^2 \Pi_\Abold^2$ imply that the right side of the above equation is bounded by
\begin{align*}
C \, |Z|^4 \bigl[ \Pi_\Abold^4 + \, |\curl \Abold|^2 + \varepsilon \, \Pi_\Abold^2 + \varepsilon^{-1} \bigl( |\curl(\curl \Abold)|^2  + |\nabla (\curl \Abold)|^2 \bigr)\bigr].
\end{align*}
This proves part (a). 

The proof of part (b) is a direct consequence of part (a) with the choice $\varepsilon = h^2$, and is therefore left to the reader. This proves Lemma~\ref{ZPiX_inequality_quartic}.
%
\end{proof}


\begin{proof}[Proof of Proposition \ref{MTB3}] We use the definition of $M_{T, \Abold}^{(3)}$ to write
\begin{align}
\langle \Delta, M_{T, \Abold}^{(3)} \Delta \rangle &= 4 \iiint_{\Rbb^3\times \Rbb^3\times \Rbb^3} \dd r\dd s\dd Z \; V\alpha_*(r) V\alpha_*(s)\, k_T(Z, r-s) \; \langle \Psi , \Rcal(Z\cdot \Pi_{\Abold_h})\Psi\rangle. \label{MTB3_1}
\end{align}
The function $\Rcal(x) = \cos(x) - 1 + \frac{x^2}{2}$ satisfies the bound $0\leq\Rcal(x) \leq \frac{1}{24} x^4$, and hence an application of Lemma~\ref{ZPiX_inequality_quartic} shows
\begin{align}
\langle \Psi, \Rcal(Z\cdot \Pi_{\Abold_h}) \Psi\rangle &\leq C\; h^6 \; |Z|^4 \; \Vert \Psi\Vert_{\Hmag^2(Q_h)}^2. \label{MTB3_2}
\end{align}
When we apply the estimate $|Z|^4 \leq | Z + \frac{r}{2} |^4 + | Z - \frac{r}{2} |^4$, we see that
\begin{align}
\int_{\Rbb^3} \dd Z\; |Z|^4 \; |k_T(Z,r)| &\leq F_T^4(r) \label{MTB3_3}
\end{align}
holds with $F_T^4$ defined in \eqref{LtildeTA-MtildeTA_FT_definition}. But this also shows
\begin{equation*}
	|\langle \Delta, M_{T, \Abold}^{(3)} \Delta \rangle| \leqslant C\; h^6 \; \Vert \Psi\Vert_{\Hmag^2(Q_h)}^2 \; \Vert (V \alpha_*) F_T^4 \ast (V \alpha_*) \Vert_1.
\end{equation*}
In combination with the $L^1(\Rbb^3)$-norm bound for $F_T^4$ in \eqref{LtildeTA-MtildeTA_FTGT}, this proves the claim.
\end{proof}


\subsubsection{A representation formula for the operator \texorpdfstring{$\Woperator_{T,\Abold}$}{WTA}}
\label{LTAW_action_Section}

In the next five subsections we study the operator $\Woperator_{T, \Abold}$ in \eqref{LTA^W_definition}. In particular, we extract the term in the GL functional that is proportional to $W$ from $\langle \Delta, \Woperator_{T, \Abold} \Delta \rangle$. The operator $\Woperator_{T, 0}$ has previously been studied in \cite{ProceedingsSpohn}. After the magnetic field has been removed, our analysis mostly follows ideas in this reference. Because of this and because several ideas of the previous sections appear again, we keep our presentation rather short and only mention the main ideas. As in the case of $L_{T, \Abold}$, we start our analysis with a representation formula for $\Woperator_{T, \Abold}$ in terms of relative and center-of-mass coordinates.

\begin{lem}
\label{LTAW_action}
The operator $\Woperator_{T, \Abold} \colon \Lsymm \ra \Lsymm$ in \eqref{LTA^W_definition} acts as
\begin{align}
\Woperator_{T, \Abold} \alpha(X, r) &= \iint_{\Rbb^3 \times \Rbb^3} \dd Z \dd s \; k_{T, \Abold, W}(X, Z, r, s) \; (\e^{\i Z \cdot (-\i \nabla_X)} \alpha)(X, s),
\label{eq:Andinew2}
\end{align}
where
\begin{align}
k_{T, \Abold, W} (X, Z, r, s) &\coloneqq  \smash{\frac 2\beta \sum_{n\in \Zbb} \int_{\Rbb^3} \dd Y} \; W_h(X + Y) \bigl[  k_{T, \Abold, +}^n(X, Y, Z, r, s) \; \e^{\i \Theta_{\Abold_h}^+(Y, Z, r, s)} \notag \\
&\hspace{80pt} + k_{T, \Abold, -}^n(X, Y, Z, r, s) \; \e^{\i  \Theta_{\Abold_h}^-(Y, Z, r, s)} \bigr], \label{kTAW_definition}
\end{align}
where
\begin{align}
k_{T, \Abold, \pm}^n(X, Y, Z, r, s) &\coloneqq  g_h^{\pm \i \omega_n} \bigl( X \pm \frac r2, X + Y\bigr) \; g_h^{\pm \i \omega_n} \bigl( X + Y, X + Z \pm \frac s2\bigr)\notag \\
&\hspace{120pt}  \times g_h^{\mp \i \omega_n} \bigl( X \mp \frac r2, X + Z \mp \frac s2\bigr). \label{kTAWn_definition}
\end{align}
The function $g_h^z$ is defined in \eqref{ghz_definition} and
\begin{align}
\Theta_\Abold^\pm (Y, Z,r, s) &\coloneqq  \Phi_\Abold \bigl( X \pm \frac r2, X + Y\bigr) + \Phi_\Abold \bigl( X + Y, X + Z \pm \frac s2\bigr)\notag \\
&\hspace{120pt} + \Phi_\Abold \bigl( X \mp \frac r2, X + Z \mp \frac s2\bigr). \label{LTA^W_Phi_definition}
\end{align}

\end{lem}

\begin{proof}
The proof that $\Woperator_{T, \Abold}$ is a bounded linear map on $\Lsymm$ goes along the same lines as that of Lemma~\ref{lem:propsLN}. The proof of the representation formula is analogous to the proof of Lemma~\ref{LTA_action}. We use \eqref{GAz_Kernel_of_complex_conjugate} for $W=0$ to write
\begin{align}
\Woperator_{T, \Abold} \alpha(x, y) &=  \frac 2\beta \sum_{n\in \Zbb} \iiint_{\Rbb^9} \dd u \dd v \dd w \; \bigl[ G_h^{\i \omega_n} (x, u) \, W_h(u) \, G_h^{\i\omega_n} (u,v) \, \alpha(v, w) \, G_h^{-\i\omega_n}(y, w) \notag \\
&\hspace{60pt} + G_{\Abold_h}^{\i\omega_n} (x, v) \, \alpha(v, w) \, G_h^{-\i\omega_n} (u,w) \, W_h(u) \, G_h^{-\i\omega_n}(y, u)\bigr].
\label{eq:Andinew}
\end{align}
When we define the coordinates $X = \frac{x + y}{2}$ and $r = x-y$, apply the change of variables
\begin{align*}
u &= X + Y, & v &= X + Z + \frac s2, & w &= X + Z - \frac s2,
\end{align*}
and use \eqref{ghz_definition}, this yields \eqref{eq:Andinew2}. We highlight that, by a slight abuse of notation, we denoted the function $\alpha$ depending on the original coordinates in \eqref{eq:Andinew} and the function depending on relative and center-of-mass coordinates in \eqref{eq:Andinew2} by the same symbol.
\end{proof}

\subsubsection{Approximation of the operator \texorpdfstring{$\Woperator_{T, \Abold}$}{WTA}}
\label{Approximation_of_LTA^W_Section}

The operator $\Woperator_{T, \Abold}$ will be analyzed in three steps. More precisely, we write
\begin{align}
\Woperator_{T, \Abold} = \bigl( \Woperator_{T, \Abold} - \tilde \Woperator_{T, \Abold} \bigr) + \bigl( \tilde \Woperator_{T, \Abold} - \Woperator_T \bigr) + \Woperator_T, 
\label{eq:Andinew3}
\end{align}
where $\tilde \Woperator_{T, \Abold}$ and $\Woperator_T$ are operators of increasing simplicity in their dependence on $W$ and $\Abold$. They are defined below in \eqref{LtildeTAW_definition} and \eqref{MTW_definition}, respectively. The term in the Ginzburg--Landau functional that is proportional to $W$ will be extracted from the expectation of the operator $\Woperator_T$ with respect to $\Delta$. The expectation of the first two terms in \eqref{eq:Andinew3} will be shown to be negligible.

\paragraph{The operator $\tilde \Woperator_{T, \Abold}$.}

We define the operator 
\begin{align}
\tilde \Woperator_{T, A} \alpha(X, r) &\coloneqq  W_h(X) \iint_{\Rbb^3 \times \Rbb^3} \dd Z \dd s \; k_{T, \Abold, 0}(X, Z, r, s) \; (\e^{\i Z \cdot (-\i \nabla_X)} \alpha)(X, s), \label{LtildeTAW_definition}
\end{align}
where $k_{T, \Abold, W}$ is defined in \eqref{kTAW_definition}. The following proposition allows us to estimate the expectation of the first term in \eqref{eq:Andinew3} with respect to $\Delta$.

\begin{prop}
\label{LTAW-LtildeTAW}
Let $| \cdot|^k  V\alpha_* \in L^2(\Rbb^3)$ for $k \in \{ 0,1 \}$, let $A\in W^{3,\infty}(\mathbb{R}^3,\mathbb{R}^3)$ and $W \in W^{1,\infty}(\mathbb{R}^3,\mathbb{R})$ be periodic, assume $\Psi\in \Hmag^1(Q_h)$, and denote $\Delta \equiv \Delta_\Psi$ as in \eqref{Delta_definition}. For any $T_0>0$ there is $h_0>0$ such that for any $T\geq T_0$ and any $0 < h \leq h_0$ we have
\begin{align*}
|\langle \Delta, \Woperator_{T, \Abold} \Delta - \tilde \Woperator_{T, \Abold} \Delta \rangle| \leq C\,  h^5\,  \max_{k=0,1} \Vert \, |\cdot|^k \ V\alpha_*\Vert_2^2 \, \Vert \Psi\Vert_{\Hmag^1(Q_h)}^2.
\end{align*}
\end{prop}

\begin{proof}
We write
\begin{align}
|\langle \Delta, \Woperator_{T, \Abold}\Delta- \tilde \Woperator_{T, \Abold} \Delta \rangle | & \notag \\
&\hspace{-110pt}\leq 4 \, \Vert \Psi\Vert_2^2 \iiint_{\Rbb^9} \dd Z \dd r \dd s \; \esssup_{X\in \Rbb^3} |(k_{T, \Abold, W} - k_{T, \Abold, 0})(X, Z, r, s)| \; |V\alpha_*(r)| \, |V\alpha_*(s)| \label{LTAW-LtildeTAW_2}
\end{align}
and note that
\begin{align}
|(k_{T, \Abold, W} - k_{T, \Abold, W})(X, Z, r, s)| & \notag\\
&\hspace{-140pt} \leq \frac 2\beta \sum_{n\in \Zbb} \int_{\Rbb^3} \dd Y \; |W_h(X + Y) - W_h(X)| \; \bigl( |k_{T, \Abold, +}^n| + |k_{T, \Abold, -}^n|\bigr)(X, Y, Z, r, s). 
%
\label{LTAW-LtildeTAW_1}
\end{align}
When we use \eqref{LTAW-LtildeTAW_1}, $|W(X + Y) - W(X)|\leq \Vert \nabla W\Vert_\infty \, |Y|$, $|Y| \leq |Y \pm \frac r2| + |\frac r 2|$, and Proposition~\ref{gh-g_decay}, we see that the right side of \eqref{LTAW-LtildeTAW_2} is bounded by a constant times
\begin{align*}
\Vert \Psi \Vert_2^2 \; \Vert \nabla W_h \Vert_{\infty} \; \Vert (1+|\cdot|) (V \alpha_*) \tilde F_{T}^1 \ast | V \alpha_* | \ \Vert_1, 
\end{align*}
where 
\begin{align*}
\tilde F_{T}^a \coloneqq  \frac 2\beta \sum_{n\in \Zbb} \sum_{b=0}^a \left[ \bigl( |\cdot|^b \, \gzfunction^{ \i\omega_n}\bigr) * \gzfunction^{\i\omega_n} * \gzfunction^{- \i \omega_n} + \bigl( |\cdot|^b \, \gzfunction^{- \i\omega_n}\bigr) * \gzfunction^{- \i\omega_n} * \gzfunction^{ \i \omega_n} \right].
\end{align*}
With Proposition~\ref{gh-g_decay} and \eqref{g0_decay_f_estimate1} we show that $\Vert \tilde F_{T}^a\Vert_1 \leq C$. In combination with the bound $\Vert \nabla W_h\Vert_\infty \leq Ch^3$, this proves the claim.
%
\end{proof}

\paragraph{The operator $\Woperator_T$.}

We define the operator $\Woperator_T$ by
\begin{align}
\Woperator_T \alpha(X, r) &\coloneqq  W_h(X) \iint_{\Rbb^3 \times \Rbb^3} \dd Z \dd s \; k_T(Z, r-s) \; \alpha(X, s) \label{MTW_definition}
\end{align}
%
where
\begin{align}
k_T(Z, r) &\coloneqq  \frac 2\beta \sum_{n\in \Zbb} \bigl( k_{T, +}^n (Z, r) + k_{T,-}^n(Z, r) \bigr) \label{kTW_definition}
\end{align}
and
\begin{align}
k_{T, \pm}^n(Z, r) &\coloneqq  (g_0^{\pm \i \omega_n} * g_0^{\pm \i \omega_n})\bigl( Z \mp \frac r2\bigr)\;  g_0^{\mp \i \omega_n} \bigl( Z \pm \frac r2\bigr). \label{kTWn_definition}
\end{align}

\begin{prop}
\label{MtildeTAW-MTW}
Let $| \cdot|^k  V\alpha_* \in L^2(\Rbb^3)$ for $k \in \{ 0,1 \}$, let $A\in W^{3,\infty}(\mathbb{R}^3,\mathbb{R}^3)$ and $W \in L^{\infty}(\mathbb{R}^3,\mathbb{R})$ be periodic, assume $\Psi\in \Hmag^1(Q_h)$, and denote $\Delta \equiv \Delta_\Psi$ as in \eqref{Delta_definition}. For any $T_0>0$ there is $h_0>0$ such that for any $T\geq T_0$ and any $0 < h \leq h_0$ we have
\begin{align*}
|\langle \Delta, \tilde \Woperator_{T, \Abold} \Delta - \Woperator_T \Delta \rangle| \leq C \,  h^5 \, \max_{k=0,1} \Vert \, |\cdot|^k \ V\alpha_*\Vert_2^2 \; \Vert \Psi\Vert_{\Hmag^1(Q_h)}^2.
\end{align*}
\end{prop}

\begin{proof}[Sketch of proof]
The proof goes along the same lines as that of Propositions~\ref{LTA-LtildeTA}, \ref{LtildeTA-MtildeTA}, and \ref{MtildeTA-MTA} with the notable simplification that we only need to prove bounds for the quadratic form. We therefore only mention the main steps that need to be carried out and leave the details to the reader. In the first step $g_h^z$ is replaced by $g_0^z$. In the second step the part of the magnetic phase coming from $\Abold_\Bbold$ is split off. A careful analysis shows that
\begin{align*}
\Theta_{\Abold_\Bbold}^\pm(X, Y, Z, r, s) = Z \cdot (\Bbold \wedge X) + \frac \Bbold 2 \cdot \theta_\pm(Y, Z, r, s),
\end{align*}
where
\begin{align}
\theta_\pm(Y, Z, r, s) &\coloneqq  \pm \frac r2\wedge \bigl(Y \mp \frac r2\bigr) + \bigl( Y \mp \frac r2\bigr) \wedge \bigl( Z - Y \pm \frac s2\bigr) \notag\\
&\hspace{60pt} \pm \frac r2\wedge \bigl( Z - Y\pm \frac s2\bigr) \mp \frac r2 \wedge \bigl( Z\pm \frac{r-s}{2}\bigr). \label{LTA^W_Phi_definition}
\end{align}
The phase $\exp(\i Z \cdot (\Bbold \wedge X))$ and $\exp(\i Z \cdot (-\i \nabla_X))$ are combined and give $\exp(\i Z \cdot \Pi)$, see \eqref{Magnetic_Translation_constant_Decomposition}. In the third step the magnetic phases coming from $\theta_\pm$ and from the periodic vector potential $\Asolo$, that is, from $\Theta_{\Asolo}$, are removed. Afterwards, the emergent symmetry of the integrand under the transformation $Z \rightarrow -Z$ is used to replace the operator $\exp(\i Z \cdot \Pi)$ by $\cos(Z\cdot \Pi)$. In the final step, we apply the estimate $1 - \cos(Z\cdot \Pi)\leq C |Z|^2 \, \Pi^2$. This ends our sketch of proof. 
\end{proof}


\subsubsection{Analysis of \texorpdfstring{$\Woperator_T$}{WT} and calculation of the quadratic \texorpdfstring{$W$}{W}-term}
\label{Analysis_of_MTW_Section}

\begin{prop}
\label{MTW}
Let $V\alpha_* \in L^2(\Rbb^3)$, let $W \in L^{\infty}(\mathbb{R}^3,\mathbb{R})$ be periodic, assume that  $\Psi\in \Hmag^1(Q_h)$, and denote $\Delta \equiv \Delta_\Psi$ as in \eqref{Delta_definition}. There is $h_0>0$ such that for any $0< h \leqslant h_0$ we have
\begin{align}
\langle \Delta, \Woperator_\Tc \Delta\rangle = -4\; \Lambda_1 \; \langle \Psi, W_h \Psi\rangle  \label{MTW_1}
\end{align}
with $\Lambda_1$ in \eqref{GL-coefficient_W}. Moreover, for any $T \geq T_0 > 0$ we have
\begin{align}
|\langle \Delta,  \Woperator_T \Delta - \Woperator_\Tc \Delta\rangle| \leq C\; h^4 \; |T - \Tc| \; \Vert V\alpha_*\Vert_2^2 \; \Vert \Psi\Vert_{\Hmag^1(Q_h)}^2. \label{MTW_2}
\end{align}
\end{prop}

\begin{proof}
Using the definition of $k_T(Z,r)$ in \eqref{kTW_definition}, we write
\begin{align*}
k_T(Z, r) &= - \frac 2\beta \sum_{n\in \Zbb} \int_{\Rbb^3} \frac{\dd p}{(2\pi)^3} \int_{\Rbb^3} \frac{\dd q}{(2\pi)^3} \; \Bigl[ \frac{\e^{\i Z\cdot (p+q)} \e^{\i \frac r2\cdot (q-p)}}{(\i \omega_n + \mu - p^2)^2 (\i\omega_n - \mu + p^2)} \\
&\hspace{150pt} - \frac{\e^{\i Z\cdot (p+q)} \e^{\i \frac r2\cdot (p-q)}}{(\i\omega_n - \mu + p^2)^2 (\i\omega_n + \mu - p^2)}\Bigr]
\end{align*}
as well as
\begin{align*}
\int_{\Rbb^3} \dd Z \; k_T(Z, r) &= -\frac 4\beta \sum_{n\in \Zbb} \int_{\Rbb^3} \frac{\dd p}{(2\pi)^3} \; \e^{\i r \cdot p} \, \frac{p^2 -\mu }{(\i \omega_n + \mu - p^2)^2 (\i\omega_n - \mu + p^2)^2}.
\end{align*}
With the Mittag-Leffler series expansion in \eqref{tanh_Matsubara}, we also check that
\begin{align}
\frac{\beta}{2} \frac{1}{\cosh^2(\frac \beta 2z)} = \frac{\dd}{\dd z} \tanh\bigl( \frac \beta 2 z\bigr) = - \frac{2}{\beta} \sum_{n\in \Zbb} \frac{1}{(\i\omega_n - z)^2} \label{cosh2_Matsubara}
\end{align}
holds. We use \eqref{cosh2_Matsubara} and the partial fraction expansion
\begin{align*}
	\frac{1}{(\i\omega_n - E)^2(\i\omega_n + E)^2} = \frac{1}{4E^2} \Bigl[ \frac{1}{(\i\omega_n - E)^2} + \frac{1}{(\i\omega_n + E)^2}\Bigr] - \frac{1}{4E^3} \Bigl[ \frac{1}{\i\omega_n - E} - \frac{1}{\i\omega_n + E}\Bigr]
\end{align*}
to see that
\begin{align*}
\frac{4}{\beta} \sum_{n\in \Zbb} \frac{E}{(\i\omega_n - E)^2(\i\omega_n + E)^2} = \beta^2 \; g_1(\beta E)
\end{align*}
with the function $g_1$ in \eqref{XiSigma}. Therefore,
\begin{align*}
\langle \Delta, M_\Tc^W \Delta\rangle &= -\betac^2 \int_{\Rbb^3} \frac{\dd p}{(2\pi)^3} \; |(-2)\hat{V\alpha_*}(p)|^2 \, g_1(\betac (p^2-\mu)) \; \langle \Psi, W_h \Psi\rangle \\
&= - 4\, \Lambda_1 \, \langle \Psi, W_h\Psi\rangle.
\end{align*}
This proves \eqref{MTW_1}. 

To obtain the bound in \eqref{MTW_2}, we argue as in the proof of \eqref{MTA2_2}. 
\end{proof}


\subsubsection{Summary: The quadratic terms}
\label{Summary_quadratic_terms_Section}

In this section, we summarize our results concerning the quadratic terms (in $\Delta$) that are relevant for the proof of Theorem~\ref{Calculation_of_the_GL-energy}. We also use our results to prove another statement (Proposition~\ref{Rough_bound_on_BCS energy} below), which will later be used in the proof of Proposition~\ref{Lower_Tc_a_priori_bound}. We start by summarizing our findings.

Let the assumptions of Theorem~\ref{Calculation_of_the_GL-energy} hold and recall the definition of $\Rcal_{T, \Abold, W}^{(2)}$ in \eqref{RTAW2_definition}. An application of H\"older's inequality in \eqref{Schatten-Hoelder} and the bound $\Vert ( \i \omega_n - \hfrak_\Abold )^{-1} \Vert_{\infty} \leqslant |\omega_n|^{-1}$ show that 
\begin{equation*}
	| \langle \Delta, \Rcal_{T, \Abold, W}^{(2)}\Delta \rangle | \leqslant C \; \Vert \Delta \Vert_2^2 \; \Vert W_h \Vert^2_{\infty} \leqslant C \; h^4 \; \Vert \Psi\Vert_{2}^2 \leqslant C \; h^6 \; \Vert \Psi\Vert_{\Hmag^2(Q_h)}^2.
\end{equation*} 
We combine \eqref{LTAW_decomposition}, this bound, and the results of Propositions~\ref{MT1}, \ref{MTB2}, \ref{MTB3}, \ref{LTAW-LtildeTAW}, \ref{MtildeTAW-MTW}, and \ref{MTW}, to see that for $T = \Tc(1-Dh^2)$ with $D \in \Rbb$ the identity
\begin{align}
	-\frac{1}{4} \langle \Delta, L_{T,\Abold, W} \Delta \rangle + \Vert \Psi \Vert_2^2 \, \langle \alpha_*, V \alpha_* \rangle & \notag \\
	&\hspace{-80pt} = \Lambda_0 \; \Vert \Pi_{\Abold_h}\Psi\Vert_2^2 + \Lambda_1 \; \langle \Psi, W_h\Psi\rangle - Dh^2 \; \Lambda_2 \; \Vert \Psi\Vert_2^2  +  R_2(\Delta) \label{eq:A15}
\end{align}
holds. The remainder term $R_2(\Delta)$ obeys the estimate
\begin{align*}
	| R_2(\Delta) | \leq C\;  \bigl( h^5 \; \Vert \Psi\Vert_{\Hmag^1(Q_h)}^2 + h^6 \; \Vert \Psi\Vert_{\Hmag^2(Q_h)}^2\bigr).
\end{align*}
This concludes the computation of the quadratic terms in the Ginzburg--Landau functional. It remains to compute the term that is proportional to $| \Psi |^4$, which is the content of the remaining part of Section~\ref{Calculation_of_the_GL-energy_proof_Section}.

Before we continue with the proof of Theorem~\ref{Calculation_of_the_GL-energy}, we state and prove the following statement, which will later be used in the proof of Proposition~\ref{Lower_Tc_a_priori_bound}. It is a straightforward consequence of our results for the quadratic terms, and we therefore prove it here.
\begin{prop}
\label{Rough_bound_on_BCS energy}
Let $| \cdot|^k  V\alpha_* \in L^2(\Rbb^3)$ for $k \in \{ 0,1,2 \}$, let $A\in W^{3,\infty}(\mathbb{R}^3,\mathbb{R}^3)$ and $W \in W^{1,\infty}(\mathbb{R}^3,\mathbb{R})$ be periodic, assume $\Psi\in \Hmag^1(Q_h)$, and denote $\Delta \equiv \Delta_\Psi$ as in \eqref{Delta_definition}. For any $T_0>0$ there is $h_0>0$ such that for any $T\geq T_0$ and any $0 < h \leq h_0$ we have
\begin{align}
- \frac 14 \langle \Delta, L_{T,\Abold, W} \Delta\rangle + \Vert \Psi\Vert_2^2 \; \langle \alpha_*, V\alpha_*\rangle & \leq c  \, \frac{T - \Tc}{\Tc}\, \Vert \Psi\Vert_2^2 + C h^4  \, \Vert \Psi\Vert_{\Hmag^1(Q_h)}^2. \label{Rough_bound_on_BCS energy_eq1}
\end{align}
\end{prop}

\begin{proof}
We write
\begin{align}
-\frac 14 \langle \Delta, L_{T, \Abold, W} \Delta \rangle  = - \frac 14 \langle \Delta, L_{T,\Abold} \Delta\rangle - \frac 14 \langle \Delta, L_{T, \Abold, W} \Delta - L_{T, \Abold} \Delta\rangle \label{Rough_bound_on_BCS energy_proof_2}
\end{align}
and use the resolvent identity in \eqref{Resolvent_Equation} to write one of the operators on the right side as
\begin{align}
L_{T, \Abold, W} \Delta - L_{T, \Abold} \Delta &= -\frac 2\beta \sum_{n\in \Zbb} \frac{1}{\i \omega_n - \hfrak_\Abold} W_h \frac{1}{\i \omega_n - \hfrak_{\Abold, W}} \Delta \frac{1}{\i \omega_n  + \ov{\hfrak_{\Abold, W}}} \notag \\
&\hspace{80pt} - \frac{1}{\i \omega_n - \hfrak_\Abold} \Delta \frac{1}{\i \omega_n + \ov{\hfrak_{\Abold, W}}} W_h \frac{1}{\i \omega_n + \ov{\hfrak_\Abold}}. \label{RTAW2_estimate_1}
\end{align}
An application of Hölder's inequality therefore implies the bound
\begin{align*}
|\langle \Delta, L_{T, \Abold, W} \Delta - L_{T, \Abold} \Delta \rangle| \leq C \, h^4 \, \Vert \Psi\Vert_{\Hmag^1(Q_h)}^2.
\end{align*}
When we additionally use the decomposition of $L_{T,\Abold}$ in \eqref{LTA_decomposition}, Propositions~\ref{LTA-LtildeTA}, \ref{LtildeTA-MtildeTA}, and \ref{MtildeTA-MTA}, we find
\begin{align}
- \frac 14 \langle \Delta, L_{T,\Abold} \Delta\rangle + \Vert \Psi\Vert_2^2 \; \langle \alpha_*, V\alpha_*\rangle &\notag\\
&\hspace{-100pt} = - \frac 14 \langle \Delta, M_T^{(1)}\Delta- M_{\Tc}^{(1)} \Delta\rangle - \frac 14 \langle \Delta, M_{T,\Abold} \Delta - M_T^{(1)}\Delta\rangle + R_1(\Delta), \label{Rough_bound_on_BCS energy_proof_1}
\end{align}
with a remainder $R_1(\Delta)$ obeying the bound
\begin{align*}
|R_1(\Delta) | \leq C \; h^5 \; \Vert \Psi\Vert_{\Hmag^1(Q_h)}^2.
\end{align*}
From Proposition~\ref{MT1} we know that 
\begin{align*}
-\frac 14 \langle \Delta, M_T^{(1)}\Delta- M_{\Tc}^{(1)} \Delta\rangle \leq  c \; \frac{T - \Tc}{\Tc} \; \Vert \Psi \Vert_2^2.
\end{align*}
We also claim that the bound
\begin{align}
|\langle \Delta, M_{T,\Abold}\Delta - M_T^{(1)}\Delta\rangle| &\leq C\; h^4 \; \Vert V\alpha_*\Vert_2^2 \; \Vert \Psi\Vert_{\Hmag^1(Q_h)}^2 \label{MTB-MT1_eq1}
\end{align}
holds. Its proof can be achieved with the same methods that have been used to prove Proposition~\ref{MtildeTA-MTA}. The main point is that we have to use the bound
\begin{align}
	|\langle \Psi, [\cos(Z\cdot\Pi_\Abold) - 1] \Psi \rangle| &\leq C\; h^4 \; |Z|^2 \; \Vert \Psi\Vert_{\Hmag^1(Q_h)}^2, \label{MTB-MT1_1}
\end{align} 
as well as the operator inequality in \eqref{ZPiX_inequality} for $(Z\cdot \Pi_\Abold)^2$. Since no additional difficulties occur, we leave the details to the reader. This completes the proof of \eqref{Rough_bound_on_BCS energy_eq1}.
\end{proof}



\subsubsection{A representation formula for the operator \texorpdfstring{$N_{T, \Abold}$}{NTA}}
\label{sec:NT}

In this and the following sections we investigate the nonlinear operator
\begin{align}
	N_{T, \Abold} \coloneqq  N_{T, \Abold, 0} \label{NTA_definition}
\end{align}
with $N_{T, \Abold, W}$ in \eqref{NTAW_definition}. In particular, we show that the quartic term in the GL functional emerges from $\langle \Delta, N_{T, \Abold} (\Delta) \rangle$. In Section~\ref{sec:quarticterms} we show that $\langle \Delta, N_{T, \Abold}(\Delta) - N_{T, \Abold, W} (\Delta) \rangle$ yields a negligible contribution. In this section we will also collect all previous results and finish the proof of Theorem~\ref{Calculation_of_the_GL-energy}. 

Before we state a representation formula for $N_{T, \Abold}$ in terms of relative and center-of-mass coordinates, we introduce the notation $\Zbold$ to denote the vector $(Z_1,Z_2,Z_3)$ with $Z_1, Z_2, Z_3 \in \Rbb^3$ as well as $\dd \Zbold = \dd Z_1 \dd Z_2 \dd Z_3$.
%
%
\begin{lem}
\label{NTA_action}
The operator $N_{T, \Abold}$ in \eqref{NTAW_definition} acts as
\begin{align*}
N_{T, \Abold}(\alpha) (X, r) &= \iiint_{\Rbb^9} \dd \Zbold \iiint_{\Rbb^9} \dd \sbold \; \ell_{T, \Abold}(X, \Zbold, r, \sbold)\; \Acal(X, \Zbold, \sbold),
\end{align*}
where
\begin{align}
\Acal(X, \Zbold, \sbold) &\coloneqq  \e^{\i Z_1\cdot (-\i \nabla_X)} \alpha(X, s_1) \; \ov{\e^{\i Z_2\cdot (-\i \nabla_X)} \alpha(X, s_2)} \; \e^{\i Z_3\cdot (-\i \nabla_X)} \alpha(X,s_3) \label{NTA_alpha_definition}
\end{align}
and
\begin{align}
\ell_{T, \Abold}(X, \Zbold, r, \sbold) &\coloneqq  \frac{2}{\beta} \sum_{n\in \Zbb} \ell_{T, \Abold}^n(X, \Zbold, r, \sbold) \; \e^{\i  \Upsilon_{\Abold_h}(X, \Zbold, r, s)}. \label{lTA_definition}
\end{align}
Here, 
\begin{align}
\ell_{T, \Abold}^n(X, \Zbold, r, \sbold) &\coloneqq   g_h^{\i\omega_n}\bigl(X + \frac r2 \, , \, X + Z_1 + \frac{s_1} 2\bigr) \, g_h^{-\i\omega_n} \bigl( X + Z_2 + \frac{s_2}{2} \, , \, X + Z_1 - \frac{s_1}{2}\bigr)  \notag\\
&\hspace{-50pt} \times g_h^{\i\omega_n}\bigl( X + Z_2 - \frac{s_2}{2} \, , \, X + Z_3 + \frac{s_3}{2}\bigr) \, g_h^{-\i\omega_n}\bigl(X - \frac{r}{2} \, , \, X + Z_3 - \frac{s_3}{2}\bigr), \label{lTAn_definition}
\end{align}
with $g_h^z$ in \eqref{ghz_definition} and
\begin{align}
\Upsilon_\Abold(X, \Zbold, r, \sbold) &\coloneqq  \Phi_\Abold \bigl(X + \frac r2 \, , \, X + Z_1 + \frac{s_1} 2\bigr) + \Phi_\Abold \bigl( X + Z_2 + \frac{s_2}{2} \, , \, X + Z_1 - \frac{s_1}{2}\bigr) \notag\\
&\hspace{-30pt} + \Phi_\Abold \bigl( X + Z_2 - \frac{s_2}{2} \, , \, X + Z_3 + \frac{s_3}{2}\bigr) + \Phi_\Abold \bigl(X - \frac{r}{2} \, , \, X + Z_3 - \frac{s_3}{2}\bigr). \label{NTA_PhitildeA_definition}
\end{align}

\end{lem}

\begin{bem}
	The above representation formula for $N_{T,\Abold}$ should be compared to that in the case of a constant magnetic field in \cite[Lemma~4.16]{DeHaSc2021} and to the representation formula for $L_{T,\Abold}$ in \ref{LTA_action}. The following two properties are relevant for us: (a) The functions $\alpha$ are multiplied by translation operators that can later be completed with appropriate phase factors to give magnetic translation operators. (b) The coordinates appearing in $\Upsilon_{\Abold}$ in \eqref{NTA_PhitildeA_definition} equal those in the different factors in the definition of $\ell_{T, \Abold}^n(X, \Zbold, r, \sbold)$ in \eqref{lTAn_definition}. When proving bounds, this allows us to find a similar structure of nested convolutions as the one we already encountered in the analysis of $L_{T,\Abold}$. The center-of-mass part of $\alpha$ never participates in these convolutions.
\end{bem}

\begin{proof}[Proof of Lemma \ref{NTA_action}]
When we compute the integral kernel of $N_{T, \Abold}$ using \eqref{GAz_Kernel_of_complex_conjugate} in the case $W =0$, we get
\begin{align}
N_{T, \Abold} (\alpha)(x,y) &= \smash{\frac 2\beta \sum_{n\in \Zbb} \iiint_{\Rbb^{9}} \dd \mathbf u \iiint_{\Rbb^9}\dd \mathbf v} \; G_h^{\i\omega_n} (x, u_1)\, \alpha(u_1,v_1)\, G_h^{-\i\omega_n} (u_2,v_1) \, \ov{\alpha(u_2,v_2)} \nonumber \\
&\hspace{120pt} \times G_h^{\i\omega_n}(v_2,u_3)\, \alpha(u_3,v_3)\,  G_h^{-\i\omega_n}(y, v_3),
\label{eq:newAndi4}
\end{align}
We highlight that, by a slight abuse of notation, $\alpha$ and $N_{T, \Abold} (\alpha)$ in the above equation are functions of the original coordinates, while they are functions of relative and center-of-mass coordinates in \eqref{NTA_alpha_definition}. Let us denote $X = \frac{x+y}{2}$, $r=x-y$ and let us also introduce the relative coordinate $\sbold$ and the center-of-mass coordinate $\Zbold$ by
\begin{align*}
u_i &= X + Z_i + \frac {s_i}{ 2 }, & v_i &= X + Z_i - \frac {s_i}{ 2 }, & i&=1,2,3. 
\end{align*}
When we express the integration in \eqref{eq:newAndi4} in terms of these coordinates and use \eqref{ghz_definition}, we see that the claimed formula holds.
\end{proof}

As the operator $L_{T,\Abold}$, we analyze the operator $N_{T, \Abold}$  in four steps. More precisely, we decompose $N_{T, \Abold}$ as
\begin{align}
N_{T, \Abold} = (N_{T, \Abold} - \tilde N_{T, \Abold}) + (\tilde N_{T, \Abold} - N_{T, \Bbold}^{(1)}) + (N_{T, \Bbold}^{(1)} - N_T^{(2)}) + N_{T}^{(2)} \label{NTB_decomposition}
\end{align}
with $\tilde N_{T, \Abold}$ defined below in \eqref{NtildeTB_definition}, $N_{T, \Bbold}^{(1)}$ in \eqref{NTB1_definition}, and $N_T^{(2)}$ in \eqref{NT2_definition}. To obtain the map $\tilde N_{T, \Abold}$ from $N_{T, \Abold}$ we need to replace $g_{h}^z$ by $g_0^z$. The operator $N_{T,\Bbold}^{(1)}$ emerges when we use a part of the phase $\exp(\i \Upsilon_{\Abold_h}(X, \Zbold, r, s))$ in the definition of $\ell_{T, \Abold}$ in \eqref{lTA_definition} to replace the translation operators $\exp(\i Z\cdot P_X)$ in front the $\alpha$ factors by magnetic translation operators. The part of the phase factor that is not needed during this procedure is shown to yield a negligible contribution. Finally, the operator $N_T^{(2)}$ is obtained when we replace the just found magnetic translations by $1$. The above decomposition of $N_{T, \Abold}$ should be compared to that in \cite[Eq.~(4.120)]{DeHaSc2021}. In Section~\ref{sec:approxNTA} we show that the terms in brackets in \eqref{NTB_decomposition} only yield negligible contributions. Afterwards, we extract in Section~\ref{sec:compquarticterm} the quartic term in the Ginzburg--Landau functional from $\langle \Delta, N_{T}^{(2)}(\Delta) \rangle$. In Section~\ref{sec:quarticterms} we summarize our findings.

\subsubsection{Approximation of \texorpdfstring{$N_{T, \Abold}$}{NTA}}
\label{sec:approxNTA}

\paragraph{The operator $\tilde N_{T, \Abold}$.} We define the operator $\tilde N_{T, \Abold}$ by
\begin{align}
\tilde N_{T, \Abold}(\alpha) (X,r) &\coloneqq  \iiint_{\Rbb^{9}} \dd \Zbold \iiint_{\Rbb^{9}} \dd \sbold \; \tilde \ell_{T, \Abold} (X, \Zbold, r, \sbold) \; \Acal(X, \Zbold , \sbold) \label{NtildeTB_definition}
\end{align}
with $\Acal$ in \eqref{NTA_alpha_definition} and
\begin{align*}
\tilde \ell_{T, \Abold}(X, \Zbold, r, \sbold) &\coloneqq  \frac 2\beta \sum_{n\in\Zbb} \ell_T^n (\Zbold, r, \sbold) \; \e^{\i \Upsilon_{\Abold_h}(X, \Zbold, r, \sbold)},
\end{align*}
where $\Upsilon_\Abold$ has been defined in \eqref{NTA_PhitildeA_definition}, and
\begin{align}
\ell_T^n(\Zbold, r, \sbold) &\coloneqq  g_0^{\i\omega_n} \bigl(Z_1 - \frac{r-s_1}{2}\bigr) \; g_0^{-\i\omega_n} \bigl( Z_1 - Z_2 - \frac{s_1 + s_2}{2}\bigr) \notag \\
&\hspace{50pt} \times g_0^{\i\omega_n} \bigl( Z_2 - Z_3 - \frac{s_2 + s_3}{2} \bigr) \; g_0^{-\i\omega_n} \bigl( Z_3 + \frac{r-s_3}{2}\bigr). \label{lTn_definition}
\end{align}
In our calculation of the BCS energy we can replace $N_{T, \Abold}(\Delta)$ by $\tilde N_{T, \Abold}(\Delta)$ because of the following error bound.

\begin{prop}
\label{NTA-NtildeTA}
Let $V\alpha_*\in L^{\nicefrac 43}(\Rbb^3)$, let $A\in W^{3,\infty}(\mathbb{R}^3,\mathbb{R}^3)$ be periodic, assume that $\Psi\in \Hmag^1(Q_h)$, and denote $\Delta \equiv \Delta_\Psi$ as in \eqref{Delta_definition}. For any $T_0>0$ there is $h_0>0$ such that for any $T\geq T_0$ and any $0 < h \leq h_0$ we have
\begin{align*}
|\langle \Delta,  N_{T, \Abold}(\Delta) - \tilde N_{T, \Abold}(\Delta)\rangle| &\leq C \; h^6 \; \Vert V\alpha_*\Vert_{\nicefrac 43}^4 \; \Vert \Psi\Vert_{\Hmag^1(Q_h)}^4.
\end{align*}
\end{prop}

The function
\begin{align}
J_{T,\Abold} &\coloneqq  \smash{\frac 2\beta \sum_{n\in \Zbb}} \; \gzfunctiondiff^{\i\omega_n} * \gzfunction^{-\i \omega_n} * \gzfunction^{\i \omega_n} * \gzfunction^{-\i \omega_n} + |g_0^{\i\omega_n}| * \gzfunctiondiff^{-\i \omega_n} * \gzfunction^{\i \omega_n} * \gzfunction^{-\i \omega_n} \notag \\
&\hspace{50pt} + |g_0^{\i\omega_n}| * |g_0^{-\i\omega_n} | * \gzfunctiondiff^{\i \omega_n} * \gzfunction^{-\i \omega_n} + |g_0^{\i\omega_n}| * |g_0^{-\i\omega_n} | * |g_0^{\i\omega_n}| * \gzfunctiondiff^{-\i \omega_n}. \label{NTA-NtildeTA_FTA_definition}
\end{align}
plays a prominent role in the proof of Proposition \ref{NTA-NtildeTA}. Using Lemmas~\ref{gh-g_decay} and \ref{g_decay} as well as \eqref{g0_decay_f_estimate1}, we see that for any $T \geq T_0 > 0$ there is a constant $C>0$ such that
\begin{align}
	\Vert J_{T, \Abold_h}\Vert_1 \leq C \; h^3 \label{NTA-NtildeTA_FTA_estimate}
\end{align}
holds.

\begin{proof}[Proof of Proposition \ref{NTA-NtildeTA}]
The function $|\Psi|$ is periodic, and hence \eqref{Magnetic_Sobolev} implies
\begin{align}
\Vert \e^{\i Z \cdot (-\i \nabla_X)}\Psi\Vert_6^2 = \Vert \Psi\Vert_6^2 \leq C\,  h^2 \, \Vert \Psi\Vert_{\Hmag^1(Q_h)}^2. \label{NTB-NtildeTB_3}
\end{align}
In particular, we have
\begin{align}
\fint_{Q_h} \dd X \; |\Psi(X)|\; \prod_{i=1}^3  |\e^{\i Z_i\cdot (-\i \nabla_X)}\Psi(X)| &\leq \Vert \Psi\Vert_2 \; \prod_{i=1}^3  \Vert \e^{\i Z_i\cdot (-\i \nabla_X)}\Psi\Vert_6 \leq C \, h^4 \, \Vert \Psi\Vert_{\Hmag^1(Q_h)}^4 \label{NTA-NtildeTBA_2}
\end{align}
as well as
\begin{align}
|\langle \Delta,  N_{T, \Abold}(\Delta)- \tilde N_{T, \Abold}(\Delta)\rangle| & \notag\\
&\hspace{-100pt}\leq C\, h^4 \, \Vert \Psi\Vert_{\Hmag^1(Q_h))}^4  \int_{\Rbb^3} \dd r \,  \iiint_{\Rbb^9} \dd \sbold\; |V\alpha_*(r)|\;  |V\alpha_*(s_1)| \; |V\alpha_*(s_2)| \; |V\alpha_*(s_3)|  \notag\\
&\hspace{-20pt}\times \iiint_{\Rbb^9} \dd \Zbold \; \esssup_{X\in \Rbb^3} \bigl|(\ell_{T, \Abold} - \tilde \ell_{T, \Abold})(X, \Zbold, r, \sbold)\bigr|.  \label{NTB-NtildeTB_1}
\end{align}
Next, we define the variables $Z_1',Z_2',Z_3'$ via the equation
\begin{align}
Z_1' - Z_2' &\coloneqq  Z_1 - Z_2 - \frac{s_1 +s_2}{2}, & Z_2' - Z_3' &\coloneqq  Z_2 - Z_3 - \frac{s_2 + s_3}{2}, & Z_3' &\coloneqq  Z_3 + \frac{r-s_3}{2}, \label{NTA_change_of_variables_1}
\end{align}
which implies
\begin{align}
Z_1 - \frac{r-s_1}{2} = Z_1' - (r - s_1 - s_2 - s_3). \label{NTA_change_of_variables_2}
\end{align}
We argue as in the proof of \eqref{LTA-LtildeTBA_5} to see that
\begin{align*}
\iiint_{\Rbb^9} \dd \Zbold\; \esssup_{X\in \Rbb^3} \bigl|(\ell_{T, \Abold} - \tilde \ell_{T, \Abold})(X, \Zbold, r, \sbold)\bigr| \leq J_{T, \Abold_h}(r - s_1 - s_2 - s_3)
\end{align*}
holds with $J_{T, \Abold}$ in \eqref{NTA-NtildeTA_FTA_definition}. When insert this bound into \eqref{NTB-NtildeTB_1} and use
\begin{align*}
\bigl\Vert V\alpha_* \; \bigl( V\alpha_* * V\alpha_* * V\alpha_* * J_{T, \Abold}\bigr) \bigr\Vert_1 &\leq C \; \Vert V\alpha_*\Vert_{\nicefrac 43}^4 \; \Vert J_{T, \Abold}\Vert_1,
\end{align*}
as well as \eqref{NTA-NtildeTA_FTA_estimate}, this finishes the proof.
\end{proof}

\paragraph{The operator $N_{T, \Bbold}^{(1)}$.} We define the operator $N_{T, \Bbold}^{(1)}$ by
\begin{align}
N_{T, \Bbold}^{(1)}(\alpha)(X,r) &\coloneqq  \iiint_{\Rbb^9} \dd \Zbold \iiint_{\Rbb^9} \dd \sbold \; \ell_T (\Zbold, r, \sbold) \; \Acal_\Bbold (X, \Zbold , \sbold), \label{NTB1_definition}
\end{align}
where
\begin{align}
\ell_T(\Zbold, r, \sbold) &\coloneqq  \ell_{T,0}(0, \Zbold, r, \sbold), \label{lT_definition}
\end{align}
with $\ell_{T,0}$ in \eqref{lTA_definition} and
\begin{align}
\Acal_\Bbold(X, \Zbold, \sbold) &\coloneqq  \e^{\i Z_1\cdot \Pi} \alpha(X, s_1) \; \ov{\e^{\i Z_2\cdot \Pi} \alpha(X, s_2)} \; \e^{\i Z_3\cdot \Pi} \alpha(X,s_3). \label{NTB1_alphaB_definition}
\end{align}
The following bound allows us to replace $\langle \Delta, \tilde N_{T, \Abold}(\Delta) \rangle$ by $\langle \Delta, N_{T, \Bbold}^{(1)}(\Delta) \rangle$ in our computation of the energy.

\begin{prop}
\label{NtildeTBA-NTB1}
Let $| \cdot| V\alpha_*\in L^{\nicefrac 43}(\Rbb^3)$ for $k\in \{0,1\}$, let $A\in L^{\infty}(\mathbb{R}^3,\mathbb{R}^3)$ be periodic, assume $\Psi\in \Hmag^1(Q_h)$, and denote $\Delta \equiv \Delta_\Psi$ as in \eqref{Delta_definition}. For any $T_0>0$ there is $h_0>0$ such that for any $T\geq T_0$ and any $0 < h \leq h_0$ we have
\begin{align*}
|\langle \Delta, \tilde N_{T, \Abold}(\Delta) - N_{T, \Bbold}^{(1)}(\Delta)\rangle| &\leq C \; h^5 \; \max_{k=0,1} \Vert \ |\cdot|^k \ V\alpha_*\Vert_{\nicefrac 43}^4 \; \Vert \Psi\Vert_{\Hmag^1(Q_h)}^4.
\end{align*}
\end{prop}

Before we give the proof of Proposition~\ref{NtildeTBA-NTB1} we define the functions
\begin{align}
	J_T^{(1)} &\coloneqq  \smash{\frac 2\beta \sum_{n\in \Zbb}} \, |g_0^{\i\omega_n}| * \bigl(|\cdot|\, |g_0^{-\i\omega_n}|\bigr) * \bigl(|\cdot|\, |g_0^{\i\omega_n}|\bigr) * |g_0^{-\i\omega_n}| \notag \\
	&\hspace{100pt}+ |g_0^{\i\omega_n}| * \bigl(|\cdot|\, |g_0^{-\i\omega_n}|\bigr) * |g_0^{\i\omega_n}| * \bigl(|\cdot|\, |g_0^{-\i\omega_n}|\bigr) \notag \\
	&\hspace{100pt}+ |g_0^{\i\omega_n}| * |g_0^{-\i\omega_n}| * \bigl(|\cdot|\, |g_0^{\i\omega_n}|\bigr) * \bigl(|\cdot|\, |g_0^{-\i\omega_n}|\bigr) \label{NtildeTB-NTB1_FT1_definition}
\end{align}
and
\begin{align}
J_T^{(2)} &\coloneqq  \smash{\frac 2\beta \sum_{n\in \Zbb}} \, \bigl( |\cdot| \, |g_0^{\i\omega_n}|\bigr) *  |g_0^{-\i\omega_n}| * |g_0^{\i\omega_n}| * |g_0^{-\i\omega_n}| + |g_0^{\i\omega_n}| * \bigl(|\cdot|\, |g_0^{-\i\omega_n}|\bigr) * |g_0^{\i\omega_n}| * |g_0^{-\i\omega_n}| \notag\\
&\hspace{30pt}+ |g_0^{\i\omega_n}| * |g_0^{-\i\omega_n}| * \bigl(|\cdot|\, |g_0^{\i\omega_n}|\bigr) * |g_0^{-\i\omega_n}| + |g_0^{\i\omega_n}| * |g_0^{-\i\omega_n}| * |g_0^{\i\omega_n}| * \bigl(|\cdot|\, |g_0^{-\i\omega_n}|\bigr). \label{NtildeTB-NTB1_FT2_definition}
\end{align}
Using Lemma~\ref{g_decay} and \eqref{g0_decay_f_estimate1}, we show that for any $T_0 > 0$ there is a constant $C>0$ such that for $T \geqslant T_0$ we have
\begin{align}
	\Vert J_T^{(1)} \Vert_1 + \Vert J_T^{(2)} \Vert_1 \leq C. \label{NtildeTB-NTB1_FT1-2_estimate}
\end{align}

\begin{proof}[Proof of Proposition \ref{NtildeTBA-NTB1}]
We recall the definition of the phase $\Upsilon_\Abold$ in \eqref{NTA_PhitildeA_definition}. A tedious but straightforward computation shows that
\begin{align*}
\Upsilon_{\Abold_{\Bbold}} (X, \Zbold, r, \sbold) &= Z_1 \cdot (\Bbold \wedge X) - Z_2 \cdot (\Bbold \wedge X) + Z_3 \cdot (\Bbold \wedge X) + \frac \Bbold 2 \cdot I(\Zbold, r, \sbold),
\end{align*}
where
\begin{align}
I(\Zbold, r, \sbold) &\coloneqq  \frac r2 \wedge \bigl( Z_1 - \frac{r - s_1}{2}\bigr) + \frac r2 \wedge \bigl( Z_3 + \frac{r-s_3}{2}\bigr) \notag \\
&\hspace{-35pt} + \bigl( Z_2 - Z_3 - \frac{s_2 + s_3}{2}\bigr) \wedge \bigl( Z_1 - Z_2 - \frac{s_1 + s_2}{2}\bigr) \notag \\
&\hspace{-35pt}+ \bigl( Z_3 + \frac{r - s_3}{2}\bigr) \wedge \bigl( Z_1 - Z_2 - \frac{s_1 + s_2}{2}\bigr)+ \bigl( s_2 + s_3 - \frac r2\bigr) \wedge \bigl( Z_1 - Z_2 - \frac{s_1 + s_2}{2}\bigr) \notag \\
&\hspace{-35pt}+ \bigl( Z_3 + \frac{r - s_3}{2} \bigr) \wedge \bigl( Z_3 - Z_2 + \frac{s_2 + s_3}{2}\bigr)+ \bigl( s_3 - \frac r2\bigr) \wedge \bigl( Z_3 - Z_2 + \frac{s_2 + s_3}{2}\bigr). \label{PhiB_definition}
\end{align}
By \eqref{Magnetic_Translation_constant_Decomposition}, the operator $\tilde N_{T, \Abold}$ can therefore be rewritten as
\begin{align*}
\tilde N_{T, \Abold}(\alpha) &= \iiint_{\Rbb^9} \dd \Zbold \iiint_{\Rbb^9} \dd s \; \ell_T(\Zbold, r, \sbold) \, \e^{\i \Upsilon_{\Asolo_h}(X, \Zbold, r, \sbold)} \e^{\i \frac \Bbold 2 \cdot I(\Zbold, r, \sbold)} \; \Acal_\Bbold(X, \Zbold, \sbold).
\end{align*}
This formula and the estimate in \eqref{NTA-NtildeTBA_2} imply the bound
\begin{align}
|\langle \Delta, \tilde N_{T, \Abold}(\Delta) - N_{T, \Bbold}^{(1)}(\Delta)\rangle| &  \notag\\
&\hspace{-120pt}\leq C\; h^4 \, \Vert \Psi\Vert_{\Hmag^1(Q_h)}^4 \int_{\Rbb^3} \dd r  \iiint_{\Rbb^9} \dd \sbold \; |V\alpha_*(r)| \; |V\alpha_*(s_1)| \; |V\alpha_*(s_2)| \; |V\alpha_*(s_3)|  \notag \\
&\hspace{-100pt}\times \frac{2}{\beta} \sum_{n\in \Zbb} \iiint_{\Rbb^9} \dd \Zbold \; |\ell_T^n(\Zbold, r, \sbold)| \; \sup_{X\in \Rbb^3} \bigl| \e^{\i \Upsilon_{\Asolo_h}(X, \Zbold, r, \sbold)} \, \e^{\i \frac \Bbold 2 \cdot I(\Zbold, r,\sbold)} - 1\bigr| \label{NtildeTBA-NTB1_1}
\end{align}
with $\Upsilon_A$ in \eqref{NTA_PhitildeA_definition} and $I$ in \eqref{PhiB_definition}. In terms of the coordinates in \eqref{NTA_change_of_variables_1} and with the help of \eqref{NTA_change_of_variables_2}, the phase function $I$ can be written as
\begin{align}
I(\Zbold, r,\sbold) &=  (Z_2' - Z_3') \wedge (Z_1' - Z_2') + Z_3' \wedge (Z_1' - Z_2') +  Z_3' \wedge (Z_3'- Z_2')  \notag\\
& \hspace{50pt} + \frac r2 \wedge \bigl( Z_1' - (r - s_1 - s_2 - s_3)\bigr) + \bigl( s_2 + s_3 - \frac r2\bigr) \wedge (Z_1 ' - Z_2')\notag \\
&\hspace{50pt} + \bigl( s_3 - \frac r2\bigr) \wedge (Z_3' - Z_2') + \frac r2 \wedge Z_3'.\label{NtildeTBA-NTB1_2}
\end{align}
Moreover, using the definition of $\Phi_\Abold$ in \eqref{PhiA_definition}, the definition of $\Upsilon_{\Asolo}(X, \Zbold, r, \sbold)$ in \eqref{NTA_PhitildeA_definition}, \eqref{NTA_change_of_variables_1}, and \eqref{NTA_change_of_variables_2}, we obtain the bound
\begin{align*}
|\Upsilon_{\Asolo}(X, \Zbold, r, \sbold)| &\leq \Vert \Asolo\Vert_\infty \bigl( (Z_1' - (r - s_1 - s_2 - s_3)) + (Z_1' - Z_2') + (Z_2' - Z_3') + Z_3'\bigr).
\end{align*}
In combination with \eqref{NtildeTBA-NTB1_1}, \eqref{NtildeTBA-NTB1_2}, and an argument that is similar to the one used to obtain \eqref{MtildeTA-MTA_5}, we find
\begin{align*}
&\frac{2}{\beta} \sum_{n\in \Zbb} \iiint_{\Rbb^9} \dd \Zbold \; |\ell_{T,0}^n(\Zbold, r, \sbold)| \; \bigl| \e^{\i \Upsilon_{\Asolo_h}(X, \Zbold, r, \sbold)}\, \e^{\i \frac \Bbold 2 \cdot I(\Zbold, r, \sbold)} - 1\bigr| \\
&\hspace{20pt} \leq Ch \; \bigl[ J_T^{(1)} (r - s_1 - s_2 - s_3) + J_T^{(2)}(r - s_1 - s_2 - s_3) \; \bigl(1 + |r| + |s_1| + |s_2| + |s_3|\bigr)\bigr]
\end{align*}
with the functions $F_T^{(1)}$ and $F_T^{(2)}$ in \eqref{NtildeTB-NTB1_FT1_definition} and \eqref{NtildeTB-NTB1_FT2_definition}, respectively. Accordingly, an application of Young's inequality shows that 
\begin{align*}
|\langle \Delta, \tilde N_{T, \Abold}(\Delta) - N_{T, \Bbold}^{(1)}(\Delta)\rangle| &\\
&\hspace{-100pt}\leq C\; h^5 \, \Vert \Psi\Vert_{\Hmag^1(Q_h)}^4  \; \bigl( \Vert V\alpha_*\Vert_{\nicefrac 43}^4 + \Vert \, |\cdot| V\alpha_*\Vert_{\nicefrac 43}^4 \bigr) \bigl( \Vert J_T^{(1)}\Vert_1 + \Vert J_T^{(2)}\Vert_1 \bigr).
\end{align*}
The claim of the proposition follows when we apply \eqref{NtildeTB-NTB1_FT1-2_estimate} on the right side of the above equation.
\end{proof}

\paragraph{The operator $N_T^{(2)}$.} We define the operator $N_{T}^{(2)}$ by
\begin{align}
N_T^{(2)}(\alpha) (X, r) &\coloneqq  \iiint_{\Rbb^9} \dd \Zbold \iiint_{\Rbb^9} \dd \sbold \; \ell_{T} (\Zbold, r, \sbold) \, \prod_{i=1}^3 \alpha(X,s_i) \label{NT2_definition}
\end{align}
with $\ell_{T}$ in \eqref{lT_definition}.

In the computation of the BCS energy we can replace $\langle \Delta, N_{T, \Bbold}^{(1)}(\Delta) \rangle$ by $\langle \Delta, N_{T}^{(2)}(\Delta) \rangle$ with the help of the following error bound. Its proof can be found in \cite[Proposition 4.20]{DeHaSc2021}. We highlight that the $\Hmag^2(Q_h)$-norm of $\Psi$ is needed once more.

\begin{prop}
\label{NTB1-NT2}
Assume that $|\cdot|^kV\alpha_* \in L^{\nicefrac 43}(\Rbb^3)$ for $k\in \{0,1,2\}$, let $\Psi \in \Hmag^2(Q_h)$, and $\Delta\equiv \Delta_\Psi$ as in \eqref{Delta_definition}. For any $T \geq T_0 >0$ there is $h_0 > 0$ such that for $0 < h \leqslant h_0$ we have
\begin{align*}
|\langle \Delta, N_{T, \Bbold}^{(1)}(\Delta) - N_{T}^{(2)}(\Delta) \rangle|  &\leq C \; h^6 \; \max_{k=0,1,2} \Vert \ |\cdot|^k \ V\alpha_*\Vert_{\nicefrac 43}^4 \, \Vert \Psi\Vert_{\Hmag^1(Q_h)}^3 \; \Vert \Psi\Vert_{\Hmag^2(Q_h)}.
\end{align*}
\end{prop}

\subsubsection{Calculation of the quartic term in the Ginzburg--Landau functional}
\label{sec:compquarticterm}

The quartic term in the Ginzburg--Landau functional in \eqref{Definition_GL-functional} is contained in $\langle \Delta, N_T^{(2)}(\Delta) \rangle$. It can be extracted with the following proposition, whose proof can be found in \cite[Proposition 4.21]{DeHaSc2021}. 

\begin{prop}
\label{NTc2}
Assume $V\alpha_* \in L^{\nicefrac 43}(\Rbb^3)$ and let $\Psi\in \Hmag^1(Q_h)$ as well as $\Delta \equiv \Delta_\Psi$ as in \eqref{Delta_definition}. For any $h>0$, we have
\begin{align*}
\langle \Delta, N_{\Tc}^{(2)}(\Delta)\rangle = 8\; \Lambda_3 \; \Vert \Psi\Vert_4^4
\end{align*}
with $\Lambda_3$ in \eqref{GL_coefficient_3}. Moreover, for any $T \geq T_0 > 0$, we have
\begin{align*}
|\langle \Delta,  N_T^{(2)}(\Delta) - N_{\Tc}^{(2)}(\Delta)\rangle| &\leq C\; h^4  \; |T - \Tc| \; \Vert V\alpha_*\Vert_{\nicefrac 43}^4 \; \Vert \Psi\Vert_{\Hmag^1(Q_h)}^4.
\end{align*}
\end{prop}

\subsubsection{Summary: The quartic term and proof of Theorem~\ref{Calculation_of_the_GL-energy}}
\label{sec:quarticterms}

%

Let the assumptions of Theorem~\ref{Calculation_of_the_GL-energy} hold. We use the resolvent identity in \eqref{Resolvent_Equation} to decompose the operator $N_{T, \Abold, W}$ in \eqref{NTAW_definition} as
\begin{align}
N_{T, \Abold, W} = N_{T, \Abold} + \Rcal_{T, \Abold,W}^{(3)} \label{NTAW_decomposition}
\end{align}
with $N_{T, \Abold}$ in \eqref{NTA_definition} and
\begin{align}
\Rcal_{T, \Abold, W}^{(3)}(\Delta) & \notag\\
&\hspace{-40pt} \coloneqq  \frac 2\beta \sum_{n\in \Zbb} \Bigl[ \frac{1}{\i\omega_n - \hfrak_\Abold} \, W_h \,  \frac{1}{\i \omega_n -\hfrak_{\Abold, W}} \, \Delta \, \frac{1}{\i\omega_n + \ov{\hfrak_{\Abold, W}}} \, \ov \Delta \, \frac{1}{\i \omega_n - \hfrak_{\Abold, W}} \, \Delta \, \frac{1}{\i \omega_n + \ov{\hfrak_{\Abold, W}}} \notag \\
&\hspace{10pt} - \frac{1}{\i\omega_n - \hfrak_\Abold} \, \Delta \, \frac{1}{\i \omega_n + \ov{\hfrak_\Abold}} \, W_h \, \frac{1}{\i\omega_n + \ov{\hfrak_{\Abold, W}}} \, \ov \Delta \,  \frac{1}{\i \omega_n - \hfrak_{\Abold, W}} \, \Delta \, \frac{1}{\i \omega_n + \ov{\hfrak_{\Abold, W}}} \notag \\
&\hspace{10pt} + \frac{1}{\i\omega_n - \hfrak_\Abold} \, \Delta \, \frac{1}{\i \omega_n + \ov{\hfrak_\Abold}} \, \ov \Delta \, \frac{1}{\i\omega_n - \hfrak_\Abold} \, W_h \,  \frac{1}{\i \omega_n - \hfrak_{\Abold, W}} \, \Delta \, \frac{1}{\i \omega_n + \ov{\hfrak_{\Abold, W}}} \notag \\
&\hspace{10pt} - \frac{1}{\i\omega_n - \hfrak_\Abold} \, \Delta \, \frac{1}{\i \omega_n + \ov{\hfrak_\Abold}} \, \ov \Delta \, \frac{1}{\i\omega_n - \hfrak_\Abold} \, \Delta \,  \frac{1}{\i \omega_n + \ov{\hfrak_\Abold}} \, W_h \, \frac{1}{\i \omega_n + \ov{\hfrak_{\Abold, W}}} \Bigr]. \label{RTAW3_definition}
\end{align}
We claim that the operator $\Rcal_{T, \Abold, W}^{(3)}$ satisfies the bound
\begin{align*}
\Vert \Rcal_{T, \Abold, W}^{(3)} (\Delta)\Vert_{\Lsymm} &\leq C\, T^{-5} \, h^{5} \, \Vert \Psi\Vert_{\Hmag^1(Q_h)}^3.
\end{align*}
This is a direct consequence of Hölder's inequality in \eqref{Schatten-Hoelder} for the trace per unit volume, which implies that the Hilbert-Schmidt norm per unit volume of the terms in the sum in \eqref{RTAW3_definition} are bounded by $C\, |\omega_n|^{-5} \, \Vert W_h\Vert_\infty \,  \Vert \Delta\Vert_6^3$. Moreover, an application of Lemma~\ref{Schatten_estimate} and \eqref{Magnetic_Sobolev} show that this expression is bounded by $C\, |2n+1|^{-5} \, T^{-5} \, h^5 \Vert \Psi\Vert_{\Hmag^1(Q_h)}^3$, which proves our claim. 

When we combine this bound, Lemma~\ref{NTA_action}, and Propositions~\ref{NTA-NtildeTA}-\ref{NTc2}, we find 
\begin{align}
\frac{1}{8} \langle \Delta, N_{T, \Abold, W}(\Delta) \rangle = \; \Lambda_3 \; \Vert \Psi\Vert_4^4 + R_4(h), \label{eq:A28}
\end{align}
where the remainder $R(h)$ satisfies the bound
\begin{align*}
| R_4(h) | \leq C \; \Vert \Psi \Vert_{\Hmag^1(Q_h)}^3 \, \bigl( h^5 \; \Vert \Psi \Vert_{\Hmag^1(Q_h)} + h^6 \; \Vert \Psi \Vert_{\Hmag^2(Q_h)}\bigr).
\end{align*}
The statement in Theorem~\ref{Calculation_of_the_GL-energy} is a direct consequence of \eqref{eq:A28} and \eqref{eq:A15}. 


\subsection{Proof of Proposition \ref{Structure_of_alphaDelta}}
\label{sec:proofofadmissibility}

We assume that the assumptions of Proposition~\ref{Structure_of_alphaDelta} hold and recall the definition of $\Gamma_\Delta$ in \eqref{GammaDelta_definition}. Using the resolvent equation in \eqref{Resolvent_Equation} and \eqref{alphaDelta_decomposition_1}, we write $\alpha_\Delta = [\Gamma_\Delta]_{12}$ as
\begin{align*}
\alpha_\Delta &= [\Ocal]_{12} + \Rcal_{T, \Abold, W}^{(4)}(\Delta).
\end{align*}
Here, $\Ocal = \frac 1\beta \sum_{n\in \Zbb} \frac{1}{ \i \omega_n - H_0} \delta \frac{1}{ \i \omega_n - H_0}$, see \eqref{alphaDelta_decomposition_2}, with $\delta$ in \eqref{Delta_definition} and
\begin{align}
\Rcal_{T,\Abold, W}^{(4)}(\Delta) &\coloneqq   \frac 1\beta \sum_{n\in \Zbb} \Bigl[ \frac{1}{ \i \omega_n - H_0} \delta\frac{1}{ \i \omega_n - H_0} \delta\frac{1}{ \i \omega_n - H_\Delta} \delta \frac{1}{ \i \omega_n - H_0}\Bigr]_{12}. \label{RTAW4_definition}
\end{align}
Moreover, we have $[\Ocal]_{12} = -\frac 12 L_{T, \Abold, W}\Delta$ with $L_{T, \Abold, W}$ in \eqref{LTAW_definition}. Using the decomposition of $L_{T, \Abold, W}$ in \eqref{LTAW_decomposition}, we define
\begin{align}
\eta_0(\Delta) &\coloneqq  \frac 12 \bigl( L_{T, \Abold, W} \Delta - L_{T, \Abold}\Delta\bigr) + \frac 12 \bigl(L_{T, \Abold}\Delta - M_{T, \Abold}\Delta\bigr) + \frac 12 \bigl( M_T^{(1)}\Delta - M_{\Tc}^{(1)}\Delta\bigr) \notag \\
&\hspace{60pt} + \frac 12 \bigl( M_{T, \Abold}\Delta - M_{T, \Abold_{e_3}}\Delta\bigr) + \Rcal_{T, \Abold, W}^{(4)}(\Delta), \notag \\
\eta_\perp(\Delta) &\coloneqq  \frac 12 \bigl( M_{T, \Abold_{e_3}}\Delta - M_{T}^{(1)}\Delta\bigr), \label{eta_perp_definition}
\end{align}
with $M_{T, \Abold}$ in \eqref{MTA_definition}, $L_{T, \Abold}^W$ in \eqref{LTA^W_definition}, and $M_T^{(1)}$ in \eqref{MT1_definition}. From Proposition~\ref{MT1} we know that $-\frac 12 M_{\Tc}^{(1)} \Delta = \Psi\alpha_*$, which allows us to write $\alpha_{\Delta}$ as in \eqref{alphaDelta_decomposition_eq1}. The operator $M_{T, \Abold_{e_3}}$ equals $M_{T, \Abold}$ in \eqref{MTA_definition} with $\Abold$ replaced by $\Abold_{e_3}$. The contribution from this operator needs to be carefully isolated for the orthogonality property in \eqref{alphaDelta_decomposition_eq4} to hold. This should be compared to part~(c) of \cite[Proposition~3.2]{DeHaSc2021}. In the following, we will establish the properties of $\eta_0$ and $\eta_\perp$ that are stated in Proposition~\ref{Structure_of_alphaDelta}.

We will first prove \eqref{alphaDelta_decomposition_eq2}, and start by noting that
\begin{align*}
\Rcal_{T, \Abold, W}^{(4)}(\Delta) &= \frac 1\beta \sum_{n\in\Zbb} \frac 1{\i\omega_n - \hfrak_{\Abold, W}} \, \Delta \,  \frac 1{\i\omega_n + \ov{\hfrak_{\Abold, W}}}\,  \ov \Delta\,  \Bigl[ \frac{1}{\i \omega_n - H_\Delta}\Bigr]_{11}\,  \Delta \, \frac 1{\i\omega_n + \ov{\hfrak_{\Abold, W}}}.
\end{align*}
An application of Hölder's inequality shows $\Vert \Rcal_{T, \Abold, W}^{(4)}(\Delta)\Vert_2 \leq C \beta^{3} \Vert \Delta\Vert_6^3$. With the operator $\pi = -\i \nabla + \Abold_{\Bbold}$ understood to act on the $x$-coordinate of the integral kernel of $\Rcal_{T, \Abold, W}^{(4)}(\Delta)$ we also have
\begin{equation*}
	\Vert \pi \Rcal_{T, \Abold, W}^{(4)}(\Delta) \Vert_2 \leq \frac 1\beta \sum_{n\in\Zbb} \Bigl\Vert \pi \frac 1{\i\omega_n - \hfrak_{\Abold, W}} \Bigr\Vert_{\infty} \Bigl\Vert \frac 1{\i\omega_n + \ov{\hfrak_{\Abold, W}}} \Bigr\Vert_{\infty}^2 \Bigl\Vert \Bigl[ \frac{1}{\i \omega_n - H_\Delta}\Bigr]_{11} \Bigr\Vert_{\infty} \Vert \Delta \Vert_6^3.
\end{equation*}
An application of Cauchy--Schwarz shows
\begin{equation}
	(- \i \nabla + \Abold_{\Bbold} + \Asolo)^2 + W_h \geqslant \frac{1}{2} (- \i \nabla + \Abold_{\Bbold} )^2 - C h^2. 
	\label{eq:Andinew7}
\end{equation}
Accordingly, we have 
\begin{align*}
\Bigl\Vert \pi \frac 1{\i\omega_n - \hfrak_{\Abold, W}} \Bigr\Vert_{\infty} &= \Bigl\Vert \frac 1{\i\omega_n + \hfrak_{\Abold, W}} \pi^2 \frac 1{\i\omega_n - \hfrak_{\Abold, W}} \Bigr\Vert_{\infty}^{\nicefrac 12} \\
&\leqslant \Bigl\Vert \frac 1{\i\omega_n + \hfrak_{\Abold, W}} \left( \hfrak_{\Abold, W} + \mu + C h^2 \right) \frac 1{\i\omega_n - \hfrak_{\Abold, W}} \Bigr\Vert_{\infty}^{\nicefrac 12} \leq C \,  |\omega_n|^{-\nicefrac 12}.
\end{align*}
It follows that
\begin{align}
\Vert \pi \Rcal_{T, \Abold, W}^{(4)}(\Delta) \Vert_2 \leq C\, \Vert \Delta \Vert_6^3. \label{eq:A25}
\end{align}
The same argument with obvious adjustments also shows that $\Vert \Rcal_{T, \Abold, W}^{(4)}(\Delta)\pi \Vert_2$ is bounded by the right side of \eqref{eq:A25}, too. Finally, \eqref{Norm_equivalence_2}, an application of Lemma~\ref{Schatten_estimate}, and \eqref{Magnetic_Sobolev} allow us to conclude that
\begin{align}
\Vert \Rcal_{T, \Abold, W}^{(4)}(\Delta) \Vert_{\Hsymm}^2 \leq C \; h^6 \; \Vert \Psi \Vert_{\Hmag^1(Q_h)}^6 \label{eq:A23} 
\end{align}
holds.

To control $M_{T, \Abold}\Delta - M_{T, \Abold_{e_3}}\Delta$, we need the following proposition.

\begin{prop}
\label{MTA-MTAe3}
Let $| \cdot| V\alpha_*\in L^{2}(\Rbb^3)$ for $k\in \{0,1\}$, let $A\in W^{1,\infty}(\mathbb{R}^3,\mathbb{R}^3)$ be periodic, assume $\Psi\in \Hmag^1(Q_h)$, and denote $\Delta \equiv \Delta_\Psi$ as in \eqref{Delta_definition}. For any $T_0>0$ there is $h_0>0$ such that for any $T\geq T_0$ and any $0 < h \leq h_0$ we have
\begin{align}
\Vert M_{T, \Abold} \Delta - M_{T, \Abold_{e_3}} \Delta\Vert_\Hsymm^2 &\leq C \, h^5 \, \max_{k=0,1} \Vert \ | \cdot |^k \ V\alpha_*\Vert_2^2 \, \Vert \Psi\Vert_{\Hmag^1(Q_h)}^2. \label{MTA-MTAe3_eq1}
\end{align}
\end{prop}

\begin{proof}
Let us define the operator
\begin{align*}
\Qcal_{T, \Bbold, \Asolo}\alpha(X, r) \coloneqq  \iint_{\Rbb^3\times \Rbb^3} \dd Z \dd s \; k_T(Z, r-s) \; \e^{2\i \Asolo_h(X) \cdot Z} \; (\e^{\i Z\cdot \Pi} \alpha)(X,s),
\end{align*}
where $k_T(Z,r) \coloneqq  k_{T, 0}(0,Z, r, 0)$ with $k_{T, 0}$ in \eqref{kTA_definition} and $\Pi = -\i \nabla + 2 \Abold_{\Bbold}$ is understood to act on the center-of-mass coordinate $X$ of $\alpha$. We start our analysis by writing
\begin{align}
M_{T, \Abold} \Delta - M_{T, \Abold_{e_3}} \Delta = \bigl( M_{T, \Abold} \Delta - \Qcal_{T, \Bbold, \Asolo} \Delta \bigr) + \bigl( Q_{T, \Bbold, \Asolo} \Delta - M_{T, \Abold_{e_3}} \Delta\bigr). \label{MTA-MTAe3_1}
\end{align}
In the following we derive bounds on the $\Hsymm$-norms of the two terms on the right side of \eqref{MTA-MTAe3_eq1}. When we use that the integrand in the definition of $M_{T, \Abold}$ is symmetric with respect to the transformation $Z \mapsto -Z$ and apply Lemma \ref{Magnetic_Translation_Representation}, we see that
\begin{align*}
(M_{T, \Abold} \alpha - \Qcal_{T, \Bbold, \Asolo} \alpha)(X, r) &\\
&\hspace{-100pt} = \iint_{\Rbb^3\times \Rbb^3} \dd Z \dd s\; k_T(Z, r-s) \, \bigl[ \e^{\i \Phi_{2A_h}(X, X+Z)} - \e^{2\i A_h(X) \cdot Z}\bigr] \, (\e^{\i Z\cdot \Pi} \alpha)(X, s)
\end{align*}
holds. Let us also recall that $\Phi_{\Abold}$ is defined in \eqref{PhiA_definition}.

We have
\begin{align*}
\Phi_{2\Asolo}(X, X+ Z) - 2\Asolo(X)\cdot Z = 2\int_0^1 \dd t \; \bigl[ \Asolo(X + (1-t) Z) - \Asolo(X)\bigr] \cdot Z,
\end{align*}
and hence
\begin{align}
\bigl| \e^{\i \Phi_{2\Asolo}(X, X+Z)} - \e^{2\i \Asolo(X)\cdot Z} \bigr| \leq \Vert DA\Vert_\infty \; |Z|^2. \label{MTA-MTAe3_2}
\end{align}
When we apply this bound and $|Z|^2 \leq
| Z + \frac{r}{2}|^2 + | Z - \frac{r}{2}|^2$, it follows that
\begin{align*}
\Vert M_{T, \Abold}\Delta - \Qcal_{T, \Bbold, \Asolo}\Delta\Vert_2^2 &\leq C\, \Vert \Psi\Vert_2^2 \, \Vert D\Asolo_h \Vert_\infty^2 \, \Vert F_T^{2} \Vert_1 \, \Vert V\alpha_*\Vert_2^2
\end{align*}
with $F_T^{2}$ in \eqref{LtildeTA-MtildeTA_FT_definition}. Using the $L^1$-norm bound for $F_T^{2}$ in \eqref{LtildeTA-MtildeTA_FTGT}, we conclude the claimed estimate for this term. 

To obtain a bound for the first gradient term, we start by noting that
\begin{align*}
\Vert \Pi (M_{T, \Asolo}\Delta - \Qcal_{T, \Bbold, \Asolo}\Delta )\Vert_2^2 &\leq C \, \Vert \Psi \Vert_2^2  \int_{\Rbb^3} \dd r \; \Bigl| \iint_{\Rbb^3\times \Rbb^3} \dd Z \dd s\; |k_T(Z, r-s)| \, |V\alpha_*(s)| \\
& \hspace{1cm} \times \sup_{X\in \Rbb^3} \bigl| \nabla_X \e^{\i \Phi_{2\Asolo_h}(X, X + Z)} - \nabla_X \e^{2\i \Asolo_h(X)\cdot Z} \bigr| \, \bigr|^2  \\
&+ C \, \Vert \Pi \e^{\i Z\cdot \Pi} \Psi\Vert_2 \, \int_{\Rbb^3} \dd r \; \Bigl| \iint_{\Rbb^3\times \Rbb^3} \dd Z \dd s\; |k_T(Z, r-s)| \, |V\alpha_*(s)| \\ 
&\hspace{1cm} \times\sup_{X\in \Rbb^3} \bigl| \e^{\i \Phi_{2\Asolo_h}(X, X + Z)} - \e^{2\i \Asolo_h(X)\cdot Z}\bigr| \, \bigr|^2,
\end{align*}
where $\Pi = -\i \nabla + 2 \Abold_{\Bbold}$ is understood to act on the center-of-mass coordinate. When we additionally use
\begin{align*}
\bigl| \nabla_X \e^{\i \Phi_{2\Asolo}(X, X + Z)} - \nabla_X\e^{2\i \Asolo(X)\cdot Z} \bigr| &\leq \bigl| \nabla_X \Phi_{2\Asolo}(X, X + Z) - 2\nabla_X \Asolo(X)\cdot Z \bigr| \\
&\hspace{-20pt} + \bigl|\Phi_{2\Asolo}(X, X+Z) - 2\Asolo(X)\cdot Z\bigr| \, |\nabla_X\Asolo(X)\cdot Z| \\
&\leq \bigl[ \Vert D^2\Asolo\Vert_\infty + \Vert D\Asolo\Vert_\infty^2 \bigr] \bigl[ |Z|^2 + |Z|^3\bigr],
\end{align*}
as well as 
\begin{equation}
	|Z|^a \leq \left| Z + \frac{r}{2} \right|^a + \left| Z - \frac{r}{2} \right|^a, \quad a \geq 0
	\label{Z_estimate}
\end{equation}
for the choices $a=2,3$, $\Vert D^2\Asolo_h\Vert_\infty \leq Ch^3$, $\Vert D\Asolo_h\Vert_\infty^2 \leq Ch^4$, \eqref{PiXcosPiA_estimate}, and \eqref{MTA-MTAe3_2}, we see that
\begin{align*}
\Vert \Pi (M_{T, \Abold} \Delta - \Qcal_{T, \Bbold, \Asolo} \Delta)\Vert_2^2 &\leq C \, h^5 \, \Vert \Psi\Vert_{\Hmag^1(Q_h)}^2 \, \Vert V\alpha_*\Vert_2^2 \, \left( \Vert F_T^{2} \Vert_1^2 + \Vert F_T^{3} \Vert_1^2 \right)
\end{align*}
holds with $F_T^{a}$ in \eqref{LtildeTA-MtildeTA_FTGT}. An application of \eqref{LtildeTA-MtildeTA_FTGT} proves the claimed bound for this term. 

In the last step we consider
\begin{align*}
\Vert \tilde \pi (M_{T, \Abold} \Delta - \Qcal_{T, \Bbold, \Asolo}\Delta) \Vert_2^2 &\leq C \, \Vert \Psi\Vert_2^2 \\
&\hspace{-130pt} \times \int_{\Rbb^3} \dd r \, \Bigl| \iint_{\Rbb^3\times \Rbb^3} \dd Z\dd s\; |\tilde \pi k_T(Z, r-s)| \, |V\alpha_*(s)| \, \sup_{X\in \Rbb^3} \bigl| \e^{\i \Phi_{2\Asolo_h}(X, X + Z)} - \e^{2\i \Asolo_h(X)\cdot Z}\bigr|\Bigr|^2.
\end{align*}
The estimate in \eqref{Z_estimate} and $\frac 14 |\Bbold \wedge r| \leq \frac 14 |\Bbold| ( |r-s| + |s| )$ allow us to prove the bound
\begin{align}
\int_{\Rbb^3} \dd Z \; |\tilde \pi k_T(Z, r-s)| \, |Z|^2 \leq F_T^3(r-s) \, (1 + |s| ) + G_T^2(r-s) \label{MTA-MTAe3_3}
\end{align}
with $F_T^a$ in \eqref{LtildeTA-MtildeTA_FT_definition} and $G_T^a$ in \eqref{LtildeTA-MtildeTA_GT_definition}. In combination with \eqref{MTA-MTAe3_2}, this proves the claimed bound for this term. It also ends the proof of the claimed bound for the first term on the right side of \eqref{MTA-MTAe3_1}. It remains to consider the second term. 

A short computation that uses $k_T(-Z,r-s) = k_T(Z,r-s)$ and $\cos(x) - 1 = -2\sin^2(\frac x2)$ shows
\begin{align*}
(\Qcal_{T, \Bbold, \Asolo}\alpha - M_{T, \Abold_{e_3}}\alpha)(X, r) &\\
&\hspace{-60pt} = -2 \iint_{\Rbb^3\times \Rbb^3} \dd Z \dd s \; k_T(Z, r-s) \, \sin^2(\Asolo_h(X)\cdot Z) (\cos(Z\cdot \Pi) \alpha)(X, s) \\
&\hspace{-30pt}- \int_{\Rbb^3\times \Rbb^3} \dd Z \dd s \; k_T(Z, r-s) \, \sin(2\Asolo_h(X) \cdot Z) \, (\sin(Z\cdot \Pi) \alpha)(X, s).
\end{align*}
From this, we check that
\begin{align*}
\Vert \Qcal_{T, \Bbold, \Asolo}\Delta - M_{T, \Abold_{e_3}}\Delta\Vert_2^2 &\leq C\, \bigl[ \Vert\Psi\Vert_2^2 \, \Vert \Asolo_h\Vert_\infty^4 + \Vert \Pi\Psi\Vert_2^2 \Vert \Asolo_h\Vert_\infty^2 \bigr] \, \Vert F_T^{2} \Vert_1^2 \, \Vert V\alpha_*\Vert_2^2
\end{align*}
holds with $F_T^{a}$ in \eqref{LtildeTA-MtildeTA_FT_definition}. When we use \eqref{MTA-MTAe3_2} to obtain a bound for the $L^1$-norm of $F_T^{2}$, this proves the claimed bound for this term.

Next, we note that
\begin{align}
&\Vert \Pi ( \Qcal_{T, \Bbold, \Asolo}\Delta - M_{T, \Abold_{e_3}}\Delta)\Vert_2^2 \leq C \, \int_{\Rbb^3} \dd r \, \Bigl| \iint_{\Rbb^3\times \Rbb^3} \dd Z\dd s \; |k_T(Z, r-s)| \, |V\alpha_*(s)| \notag \\%
& \times \Bigl[ \sup_{X \in \mathbb{R}^3} | \nabla_X \sin^2 (\Asolo_h(X)\cdot Z) | \, \Vert \cos(Z\cdot \Pi)\Psi\Vert_2 + \sup_{X \in \mathbb{R}^3} | \sin^2(\Asolo_h(X)\cdot Z) | \, \Vert \Pi \cos(Z\cdot\Pi)\Psi\Vert_2 \notag \\
&+ \sup_{X \in \mathbb{R}^3} | \sin(2\Asolo_h(X)\cdot Z) | \, \Vert \sin(Z\cdot \Pi) \Psi\Vert_2 + \sup_{X \in \mathbb{R}^3} | \sin(2\Asolo_h(X)\cdot Z) | \, \Vert \Pi\sin(Z\cdot \Pi) \Psi\Vert_2\Bigr]\Bigr|^2. \label{alphaDelta_decomposition_5}
\end{align}
We have
\begin{align*}
\Vert \sin(Z\cdot \Pi)\Psi\Vert_2 \leq C \, h^2 \, \Vert \Psi\Vert_{\Hmag^1(Q_h)} \, |Z|.
\end{align*}
Furthermore, from a straight forward computation or from \cite[Lemma 5.12]{DeHaSc2021}, we know that
\begin{align*}
\Pi_X \, \sin(Z\cdot \Pi_X) = \sin(Z\cdot \Pi_X) \;\Pi_X + 2\i \, \cos (Z\cdot \Pi_X) \; \Bbold \wedge Z,
\end{align*}
and hence
\begin{align*}
\Vert \Pi \sin(Z\cdot \Pi)\Psi\Vert_2 &\leq C \, h^2 \Vert \Psi\Vert_{\Hmag^1(Q_h)} \, (1 + |Z|).
\end{align*}
To obtain the bound we used that $|\Bbold| = h^2$. For $\Pi\cos(Z\cdot \Pi)\Psi$ a similar estimate was obtained in \eqref{MtildeTA-MTA_Lemma}. Putting these bounds together, we find that the term on the left side of \eqref{alphaDelta_decomposition_5} is bounded by a constant times $h^6 \Vert V \alpha_* \Vert_2^2 \Vert \Psi\Vert_{\Hmag^1(Q_h)}^2 $. It remains to consider the term proportional to $\tilde \pi$.

A straightforward computation shows that
\begin{align*}
\Vert \tilde \pi(\Qcal_{T, \Bbold, \Asolo}\Delta - M_{T, \Abold_{e_3}}\Delta)\Vert_2^2 &\leq C\bigl[ \Vert \Asolo_h\Vert_\infty^4 \Vert \Psi\Vert_2^2 + \Vert \Asolo_h\Vert_\infty^2 \Vert \Pi\Psi\Vert_2^2 \bigr] \\
&\hspace{20pt} \times \int_{\Rbb^3} \dd r \, \Bigl| \iint_{\Rbb^3\times \Rbb^3} \dd Z \dd s \; |\tilde \pi k_T(Z, r-s)| \, |V\alpha_*(s)| \, |Z|^2\Bigr|^2.
\end{align*}
We use \eqref{MTA-MTAe3_3} to see that the term on the left side is bounded by a constant times $h^6 \max_{k=0,1} \Vert \ | \cdot |^k \ V\alpha_*\Vert_2^2 \, \Vert \Psi\Vert_{\Hmag^1(Q_h)}^2$. This proves Proposition~\ref{MTA-MTAe3}.
\end{proof}

The next lemma provides us with a bound for the term in \eqref{eta_perp_definition} that is proportional to $L_{T, \Abold, W} - L_{T, \Abold}$.

\begin{lem}
\label{RTAW2_estimate}
Let $V\alpha_*\in L^{2}(\Rbb^3)$, let $W \in L^{\infty}(\mathbb{R}^3)$ and $A\in L^{\infty}(\mathbb{R}^3,\mathbb{R}^3)$ be periodic, assume $\Psi\in \Hmag^1(Q_h)$, and denote $\Delta \equiv \Delta_\Psi$ as in \eqref{Delta_definition}. For any $T_0>0$ there is $h_0>0$ such that for any $T\geq T_0$ and any $0 < h \leq h_0$ we have	
\begin{align*}
\Vert L_{T, \Abold, W} \Delta - L_{T, \Abold} \Delta \Vert_{\Hsymm}^2 &\leq C \, h^6 \, \Vert V \alpha_* \Vert_2^2 \, \Vert \Psi\Vert_{\Hmag^1(Q_h)}^2.
\end{align*}
\end{lem}

\begin{proof}
To prove the lemma, we write $L_{T, \Abold, W} \Delta - L_{T, \Abold} \Delta$ as in \eqref{RTAW2_estimate_1} and use the representation of the $H^1$-norm in \eqref{Norm_equivalence_2}. The details are a straight forward application of arguments that have been used already several times above, and are therefore left to the reader.
%
\end{proof}

When we combine Propositions~\ref{LTA-LtildeTA}, \ref{LtildeTA-MtildeTA}, \ref{MtildeTA-MTA}, \ref{MT1} as well as \eqref{eq:A23} and Lemma~\ref{RTAW2_estimate}, we obtain the claimed bound for $\eta_0(\Delta)$ in \eqref{alphaDelta_decomposition_eq2}, that is, part (a) of Proposition~\ref{Structure_of_alphaDelta}. The proofs of parts (b) and (c) can be found in \cite{DeHaSc2021}, see the proofs of Proposition 3.2 (b), (c). This ends our proof of Proposition~\ref{Structure_of_alphaDelta}.

\subsection{Proof of Proposition \ref{Lower_Tc_a_priori_bound}}
\label{Lower_Tc_a_priori_bound_proof_Section}

Let the assumptions of Proposition~\ref{Lower_Tc_a_priori_bound} hold. In the following we prove that there are constants $D_0>0$ and $h_0>0$ such that for $0 < h \leq h_0$ and
\begin{align*}
	0 < T_0 \leq T < \Tc (1 - D_0 h^2)
\end{align*}
there is a function $\Psi \in \Hmag^2(Q_h)$, such that the energy of the Gibbs state $\Gamma_\Delta$ in \eqref{GammaDelta_definition} with gap function $\Delta(X,r) = -2 V\alpha_*(r) \Psi(X)$ satisfies \eqref{Lower_critical_shift_2}.

Let $\psi \in \Hmag^2(Q_1)$ with $\Vert \psi\Vert_{\Hmag^2(Q_h)}=1$ and define $\Psi(X) = h \psi(hX)$. The function $\Psi$ satisfies $\Vert \Psi\Vert_{\Hmag^2(Q_h)}=1$. When we apply Propositions~\ref{Structure_of_alphaDelta}, \ref{BCS functional_identity}, \ref{Rough_bound_on_BCS energy}, as well as \eqref{Magnetic_Sobolev} and \eqref{eq:A28}, we find
\begin{align*}
	\FBCS(\Gamma_\Delta) - \FBCS(\Gamma_0) &< h^2 \, \bigl( - cD_0 \, \Vert \psi\Vert_2^2 + C \bigr)
\end{align*}
for $h$ small enough.
The proof of Proposition~\ref{Lower_Tc_a_priori_bound} is completed when we choose $D_0 = \frac{C}{c \Vert \psi\Vert_2^2}$. 

\section{The Structure of Low-Energy States}
\label{Lower Bound Part A}

In this section we prove a priori bounds for low-energy states of the BCS functional in the sense of \eqref{Second_Decomposition_Gamma_Assumption} below. The goal is to show that their Cooper pair wave function has a structure similar to that of the trial state we use in the proof of the upper bound in Section~\ref{Upper_Bound}. These bounds and the trial state analysis in Section~\ref{Upper_Bound} are the main technical ingredients for the proof of the lower bound in Section~\ref{Lower Bound Part B}. To prove the a priori bounds, we show that the periodic external potentials $W_h$ and $A_h$ can be treated as a perturbation, which reduces the problem to proving a priori bounds for the case of a constant magnetic field. The solution of this problem has been the main novelty in \cite[Theorem~5.1]{DeHaSc2021} and we apply it here. In case of a magnetic field with zero flux through the unit cell such bounds have been proved for the first time in \cite{Hainzl2012}. The idea to reduce the problem to the case of a constant magnetic field is inspired by a similar perturbative analysis in \cite{Hainzl2012}.

%
%
We recall the definition of the generalized one-particle density matrix $\Gamma$ in \eqref{Gamma_introduction}, its Cooper pair wave function $\alpha = \Gamma_{12}$, as well as the normal state $\Gamma_0$ in \eqref{Gamma0}.

\begin{thm}[Structure of low-energy states]
\label{Structure_of_almost_minimizers}
Let Assumptions \ref{Assumption_V} and \ref{Assumption_KTc} hold. For given $D_0, D_1 \geq 0$, there is a constant $h_0>0$ such that for all $0 <h \leq h_0$ the following holds: If $T>0$ obeys $T - \Tc \geq -D_0h^2$ and if $\Gamma$ is a gauge-periodic state with low energy, that is,
\begin{align}
\FBCS(\Gamma) - \FBCS(\Gamma_0) \leq D_1h^4, \label{Second_Decomposition_Gamma_Assumption}
\end{align}
then there are $\Psi\in \Hmag^1(Q_h)$ and $\xi\in \Hsymm$ such that
\begin{align}
\alpha(X,r) = \alpha_*(r) \Psi(X) + \xi(X,r), \label{Second_Decomposition_alpha_equation}
\end{align}
where
\begin{align}
\sup_{0< h\leq h_0} \Vert \Psi\Vert_{\Hmag^1(Q_h)}^2 &\leq C, &  \Vert \xi\Vert_\Hsymm^2 &\leq Ch^4 \bigl( \Vert \Psi\Vert_{\Hmag^1(Q_h)}^2 + D_1\bigr). \label{Second_Decomposition_Psi_xi_estimate}
\end{align}
\end{thm}

\begin{varbems}
\begin{enumerate}[(a)]
\item Equation~\eqref{Second_Decomposition_Psi_xi_estimate} shows that $\Psi$ is a macroscopic quantity in the sense that its $\Hmag^1(Q_h)$-norm scales as that of the function in \eqref{GL-rescaling}. It is important to note that the $\Hmag^1(Q_h)$-norm is scaled with $h$, see \eqref{Periodic_Sobolev_Norm}. The unscaled $\Lmag^2(Q_h)$-norm of $\Psi$ is of the order $h$, and therefore much larger than that of $\xi$, see \eqref{Second_Decomposition_Psi_xi_estimate}.



\item Theorem~\ref{Structure_of_almost_minimizers} has been proven in \cite[Theorem 5.1]{DeHaSc2021} for the case of a constant external magnetic field, where $A_h =0$ and $W_h = 0$. Our proof of Theorem~\ref{Structure_of_almost_minimizers} for general external fields reduces the problem to that case.
\end{enumerate}
\end{varbems}

Although Theorem~\ref{Structure_of_almost_minimizers} contains the natural a priori bounds for low-energy states, we need a slightly different version of it in our proof of the lower bound for the BCS free energy in Section~\ref{Lower Bound Part B}. The main reason is that we intend to use the function $\Psi$ from the decomposition of the Cooper pair wave function of a low-energy state in \eqref{Second_Decomposition_alpha_equation} to construct a Gibbs state $\Gamma_{\Delta}$ as in \eqref{GammaDelta_definition}. In order to be able to justify the relevant computations with this state, we need $\Psi \in \Hmag^2(Q_h)$, which is not guaranteed by Theorem~\ref{Structure_of_almost_minimizers} above, see also Remark~\ref{rem:alpha}. The following corollary provides us with a decomposition of $\alpha$, where the center-of-mass wave function $\Psi_\leq$ has the required $\Hmag^2(Q_h)$-regularity. A decomposition with a cut-off function of the form in the corollary has also been used in \cite{Hainzl2012,Hainzl2014,Hainzl2017,DeHaSc2021}.



\begin{kor}
\label{Structure_of_almost_minimizers_corollary}
Let the assumptions of Theorem~\ref{Structure_of_almost_minimizers} hold and let $\varepsilon \in [h^2, h_0^2]$. Let $\Psi$ be as in 
\eqref{Second_Decomposition_alpha_equation} and define
\begin{align}
	\Psi_\leq &\coloneqq  \Idbb_{[0,\varepsilon]}(\Pi^2) \Psi, &  \Psi_> &\coloneqq  \Idbb_{(\varepsilon,\infty)}(\Pi^2) \Psi. \label{PsileqPsi>_definition}
\end{align}
Then, we have
\begin{align}
	\Vert \Psi_\leq\Vert_{\Hmag^1(Q_h)}^2 &\leq \Vert \Psi\Vert_{\Hmag^1(Q_h)}^2, \notag \\ 
	\Vert \Psi_\leq \Vert_{\Hmag^k(Q_h)}^2 &\leq C\, (\varepsilon h^{-2})^{k-1} \, \Vert \Psi\Vert_{\Hmag^1(Q_h)}^2, \qquad k\geq 2,  \label{Psileq_bounds}
\end{align}
as well as 
\begin{align}
\Vert \Psi_>\Vert_2^2 &\leq C \varepsilon^{-1}h^4 \, \Vert \Psi\Vert_{\Hmag^1(Q_h)}^2, & \Vert \Pi\Psi_>\Vert_2^2 &\leq Ch^4 \, \Vert \Psi\Vert_{\Hmag^1(Q_h)}^2. \label{Psi>_bound}
\end{align}
Furthermore,
\begin{align}
	\sigma_0(X,r) \coloneqq  \alpha_*(r) \Psi_>(X) \label{sigma0}
\end{align}
satisfies
\begin{align}
	\Vert \sigma_0\Vert_{H^1_\symm(Q_h\times \Rbb^3)}^2 &\leq C\varepsilon^{-1}h^4 \, \Vert \Psi\Vert_{\Hmag^1(Q_h)}^2 \label{sigma0_estimate}
\end{align}
and, with $\xi$ in \eqref{Second_Decomposition_alpha_equation}, the function
\begin{align}
	\sigma \coloneqq   \xi + \sigma_0 \label{sigma}
\end{align}
obeys
\begin{align}
	\Vert \sigma\Vert_{H^1_\symm(Q_h\times \Rbb^3)}^2 \leq Ch^4 \bigl( \varepsilon^{-1}\Vert \Psi\Vert_{\Hmag^1(Q_h)}^2 + D_1\bigr). \label{Second_Decomposition_sigma_estimate}
\end{align}
In terms of these functions, the Cooper pair wave function $\alpha$ of the low-energy state $\Gamma$ in \eqref{Second_Decomposition_Gamma_Assumption} admits the decomposition
\begin{align}
	\alpha(X,r) = \alpha_*(r) \Psi_\leq (X) + \sigma(X,r). \label{Second_Decomposition_alpha_equation_final}
\end{align}
\end{kor}
For a proof of the corollary we refer to the proof of Corollary~5.2 in \cite{DeHaSc2021}.


\subsection{A lower bound for the BCS functional}

We start the proof of Theorem \ref{Structure_of_almost_minimizers} with the following lower bound on the BCS functional, whose proof is literally the same as that of the comparable statement in \cite{Hainzl2012}.


\begin{lem}
Let $\Gamma_0$ be the normal state in \eqref{Gamma0}. We have the lower bound
\begin{align}
\FBCS(\Gamma) - \FBCS(\Gamma_0) \geq  \Tr\bigl[ (K_{T,\Abold, W} - V) \alpha  \alpha^*\bigr] + \frac{4T}{5} \Tr\bigl[ (\alpha^* \alpha)^2\bigr], \label{Lower_Bound_A_3}
\end{align}
where
\begin{align}
K_{T, \Abold, W} = \frac{(-\i \nabla + \Abold_h)^2 + W_h- \mu}{\tanh (\frac{(-\i \nabla + \Abold_h)^2 + W_h - \mu}{2T})} \label{KTAW_definition}
\end{align}
and $V\alpha(x,y) = V(x-y) \alpha(x,y)$.
\end{lem}

In Proposition~\ref{prop:gauge_invariant_perturbation_theory} in Appendix~\ref{KTV_Asymptotics_of_EV_and_EF_Section} we show that the external electric and magnetic fields can lower the lowest eigenvalue zero of $K_{\Tc} - V$ at most by a constant times $h^2$. We use this in the next lemma to show that $K_{T, \Abold, W} - V$ is bounded from below by a nonnegative operator, up to a correction of the size $Ch^2$. 
\begin{lem}
\label{KTB_Lower_bound}
Let Assumptions \ref{Assumption_V} and \ref{Assumption_KTc} be true. For any $D_0 \geq 0$, there are constants $h_0>0$ and $T_0>0$ such that for $0< h\leq h_0$ and $T>0$ with $T - \Tc \geq -D_0h^2$, the estimate
\begin{align}
K_{T, \Abold, W} - V &\geq c \; (1 - P) (1 + \pi^2) (1- P) + c \, \min \{ T_0, (T - \Tc)_+\} - Ch^2 \label{KTB_Lower_bound_eq}
\end{align}
holds. Here, $P = |\alpha_*\rangle\langle \alpha_*|$ is the orthogonal projection onto the ground state $\alpha_*$ of $K_{\Tc} - V$ and $\pi = -\i \nabla + \Abold_\Bbold$.
\end{lem}

\begin{proof}
Since $W \in L^{\infty}(\mathbb{R}^3)$ we can use Lemma \ref{KT_integral_rep} to show that $K_{T, \Abold, W} \geq K_{T, \Abold, 0} - C h^2$ holds. The rest of the proof goes along the same lines as that of \cite[Lemma 5.4]{DeHaSc2021} with the obvious replacements. In particular, \cite[Proposition~A.1]{DeHaSc2021} needs to be replaced by Proposition~\ref{prop:gauge_invariant_perturbation_theory}. We omit the details.
\end{proof}

We deduce two corollaries from \eqref{Lower_Bound_A_3} and Lemma \ref{KTB_Lower_bound}. The first statement is an a priori bound that will be used in the proof of Theorem \ref{Main_Result_Tc}~(b). Its proof goes along the same lines as that of \cite[Corollary 5.5]{DeHaSc2021}.
\begin{kor}
\label{TcB_First_Upper_Bound}
Let Assumptions \ref{Assumption_V} and \ref{Assumption_KTc} be true. Then, there are constants $h_0>0$ and $C>0$ such that for all $0 < h \leq h_0$ and all temperatures $T\geq \Tc(1 + Ch^2)$, we have $\FBCS(\Gamma) - \FBCS(\Gamma_0) >0$ unless $\Gamma = \Gamma_0$.
\end{kor}


The second corollary provides us with an inequality for Cooper pair wave functions of low-energy BCS states in the sense of \eqref{Second_Decomposition_Gamma_Assumption}. The left side of \eqref{Lower_Bound_A_2} appears as a lower bound for the full BCS functional. Despite of its apparent simplicity, it still contains all the information needed for a proof of Theorem~\ref{Structure_of_almost_minimizers}. Before we state the corollary, let us define the operator
\begin{align}
U  &\coloneqq  \e^{-\i \frac r2 \Pi_X}, \label{U_definition}
\end{align}
with $\Pi_X$ in \eqref{Magnetic_Momenta_COM}, which acts on the relative coordinate $r = x-y$ as well as on the center-of-mass coordinate $X = \frac{x+y}{2}$ of a function $\alpha(X,r)$.

\begin{kor}
\label{cor:lowerbound}
Let Assumptions \ref{Assumption_V} and \ref{Assumption_KTc} be true. For any $D_0, D_1 \geq 0$, there is a constant $h_0>0$ such that if $\Gamma$ is a low-energy state in the sense that it satisfies \eqref{Second_Decomposition_Gamma_Assumption}, if $0 < h\leq h_0$, and if $T$ is such that $T - \Tc \geq -D_0h^2$, then $\alpha = \Gamma_{12}$ obeys
\begin{align}
&\langle \alpha, [U(1 - P)(1 + \pi^2)(1 - P)U^* + U^*(1 - P)(1 + \pi^2)(1 - P)U] \alpha \rangle \notag\\
&\hspace{200pt} + \Tr\bigl[(\alpha^* \alpha)^2\bigr] \leq C h^2 \Vert \alpha\Vert_2^2 + D_1h^4, \label{Lower_Bound_A_2}
\end{align}
where $P = | \alpha_* \rangle \langle \alpha_* |$ and $\pi = -\i \nabla_r + \Abold_\Bbold(r)$ both act on the relative coordinate $r$ of $\alpha(X,r)$.
\end{kor}

In the statement of the corollary and in the following, we refrain from equipping the operator $\pi$ and the projection $P = |\alpha_*\rangle \langle \alpha_*|$ with an index $r$ although it acts on the relative coordinate. This does not lead to confusion and keeps the formulas readable. The proof of the corollary is inspired by the proof of \cite[Proposition~23]{Hainzl2017}.

\begin{proof}[Proof of Corollary~\ref{cor:lowerbound}]
In the following we use the notation $\pi_{\Abold}^x = -\i \nabla_x + \Abold(x)$. We claim that
\begin{align}
\pi_{\Abold_h}^x &= U \pi_{\Abold^+_h}^r U^*, & -\pi_{\Abold_h}^y &= U^* \pi_{\Abold^-_h}^r U, \label{cor:lowerbound_1}
\end{align}
where $\pi_{\Abold^\pm}^r = -\mathrm{i} \nabla_r + \Abold^{\pm}(r)$ with
\begin{align*}
\Abold^\pm(r) &\coloneqq \Abold_{e_3}(r) \pm A(X \pm r).
\end{align*}
To obtain \eqref{cor:lowerbound_1}, we denote $P_X = -\mathrm{i} \nabla_X$ and note that $[r \cdot P_X, r \cdot (\Bbold \wedge X)] = 0$ implies $U = \e^{\i \frac{\Bbold}{2} (X \wedge r)} \e^{- \i \frac{r}{2} P_X}$. Using this identity we conclude that
\begin{align*}
	U \; (-\mathrm{i} \nabla_r + \Abold_{\Bbold}(r)) \,  U^* &= -\mathrm{i} \nabla_r + \frac{1}{2} \Abold_{\Bbold}(r) - \frac{1}{2} \Pi_X, \\
	U^* (-\mathrm{i} \nabla_r + \Abold_{\Bbold}(r)) \, U \; &= -\mathrm{i} \nabla_r + \frac{1}{2} \Abold_{\Bbold}(r) + \frac{1}{2} \Pi_X,
\end{align*}  
where $\Pi = -\i \nabla + 2 \Abold_{\Bbold}$. Eq.~\eqref{cor:lowerbound_1} is a direct consequence of these two identities. 

We also have
\begin{align*}
W(x) &= U \, W^+(r) \, U^*, & W(y) &= U^* \, W^-(r) \, U, & W^\pm(r) &\coloneqq  W(X \pm r).
\end{align*}
Consequently, if $K_{T, \Abold, W}^x$ and $K_{T, \Abold, W}^y$ denote the operators $K_{T, \Abold, W}$ acting on the $x$ and $y$ coordinate, respectively, we infer
\begin{align}
K_{T, \Abold, W}^x - V(r) &= U^* ( K_{T, \Abold^+, W^+}^r - V(r) ) \, U, \notag \\ 
K_{T, \Abold, W}^y - V(r) &= U \; ( K_{T, \Abold^-, W^-}^r - V(r) ) \, U^*. \label{eq:idK_T^r} 
\end{align}
We highlight that $\Abold^\pm$ and $W^\pm$ depend on $X$.

The operator $V$ in \eqref{Lower_Bound_A_3} acts by multiplication with the function $V(x-y)$. We use the symmetry $\alpha(x,y) = \alpha(y,x)$ to deduce
\begin{align}
\Tr \bigl[ (K_{T, \Abold, W} - V) \alpha \alpha^*\bigr] & \notag\\
&\hspace{-80pt} = \frac 12 \fint_{Q_h} \dd x \int_{\Rbb^3} \dd y \; \overline{\alpha(x,y)} \bigl[ (K_{T, \Abold, W}^x - V) + (K_{T, \Abold, W}^y - V) \bigr] \alpha(x,y). \label{Lower_Bound_A_4}
\end{align}
In combination, \eqref{Second_Decomposition_Gamma_Assumption}, \eqref{Lower_Bound_A_3}, \eqref{eq:idK_T^r}, and \eqref{Lower_Bound_A_4} therefore prove the bound
\begin{align*}
\frac 12\langle \alpha, [U^* ( K_{T, \Abold^+, W^+}^r - V(r) ) \, U + U \; ( K_{T, \Abold^-, W^-}^r - V(r) ) \, U^*]\alpha\rangle + c \Tr \bigl[ (\alpha^* \alpha)^2\bigr] \leq D_1 h^4.
\end{align*}
An application of Lemma \ref{KTB_Lower_bound} on the left side establishes \eqref{Lower_Bound_A_2}.
\end{proof}

%
%

\subsection{Proof of Theorem \ref{Structure_of_almost_minimizers}}

The statement in Theorem \ref{Structure_of_almost_minimizers} is a direct consequence of Corollary~\ref{cor:lowerbound} above and the results in \cite{DeHaSc2021}. More precisely, we need to combine Corollary~\ref{cor:lowerbound}, \cite[Proposition~5.7]{DeHaSc2021}, \cite[Lemma~5.14]{DeHaSc2021}, and the arguments in \cite[Section~5.4]{DeHaSc2021}.

\section{The Lower Bound on \texorpdfstring{(\ref{ENERGY_ASYMPTOTICS})}{(\ref{ENERGY_ASYMPTOTICS})} and Proof of Theorem \ref{Main_Result_Tc} (b)}
\label{Lower Bound Part B}

\subsection{The BCS energy of low-energy states}

In this section, we complete the proofs of Theorems \ref{Main_Result} and \ref{Main_Result_Tc}, which amounts to providing the lower bound on \eqref{ENERGY_ASYMPTOTICS}, the bound in \eqref{GL-estimate_Psi}, and the proof of Theorem~\ref{Main_Result_Tc}~(b). Since these proofs mostly go along the same lines as those in \cite{DeHaSc2021}, we only mention the differences and keep the presentation to a minimal length. Once the a priori estimates in \cite[Theorem~5.1]{DeHaSc2021} are proved, the proofs of the lower bound for the free energy, the decomposition of the Cooper pair wave function of an approximate minimizer, and the upper bound for the critical temperature shift in \cite{DeHaSc2021} follow the same strategy as the related proofs in \cite{Hainzl2012,Hainzl2014}. In the following we will, however, only refer to \cite{DeHaSc2021} because our presentation is closer to the analysis in this reference than to those in \cite{Hainzl2012,Hainzl2014}.

Let $D_1\geq 0$ and $D\in \Rbb$ be given, choose $T = \Tc(1 - Dh^2)$, and assume that $\Gamma$ is a gauge-periodic state that satisfies \eqref{Second_Decomposition_Gamma_Assumption}.  Corollary~\ref{Structure_of_almost_minimizers_corollary} guarantees a decomposition of the Cooper pair wave function $\alpha = [\Gamma]_{12}$ in terms of $\Psi_{\leq}$ in \eqref{PsileqPsi>_definition} and $\sigma$ in \eqref{sigma}. The function $\Psi_{\leq}$ satisfies the bounds
\begin{align}
	\Vert \Psi_{\leq} \Vert_{\Hmag^1(Q_h)}^2 &\leq \Vert \Psi \Vert_{\Hmag^1(Q_h)}^2 \leq C, &  \Vert \Psi_{\leq} \Vert_{\Hmag^2(Q_h)}^2 &\leq C \varepsilon h^{-2} \Vert \Psi \Vert_{\Hmag^1(Q_h)}^2, \label{eq:lowerboundB1}
\end{align}
with $\Psi$ in \eqref{Second_Decomposition_alpha_equation}. Let us define the state $\Gamma_{\Delta}$ as in \eqref{GammaDelta_definition} with $\Delta(X,r) = -2 V \alpha_* (r) \Psi_{\leq}(X)$. We apply Proposition~\ref{BCS FUNCTIONAL_IDENTITY} and Theorem~\ref{CALCULATION_OF_THE_GL-ENERGY} to obtain the following lower bound for the BCS energy of $\Gamma$:
\begin{align}
	\FBCS(\Gamma) - \FBCS(\Gamma_0) &\geq h^4\; \EGL(\Psi_{\leq}) - C \left( h^5 + \varepsilon h^4 \right) \Vert \Psi \Vert_{\Hmag^1(Q_h)}^2 \nonumber \\
	&\hspace{20pt}+ \frac{T}{2} \Hcal_0(\Gamma, \Gamma_\Delta) - \fint_{Q_h} \dd X \int_{\Rbb^3} \dd r \; V(r) \, | \sigma(X,r) |^2. \label{eq:lowerboundB2}
\end{align}
In the next section we prove a lower bound for the terms in the second line of \eqref{eq:lowerboundB2}.

\subsection{Estimate on the relative entropy}

The arguments in \cite[Eqs.~(6.1)-(6.14)]{DeHaSc2021} apply in literally the same way here, too. We obtain the correct bounds when we replace $B$ by $h^2$ in all formulas. This, in particular, applies to the statement of \cite[Lemma~6.2]{DeHaSc2021}. The only difference is that \cite[Eq.~(6.10)]{DeHaSc2021} is now given by
\begin{equation*}
	| \langle \eta_{0},  K_{T_{\mathrm{c}},\Abold,W} \sigma \rangle | \leq \ C \varepsilon^{-\nicefrac{1}{2}} h^{\nicefrac{9}{2}} \Vert \Psi \Vert_{\Hmag^1(Q_h)} \bigl( \Vert \Psi\Vert_{\Hmag^1(Q_h)}^2 + D_1\bigr)^{\nicefrac{1}{2}},
\end{equation*}
which is due to the reason that the bound for the $L^2$-norm of $\eta_0$ in Proposition~\ref{Structure_of_alphaDelta} is worse than the comparable bound we obtained in \cite[Proposition~3.2]{DeHaSc2021}. This, however, does not change the size of the remainder in the final bound because other error terms come with a worse rate. 

With the choice $\varepsilon = h^{\nicefrac 13}$, we therefore obtain the bound
\begin{align}
	&\FBCS(\Gamma) - \FBCS(\Gamma_0) \notag \\
	&\hspace{2cm}\geq h^4 \, \bigl(\EGL(\Psi_\leq) - C \,  h^{\nicefrac {1}{6}} \Vert \Psi\Vert_{\Hmag^1(Q_h)} \; \bigl(  \Vert \Psi\Vert_{\Hmag^1(Q_h)}^2 + D_1\bigr)^{\nicefrac 12} \bigr), \label{Lower_Bound_B_5}
\end{align}
which is the equivalent of \cite[Eq~(6.14)]{DeHaSc2021}.

\subsection{Conclusion}

The arguments in \cite[Section~6.3]{DeHaSc2021} apply in the same way also here and we obtain the correct formulas when we replace $B^{\nicefrac{1}{2}}$ by $h$. This concludes the proof of Theorem~\ref{Main_Result} and Theorem~\ref{MAIN_RESULT_TC}.

\subsection{Proof of the equivalent of \texorpdfstring{\cite[Lemma~6.2]{DeHaSc2021}}{Lemma~6.2 in Deuchert, Hainzl, Maier (2021)} in our setting}

To obtain a proof of the equivalent of \cite[Lemma~6.2]{DeHaSc2021} in our setting, we follow the proof strategy in \cite{DeHaSc2021}. The additional terms coming from the external electric potential are not difficult to bound because $W$ is a bounded function. To obtain bounds of the correct size in $h$ for the terms involving the periodic vector potential $A_h$, we need to use that $A(0) = 0$, which is guaranteed by Assumption~\ref{Assumption_V}. This is relevant for example when we estimate our equivalent of the term on the left side of \cite[Eq.~(6.24)]{DeHaSc2021}, that is, 
\begin{equation*}
	\Vert [ \pi_{\Abold_h}^2 + W_h(r) - p_r^2 ] \sigma_0 \Vert_2   
\end{equation*}
with $p_r = - \mathrm{i} \nabla_r$ and $\sigma_0$ in \eqref{sigma0}. We write the operator multiplying $\sigma_0$ as
\begin{equation}
	\pi_r^2 - p_r^2 + W(r) + A(r) \cdot \pi_r + \pi_r \cdot A(r) + |A(r)|^2,
	\label{eq:Andi17}
	\end{equation}
where $\pi_r = -\i \nabla_r + \Abold_{\Bbold}(r)$. When we use \eqref{Psi>_bound}, we see that the terms involving $|A_h|^2$ and $W_h$ are bounded by 
\begin{align}
	\bigl( \Vert A_h\Vert_\infty^2 + \Vert W_h \Vert_\infty \bigr)  \ \Vert \sigma_0 \Vert_2 \leq C \varepsilon^{-\nicefrac{1}{2}} h^4 \Vert \Psi\Vert_{\Hmag^1(Q_h)}.
	\label{eq:proofoflemma32b}
\end{align}
Moreover, from \cite[Eq.~(6.24)]{DeHaSc2021} we know that 
\begin{equation*}
	 \Vert [\pi_r^2 - p_r^2]\sigma_0\Vert_2 \leq C \varepsilon^{-\nicefrac{1}{2}} h^4 \Vert \Psi\Vert_{\Hmag^1(Q_h)}.
\end{equation*}
%
%
%
%

%
%
%
%
To obtain a bound for the contribution from the fourth and the fifth term on the right side of \eqref{eq:Andi17}, we write
\begin{equation*}
	A_h(r) = h^2 \int_0^1 \dd t \ (D A)(h r t) \cdot r, 
\end{equation*}
where $DA$ denotes the Jacobi matrix of $A$. Hence,
\begin{equation*}
	\Vert A_h(r) \cdot \pi_r \, \sigma_0 \Vert_2 \leq \ h^2 \Vert DA \Vert_\infty \, \Vert \ | \cdot | \pi \alpha_* \Vert_2 \, \Vert \Psi_> \Vert_2 \leq C h^4 \varepsilon^{-\nicefrac{1}{2}} \Vert \Psi\Vert_{\Hmag^1(Q_h)}.
\end{equation*}
The term involving $\pi_r \cdot A_h(r)$ can be treated similarly when we commute $\pi_r$ to the right. In combination, the above considerations show
\begin{equation*}
	\Vert [ \pi_{\Abold_h}^2 + W_h(r) - p_r^2 ] \sigma_0 \Vert_2 \leq C h^4 \varepsilon^{-\nicefrac{1}{2}} \Vert \Psi\Vert_{\Hmag^1(Q_h)}.
\end{equation*}
All other bounds in the proof of the equivalent of \cite[Lemma~6.2]{DeHaSc2021} in our setting that involve $W_h$ or $A_h$ can be estimated with similar ideas. We therefore omit further details.

\appendix
\begin{center}
\huge \textsc{--- Appendix ---}
\end{center}

\section{Gauge-Invariant Perturbation Theory for \texorpdfstring{$K_{\Tc, \Abold} -V$}{KTcA-V}}
\label{KTV_Asymptotics_of_EV_and_EF_Section}

In this appendix, we discuss the behavior of the eigenvalues below the essential spectrum of the operator $K_{\Tc, \Abold} - V$ for small $h >0$, where $K_{\Tc, \Abold} \coloneqq  K_{\Tc, \Abold, 0}$ with $K_{T, \Abold, 0}$ defined in \eqref{KTAW_definition}. We recall that the full magnetic vector potential is given by $\Abold \coloneqq  \Abold_{e_3} + A$ with the magnetic vector potential $\Abold_{e_3}(x) = \frac 12 e_3\wedge x$ of a constant magnetic field and a periodic vector potential $A$ that satisfies $A(0) =0$. We also recall the notation $\Abold_h (x) \coloneqq  h\Abold(hx)$. The goal of this appendix is to prove the following proposition. 

\begin{prop}
	\label{prop:gauge_invariant_perturbation_theory}
	Assume $(1+|\cdot|^2) V \in L^{2}(\mathbb{R}^3) \cap L^{\infty}(\mathbb{R}^3)$ and let $A \in W^{3, \infty}(\Rbb^3; \Rbb^3)$ be a periodic function. Then there is $h_0>0$ such that for $0 < h \leq h_0$ the following statements hold:
	\begin{enumerate}[(a)]
		\item Let $\lambda$ be an isolated eigenvalue of multiplicity $m \in \mathbb{N}$ of the operator $K_\Tc - V$ with spectral projection $P$. Then there are $m$ eigenvalues $\lambda_1(h), \ldots ,\lambda_m(h)$ of the operator $K_{\Tc, \Abold}-V$ with spectral projection $P(h)$ such that  
		\begin{align}
			\max_{i=1,\ldots ,m} | \lambda_i(h) - \lambda | &\leq C h^2, & \Vert P(h) - P \Vert_{\infty} &\leq  C h^2.
			\label{eq:app1}
		\end{align}
		\item Assume that $K_\Tc-V$ has a simple lowest eigenvalue with eigenfunction $\alpha_*$ and denote by $\alpha_*^h$ the eigenfunction to the lowest eigenvalue of $K_{\Tc, \Abold} -V$, which is normalized such that $\langle \alpha_*, \alpha_*^h \rangle \geq 0$ holds. Then we have the bound
		\begin{align}
			\Vert (1+\pi^2) (\alpha_*^h - \alpha_*) \Vert_2 \leq C h^2,
			\label{eq:app2}
		\end{align}
		where $\pi^2 = (-\i\nabla + \Abold_{\Bbold})^2$ denotes the magnetic Laplacian.
	\end{enumerate}
\end{prop}

\begin{bem}
	\label{rem:CombesThomas}
	The above Proposition should be compared to \cite[Proposition~A.1]{DeHaSc2021}, where a similar statement is proved in the special case, where $V \geq 0$ and where $\Abold(x) = \frac{1}{2} e_3 \wedge x$. In the reference the assumption for $V$ is needed to assure that the Birman--Schwinger operator related to $K_{T,\Abold} - V$ is self-adjoint. To prove the above proposition we investigate the resolvent kernel of $K_{T,\Abold} - V$ rather than the Birman--Schwinger operator, which is the reason why we do not need the positivity assumption for $V$. The main technical ingredient of our proof is the estimate for the integral kernel of the resolvent of $K_{T} - V$ in Lemma~\ref{lem:App2} below. Afterwards, we apply a standard version of gauge-invariant perturbation theory. As has been shown in \cite[Chapter~6]{Diss_Marcel} one can also obtain exponential decay estimates for the resolvent kernel of $K_{T} - V$ via Combes--Thomas estimates. The proof, however, requires more effort than for the case of the Laplacian because $K_{T}$ has a more complicated structure. Since these estimates are not needed for the proof of Proposition~\ref{prop:gauge_invariant_perturbation_theory},we refrain from presenting them here.
\end{bem}

We denote the resolvent of $K_{T, \Abold} - V$ at $z\in \rho(K_{T, \Abold} - V)$ by $\Rcal_\Abold^{z, V} \coloneqq  (z - (K_{\Tc, \Abold} - V))^{-1}$. The integral kernel of $\Rcal_\Abold^{z, V}$ is denoted by $\Gcal_\Abold^{z, V}(x,y)$. We highlight that we use another symbol for the resolvent and for its integral kernel in this section. Since we frequently work with $L^p$-norms of integral kernels this simplifies our notation. If $\Abold =0$, we write $\Rcal^{z, V}$ for $\Rcal_0^{z, V}$ and $\Gcal^{z, V}$ for $\Gcal_0^{z, V}$. Similarly, $\Gcal^z$ stands for $\Gcal^{z, 0}$. Before we give the proof of the above proposition, we state and prove four preparatory lemmas.

\subsection{Preparatory lemmas}

The first lemma concerns the regularity of the kernel $\Gcal^z$.

\begin{lem}
	\label{lem:App1}
	There is a continuous function $a \colon \mathbb{C} \backslash [2\Tc,\infty) \to \mathbb{R}_+$ such that
	\begin{align}
		\int_{\mathbb{R}^3} \dd x \ (1+x^2) | \nabla \Gcal^z(x) |  \leq a(z).
		\label{eq:Prep1}
	\end{align}
\end{lem}
\begin{proof}
	We use the resolvent identity $\Rcal^z = \Rcal^0 + z \Rcal^0 \Rcal^z$ to write $\Gcal^z$ as
	\begin{align}
		\Gcal^z(x) = \Gcal^0(x) + z \int_{\mathbb{R}^3} \dd v \; \Gcal^0(x-v) \, \Gcal^z(v),
		\label{eq:Prep2}
	\end{align}
	which implies
	\begin{align}
		\Vert (1+|\cdot|^2) \nabla \Gcal^z \Vert_1 \leq \Vert (1+2 |\cdot|^2) \nabla \Gcal^0 \Vert_1 \left( 1 + |z|\,  \Vert (1+ |\cdot|^2) \Gcal^z \Vert_1 \right). 
		\label{eq:Prep3}
	\end{align}
	The second $L^1(\mathbb{R}^3)$-norm on the right side of \eqref{eq:Prep3} is bounded by
	\begin{align}
		\Vert (1+|\cdot|^2) \Gcal^z \Vert_1 \leq C \ \Vert (1+|\cdot|^2)^2 \Gcal^z \Vert_2 = C \, \Bigl( \int_{\mathbb{R}^3} \dd p \ \Bigl| (1-\Delta_p)^2 \frac{1}{z - K_\Tc(p)} \Bigr|^2 \Bigr)^{\nicefrac 12}, 
		\label{eq:Prep4}
	\end{align}
	which, when multiplied with $z$, meets the requirements of the lemma. It therefore remains to consider $\Vert (1+|\cdot|^2) \nabla \Gcal^0 \Vert_1$.
	
	From \cite[Eq.~(A.6)]{Hainzl2012} and \cite[Theorem~6.23]{LiebLoss} we know that $\Gcal^0(x)$ can be written as
	\begin{align}
		\Gcal^0(x) &= \frac{2}{\pi} \sum_{n=1}^{\infty} \frac{1}{n - \frac{1}{2}} \text{Im} \ g_0^{\i (n-\frac{1}{2}) 2 \pi \Tc}(x) \label{eq:Prep5} \\
		&= \frac{1}{2 \pi^2 |x|} \sum_{n=1}^{\infty} \frac{1}{n - \frac{1}{2}} \text{Im} \exp \Bigl( \i \sqrt{\mu+\i \left(n-\frac{1}{2} \right) 2 \pi \Tc} \, |x| \Bigr), \nonumber
	\end{align}
	where $g_0^z$ is the resolvent kernel of the Laplacian in \eqref{ghz_definition} and $\sqrt{\cdot}$ denotes the principal square root. We use $\Im \e^{\i z} = \sin(\Re z) \exp(- \Im z)$ for $z \in \mathbb{C}$ and 
	\begin{align}
		\sqrt{a+ \i b} = \frac{1}{\sqrt{2}} \sqrt{\sqrt{a^2+b^2}+a} +  \frac{\i \sgn(b)}{\sqrt{2}} \sqrt{\sqrt{a^2+b^2}-a} 
		\label{eq:Prep5b} 
	\end{align}
	for $a, b \in \mathbb{R}$ to see that
	\begin{align}
		\text{Im} \exp\Bigl( \i \sqrt{\mu+\i \left(n-\frac{1}{2} \right) 2 \pi \Tc} \, |x| \Bigr) = \sin\left( |x| c_n^{+} \right)  \exp\left( -|x| c_n^{-} \right), \label{eq:Prep6}
	\end{align}
	where 
	\begin{align}
		c_n^{\pm} = \frac{ 1 }{ \sqrt{2} } \sqrt{ \sqrt{\mu^2 + ((n-1/2) 2 \pi \Tc )^2} \pm \mu}.
		\label{eq:Prep6b}
	\end{align}
	In particular,
	\begin{align}
		\nabla \Gcal^0(x) &= -\frac{x}{2 \pi^2 |x|^3} \sum_{n=1}^{\infty} \frac{1}{n - \frac{1}{2}} \sin\left( |x| c_n^{+} \right)  \exp\left( -|x| c_n^{-} \right) \nonumber\\
		&\hspace{50pt}+ \frac{x}{2 \pi^2 |x|^2} \sum_{n=1}^{\infty} \frac{c_n^{+}}{n - \frac{1}{2}} \cos\left( |x| c_n^{+} \right)  \exp\left( -|x| c_n^{-} \right)  \nonumber \\
		&\hspace{50pt}- \frac{x}{2 \pi^2 |x|^2} \sum_{n=1}^{\infty} \frac{c_n^{-}}{n - \frac{1}{2}} \sin\left( |x| c_n^{+} \right)  \exp\left( -|x| c_n^{-} \right).  \label{eq:Prep6c}
	\end{align}
	The above formula implies the bound
	\begin{align}
		\Vert (1+|\cdot|^2) \nabla \Gcal^0 \Vert_1 &\leq \sum_{n=1}^{\infty} \frac{C}{n} \int_{0}^{\infty} \dd r \left( 1+r^2 + (r+r^3) \sqrt{1+n} \right) \exp\left( -r c_n^{-} \right) \leq C \sum_{n=1}^{\infty} n^{-\frac 32}. \label{eq:Prep6d} 
	\end{align}
	This proves the claim of the lemma.
\end{proof}

The second lemma concerns bounds for the operator norm of $(z- (K_\Tc - V))^{-1}$ and commutators of this operator with $x$, when viewed as maps from $L^2(\mathbb{R}^3)$ to $L^{\infty}(\mathbb{R}^3)$. Here and in the following we denote by $\Vert A \Vert_{2;\infty}$ the norm of a bounded operator $A$ from $L^2(\mathbb{R}^3) \ra L^{\infty}(\mathbb{R}^3)$. From \cite[Corollary~A.1.2]{Simon82} we know that if such an operator is also a bounded from $L^2(\Rbb^3)$ to $L^2(\Rbb^3)$ then it has an integral kernel given by a measurable function $A(x,y)$, which obeys
\begin{align}
	\esssup_{x \in \mathbb{R}^3} \left( \int_{\mathbb{R}^3} \dd y \; | A(x,y) |^2 \right)^{\nicefrac 12} < \infty.	\label{eq:Prep6e}
\end{align}
Moreover, the norm $\Vert A \Vert_{2;\infty}$ equals the norm of the integral kernel of $A$ in \eqref{eq:Prep6e}.

The following Lemma \ref{lem:App2} implies that the resolvent kernel $\Gcal^{z, V}$ satisfies
\begin{align}
\esssup_{x\in \Rbb^3} \Bigl( \int_{\Rbb^3} \dd y \; \bigl( 1 + |x-y|^4\bigr) |\Gcal^{z, V}(x,y)|^2 \Bigr)^{\nicefrac 12} < \infty. \label{eq:App_decay_resolvent_kernel}
\end{align}
In passing we note that our assumptions on $V$ would allow for more: it can be shown that $\Gcal^{z, V}$ is exponentially decaying in the sense that \eqref{eq:App_decay_resolvent_kernel} holds with $1 + |x-y|^4$ replaced by $\e^{\delta |x-y|}$ for some $\delta >0$ depending on the distance of $z$ to the spectrum of $K_\Tc- V$. For more information we refer to Remark~\ref{rem:CombesThomas} above.

\begin{lem}
	\label{lem:App2}
	Assume that $V \in L^{\infty}(\mathbb{R}^3)$ vanishes at infinity. Then there is a continuous function $a \colon \rho(K_\Tc-V) \to \mathbb{R}_+$ such that
	\begin{align}
		\Vert \Rcal^{z, V}  \Vert_{2;\infty} + \bigl\Vert [x, \Rcal^{z, V} ] \bigr\Vert_{2;\infty} + \bigl\Vert [x,[x, \Rcal^{z, V} ]] \bigr\Vert_{2;\infty} \leq a(z). 
		\label{eq:Prep7}
	\end{align}
\end{lem}

\begin{proof}
	We start by proving the bound for the first term on the right side of \eqref{eq:Prep7}. We use the fact that $(1-\Delta)^{-1}$ is a bounded linear map from $L^2(\mathbb{R}^3)$ to $L^{\infty}(\mathbb{R}^3)$ and the resolvent identity to estimate
	\begin{align}
		\Vert \Rcal^{z, V} \Vert_{2;\infty} \leq C \Vert (1-\Delta) \Rcal^{z, V} \Vert_\infty \leq C \Vert (1-\Delta) \Rcal^z \Vert_\infty \left( 1 + \Vert V \Vert_\infty \Vert \Rcal^{z, V} \Vert_\infty \right). 
		\label{eq:Prep8}
	\end{align}
	Our assumptions on $V$ imply $\rho(K_\Tc - V) \subseteq \rho(K_\Tc)$, whence both $z$-dependent terms on the right side meet the requirements of the lemma.
	
	To obtain a bound for the second term on the right side of \eqref{eq:Prep7}, we note that
	\begin{align}
		[x, \Rcal^{z, V} ]  &= \Rcal^{z, V} [K_\Tc,x] \Rcal^{z, V}, & [K_\Tc,x] &= -\i (\nabla f)(-\i\nabla),
		\label{eq:Prep9}
	\end{align}
	where $f(p) \coloneqq  K_\Tc(p)$ is the symbol in \eqref{KT-symbol}. Using this, we estimate
	\begin{align}
		\Vert [x,\Rcal^{z, V}] \Vert_{2;\infty} \leq \Vert \Rcal^{z, V} \Vert_{2;\infty} \, \Vert (\nabla f)(-\i\nabla) \Rcal^{z, V} \Vert_\infty.
	\end{align}
	A bound for the first factor on the right side was obtained in \eqref{eq:Prep8}. Using the resolvent identity again, we bound the second factor by
	\begin{align}
		\Vert (\nabla f)(-\i\nabla) \Rcal^{z, V} \Vert_\infty \leq \Vert (\nabla f)(-\i\nabla) \Rcal^z \Vert_\infty \left( 1 + \Vert V \Vert_\infty \Vert \Rcal^{z, V} \Vert_\infty \right),
	\end{align}
	which proves the claim for the second term on the right side of \eqref{eq:Prep7}. A bound for the third term can be derived similarly, and we therefore leave the remaining details to the reader. This proves the claim.
\end{proof}

\begin{lem}
	\label{lem:regEF}
	Assume $(1+|\cdot|^2)V \in L^{2}(\mathbb{R}^{3}) \cap L^{\infty}(\mathbb{R}^{3})$ and let $\alpha$ be an eigenfunction of the operator $K_\Tc - V$ with eigenvalue $\lambda < 2 \Tc$. Then we have
	\begin{align}
		\Vert \, |\cdot| \nabla \alpha \Vert_2 + \Vert \, |\cdot|^2 \alpha \Vert_2 < \infty.
		\label{eq:reg1}
	\end{align}
\end{lem}

\begin{proof}
	We use the eigenvalue equation to write the Fourier transform of $\alpha$ as 
	\begin{align}
		\hat{\alpha}(p) = -\frac{1}{\lambda - K_\Tc(p)} \,  (\hat{V} \ast \hat{\alpha})(p).
		\label{eq:reg2}
	\end{align}
	Using Young's inequality, we see that this implies
	\begin{align}
		\Vert \, |\cdot|^2 \alpha \Vert &= \Bigl( \int_{\mathbb{R}^3} \dd p \; \Bigl| \Delta_p \frac{1}{\lambda - K_\Tc(p)} \, (\hat{V} \ast \hat{\alpha}) (p) \Bigr|^2  \Bigr)^{\nicefrac 12} \nonumber \\
		&\leq C \left( \Vert \hat{V} \ast \hat{\alpha} \Vert_\infty + \Vert \Delta (\hat{V} \ast \hat{\alpha}) \Vert_\infty \right) \leq C \, \Vert (1+|\cdot|^2) V \Vert_\infty. \label{eq:reg3} 
	\end{align}
	To prove the other bound, we use the resolvent identity to write \eqref{eq:reg2} as
	\begin{align}
		\hat{\alpha}(p) = - \frac{1}{K_\Tc(p)} \, (\hat{V} \ast \hat{\alpha})(p) + \frac{\lambda}{K_\Tc(p)} \frac{1}{\lambda - K_\Tc(p)} \, (\hat{V} \ast \hat{\alpha})(p). \label{eq:reg4}
	\end{align}
	We argue as in \eqref{eq:reg3} to see that the $L^2(\mathbb{R}^3)$-norm of $\nabla_p p \frac{\lambda}{K_\Tc(p)} \frac{1}{\lambda - K_\Tc(p)} (\hat{V} \ast \hat{\alpha})(p)$ is bounded by a constant times $\Vert \, |\cdot| V \Vert_\infty$. To treat the other term, we go back to position space and note that
	\begin{align}
		\Bigl( \int_{\mathbb{R}^3} \dd x \left| \int_{\mathbb{R}^3} \dd y \; |x| \, |\nabla \Gcal^0(x-y)| \, |V\alpha(y)| \right|^2 \Bigr)^{\nicefrac 12} \leq \Vert (1+|\cdot|) \nabla \Gcal^0 \Vert_1 \, \Vert (1+|\cdot|) V \Vert_{\infty}. 
		\label{eq:reg5}
	\end{align} 
In combination with Lemma~\ref{lem:App1}, these considerations prove the claim.
\end{proof}

The last lemma provides us with a convenient representation for the operator $K_{T,\Abold}$. Its proof can be found in \cite[Lemma~6.4]{DeHaSc2021}. Before we can state the lemma we need the following definition.

\begin{defn}[Speaker path]
	\label{speaker path}
	Let $R>0$, assume that $\mu \geq -1$ and define the following complex paths 
	\begin{align*}
		\begin{split}
			u_1(t) &\coloneqq  \frac{\pi\i}{2\betac} + (1 + \i)t\\
			u_2(t) &\coloneqq  \frac{\pi\i}{2\betac} - (\mu + 1)t\\
			u_3(t) &\coloneqq  -\frac{\pi\i}{2\betac}t - (\mu + 1)\\
			u_4(t) &\coloneqq  -\frac{\pi\i}{2\betac} - (\mu + 1)(1-t) \\
			u_5(t) &\coloneqq  -\frac{\pi\i }{2\betac} + (1 - \i)t
		\end{split}
		&
		\begin{split}
			\phantom{ \frac \i\betac }t&\in [0,R], \\
			\phantom{ \frac \i\betac }t &\in [0,1], \\
			\phantom{ \frac \i\betac }t&\in [-1,1],\\
			\phantom{ \frac \i\betac }t &\in [0,1],\\
			\phantom{ \frac \i \betac }t&\in [0,R].
		\end{split}
		& \begin{split} 
			\text{\includegraphics[width=6cm]{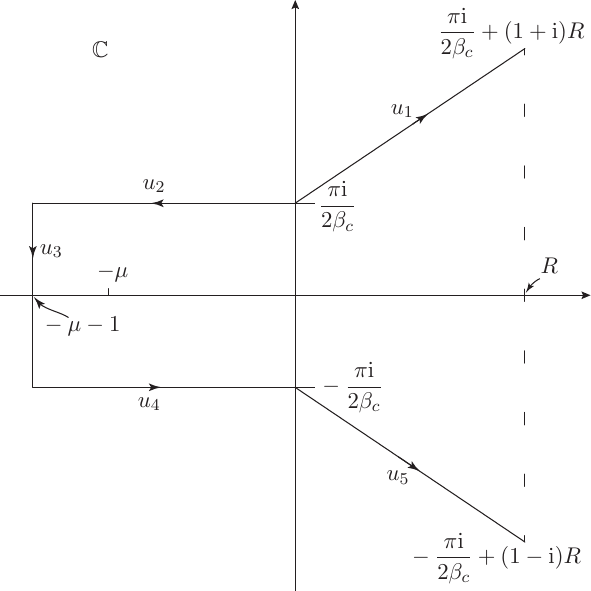}}
		\end{split} 
	\end{align*}
	The speaker path is defined as the union of paths $u_i$, $i=1, \ldots, 5$, with $u_1$ taken in reverse direction, i.e.,
	\begin{align*}
		\speaker_R \coloneqq  \mathop{\dot -}u_1 \mathop{\dot +} u_2 \mathop{\dot +} u_3 \mathop{\dot +} u_4 \mathop{\dot +} u_5.
	\end{align*}
	If $\mu < -1$ we choose the same path as in the case $\mu = -1$.
\end{defn} 

\begin{lem}
	\label{KT_integral_rep}
	Let $H\colon \Dcal(H)\ra \Hcal$ be a self-adjoint operator on a separable Hilbert space~$\Hcal$ with $H\geq -\mu$ and let $\beta > 0$. Then, we have
	\begin{align}
		\frac{H}{\tanh(\frac{\beta}{2} H)} = H + \lim_{R\to\infty} \int_{\speaker_R} \frac{\dd z}{2\pi\i} \Bigl( \frac{z}{\tanh(\frac{\beta}{2} z)} - z \Bigr) \frac{1}{z - H}, \label{DHS1:KT_integral_rep_eq}
	\end{align}
	with the speaker path $\speaker_R$ in Definition~\ref{speaker path}. The above integral including the limit is understood as an improper Riemann integral with respect to the uniform operator topology.  
\end{lem}

In the following, we use the symbol $\int_{\speaker}$ to denote the integral on the right side of \eqref{DHS1:KT_integral_rep_eq} including the limit and we denote $\speaker = \bigcup_{R > 0} \speaker_R$.

\subsection{Proof of Proposition~\ref{prop:gauge_invariant_perturbation_theory}}

The proof of Proposition~\ref{prop:gauge_invariant_perturbation_theory} is based on an adaption of gauge-invariant perturbation theory for Schr\"odinger operators as introduced in \cite{Nenciu2002} to our setting.  The core of the argument is contained in the following lemma.

\begin{lem}
	\label{lem:Appmain}
	Assume that $(1+|\cdot|^2) V \in L^{2}(\mathbb{R}^3) \cap L^{\infty}(\mathbb{R}^3)$. There is a continuous function $a \colon  \rho(K_\Tc-V) \to \mathbb{R}_+$ such that the following holds: For every compact set $K \subset \rho(K_\Tc-V)$ there is a constant $h(K) > 0$ such that for $0 < h < h(K)$ we have
	\begin{align}
		\Rcal_\Abold^{z, V} = \Scal_{\Abold_h}^{z, V} + h^2 \, \eta_h(z) \quad \text{ with } \quad \Vert (1+\pi^2) \eta_h(z) \Vert_\infty \leq a(z).
	\end{align} 
	Here $\Scal_\Abold^{z, V}$ denotes the operator defined by the kernel
	\begin{align}
		\Scal_\Abold^{z, V}(x,y) &\coloneqq  \e^{\i \Phi_{\Abold}(x,y)} \, \Gcal^{z, V}(x,y) \label{eq:app4}
	\end{align}
	with the phase factor $\Phi_\Abold(x, y)$ in \eqref{PhiA_definition}.
\end{lem}

\begin{proof}
We employ \eqref{PhiA_Magnetic_Momentum_Action} and apply the integral representation in Lemma~\ref{KT_integral_rep} to $K_{\Tc, \Abold}$ to see that
\begin{align}
	K_{\Tc, \Abold}^x \, \e^{\i \Phi_{\Abold_h}(x,y)} = \e^{\i \Phi_{\Abold_h}(x,y)} \, K_{\Tc, \Abold_y}^x
	\label{eq:app6}
\end{align}
holds. Here $K_{\Tc, \Abold}^x$ is understood to act on the $x$-coordinate and $\Abold_y(x) \coloneqq  \tilde \Abold(x,y)$ denotes the vector potential in transversal Poincar\'e gauge relative to the point $y$, defined in \eqref{Atilde_definition}. Using Lemma~\ref{KT_integral_rep} again, we also find the identity
\begin{align}
	K_{\Tc, \Abold_y} - K_\Tc &= \left( -2\i (\Abold_h)_y(x) \nabla_x -\i \divv (\Abold_h)_y(x) + |(\Abold_h)_y(x)|^2 \right) \label{eq:app8} \\
	&\hspace{20pt} + \int_{\speaker} \frac{\dd z}{2 \pi \i} \; \varphi(z) \; \frac{1}{z + (-\i \nabla + (\Abold_h)_y)^2 + \mu} \notag \\
&\hspace{40pt} \times \left( -2\i (\Abold_h)_y(x) \nabla_x -\i \divv (\Abold_h)_y(x) + |(\Abold_h)_y(x)|^2 \right) \frac{1}{z + \Delta + \mu}, \nonumber
\end{align}
where $\varphi(z) = (z/\tanh( z/(2\Tc) ) - z )$. Let us define the operator $\Tcal_\Abold^{z, V}$ via the equation
\begin{align}
	\left( z - (K_{\Tc, \Abold} - V) \right) \Scal_{\Abold_h}^{z, V} = 1 - \Tcal_{\Abold_h}^{z, V}.
	\label{eq:app5}
\end{align}
Using \eqref{eq:app6}, \eqref{eq:app8}, and \eqref{eq:app5}, we write the integral kernel of $\Tcal_\Abold^{z, V}$ as
\begin{align}
	\Tcal_\Abold^{z, V}(x,y) &= \e^{\i \Phi_{\Abold}(x,y)} \Big[ \left( -2 \i \Abold_y(x) \nabla_x - \i \divv \Abold_y(x) + |\Abold_y(x)|^2 \right) \Gcal^{z, V}(x,y) \nonumber \\
	&\hspace{20pt}+ \int_{\speaker} \frac{\dd \zeta}{2 \pi \i} \; \varphi(\zeta) \int_{\mathbb{R}^6} \dd v \dd w \; \frac{1}{\zeta - (-\i \nabla + \Abold_y)^2+\mu}(x,v)  \nonumber \\
	&\hspace{30pt} \times \left( -2\i \Abold_y(v) \nabla_v -\i \divv \Abold_y(v) + |\Abold_y(v)|^2 \right)  g_0^\zeta(v-w) \, \Gcal^{z, V}(v, y) \Big]. \label{eq:app9}
\end{align}
In the next step we use this formula to prove a bound for the operator norm of $\Tcal_\Abold^{z, V}$.

Let us denote the first and the second term on the right side of \eqref{eq:app9} by $\Tcal_\Abold^{(1)}(x,y)$ and $\Tcal_\Abold^{(2)}(x,y)$, respectively. Using \eqref{Thz_boundedness_2} and \eqref{Thz_boundedness_3}, we see that
\begin{align}
	| \Tcal_{\Abold_h}^{(1)}(x,y) | \leq Ch^2 \bigl( |x-y| \, | \nabla_x \Gcal^{z, V}(x,y) | + \bigl( |x - y|  + | x-y |^2   \bigr) | \Gcal^{z, V}(x,y) | \bigr).
	\label{eq:app10}
\end{align} 
Using the resolvent identity $\Rcal^{z, V} = \Rcal^z + \Rcal^z V \Rcal^{z, V}$, we estimate the first term on the right side of \eqref{eq:app10} by
\begin{align}
	\left| x-y \right| \left| \nabla_x \Gcal^{z, V}(x,y) \right| \leq& \left| x-y \right| \left| \nabla \Gcal^z( x-y) \right| \nonumber \\
	&+ \int_{\mathbb{R}^3} \dd w \ \left| |x-w| \, \nabla \Gcal^z (x-w) \, V( w) \, \Gcal^{z, V}(w,y) \right| \nonumber \\
	&+ \int_{\mathbb{R}^3} \dd w \ \left| \nabla \Gcal^z (x-w) \, V(w) \, |w-y| \, \Gcal^{z, V}(w,y) \right|. \label{eq:app11}
\end{align}
Eq.~\eqref{eq:app11} allows us to obtain the following bound for the operator norm of $\Tcal_\Abold^{(1)}$: 
\begin{align}
	\Vert \Tcal_{\Abold_h}^{(1)} \Vert_{\infty} &\leq Ch^2 \big[ \left\Vert |\cdot| \nabla \Gcal^z \right\Vert_1 \left( 1 + \left\Vert V \right\Vert_2 \Vert \Rcal^{z, V} \Vert_{2;\infty} \right) \nonumber \\
	&\hspace{20pt} + \left( 1+ \Vert \nabla \Gcal^z \Vert_1 \left\Vert V \right\Vert_2  \right) \Vert [x, \Rcal^{z, V} ] \Vert_{2;\infty} +  \Vert [x,[x, \Rcal^{z, V} ]] \Vert_{2;\infty} \big]. \label{eq:app12}
\end{align}
From Lemma~\ref{lem:App1}~and~\ref{lem:App2}, we know that there is a continuous $a \colon \rho(K_\Tc-V) \to \mathbb{R}_+$ such that the right side of \eqref{eq:app12} is bounded by $a(z)$. In the following we will denote by $a(z)$ a generic function with these properties whose precise form may change from line to line.

To obtain a bound for the operator norm of $\Tcal_\Abold^{(2)}(z)$, we first estimate its kernel by
\begin{align}
	&| \Tcal_{\Abold_h}^{(2)}(x,y) | \leq Ch^2 \Bigl( \int_{\speaker} \text{d}|\zeta| \; \left| \varphi(\zeta) \right|  \Bigr) \sup_{\zeta \in \speaker} \int_{\mathbb{R}^6} \dd v \dd w \; \Bigl| \frac{1}{\zeta - (-\i \nabla + (\Abold_h)_y)^2+\mu}(x,v) \Bigr| \nonumber\\
	&\hspace{0.5cm} \times \big[ | v-y | \ | \nabla g_0^{\zeta}(v-w)| + \left( |v-y| + |v-y|^2 \right) | g_0^\zeta(v-w) | \big]  |\Gcal^{z, V}(w, y)|. \label{eq:app13}
\end{align}
From Lemma~\ref{gh-g_decay} we know that the absolute value of the resolvent kernel of the magnetic Laplacian is bounded from above by a function only depending on $x-v$, whose $L^1(\mathbb{R}^3)$-norm is bounded by a constant times  $f( \Re \zeta, \Im \zeta)$ with $f$ in \eqref{gh-g_decay_f}. This, in particular, implies that this $L^1(\mathbb{R}^3)$-norm is uniformly bounded in $\zeta \in \speaker$ and $h$ as long as the latter is small enough, compare this to the bound in \cite[Eq. (6.19)]{DeHaSc2021}. We use this bound, $| v-y | \leq | v-w | + | w - y |$, and the resolvent identity for $\Rcal^{z, V}$ to bound the operator norm of $\Tcal_\Abold^{(2)}$ by
\begin{align}
	\Vert \Tcal_{\Abold_h}^{(2)} \Vert_{\infty} &\leq h^2 C \big(  \sup_{\zeta \in \speaker} \Vert (1+|\cdot|) \nabla g_0^\zeta \Vert_1 \; , \; \sup_{\zeta \in \speaker} \Vert (1+|\cdot|^2) g_0^{\zeta} \Vert_1 \; , \; \Vert (1+|\cdot|^2) \Gcal^z \Vert_1 \; , \; \nonumber\\
	&\hspace{50pt} \Vert V \Vert_2 \; , \; \Vert \Rcal^{z, V} \Vert_{2;\infty} \; , \; \Vert [x, \Rcal^{z, V} ] \Vert_{2,\infty} \; , \; \Vert [x, [x, \Rcal^{z, V}] ] \Vert_{2,\infty} \big). \label{eq:app14}
\end{align}
The constant on the right side is an affine function of each of its arguments. From Lemma~\ref{g_decay} we know that the norms involving $g_0^{\zeta}$ are finite. \ifthenelse{\equal\masterfile{Diss}}{With}{In combination with} Lemmas~\ref{lem:App1} and \ref{lem:App2}, this implies that the right side of \eqref{eq:app14} is bounded by $a(z)$. We conclude that
\begin{align}
	\Vert \Tcal_{\Abold_h}^{z, V} \Vert_\infty \leq a(z)\, h^2
	\label{eq:app15}
\end{align}
holds.

Let $K \subset \rho(K_\Tc-V)$ be compact. The above bounds allow us to find $h_0(K) > 0$ such that for $z \in K$ and as long as $0 < h < h_0(K)$ we can write the resolvent of $K_{\Tc, \Abold}-V$ as
\begin{align}
	\frac{1}{z - (K_{\Tc, \Abold} - V)} = \Scal_{\Abold_h}^{z, V} + h^2 \,  \eta_h(z) \quad \text{ with } \quad \eta_h(z) \coloneqq  h^{-2} \Scal_{\Abold_h}^{z, V} \sum_{n=1}^{\infty} (\Tcal_{\Abold_h}^{z, V})^n.
	\label{eq:app16}
\end{align}
To show that the operator norm of $(1 + \pi^2) \eta_h(z)$ is bounded by $a(z)$, we use 
\begin{align}
	\Vert (1 + \pi^2) \eta_h(z) \Vert_\infty \leq \Vert (1 + \pi^2) \Scal_{\Abold_h}^{z, V} \Vert_\infty \; \sum_{n=1}^{\infty} h^{2(n-1)} a(z)^n. \label{eq:app17}
\end{align} 
With the resolvent identity for $\Rcal^{z, V}$ and Lemma~\ref{lem:App1}, we easily see that the operator norm of $(1 + \pi^2) \Scal_{\Abold_h}^{z, V}$ is bounded by $a(z)$. This proves the claim.
\end{proof}

With the resolvent estimates in Lemma~\ref{lem:Appmain} at hand, we turn to the proof of Proposition~\ref{prop:gauge_invariant_perturbation_theory}. Let $\lambda < 2\Tc$ be an eigenvalue of the operator $K_\Tc - V$. Our assumption on $V$ guarantees that it has finite multiplicity $m \in \mathbb{N}$. We choose $\varepsilon > 0$ such that the ball $B_{\varepsilon}(\lambda) \subseteq \mathbb{C}$ contains no other point of the spectrum of $K_\Tc - V$ than $\lambda$ and define 
\begin{align}
	P(h) \coloneqq   \int_{\partial B_{\varepsilon}(\lambda)} \frac{\dd z}{2 \pi \i}  \; \Rcal_\Abold^{z, V} =  \int_{\partial B_{\varepsilon}(\lambda)} \frac{\dd z}{2 \pi \i} \; \Scal_{\Abold_h}^{z, V} + h^2 \int_{\partial B_{\varepsilon}(\lambda)} \frac{\dd z}{2 \pi \i}  \; \eta_h(z). \label{eq:app19}
\end{align}
From Lemma~\ref{lem:Appmain} we know that the operator norm of the second term on the right side is bounded by a constant times $h^2$ provided $h$ is small enough. The integral kernel of the first term is given by
\begin{align}
	\Bigl( \int_{\partial B_{\varepsilon}(\lambda)} \frac{\dd z}{2 \pi \i} \; \Scal_\Abold^{z, V} \Bigr)(x,y) = \e^{\i \Phi_{\Abold}(x,y)} \sum_{i=1}^m u_i(x) \overline{ u_i(y) },
	\label{eq:app20}
\end{align}
where the vectors $\{ u_i \}_{i=1}^m$ span the eigenspace of $\lambda$. Let us denote by $P$ the projection onto that linear space. Using \eqref{eq:app19}, \eqref{eq:app20}, and $|\Phi_{\Abold_h}(x,y)| \leq C h^2 (|x|^2 + |y|^2)$, which follows from the assumption $A(0) =0$, we obtain the bound
\begin{align}
	\Vert P(h) - P \Vert_{\infty} \leq C h^2 \, \max_{i=1, \ldots ,m} \Vert \, |\cdot|^2 u_i \Vert_2.
	\label{eq:app21}
\end{align}
In combination with Lemma~\ref{lem:regEF}, this proves $\rank P(h) = m$ for $h$ small enough as well as the second bound in \eqref{eq:app1}.

To prove the bounds for the eigenvalues we use the identity 
\begin{align}
	(K_{\Tc, \Abold} - V) P(h) =  \int_{\partial B_{\varepsilon}(\lambda)} \frac{\dd z}{2 \pi \i} \; z \, \Rcal_\Abold^{z, V} . \label{eq:app22}
\end{align}
As long as $h$ is small enough, the rank of this operator equals $m$ and its eigenvalues are given by $\lambda_1(h), \ldots , \lambda_m(h)$. Similar arguments to the above for the spectral projections allow us to conclude that
\begin{align}
	\Vert (K_{\Tc, \Abold} - V) P(h) - (K_\Tc - V) P \Vert_{\infty} \leq  C h^2 \, \max_{i=1, \ldots ,m} \Vert \, |\cdot|^2 u_i \Vert_2 \label{eq:app23}
\end{align}
holds. This proves the claimed bound for the eigenvalues. It remains to prove \eqref{eq:app2}.

Let us write $\alpha_*^h = a(h) \alpha_* + b(h) \phi_h$ with $\langle \alpha_*, \phi_h \rangle = 0$ and $|a(h)|^2 + |b(h)|^2 = 1$. Our assumptions imply $a(h) = \langle \alpha_* , \alpha_*^h \rangle \geq 0$. We rewrite the equation $P(h) \alpha_*^h = \alpha_*^h$ to see that $b(h) \phi_h = (P(h)-P) \alpha_*^h$. An application of \eqref{eq:app1} thus implies $|b(h)| \leq C h^2$. Using this, $|a(h)|^2 + |b(h)|^2 = 1$, and the fact that $a(h) \geq 0$, we see that $| a(h) - 1 | \leq C h^2$. This allows us to conclude that
\begin{align}
	\Vert (1+\pi^2) ( \alpha_*^h - \alpha_* ) \Vert_2 \leq |a(h)-1| \ \Vert (1 + \pi^2) \alpha_* \Vert_2 + | b(h) | \ \Vert (1 + \pi^2) \phi_h \Vert_2 \leq C h^2.
	\label{eq:app24}
\end{align}
To obtain the result, we used Lemma~\ref{lem:regEF} to see that $\Vert (1 + \pi^2) \alpha_* \Vert_2 < \infty$, as well as $\Vert (1 + \pi^2) \phi_h \Vert_2 \leq \Vert (1 + \pi^2) \alpha_* \Vert_2 + \Vert (1 + \pi^2) \alpha_*^h \Vert_2$, and 
\begin{align}
	\Vert (1+\pi^2) \alpha_*^h \Vert_2 \leq C \Vert K_{\Tc, \Abold} \alpha_*^h \Vert_2 \leq C \left( | \lambda(h) | + \Vert V \alpha_*^h \Vert_2 \right) \leq C \left( | \lambda(h) | + \Vert V \Vert_\infty \right).
\end{align}
This proves \eqref{eq:app2} and also finishes the proof of Proposition~\ref{prop:gauge_invariant_perturbation_theory}.


\begin{center}
\textsc{Acknowledgements}
\end{center}

A.~D. gratefully acknowledges funding from the Swiss National Science Foundation through the Ambizione grant PZ00P2 185851.

\printbibliography[heading=bibliography]


\vspace{1cm}

\setlength{\parindent}{0em}

(Andreas Deuchert) \textsc{Institut für Mathematik, Universität Zürich}

\textsc{Winterthurerstrasse 190, CH-8057 Zürich}

E-mail address: \href{mailto:  andreas.deuchert@math.uzh.ch}{\texttt{andreas.deuchert@math.uzh.ch}}

\vspace{0.3cm}

(Christian Hainzl) \textsc{Mathematisches Institut der Universität München}

\textsc{Theresienstr. 39, D-80333 München}

E-mail address: \href{mailto: hainzl@math.lmu.de}{\texttt{hainzl@math.lmu.de}}

\vspace{0.3cm}

(Marcel Maier, born Schaub) \textsc{Mathematisches Institut der Universität München}

\textsc{Theresienstr. 39, D-80333 München}

\end{document}